\documentclass[aps,rmp,twocolumn,superscriptaddress,longbibliography]{revtex4-1}

\usepackage{graphicx}
\usepackage{epsfig}
\usepackage{subfigure}
\usepackage{changes}
\usepackage{longtable}
\usepackage{soul}
\usepackage{gensymb}

\usepackage{amsmath,amsthm,amsfonts,latexsym}

\begin{document}

\title{Emergent constraints on climate sensitivities}

\author{Mark S. Williamson}
\affiliation{College of Engineering, Mathematics and Physical Sciences, University of Exeter, UK}
\affiliation{Global Systems Institute, University of Exeter, UK}
\author{Chad W. Thackeray}
\affiliation{Department of Atmospheric and Oceanic Sciences, University of California, Los Angeles, CA, USA}
\author{Peter M. Cox}
\affiliation{College of Engineering, Mathematics and Physical Sciences, University of Exeter, UK}
\author{Alex Hall}
\affiliation{Department of Atmospheric and Oceanic Sciences, University of California, Los Angeles, CA, USA}
\author{Chris Huntingford}
\affiliation{Centre for Ecology and Hydrology, Wallingford, UK}
\author{Femke J. M. M. Nijsse}
\affiliation{College of Engineering, Mathematics and Physical Sciences, University of Exeter, UK}

\date{\today}

\begin{abstract}
Despite major advances in climate science over the last 30 years, persistent uncertainties in projections of future climate change remain. Climate projections are produced with increasingly complex models which attempt to represent key processes in the Earth system, including atmospheric and oceanic circulations, convection, clouds, snow, sea-ice, vegetation and interactions with the carbon cycle. Uncertainties in the representation of these processes feed through into a range of projections from the many state-of-the-art climate models now being developed and used worldwide. For example, despite major improvements in climate models, the range of equilibrium global warming due to doubling carbon dioxide still spans a range of more than three. Here we review a promising way to make use of the ensemble of climate models to reduce the uncertainties in the sensitivities of the real climate system. The emergent constraint approach uses the model ensemble to identify a relationship between an uncertain aspect of the future climate and an observable variation or trend in the contemporary climate. This review summarises previous published work on emergent constraints, and discusses the huge promise and potential dangers of the approach. Most importantly, it argues that emergent constraints should be based on well-founded physical principles such as the fluctuation-dissipation theorem. It is hoped that this review will stimulate physicists to contribute to the rapidly-developing field of emergent constraints on climate projections, bringing to it much needed rigour and physical insights.

\end{abstract}
\maketitle


%

\tableofcontents{}


\section{Introduction}\label{sec:intro}

Numerical methods have become a standard technique to simulate complex systems. The equations governing components of such systems may be well-known. But their solutions cannot be solved analytically, creating a need for numerical approaches. Because of the discretization of time and space inherent in numerical techniques, modelling complex systems must involve ways to include effects of unresolved processes. Often there is no `first principles' approach to do this. Typically, the effects of unresolved processes are included by resorting to quasi-empirical relationships between them and explicitly-resolved variables, otherwise known as `parameterization'. There are usually multiple defensible ways to parameterize unresolved processes. Thus, independently developed models of the same complex system might incorporate different parameterization choices. The more models that are independently developed, the greater the diversity of approaches for modelling the same natural system.

A classic example of this model diversity is the use of numerical models of the atmosphere to predict hurricane development. Initial conditions are imposed on a model at some time, and it is integrated forward in time to produce simulations of critical hurricane features such as intensity and track. This approach might be replicated for multiple models incorporating different parameterization choices, producing an `ensemble' of hurricane forecasts. The spread in the forecasts is a measure of the uncertainty in the future hurricane behaviour, given the range of plausible approaches to atmospheric modelling. Over time, with enough hurricanes and associated predictions, the various models can be evaluated for their prediction skill. Certain parameterization choices may emerge as producing systematically better predictions. The models can then be rebuilt or re-calibrated with the better choices. Over time the ensemble will become more skilful, with less spread. In fact, this is approximately the process that has resulted in dramatic improvements not only in hurricane prediction, but in weather forecasting generally over the past seven decades.

Earth's climate is another example of a complex system whose governing equations can only be solved through numerical methods. (In fact, the dynamical equations for the atmospheric component of a climate model are typically almost identical to those in the hurricane models referred to above.) As expected, there are a variety of plausible approaches to parameterization in the components of Earth system models (ESMs)\footnote{State-of-the-art climate models are also commonly called general circulation models (GCMs). This was particularly true in the past when they consisted of just an atmosphere and sometimes an ocean. As time has progressed and more processes have been included the term ESM has become more common (previously ESM may have referred to models of reduced complexity featuring a carbon cycle). In this review we use the term ESM to mean a full complexity dynamical state-of-the-art climate model although this could be used interchangeably with GCM.}.  Thus, modelling groups throughout the world have built a few dozen ESMs with different approaches to parameterization. Because of these differences, these models produce different future climate states, even when the same scenario of radiative forcing is imposed (associated, for example, with an increase in greenhouse gases).

A classic climate change experiment is to double CO$_{2}$ concentrations in the atmospheric component of an ESM, and measure the surface warming that occurs after the simulation has equilibrated \cite{ref:Manabe&Wetherald75}, an important number in climate science referred to as equilibrium climate sensitivity (ECS). Reaching a true equilibrium with a full complexity ESM requires long, computationally expensive simulations of thousands of years so ECS is usually estimated from shorter duration CO$_2$ doubling experiments. When these experiments are done with contemporary ESMs, the spread in values across the ensemble of ESMs is large, between 1.5 and 6 degrees Celsius \cite{ref:Forster19}, although true equilibrium values are higher (median 17\% higher, \citet{ref:Rugenstein20}). The international climate science community has organised itself to generate scenarios of greenhouse gas (GHG) emissions that result in more realistic future radiative forcing than the CO$_{2}$ doubling experiment. These correspond to scenarios of controls (or lack thereof) on future GHG emissions \cite{ref:Moss2010,ref:Riahi17}. The scenarios are designed so that the same radiative forcing time series is imposed on each ESM. This allows for inter-comparison of ESM response at some future specified time, say the end of the 21st century. The ESM responses vary significantly across the ensemble, an indicator of deep uncertainty in Earths’ climate future, simply due to the variety of plausible ways to handle parameterization within ESM components.

If each ESM is treated as being an equally plausible analogue of the real climate system, known as `model democracy', an undesirably large factor of three difference in future response to man-made GHG response currently results. Apart from the subject of this review, there are several techniques used to reduce this uncertainty and much research effort has been dedicated to this important task. One approach is to give ESMs that simulate the real world `better' (according to some chosen metric) more weight in the future projection. There are also methods that do not use ESMs at all, by constraining future response from past observations and those that use a combination of all of these. Many more details can be found in e.g. \citet{ref:Eyring19,ref:Sherwood20} and chapters 9 and 11-14 in \citet{ref:IPCC_AR5}.

The approach taken to improve hurricane forecasting models -- to evaluate their performance and adjust parameterizations over the course of multiple prediction cycles -- is unfortunately impractical in the case of projecting future climate with ESMs. With future climate, we have only one realisation of the real system’s trajectory, and our mandate is to predict it as best we can now. To compound the problem, the climate system involves many more components than the atmosphere alone, including the ocean, the land surface, glaciers and ice sheets, and the marine and terrestrial biospheres. The engineering required to model each component and allow them to interact in a simulation is impressive. But the components can interact with one another in information flows are so complex that it is not easy to predict how behavior in one component might affect behavior in another. For example, parameterization reformulation or adjustment in one component can significantly affect the simulated state of another component for reasons that are not always clear (e.g. \citet{ref:Donner11}).

\begin{figure}
\includegraphics[width=\columnwidth]{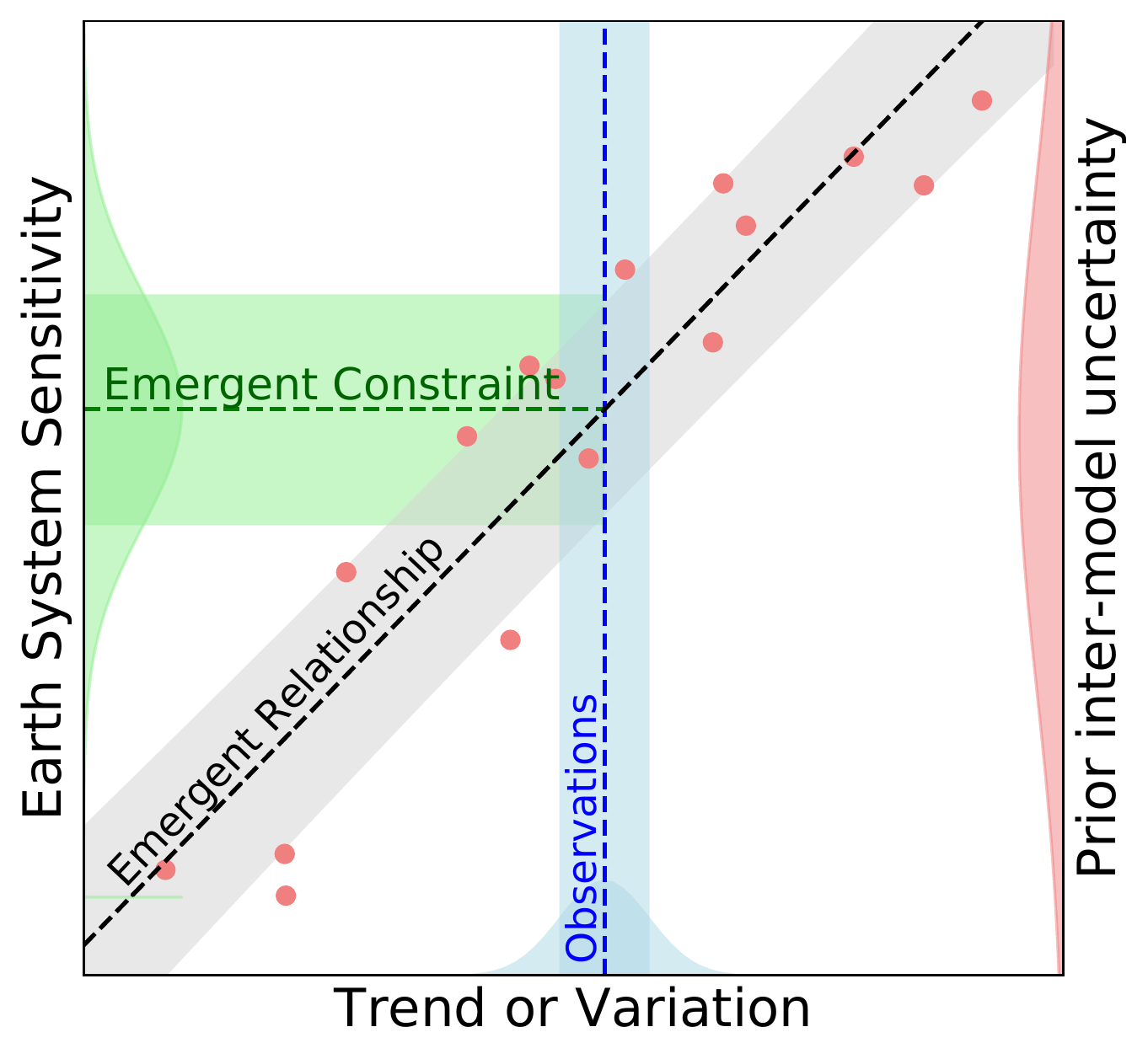}
\caption[]{Schematic showing the most common procedure used to derive emergent constraints on Earth system sensitivities. An ensemble of ESMs (each red dot is an individual ESM) running the same experiment (the pdf on the right-hand $y$-axis represents the spread in the ensemble) is used to identify an emergent relationship (black dashed line with grey uncertainty range), between an uncertain Earth system sensitivity $Y$ ($y$-axis) and an observed trend or variation $X$ ($x$-axis). An observation of the trend or variation (blue pdf on the $x$-axis) can then be combined with the model-based emergent relationship to derive an emergent constraint on the Earth system sensitivity (green pdf on the left-hand $y$-axis).}\label{sec:intro:fig:EC_schematic}
\end{figure}

To circumvent the impossibility of directly evaluating ESMs for their ability to simulate a future climate state which has not yet been observed, a new technique has emerged over the past decade and a half, known as the \emph{emergent constraint} (EC) approach (Figure \ref{sec:intro:fig:EC_schematic} shows a schematic). The basic idea is to identify an element ($X$) of the observable climate that varies significantly across the ESM ensemble, and which exhibits a statistically-significant relationship with variations in some important variable ($Y$) describing the ESM's future simulated state. If we call this relationship $f$, then $Y=f(X)+\varepsilon$, where $\varepsilon$ is a relatively small departure from $f$. Since $X$ is an element of the observable climate, it is a quantity that can be measured. The relationship $f$ may then place a useful constraint on $Y$, provided the measurement uncertainty in $X$ is small compared to the range of simulated values. This constraint is `emergent' because the \emph{emergent relationship} $f$ cannot be diagnosed from a single ESM. It only becomes apparent when the full ensemble is analysed. If the relationship $f$ arises from model physics or dynamics common to the ESMs, then reducing the spread in $X$ through reformulation or adjustment of parameterizations ought to result in spread reduction in $Y$. Corresponding reduction in the spread in other future climate variables affected by $Y$ should also occur. If the process is repeated for enough variables $X$ and $Y$, then we can imagine that overall simulated spread in multiple aspects of future climate would gradually be reduced. We also note that ECs can be used directly to ascertain the most likely values for a particular $Y$.

\begin{figure}[t]
    \centering
    \includegraphics[width=\columnwidth]{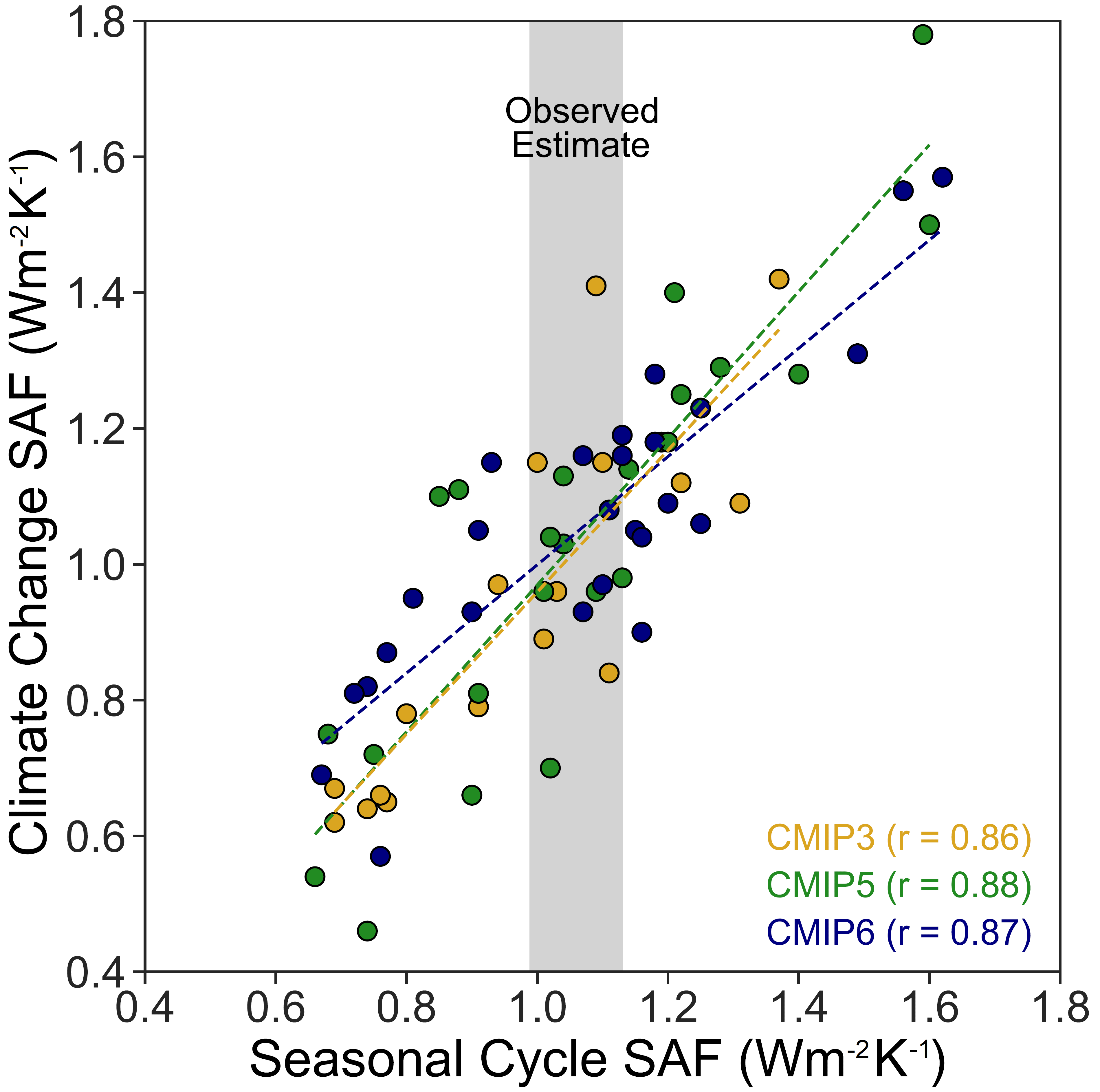}
    \caption{Emergent relationship between springtime snow albedo feedback (SAF) across Northern Hemisphere land under climate change ($Y$) and an observable snow albedo feedback associated with the current climate's seasonal cycle ($X$). An observational estimate derived from satellite data is shown as a vertical bar. Each point represents an individual climate model from the three most recent generations (CMIP3, CMIP5, and CMIP6). Methodology for calculating SAF is adapted from Qu and Hall (2014) (further details in \textcite{ref:Thackeray2020})}
    \label{fig:SAF}
\end{figure}

\begin{figure*}
\includegraphics[width=0.8\textwidth]{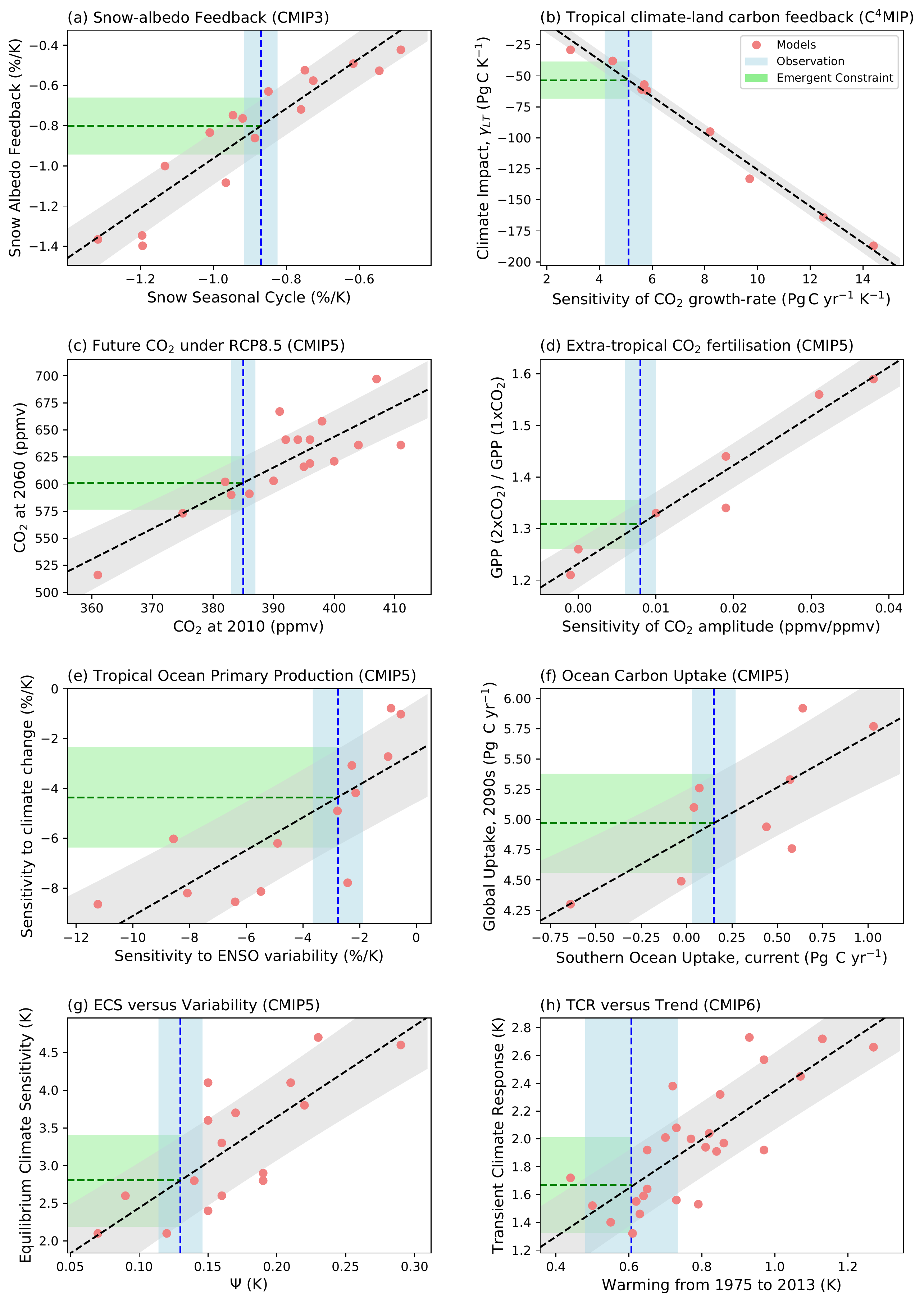}
\caption[]{Emergent constraints on Earth System sensitivities based on some key examples published in the literature: (a) snow-albedo feedback, from snow seasonal cycle \cite{ref:Hall&Qu06}; (b) sensitivity of tropical land carbon to global warming, from interannual variability in CO$_2$ \cite{ref:Cox13}; (c) atmospheric CO$_2$ concentration at 2060, from atmospheric CO$_2$ concentration at 2010 \cite{ref:Hoffman14}; (d) CO$_2$ fertilization of plant photosynthesis, from changes in the seasonal cycle of CO$_2$ \cite{ref:Wenzel16}; (e) sensitivity of tropical ocean primary production to warming, from interannual variability \cite{ref:Kwiatkowski17}; (f) global ocean carbon sink in the 2090s, from the current day carbon sink in the Southern Ocean \cite{ref:Kessler2016}; (g) Equilibrium climate sensitivity, from interannual variability of temperature \cite{ref:Cox18}; (h) Transient climate response, from increase in global mean temperature \cite{ref:Nijsse20}. In each case the emergent constraint was reconstructed from data available in the literature or provided directly by the authors. The model ensemble used in each original study is shown in the brackets after the panel title.}\label{sec:intro:fig:example_ECs}
\end{figure*}

To further illustrate the concept, the subject of the first published EC well established across multiple generations of ESMs, is used as an example \cite{ref:Hall&Qu06}. In this case, the future climate variable of interest ($Y$) is the snow albedo feedback (SAF), a climate mechanism characterized by the retreat of highly reflective snow cover under climate warming and the associated reduction in surface albedo, which amplifies warming and promotes further snow melt \cite{ref:Hall2004,ref:Bony2006,ref:Thackeray16}. However, modern ESMs disagree on the strength of this feedback, exhibiting a nearly threefold spread \cite{ref:Qu&Hall14,ref:Thackeray18}. SAF also occurs each spring in the current climate when Northern Hemisphere snow cover recedes from its winter peak to its summer minimum, enhancing seasonal warming in the process ($X$). A comparison between simulated SAF under climate change ($Y$) and in the seasonal cycle context ($X$), uncovers a strong linear relationship in three different generations of ESM ensembles (Figure \ref{fig:SAF}). The strength of this relationship tells us that seasonal SAF is likely highly predictive of climate change SAF, while the consistency across model generations illustrates robustness to out-of-sample testing. Furthermore, we can quantify the seasonal SAF using satellite-derived observations, thus allowing for model bias to be properly assessed. For example, an ESM that underestimates observed SAF in the seasonal context is likely to underestimate SAF in future climate. Producing a more accurate SAF in ESMs is more than just an academic exercise as variability in SAF can account for a significant portion of uncertainty in 21st century projections of warming across Northern Hemisphere extratropical land \cite{ref:Qu&Hall14}. Thus, by reducing variability in SAF we can expect similar reductions in projections of regional warming. Figure \ref{sec:intro:fig:example_ECs} highlights the first application of this and several other ECs from the literature (pertaining to the carbon cycle and climate sensitivity), all of which will be further discussed in Section \ref{sec:ECs}.

In this review article, we synthesise what is now understood about ECs. We discuss how relationships $f$ might emerge in ESM ensembles in section \ref{sec:emerge}, as well as the theory that underlies ECs in section \ref{sec:theory}. In section \ref{sec:wrong} we note pitfalls associated with ECs and how to guard against them. In section \ref{sec:ECs} we survey the ECs that have been found in the climate system, and describe the statistics that can be used to apply them rigorously in section \ref{sec:stats}.  Finally, we discuss how techniques similar to ECs have been or might be used in other fields, and consider the outlook for development of the technique going forward in section \ref{sec:outlook} before concluding in section \ref{sec:conclusion}.

\section{How relationships in model ensembles might `emerge'}\label{sec:emerge}

ECs are possible because of `emergent relationships' appearing in an ensemble of ESMs. An emergent relationship, $Y=f(X)+\varepsilon$, between an element of the observable climate $X$, something that can be measured now in the real world, and the unknown future response $Y$ one wants to know, can then be used to place a constraint on the real world value of $Y$ via observations of $X$ to give an EC. For example, in \textcite{ref:Cox18} the observable was the variability in global annual mean surface air temperature during roughly the last one hundred years and the unknown future response was equilibrium climate sensitivity (ECS, defined as the magnitude of the Earth's warming to doubled CO$_2$ levels), one of the oldest and most important numbers in climate science. Despite decades of research the latest ESMs still vary widely in their predictions of this number by a factor of three \cite{ref:Forster19,ref:Zelinka20}.

How might a relationship, $f$, between observable and response `emerge' in an ensemble of ESMs? In the \textcite{ref:Cox18} EC on ECS, an analytically soluble energy balance model, much reduced in complexity compared to the ESMs in the ensemble, predicted a functional form between $X$ and $Y$ for the emergent relationship (although this provoked a lively debate, see \citet{ref:Brown18,ref:Po-Chedley18,ref:Rypdal18,ref:Cox18a,ref:Williamson19}). This mechanism from high to low dimensional relationship emergence and others are given in the following subsections. First we discuss the differences in the types of climate model ensembles commonly used to find emergent relationships.

\subsection{Commonly used ESM ensembles: Multi-model (MME) and perturbed physics (PPE) ensembles}\label{sec:emerge:common_ensembles}

Model simulations are not perfect reproductions of the system they are designed to emulate. Ensembles of models are used to get a handle on (i) initial condition uncertainty, (ii) parametric uncertainty and (iii) structural uncertainty.

Uncertainty in the initial state is particularly important for numerical weather prediction. As the equations governing the weather are chaotic, forecast solutions with very small differences in their initial states, equally likely to be the `real' initial state, can diverge strongly after just a few days. This sensitivity to initial conditions means running the weather model just once, even if the model is a perfect reproduction of the real world, may produce a very different forecast from the one actually experienced. By running the same weather model many times with different, but equally plausible initial states chosen to sample the most unstable and divergent regions of the model phase space, the likelihood of experiencing a particular weather forecast can be constructed. Ensemble forecasting is now a standard tool in numerical weather prediction \cite{ref:Epstein69,ref:Leith74,ref:Toth&Kalnay93,ref:Molteni96}. Initial value ensembles in climate applications are usually used to assess natural climate variability \cite{ref:Kay15,ref:Maher19}.

Climatic variables are (generally speaking) the long term statistics of weather and their prediction is less impacted by initial state uncertainty relative to parametric and structural uncertainties \cite{ref:Hawkins&Sutton09,ref:Hawkins&Sutton11}. Parametric uncertainty arises from uncertainty in the values of constants (`parameters') in quasi-empirical relations (`parameterizations') used to model the effects of unresolved, subgrid scale processes. Parametric uncertainty can be sampled by running the same model with different values of these parameters. ESM model ensembles that do this are known as perturbed physics or perturbed parameter ensembles (PPEs, \citet{ref:Murphy04}). Examples of ESM PPEs are the climateprediction.net experiment \cite{ref:Stainforth05} and the Met Office's Quantifying Uncertainties in
Model Projections (QUMP) ensemble \cite{ref:Collins11} that both used the HadCM3 ESM.

Structural uncertainty arises from uncertainty in the functional form of equations. Although a lot of the model equations in ESMs are well known, there are multiple, equally defensible parameterization schemes for unresolved processes (e.g. convective precipitation and cloud radiative properties) and so different equations can appear in different ESMs. ESM ensembles that sample structural uncertainty are called multi-model ensembles (MMEs) and include the coupled model intercomparison project (CMIP) ensembles \cite{ref:CMIP3,ref:CMIP5,ref:CMIP6} used to inform the Intergovernmental Panel on Climate Change (IPCC) reports \cite{ref:IPCC_AR3_projections,ref:IPCC_AR4_projections,ref:IPCC_AR5_projections} and used extensively in much of the EC literature.

The CMIP MME ensembles consist of a number of different ESMs developed by different international groups thus sampling structural uncertainty. Some CMIP groups also submit multiple initial value runs and occasionally multiple physics parameter runs performed with the same model, sampling initial condition and parametric uncertainty to some degree. Numerical weather prediction ensemble forecasting has also incorporated more parametric and structural uncertainty into their forecasting as time has gone on \cite{ref:Palmer19}.

Most ECs have derived from MME ensembles to date although there are some examples from PPEs e.g. \textcite{ref:Knutti06}. ECs derived from MMEs are, in our opinion, the most believable. Each model in a MME could be thought of as that particular group's best guess of the Earth system. PPEs on the other hand, generally vary only a handful of parameters and thus sample just a few dimensions of the model's phase space. Due to the few varied dimensions, relationships hard coded in the fixed equations between model variables become easy to find (based on the authors experience with the QUMP PPE \cite{ref:Lambert13}).  These ECs need to be backed up with physical arguments and ideally tested in MMEs. \citet{ref:Collins11} found that long wave cloud feedback was highly correlated to climate sensitivity in the QUMP HadCM3 PPE ensemble, however this correlation was absent in (smaller sample size) MMEs. \citet{ref:Yokohata10} also found that strong correlations in a PPE with HadSM3 were not always present in the equivalent MIROC3.2 PPE. Further, \citet{ref:Yokohata10} found the mechanism for variations in climate sensitivity was different.

Values of parameters chosen to correctly give the proper balance between opposing processes in MMEs could be skewed to one of the processes by simultaneously varying them in a PPE thus reducing the realism of each ensemble member solution. However, PPEs can be useful for testing ECs (e.g. \citet{ref:Kamae16,ref:Wagman&Jackson18} and section \ref{sec:wrong:PPEs}) and quantifying parametric uncertainty (section \ref{sec:outlook:PPEs}).

\subsection{Null hypothesis: Emergent relationships occur by chance}\label{sec:emerge:data_mine}

A starting null hypothesis is that emergent relationships occur by chance and are not indicative of a deeper mechanistic relationship \cite{ref:Hall19}. Data mining an ESM ensemble for high correlations between pairs of variables may fall into this category (see also section \ref{sec:wrong:data_mine}). If an ESM ensemble is reasonably small and the number of variables output by each ESM is high, the expected number of variable pairs with high correlation is high purely by chance \cite{ref:Caldwell14}.

More precisely, an ESM labelled by the index $i$ in an ensemble of $n$ ESMs $i=\{1,2,\ldots,n\}$ calculates a large number $m$ of different variables $X^l$, each variable labelled with index $l$ and $l=\{1,2,\ldots,m\}$.  In this data mining scenario, a pair of variables $(X^l_i,X^k_i)$ is chosen for the prospective emergent constraint, one for the observable $X^l$ and the other, the response $X^k$ one wishes to know in the real world (previously written $Y$). In an indiscriminate data mining approach, every possible pair of variables are used to create $\tfrac{1}{2}m(m-1)$ data sets, each labelled $\mathcal{S}^{kl}$. Each dataset has $n$ elements, $\mathcal{S}^{kl}=\{(X^l_1,X^k_1),(X^l_2,X^k_2),\ldots, (X^l_n,X^k_n)\}$ with each element corresponding to one of the $n$ ESMs. If the pair of variables making up $\mathcal{S}^{kl}$ is a good candidate for an emergent relationship and we assume this relationship is linear, then the correlation $r^{kl}$ in $\mathcal{S}^{kl}$ should be high. Correlation here is defined as $r^{kl}=\text{cov}(X^l,X^k)/\sigma_{X^l} \sigma_{X^k}$ and $\sigma_X$ is the standard deviation of $X$.

The number of ESMs in a model ensemble, $n$, is typically small being around 10 to 40 in the state-of-the-art CMIP ensembles although numbers increase with each successive generation \cite{ref:CMIP3,ref:CMIP5,ref:CMIP6}. If we calculate the correlation $r^{kl}$ within every one of the $\tfrac{1}{2}m(m-1)$ possible datasets, some will have high correlations purely by chance. The likelihood increases as the number of models in the ensemble $n$ gets smaller giving less data points in each $\mathcal{S}^{kl}$. While the fraction of datasets with high correlations $|r|>|r_{high}|$, does not increase as the number of variables $m$ gets larger, the total number of datasets with correlation above $|r_{high}|$ will increase simply because there are more possible variable pairs to correlate. Thus one must be careful that any prospective emergent constraint makes physical sense, particularly in current, small ESM ensembles to avoid this pitfall.

\begin{figure}
\includegraphics[width=\columnwidth]{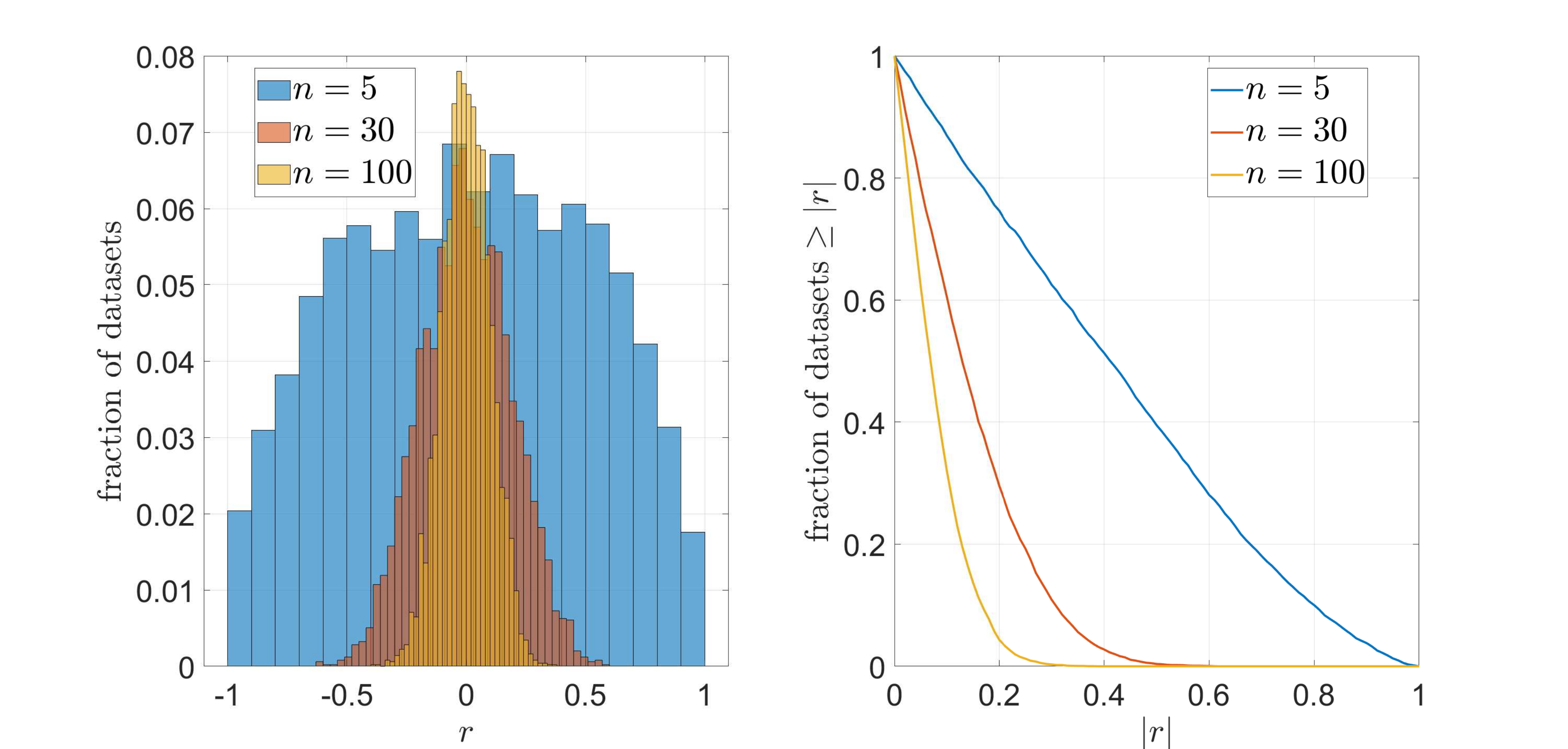}
\caption[]{High correlations between pairs of variables across small ESM ensembles are expected by chance. To illustrate this point, the fraction of all possible pairs of variables across a `toy' ESM ensemble of size $n$ plotted against correlation $r$. Each ESM is not a real ESM; it is represented by a random variable drawn from a normal distribution of unit standard deviation and each of the $n$ ESMs produces $m=100$ output variables giving 4950 different possible variable pairs to correlate with each other. This example is designed to be a worst case scenario yet still shows calculated correlation even between uncorrelated objects can be large in small enough ensembles. In the left panel, histograms of fraction of datasets with correlation falling in a particular interval are plotted. In the right hand panel, the total fraction of datasets with magnitude of correlation greater than $|r|$ is plotted. Ensemble size is varied for toy ESM ensemble sizes of $n=5$, $n=30$ and $n=100$,}\label{sec:emerge:fig:chance_relationships}
\end{figure}

To illustrate this point in a worst case type scenario where every ESM output variable is uncorrelated to any other output variable or any of the other ESMs, we can calculate the correlations expected between all possible pairs of variables as a function of ensemble size $n$. Figure \ref{sec:emerge:fig:chance_relationships} shows the fraction of all possible datasets with a particular correlation for a `toy' ESM ensemble of $n$ models where each of the toy ESMs is a modelled as a random number generator. This is a rather extreme and unrealistic example, each ESM producing $m$ random output variables, the $l$th one labelled $X^l_i$, drawn from a normal distribution with unit standard deviation, $X^l_i \in \mathcal{N}(0,1)$. Even though there are no mechanistic relationships in this example by construction, a fraction of the datasets have strong correlations, particularly in smaller ensemble sizes (small $n$). This is purely by chance.

\subsection{Low dimensional relationships `emerge' from high dimensional ESMs}

In the last example even though all $n\times m$ `toy' ESM variables in the ensemble were uncorrelated from each other by construction, high correlations were still possible. This is more likely in small ensembles. In reality \emph{there are} mechanistic relationships between $X$ and $Y$ in any particular ESM. Real ESMs are not random number generators, the output of a ESM is the numerical solution of a large set of coupled, nonlinear equations codifying relationships between variables representing the ocean and atmosphere dynamics and thermodynamics and biogeochemical feedbacks\footnote{Many of these equations are well known from other fields of physics such as the fluid dynamics Navier-Stokes equations. See for example \citet{ref:Ghil&Lucarini20} and references within for the equations that commonly feature in ESMs.}. However the exact relationship between any pair of $X$ and $Y$ is not solely a function of just these two variables, it is generally also a function of many others that prescribe the exact state of the ESM at every point in space and time on the ESM grid. The minimal number of (non-unique) variables, $d$, required to describe the exact state of the ESM at time $t$ can be thought of as a point in $d$ dimensional phase space with coordinates $(x_1,x_2,\ldots,x_d)$ and in the case of a ESM, the dimension of this space is extremely high\footnote{To give an idea of a lower bound on the dimension of the phase space required to describe a typical CMIP6 era ESM, the atmosphere is typically cut into a total number of $\sim 200\times 200 \times 100 = 4\times 10^6$ spatial cubes. A single atmospheric prognostic variable will need at least this many phase space dimensions at each point in time to specify it exactly. A lower estimate for the number of prognostic variables required just to specify the physical state of the atmosphere is $\mathcal{O}(10)$. This already gives $d\sim 10^7$ even before including the ocean, land and biogeochemisty.}. As the ESM state changes with time, a trajectory is drawn between these phase space coordinates connecting the past and present states. Variables describing the precise state of the ESM such as temperature or precipitation, the $X$ and the $Y$, may be functions of all the $d$ phase space coordinates as well as time $(d+1)$ i.e. $X=f_X(t,x_1,x_2,\ldots,x_d)$ and $Y=f_Y(t,x_1,x_2,\ldots,x_d)$.

Although the mechanistic relationship between $X$ and $Y$ is generally a function of all $d+1$ coordinates, climatic variables of interest are usually, although not exclusively, long term temporal and/or spatial averages that can be thought of as (time invariant) attractors of the phase space reducing the information relative to a precise configuration at any one time. In many circumstances the mechanistic relation between $X$ and $Y$ can be approximated to a good degree by a much smaller dimensional subset of the full phase space reducing the effective dimension.

The success of statistical mechanics and science in general\footnote{Low dimensional models can often mimic the responses of more complex ones to a good approximation and this is essentially why science is successful i.e. we can comprehend and predict the high dimension real world using (relatively) simple models.} relies on this effective dimension reduction, that certain properties of systems with many degrees of freedom can be well approximated by fewer state parameters. For example, a container of gas is, as far as we know at present, most completely described by the equations of quantum mechanics. In principle, one could solve these equations for each of the $N$ gas molecules, at each position and time in the container. The dimension of this space increases exponentially with $N$, requiring a huge number of degrees of freedom to specify the state exactly even for just a handful of molecules. This could be useful if the question one wanted to answer required detailed knowledge of the full quantum state. However, if we `only' want to know the configuration of positions and momenta of the gas molecules, one can use an approximation of the quantum theory. Treating the gas as a bunch of identical spheres each with a mass and a different position and momentum, the state can be described by a point in 6$N$ dimensional phase space, a reduction in the phase space dimension yet still a massive number of dimensions for a realistic gas.

Generally detailed knowledge of a precise system state is not desired or required. Bulk properties describing the mean overall state are a useful way of reducing information and making sense of complex systems. Going back to the container of gas, kinetic theory provides a way of mapping the many degrees of freedom from each of the molecules to the few bulk parameters of the ideal gas law, under reasonable assumptions (see for example \textcite{ref:SchekochihinNotes}). That is, the $6N$ dimensions effectively reduces to the three encoded by the relationship $pV=nRT$. The bulk variable pressure $p$ felt as a force per unit area on the container wall and is related to how often and how vigorously on average molecules with mean square velocity $\langle v^2 \rangle$ hit the sides of the container ($p=\frac{n}{3V} \langle v^2 \rangle$, $n$ is the number of moles in the container of volume $V$), whereas the temperature $T$ is a measure of the most probable kinetic energy of a molecule in the container. $R$ is the ideal gas constant. This approach works well when $N$ is large as one finds more of the $6N$ states correspond to the ideal gas law.

Effective dimension reduction in ESMs can also happen when we want answers to questions about bulk mean states of the climate. For instance, relations between bulk variables such as the annual global mean surface air temperature and the net incoming radiation are well modelled by the few degrees of freedom in a simple energy balance model (see for example \textcite{ref:Hasselmann76,ref:Wigley&Raper90,ref:Gregory00}) even though the exact state of a ESM is given by all prognostic variables at every point and multiple times on its spatial grid. If one wanted to ask detailed questions about the state of a ESM such as `How many days in March does it rain more than $x$ mm in Madagascar?' then the full ESM theory is best placed, however if the question was `How much does the global mean temperature increase from a spatially uniform doubling of atmospheric CO$_2$', a lot of that extra detail seems unnecessary. The analogy we attempt to present here is that ESMs are analogous to the solutions to the laws of quantum mechanics and the effective, reduced dimension relationship for a particular ESM $i$, $Y=f_i(X)+\epsilon$ is analogous to the ideal gas law. The analogy is far from perfect, ESMs are of course not as simple as containers of large numbers of identical molecules and there is, as yet, no such elegant route between the two solutions provided by the statistical mechanical or kinetic theory recipes. The best candidates for ESM ideal gas laws to serve as theoretical bases for emergent relationships are simplified, analytically soluble models of the climate or its subsystems such as the energy balance models previously mentioned \cite{ref:Hasselmann76,ref:Wigley&Raper90,ref:Gregory00}. These act more as testable, plausible hypotheses of the bulk behaviour of ESMs rather than derivations from the basic laws that constitute them.

So far we have discussed how a physically plausible, mechanistic, low dimensional relationship $f_i$ between $X$ and $Y$ may appear from a particular ESM $i$, $Y=f_i(X)+\varepsilon$. We have not yet discussed how low dimensional relationships, the \emph{emergent relationship}, could appear across an ensemble of ESMs. Although each ESM is different, they should be equally plausible models of the $(X,Y)$ relationship in the real world and it should not be a surprise that ESM solutions across a model ensemble should be also be mechanistically related. This is because many of equations are based on well established and tested physics (such as the Navier-Stokes equations) and are common amongst ESMs. Indeed, if the all the equations were well known and could be numerically integrated exactly, the ESMs should produce identical outputs (and the problem of climate modelling would be solved and emergent relationships would disappear).

However, some equations can not be exactly numerically integrated due to limitations on temporal and spatial resolution. One then has to parameterize subgrid scale processes and this may be a source of difference between ESM solutions, each model giving a different $(X,Y)$ pair (see section \ref{sec:emerge:common_ensembles}). Although they are our best attempts at understanding the climate, some ESM responses diverge on important questions. Such a scenario is where the emergent constraint approach may be helpful. To have a chance of finding an emergent relationship, the set of $n$ $(X_i,Y_i)$ data points should be different (i.e. $(Y_1,X_1)\neq (Y_2,X_2)$) and span a wide enough range for the emergent relationship to become apparent. The model relationship $f_i$ also needs to be shared amongst the model ensemble ($f_i(X)\approx f_j(X) \forall i,j = \{ 1,2,\ldots, n\}$). Provided such a relationship exists, the emergent relationship $f(X)$ can be determined. If this model relation is also shared by the real world, the real world's response can also be determined (along with its uncertainty) to give an emergent constraint (see section \ref{sec:emerge:ECrequirements}).



%

\subsection{Range in response due to the same physical process having a wide range across ESMs}

A related mechanism for a low dimensional relationship to emerge between $X$ and $Y$ occurs when the magnitude of the same physical process, correlated to the observable $X$, differs appreciatively across the ESM ensemble and the magnitude of this process heavily determines the size of the response $Y$. This is also a case of effective dimension reduction - the range in response is dominated by the dimensions of phase space of just that physical process.

An example of this happening was provided by \textcite{ref:Caldwell18} who evaluated ECs on equilibrium climate sensitivity (ECS) in the CMIP3 and CMIP5 ESM ensembles. That is, the targeted unknown response, $Y$, amongst these ECs was the same, ECS, although different observables $X$ were found to be well correlated to it. Examples of a few of the $X$ used in the different ECs were the strength of resolved-scale mixing between the boundary layer and lower troposphere in the tropical east Pacific and Atlantic \cite{ref:Sherwood14}, error in the distribution of cloud-top pressure and optical thickness for regions between 60$^o$N/S \cite{ref:Klein13}, fraction of tropical clouds with tops below 850 mb whose tops are also below 950 mb \cite{ref:Brient16a} and variability in global mean air temperature \cite{ref:Cox18}. \textcite{ref:Caldwell18} showed all of these observables were also highly correlated to the short wave (SW) cloud feedback, a strong and uncertain feedback on the resulting global temperature. The SW cloud feedback in turn was shown to be highly correlated to intermodel variations in ECS within the CMIP5 ensemble. To paraphrase \textcite{ref:Caldwell18}, `...intermodel variations in cloud feedback were so big that they left a strong imprint on intermodel variations in ECS. This means that fields that are strongly correlated with ECS are probably correlated with the SW cloud feedback (and vice versa).'

\subsection{What is needed for an EC?}\label{sec:emerge:ECrequirements}

In addition to having an emergent relationship in a model ensemble, an EC also ideally requires a few other things to be useful.

\subsubsection{Observable ($X$) range and uncertainty}

The range of $X$ in the ESM ensemble should be large relative to the uncertainty in the value of the real world observable. Uncertainty in each of the model $X$ should also be small relative to the range in model $X$. Ideally the real world value of the observable $X_\text{obs}$ should also fall well within the range of values in the ESM ensemble to avoid extrapolation issues.

\subsubsection{Response ($Y$) range and uncertainty}
One would like a large range in the values of the ESM ensemble responses relative to the uncertainty in each of the individual ESM responses. This reduces the uncertainty in the real world value of $Y$ derived from the EC.

\subsubsection{Relationship between $X$ and $Y$}
A particular ESM labelled $i$ should have a physically plausible relationship $f_i$ between the observable and response $Y=f_i(X)+\varepsilon$. It should be based on a simplified theory that predicts a functional response and has clear, testable assumptions that can be falsified independently. In addition, all models in the ensemble should share roughly the same relation between observable and response i.e. $f_i=f_j$. The stronger this relationship is, the more useful the emergent constraint due to the potential uncertainty reduction in $Y$. \emph{And} this relationship should also be shared by the real world.

\subsubsection{Large ensemble size $n$}
Ideally you would have a large number of independent ESMs in your ensemble (larger sample, lower error in estimates, less chance for high correlations by chance). See section \ref{sec:wrong:common_code} for further discussion of ESM ensemble independence.

\section{Underlying theory for emergent constraints based-on temporal variability}\label{sec:theory}
Many published EC studies relate the longer-term sensitivity of interest to aspects of the mean climate simulation in models - see Table I \citep{ref:Volodin08, ref:Kidston&Gerber10, ref:Massonnet12, ref:Tian15, ref:Brient16, ref:Simpson&Polvani16, ref:Lin17, ref:Lipat17, ref:Siler18, ref:Selten2020}. However, a growing number of papers are also now relating Earth System sensitivities to observable temporal variations, such as trends \citep{ref:Boe09,ref:Jimenez-de-la-Cuesta2019, ref:Nijsse20, ref:Tokarska2020}, interannual variability \citep{ref:Clement09, ref:OGorman12, ref:Cox13, ref:Wenzel14, ref:Qu15, ref:Kwiatkowski17, ref:Cox18} and seasonal cycles \citep{ref:Hall&Qu06, ref:Knutti06, ref:Qu&Hall14, ref:Zhai15, ref:Wenzel16, ref:Thackeray2019}. This section discusses the theoretical basis for emergent constraints based on such temporal variations.

 In the case of changes in relatively fast variables, such as seasonal snow-cover \cite{ref:Qu&Hall07}, or marine phytoplankton concentration \cite{ref:Kwiatkowski17}, there may be a fairly straightforward near one-to-one relationship between the short-term variability and the longer-term sensitivity, because the fast variable will be in a quasi-equilibrium state even with short-term climate variations. For slower variables (such as the forest carbon storage), short-term variations are more likely to measure fluxes (or equivalently the rate of change of the store). In this case, finding a constraint on future changes in the store requires multiplying the flux sensitivity to short-term variations by a characteristic timescale for each model. In some cases, the characteristic timescale may be similar across the model spectrum, leading to a simple emergent relationship between the short-term flux sensitivity and the long-term sensitivity of the store \cite{ref:Cox13, ref:Wenzel14}.  In general though, converting a flux sensitivity to a store sensitivity requires an independent estimate of the characteristic timescale of the store, which itself requires a theoretical basis \cite{ref:Williamson19}.

The Fluctuation-Dissipation Theorem (FDT) provides one such theoretical framework \cite{ref:Kubo66}. The FDT relates the sensitivity of a system to a small external forcing to the response of the same system to its own internally-generated fluctuations. FDT is therefore a potential theoretical basis for ECs, as it links the natural variability of a system to its sensitivity.  The size of the forcing is important because the theorem only strictly applies to near-equilibrium linear systems. Nevertheless, FDT-based approaches have had a huge impact in statistical physics, including Einstein's work on Brownian motion \cite{ref:Einstein05}, and the understanding of Johnson-Nyquist noise in electrical circuits \cite{ref:Johnson28,ref:Nyquist28} .

The first proposal to apply FDT to the climate system came almost 40 years ago \cite{ref:Leith75}, but there has been a recent resurgence of interest in this area fuelled by methodological advances  and detailed comparison of FDT-derived estimates of climate change to climate model simulations \cite{ref:Majda10}. In principle, it may be possible to use FDT to get good estimates of the response of the real climate system to small forcing (e.g. that due to doubling CO$_2$) purely from accurate long-term climate observations that reveal the full spectrum of natural fluctuations of the climate \cite{ref:Bell80, ref:Schwartz07}. Unfortunately though, the length of the detailed climate records required to achieve this are typically much longer than those available \citep{ref:Kirk-Davidoff09}, and include the contemporary period when the mean climate is changing.

In the EC technique model projections are instead used to define emergent relationships between observable variations and future climate \citep{ref:Hall19}, and then specific observations provide a selection principle to constrain the range of future climate projections from these model-defined relationships \citep{ref:Eyring19}. The EC technique has yielded many proposed constraints on aspects of the future climate \citep{ref:Hall19} and the carbon cycle \citep{ref:Cox19}.  However, this power also comes with a danger - that blind data-mining of multidimensional model outputs could lead to spurious and misleading constraints \citep{ref:Caldwell14}. Protection from this risk can come from tests of robustness across different model ensembles, and from basing the search for ECs on firm theoretical and mathematical foundations. Theory-led ECs can also be considered as hypotheses that can be tested against the ensemble of complex ESMs. The remainder of this section aims to describe the reasons why relationships between Earth System variability and sensitivity are ubiquitous, and  to provides some examples of the emergent relationships to be expected under different types of time-varying forcing (such as seasonal cycles and longer-term trends).

\subsection{Relationships between variability of fluxes and the sensitivity of stores}

We can write the time evolution of a dynamical system variable $V$ in the general form:
\begin{equation}\label{sec:theory:eq:cons}
\frac{d V}{dt} = F(V, z_i)
\end{equation}
where $F$ is the net flux into the system, and $z_i$ are the environmental variables that affect this net flux. 
Equation \ref{sec:theory:eq:cons} represents a local conservation law as it implies that the variable $V$ can only change if the net flux is non-zero.
When the environmental variables are in a steady state, $z_i=z_i(0)$, the variable $V$ has an equilibrium state $V=V(0)$, defined by:
\begin{equation}
F(V(0), z_i(0))=0
\label{sec:theory:eq:equil}
\end{equation}
In climate science we are typically interested in how a climate system variable, such as global mean temperature or land carbon storage in the tropics, varies with environmental factors, such as the atmospheric carbon dioxide concentration or the climate in the tropics.
Where the associated perturbations $\Delta z_i$ are small compared to $z_i(0)$, equation \ref{sec:theory:eq:cons} can be linearly expanded about the equilibrium state:
\begin{equation}
\frac{d \Delta V}{dt} = \frac{\partial F}{\partial V} \, \Delta V + \frac{\partial F}{\partial z_i} \, \Delta z_i
\label{sec:theory:eq:lin1}
\end{equation}
where $\Delta V$ is the resulting perturbation to the state variable, and the partial derivatives are calculated around the initial equilibrium $\left\{V(0),z_i(0)\right\}$.
The initial equilibrium is stable if $\frac{\partial F}{\partial V}<0$, and we can then define an effective timescale to perturbations of $V$ as $\tau =(-\tfrac{\partial F}{\partial V})^{-1}$. Equation \ref{sec:theory:eq:lin1} can therefore be written as: 
\begin{equation}
\frac{d \Delta V}{dt} +\frac{\Delta V}{\tau} = \frac{\partial F}{\partial z_i} \, \Delta z_i
\label{sec:theory:eq:lin2}
\end{equation}
The righthand-side of equation \ref{sec:theory:eq:lin2} can be viewed as the `external' forcing factors that produce changes in $V$ (e.g. the radiative forcing due to increasing atmospheric CO$_2$).
For a permanent time-invariant change in the environmental variable $\Delta z_i$, $V$ will be changed by an amount $\Delta V_{eq}$ where:
\begin{equation}
\Delta V_{eq} = \tau \, \frac{\partial F}{\partial z_i} \, \Delta z_i
\label{sec:theory:eq:dYeq}
\end{equation}
Defining the sensitivity of $V$ to the environmental variable $z_i$ as $\xi_i = \partial V_{eq}/\partial z_i$, we can rewrite this equation as:
\begin{equation}
\xi_i = \tau \, \mu_i 
\label{sec:theory:eq:dYeq_dx}
\end{equation}
where $\mu_i = \partial F/\partial z_i$.
Equation \ref{sec:theory:eq:dYeq_dx} implies that the sensitivity of $V$ to a step change in $z_i$ will be proportional to the sensitivity of the net flux $F$, with a constant of proportionality which is the characteristic lifetime of perturbations to $V$. In climate change research, the equilibrium  sensitivities ($\xi_i$) are often the things we would most like to constrain (e.g. Equilibrium Climate Sensitivity, ECS).


\subsection{Theoretical emergent relationships for idealised time-varying forcing}

For emergent constraints we therefore need to find relationships between sensitivities $\xi_i$ and observable variations in the climate system.
In this subsection we derive candidate emergent relationships for different time-variations in the environmental variables $z_i$. First we rewrite equation \ref{sec:theory:eq:lin2} in the form: 
\begin{equation}
\frac{d \Delta V}{dt} +\frac{\Delta V}{\tau} = \mu_i \, \Delta z_i
\label{sec:theory:eq:lin3}
\end{equation}
As we limit ourselves here to this linear model with a single timescale $\tau$, the emergent relationships presented below are intended to be illustrative, or as simple hypotheses to be tested against the outputs from complex models.  

\subsubsection{Sinusoidal forcing}

As an idealised representation of the response of our system to diurnal and seasonal forcing, we first consider sinusoidal environmental variations of angular frequency $\omega$ and amplitude $a_z$, for which $\Delta z_i=a_z\,  e^{i \omega t}$. The solution to equation \ref{sec:theory:eq:lin3} under these circumstances (after initial transients have died-down) is:
\begin{equation}
\Delta V= \mu_i \, a_z \, \frac{\tau}{\sqrt{1 + \omega^2 \tau^2}}  \, e^{i (\omega t - \phi)}
\label{sec:theory:eq:sine}
\end{equation}
where $\phi =\arctan{(\omega \tau)}$. Therefore $\Delta V$ also has a sinusoidal variation but with a phase-lag relative to the forcing of $\phi$ which asymptotes to $\pi/2$ as $\omega \tau \rightarrow \infty$. Substituting  $\Delta z_i=A_z \,  e^{i \omega t}$. The amplitude of the sinusoidal variation in $\Delta V$, $a_v$, is proportional to $a_v$ but also depends on the frequency of the sinusoidal forcing:
\begin{equation}
\frac{a_v}{a_z}= \mu_i \, \frac{\tau}{\sqrt{1 + \omega^2 \tau^2}}  
\label{sec:theory:eq:amp}
\end{equation}
Using equation \ref{sec:theory:eq:dYeq_dx} we can therefore write the sensitivity $\xi_i$ in terms of the ratio of the sinusoidal amplitudes: 
\begin{equation}
\xi_i = \frac{a_v}{a_z} \, \sqrt{1 + \omega^2 \tau^2}  
\end{equation}
In the limit of slowly-varying forcing compared to the system timescale ($\omega \tau \rightarrow 0$) this formula reduces to :
\begin{equation}
\xi_i \rightarrow \frac{a_v}{a_z}    \mbox{ \hspace*{10mm}   ; for  \hspace*{5mm}}  \omega \tau \rightarrow 0
\end{equation}
This is the limit of very fast variables that are in a quasi-equilibrium with the sinusoidal forcing, such as seasonal snow-cover \citep{ref:Qu&Hall07} or marine primary production \citep{ref:Kwiatkowski17}.
However, in the opposite high-frequency limit ($\omega \tau \gg 1$) the emergent relationship also depends linearly on the frequency of the forcing:
\begin{eqnarray}
\xi_i \rightarrow \frac{a_v}{a_z}  \, \omega \tau     \mbox{ \hspace*{10mm}   ; for  \hspace*{5mm}}   \omega \tau \gg 1 
\end{eqnarray}

\subsubsection{Linearly-increasing forcing}

Contemporary climate change is largely being driven by an approximately exponential rate of increase in atmospheric carbon dioxide above the pre-industrial level, which yields a radiative forcing that is approximately linear in time. This has motivated a long-running series of idealised climate model experiments which prescribe an exponential increase in carbon dioxide, including the \emph{1\% per year runs} which are used to define the concept of the \emph{Transient Climate Response (TCR)} . In addition, many other Earth System changes can be approximated by linear trends. In this sub-section we therefore consider emergent relationships under a linear increase in the environmental variable $z_{i}$:
\begin{equation}
\Delta z_i = \gamma \, t
\label{sec:theory:eq:trd}
\end{equation}
where $\gamma$ is the linear rate of increase $z_{i}$ which begins at time $t=0$. Under this idealised forcing the solution to equation  \ref{sec:theory:eq:lin3} is
\begin{equation}
\Delta V(t) = \tau \mu_i \, \gamma \, \left\{t - \tau (1 - e^{-t/\tau}) \right\}
\label{sec:theory:eq:trd_sol}
\end{equation}
From equation \ref{sec:theory:eq:dYeq} we note that the sensitivity $\xi_i = \tau \mu_i$, so equation \ref{sec:theory:eq:trd_sol} can be rewritten as:
\begin{equation}
\xi_i = \frac{\Delta V(t)}{\gamma} \, \frac{1}{\left\{t - \tau (1 - e^{-t/\tau}) \right\}}
\end{equation}
The exponential decay-term here is a transient response to the sudden switch-on of the linear trend at $t=0$. Once this term has died-out the transient solution lags the quasi-equilibrium solution by $\tau$ years such that:
\begin{equation}
\xi_i = \frac{\Delta V(t)}{\gamma \, (t-\tau)}
\label{sec:theory:eq:trd_sol1}
\end{equation}
This equation represents a potential emergent relationship between the sensitivity $\xi_i$ and a transient change $\Delta V(t)$. 

\subsubsection{White-noise forcing}

Emergent relationships have also been proposed between interannual variability and sensitivities \citep{ref:Schwartz07, ref:Cox13, ref:Nijsse19}, assuming that the environmental variable $z_i$ is approximately gaussian white-noise.  Under those circumstances equation \ref{sec:theory:eq:lin3} becomes the widely-used Ornstein-Uhlenbeck equation of statistical physics \cite{ref:Uhlenbeck&Ornstein30}. Standard solutions relating the variance of $V$ to the variance of $z_i$ can be derived by integrating the sinusoidal solution given by equation \ref{sec:theory:eq:sine} over all frequencies $\omega$. For interannual variability this yields the following relationship for the ratio of standard deviations:
\begin{equation}
\frac{\sigma_v}{\sigma_z} = \mu_i \, \sqrt{\frac{\tau}{2}} = \xi_i \sqrt{\frac{1}{2 \tau}}
\label{sec:theory:eq:rat_var}
\end{equation}
and an even simpler relationship for the lag-1 autocorrelation of $Y$:
\begin{equation}
\alpha_{v1} = e^{-1/\tau}
\label{sec:theory:eq:ar1}
\end{equation}
These equations can be combined to yield an equation for the sensitivity $\xi_i$ without needing to know the timescale $\tau$ \citep{ref:Cox18} :
\begin{equation}
 \xi_i = \frac{\sigma_v}{\sigma_z} \, \sqrt{\frac{2}{-\ln{(\alpha_{v1})}}}
\label{sec:theory:psi}
\end{equation}
Similar formulae can be derived for more sophisticated representations, such as two-box and semi-infinite heat diffusion models of ocean heat uptake \citep{ref:Williamson19}.

\section{How might emergent constraints go wrong and how to guard against it?}\label{sec:wrong}

Uncertainty in future projections makes adaptation planning difficult, and so there is tremendous pressure on climate researchers to provide much
more refined predictions of expected large-scale environmental change as atmospheric greenhouse gases (GHGs) rise. However, in the absence of a
full knowledge of all climate processes and their parameterizations, such deficiencies will continue to cause substantial ESM differences.
For this reason, the method of emergent constraints has attracted substantial attention. Such interest is because ECs offer a method to potentially `short cut'
current deficiencies in process understanding, by providing better estimates of change. Or, as a minimum, ECs offer more reliable estimates of
bulk aggregated parameters of the climate system of interest to policy. We believe ECs do provide a route to reduce uncertainty,
and consider it is appropriate for the technique to substantially underpin societal decisions both regarding adaptation planning and parallel
mitigation programs to reduce GHG emissions. However this brings tremendous responsibility, and so the method must be reliable and robust.
It is therefore prudent to discuss the multiple circumstances in which the method may fail, thus altering to where caution is needed. We now describe
such situations.

\subsection{Risks of purely using data mining}\label{sec:wrong:data_mine}

One approach with ECs is to simply `data mine' within climate model ensembles, such as CMIP5 and CMIP6, searching for two quantities in each model that when analyzed form a statistically significant inter-ESM regression. In these circumstances, the researcher is not led by process understanding, a hypothesis to be tested, or indeed intuition (see section \ref{sec:emerge:data_mine} for an extreme example). However, there is a logic that in some circumstances, this approach might be valid. In the climate system, there may be links between different parts of the Earth system, and including between a contemporary measurable quantity and an attribute change in a future climate state, that are not immediately obvious as being connected.

One set of examples could be similar to the on-going discovery of teleconnections in the climate system, where strong correlations are noted between variations in two parts that are a substantial geographical distance apart \cite{ref:Nigam&Baxter15}. Besides spatial distances, discovered teleconnections can contain lags, although unlike ECs that are designed to project decades ahead, the delays are months. These delays are often used to estimate expected meteorological conditions in the months ahead. For example, the status of El Ni\~{n}o, can strongly weight the probabilities of particular weather features for land regions and across the tropics and for the season ahead. Although El Ni\~{n}o is an obvious system fluctuation to investigate for its impacts elsewhere, over the last couple of decades, researchers have been investigating other candidates e.g. \textcite{ref:Feldstein00}.

Given the sometimes diverse form of teleconnections, their discovery can require a data mining approach, as comparison between two different parts of the
climate system, additionally offset in space and time, are unlikely to be found through intuition-led inspection of ESMs. If novel ECs exist, between diverse parts
of the Earth system, then such mining methods may be needed to aid their discovery. If ECs are discovered by statistical methods, then subsequent process
analysis may reveal the connection to have a strong physical basis. \textcite{ref:Hall19} refers to this as `EC confirmation', although this may take many years
to achieve, and especially if additional ESM diagnostics are needed to confirm the size of any transport equation terms between different spatial locations.

However, whilst such an approach is expected to reveal new ECs, caution is needed. The concern is that statistically-significant regressions
are likely to be found in a small number of instances, simply by chance, and the resultant EC is therefore not a robust indicator of future change (i.e. section \ref{sec:emerge:data_mine}). That is, and presented loosely, if statistical mining finds a large number of new ECs, all with non-zero regressions forming them and at a 90\% confidence level, then roughly one-in-ten will be invalid. This highlights the need to undertake process-based confirmation of ECs found through statistical means.

\subsection{The risk of $p$-hacking and overconfidence}\label{sec:wrong:phack}

The term `$p$-hacking' was first introduced by \textcite{ref:Nuzzo14}, and alerts to the risk of self-selecting only scientific findings that are statistically significant. In the context here, that would be ECs having their regressions at small $p$ values, suggestive of a low probability of an $X-Y$ relation occurring merely by chance. The concept has strong similarities to the dangers noted in data mining, where scanning across a range of potential ECs could lead investigators to only concentrate on a few with low $p$-values, with the attendant risk they occur by chance. However, $p$-hacking goes further and cautions over other decisions that researchers may make, in a rush to find a significant result. Behaviours could include a deliberate selection of one-sided tests and the stopping of sampling upon finding a low $p$-value. It could also involve adjusting the parameters, relationships between data, or time frequencies considered until lower $p$-values are discovered. In the case of the climate system, and as an example, the assessment of attributes of the global hydrological cycle offers many time scales of interest.  Precipitation statistics with potential to form different $X$ and $Y$ quantities of ECs, range from short, intense rainfall events through to seasons or even decades. To account for overconfidence \citet{ref:Bretherton2020} suggests that ECs be corrected by scaling up the unexplained variance by a user-defined factor across the board.

\subsection{Missing process in all current models, measurement errors and model compensating errors}\label{sec:consistent_bias}

ECs most frequently link a fluctuating quantity that is measurable for the contemporary period, to either a future change
(e.g. extent of polar ice sheet cover), or an invariant system attribute that describes change (e.g. climate sensitivity).
A potential concern for the validity of an EC is if there is a missing process in every model that affects the regression
forming the basis for the EC. If, for instance, that error affected the $X$-axis fluctuating quantity, but the measured quantity was accurate,
then this would introduce a bias in to the projection of the future $Y$-axis quantity.

Such a missing process could itself impact ESM performance at all modelled levels of atmospheric GHG concentrations. One potential example is that many ESMs underestimate interception loss, by vegetation, of rainfall e.g. \textcite{ref:Yang18,ref:Lian18}. Interception loss is the return of water from rainfall to the atmosphere, and that has not passed through soils but instead is temporarily held on leaves and branches instead. If an emergent constraint on, for instance, a feature of the atmospheric part of the hydrological cycle that depended on the overall land-atmosphere exchange of water, then the EC could have an overly strong dependence on plant evaporation (i.e. to compensate for the low modelled values of interception loss). If, then, the true $X$ value was taken from data of actual plant evaporation, this would introduce a bias in the estimate of the $Y$ quantity. A straightforward corollary to this is if an EC depends on an $X$ quantity that is modelled well, but measurements of it contain a bias, then this too would lead to inaccurate
estimates of the value of the $Y$ quantity.

Alternatively, other process omissions could become more important as GHG levels rise. Until recently, an example of a component frequently not modelled in ESMs is the terrestrial nitrogen cycle \cite{ref:Quinn15}. Nitrogen limitation could become an increasingly important factor as CO$_{2}$ concentrations rise, by suppressing levels of vegetation fertilisation from the increased CO$_2$ in the atmosphere. The lack of nitrogen cycle inclusion in the  land components of ESMs could therefore affect ECs linking current variations of the carbon cycle to future carbon stores, leading to potential overestimates of the capability of terrestrial ecosystems to offset future anthropogenic CO$_{2}$ emissions. That said, in \textcite{ref:Wenzel14}, two ESMs that did have the nitrogen cycle included were close to an EC regression line in an ensemble of carbon cycle only ESMs. That analysis linked contemporary fluctuations in temperature and atmospheric CO$_{2}$ concentrations to the long-term stability of terrestrial carbon stores under global warming.

A corollary to single model errors is the case where two contemporary processes are defined, but together they allow a model to perform well on some metrics as a consequence of compensating errors. There is quite substantial anecdotal evidence that many ESMs have these issues, as model developers often  discover that improvement of one component, by removing the compensation, actually results in less accurate contemporary climate simulations. Poor performance is only then resolved when the process representation of the second feature is also upgraded. Such compensations present a risk for ECs. Specifically, if the $X$-axis quantity of an EC relates to just one of the two process that compensate, then that will introduce a bias. Returning to the interception example, if we consider the deficiency recognised but accounted for by parameterizing overly strong plant evaporation amounts, any EC dependent on plant evaporation ($X$) would introduce a bias in prediction of $Y$ if actual measurements of $X$ are used.

Ultimately, climate science requires a set of ESMs that make similar projections and are highly accurate. Hence the on-going attempt to
include all relevant processes in ESMs, and to remove compensating errors, remains an appropriate activity to advance such predictive tools.
However, a co-benefit is that such ESM enhancements will likely raise the accuracy of any ECs that are used to constrain understanding where uncertainty
remains.

\subsection{System passes through a tipping point}\label{sec:wrong:tipping}

It is arguable that moderate human adjustment to the climate system is relatively linear, and so the expected meteorological changes increase proportionally with changes to the combined radiative forcing of atmospheric GHG concentrations. If this same linearity extends to higher frequency system responses such as interannual variability, then observations of contemporary fluctuations may give a strong indication of the system's response to a more permanent forcing from increased GHGs. One example is \textcite{ref:Cox13}, where the change to tropical atmospheric CO$_{2}$ concentration in response to annual temperature variation may project any expected loss of rainforest carbon stores under sustained global warming. Hence, there is an assumption of linearity present in each individual ESM, or at least a relationship that is monotonic in increasing temperature. When plotting the simultaneous variations in temperature and CO$_{2}$ ($X$) against response of future carbon stores in response to warming ($Y$), then a further linear inter-ESM relationship is found, and this is the EC itself.

A legitimate question, therefore, is what happens should the planetary system move beyond linear responses, and further, start to pass through tipping points \cite{ref:Lenton08}? Many examples of such potential nonlinear behaviours are conjectured, where a relatively small increase in radiative forcing could cause major system changes. Potential climate tipping points such as the collapse of the Atlantic meridional circulation \cite{ref:Stocker&Wright91}, Amazon dieback \cite{ref:Cox00}, Greenland ice sheet melt \cite{ref:Toniazzo04} and others \cite{ref:Drijfhout15a} are presently thought to be low probability, high risk events although there is evidence subsystems of the climate have `tipped' many times in the past \cite{ref:Bond92}. Most of these `tips' have resulted in regional rather than global climate changes and have been emulated through simpler equation systems, amenable to bifurcation analysis to characterise the nonlinearity  (e.g. \textcite{ref:Dijkstra13}).

Once past a large-scale tipping point, the Earth System might have radically different responses and feedbacks, and so behave very differently. This is likely to impact on the EC approach is two ways. First, for each individual ESM, for sufficiently changed climate,
the contemporary fluctuating quantity may no longer give accurate information in that model on future store size.
Second, if there are system tipping points, but an EC is only validated for moderate changes to the climate system
(so, for instance, the $Y$-axis quantity is only derived from ESMs operated under lower emission scenarios),
then using it to project to more severe climate alteration might be inappropriate.

A more abstract but relevant question is whether information from pre-tipping point conditions has any validity at constraining climatic conditions beyond one?
Do tipping points invalidate `information exchange' between different GHG levels when passing through one?
Two ESMs may have very similar responses up to a tipping point, and thereby appear as points near each other on a standard EC $X-Y$ plot,
but marked different responses beyond a tipping point, thus weakening an EC-based inter-ESM regression. Also in the abstract is a view that
ECs are capitalising on the `hard-wiring' implicit in ESMs, as based on model parameterizations constrained from contemporary measurement datasets.
Beyond any major tipping point represents a world that has not yet been witnessed and so not measured for equation parameterization purposes.

We hope that climate researchers can be encouraged to investigate further whether the planet passing through a tipping point invalidates ECs,
and possibly also the period just beforehand. This could be in the form of a conceptual model, or by analysis of ESMs in the CMIP5 ensemble.
ESM simulations with a tipping point could be split in to a future period pre-tipping point, and one post-tipping point for a high emissions scenario,
testing the EC on both modelled time frames. One obstacle to undertaking such analysis is that although ESMs do exhibit tipping points,
they vary markedly between models regarding their location, component of the Earth system affected, and level of global warming \cite{ref:Drijfhout15a}.
Notable is that the strong regional features of tipping points implies that any ECs that
may be affected by their presence also have to be localised in application.
Possibly of more concern is that if tipping points affect the predictive capability of ECs, there will be little inter-ESM consensus on when and how
this might occur. This lack of understanding is due to major model differences in predictions of tipping point occurrence, or even existence \cite{ref:Drijfhout15a},
and so an `EC-type' modulation factor to the original EC and common across models is unlikely to be discovered.

\subsection{Problems with common code across many models and implications for `out-of-sample' testing}\label{sec:wrong:common_code}

ECs rely on the statistics of regressions, which in turn assume an independence of data points. However, individual ESMs may not be completely independent. For parts of the climate system known to respond to well-established and well-understood physical processes, commonality in models is to be expected. The concern is for the model parts that are suspected as characterising actual processes less well, and where ECs seek to constrain this uncertainty. The lack of independence may take the form of individual research centres offering multiple model versions but at different resolutions.  Alternatively, some components of ESMs are shared between research centres, or are coded in similar ways \cite{ref:Knutti13}. The lack of independence could appear in the range of effective bulk parameters sampled, such as equilibrium climate sensitivity. Another possibility is there may be a common component existing between models, and that an EC seeks to reveal, but the ESM differences to reveal this are in the fluctuating forcings. These fluctuating quantities could have relatively low sampling due to similarities between how ESMs calculate their values.

As part of developing comprehensive process understanding of discovered ECs, \textcite{ref:Hall19} encourage `out-of-sample'
testing. Such testing is where the EC relationship is checked to be valid in additional ESMs that are not part of the
original set used to initially find the EC. New simulations from individual modelling centres can be analysed as they
become available. A more comprehensive test is to search within new ensembles, and so for instance an EC found in the
CMIP5 climate model ensemble can be checked for its presence in the newly-released CMIP6 set of ESM simulations.
However this too may not be a completely independent assessment, if modelling centres retain substantial amounts of model
code and parameterizations between ensemble contributions. We note that the out-of-sampling testing performed by
\textcite{ref:Schlund20}, that assesses if the ECs for climate sensitivity, ECS, found in the CMIP5 ensemble
remain valid for the CMIP6 models. As commented elsewhere, they find that for CMIP6, the CMIP5-based ECs have less predictive
capability, and also give generally higher ECS values. \textcite{ref:Schlund20} note that the majority of emergent constraints for ECS
are related to some extent to cloud feedbacks, and this is a major on-going area of climate research to create numerical cloud
schemes with strong predictive skill.  As many cloud schemes are currently under development, then their newness implies that this
is an example of little carrying of common code from CMIP5 to CMIP6.
From that perspective, CMIP6 models have independent features from CMIP5, validating using this more recent ensemble for out-of-sample testing.
However, if it is the new cloud schemes that cause the CMIP6-derived ECs to have less predictive capability, then this suggests
that previously estimated EC-based uncertainties on ECS from the CMIP5 models are overly narrow. This form of error differs from
that of Section~\ref{sec:consistent_bias} that considers how a missing process in all models may create a systematic bias in an EC projection.
Here, instead, the suspected missing processes associated with cloud dynamics and feedbacks are such that there likely remains
substantial uncertainty in how to model them accurately. Alternative cloud schemes may weaken ECs by causing a larger spread
around their regression lines, expanding the bounds on predictions of quantity $Y$ for quantity $X$ and for which contemporary data also exists.

\subsection{What to do when different ECs are found for the same quantity, but differ in value, or differ between
ensembles?}\label{sec:wrong:different_ECs}

In some instances, multiple and different contemporary measurements have been suggested, via ECs, to be able to predict the same quantity.
This is particularly the situation where a broad range of climate attributes have been used to evaluate equilibrium climate sensitivity (ECS).
However, and notably, cloud feedback \cite{ref:Sherwood14,ref:Brient16,ref:Brient16a,ref:Zhai15} or top-of-atmosphere (TOA) radiation
flux \cite{ref:Brown&Caldeira17} based emergent constraints tend to project higher ECS values than surface temperature based
constraints \cite{ref:Cox18}. Investigations need to continue to ascertain why these ECs disagree. Some disagreements may be
due to spurious correlations, affecting Y values. However, \textcite{ref:Caldwell18} show that these emergent relationships are highly
correlated. With high correlation, but different ECS estimates, this implies that differences may instead be due to measurement biases
in some X value observations.
Alternatively, the real world may not be sharing the same responses as the models, which could suggest a persistent error across ESMs
for at least one of the ECs.

As ECs come under increasing scrutiny due to their growing widespread use, more needs to be understood about how they operate and any limitations.
In that context, we suggest understanding the differences between projections of the same quantity will be a highly rewarding research path. However in order to
generate a single EC-based projection, \textcite{ref:Bretherton2020} present statistical methods to merge multiple ECs of the same quantity in to a single
range of uncertainty for the quantity being estimated. That analysis focuses on combining the multiple ECs used by other researchers to estimate ECS.

As noted in Section~\ref{sec:wrong:common_code}, new ensembles provide an `out-of-sample' test for existing ECs. The question then is how to
use that new value, and especially if it is substantially different to the value for an earlier ensemble. One possibility is to use the methods
of \textcite{ref:Bretherton2020} to merge the findings from two ensembles. A further possibility is to introduce a weighting to the \textcite{ref:Bretherton2020}
approach, but with a high influence for the more recent ensemble, corresponding to the hope that newer ESMs are better models.

\subsection{ECs may cause future CMIP-type climate model ensembles to have much less spread in projections}\label{sec:wrong:spread_reduction}

In the most general terms, a reduction of spread of climate models is to be highly welcomed, and especially if their convergence is on to projections that are accurate estimates of future change. Such convergence is good if what is learnt from ECs applied to previous ensembles is accurate, and this has caused climate modellers to make their new simulations achieve the constrained $Y$ values (see section \ref{sec:outlook:spread_reduction}). Convergence may also occur simply because previously uncertain parts of the Earth system have become better understood. However, the rapid development of ECs could cause a reduction in model spread that is potentially unwelcome, precluding further progress with the methodology.

As ECs are discovered, and estimates of future change are refined, this may cause climate modelling groups to, either consciously or unconsciously,
parameterize new ESMs so as to estimate changes that fall within those EC-based bounds. This convergence of projections could be because many modelling groups might not wish to estimate climatological changes that fall outside a consensus view based on previous EC-based estimates. However, this could have two detrimental side effects, and potentially cause future EC calculations to be less accurate. First, re-noting that it is almost a paradox of ECs, as designed to constrain estimates of change, that they can only work well with large inter-ESM differences.  It is a substantial spread of ESM estimates that enable `strong' regressions. Hence any clustering of projections will reduce the capability to re-test ECs for new ensembles of ESMs, and certainly make them less reliable. Second, it risks that if new processes are believed to be needed in the majority of new models, then other balancing process could be tuned, incorrectly, to balance these changes an in order to fit the earlier EC. For example, introducing the terrestrial nitrogen cycle more routinely into ESMs may suppress projections of future land carbon stores. But to remain in alignment with earlier EC estimates and as using ESMs mainly without nitrogen suppression, then instead in new ESMs this could encourage parameterization of an offsetting and incorrect overly strong CO$_{2}$ fertilisation effect for vegetation.

Despite the aim of ECs being to reduce the effects of ESM uncertainty, we suggest that climate science remains alert to these two concerns, where the
existence of an EC may overly reduce the model spread in any new ensemble (although this is not something that we have encountered to date). Such a reduction in a new ensemble, where model developers may wish to
replicate the earlier EC, risks, in particular, the propagation of errors that might be prevalent in the earlier ensemble.

\subsection{Inability to verify an EC}

In Section~\ref{sec:wrong:data_mine} we heed the dangers of ECs based on data mining, noting that \textcite{ref:Hall19} encourage that
ECs discovered statistically should first move to having potential, i.e. contain an element of intuition as their confirmation. Detailed process understanding,
maybe even as far as analytical assessment of related key climate-based differential equations, ideally then move an EC from having `potential' to being `verified'.
At the same time, we suggest a potential danger whereby it is not possible to follow this cascade of increased certainly in validity of an EC.
For example, there may be insufficient saved diagnostics from ESMs to build and study offline the process interactions that need understanding to
achieve EC verification. The risk here is that important and informative potential ECs are dismissed as their understanding cannot be developed. Although this may initially appear opposite to the suggestions in Section~\ref{sec:wrong:data_mine}, we advise such ECs are not cast aside, thereby with the danger of losing valuable insights into the climate system. Instead such potential ECs from data mining can generate
requests to climate research centres to provide related additional ESM outputs, to allow additional testing.

Any additional ESM diagnostics may also reveal the extent to which a potential EC is dependent on one major physical process, or is caused by an
amalgamation of multiple interacting effects. If more than one major process is present, then this may make it more difficult to transition from `plausible' to
`verified'.

\subsection{Lack of perturbed physics experiments with ESMs}\label{sec:wrong:PPEs}

The operation of ESMs is hugely computationally expensive and often there is only capacity in research centres to perform simulations for standardised scenarios of changes in future atmospheric greenhouse gas concentrations (e.g. the Representative Concentration Pathways, RCPs; \textcite{ref:Meinshausen11}). Unfortunately, this precludes performing simulations with varied parameters, and especially for those where there is substantial uncertainty surrounding their true value (or effective value, if grid-dependent). We suggest that this lack of model investigation restricts further testing of ECs, and in a way that may either preclude gaining extra confidence in their reliability or alternatively fails to alert to one that may not be as robust as initially believed. We suggest that PPEs may require interpretation in two ways, and that are converse to each other. If a parameter value is changed, and if it relates to a part of the Earth system believed to be independent of the underlying processes associated with a particular emergent constraint, then for that ESM the $Y$ value should not change in the EC. Such invariance both confirms EC robustness and helps illustrates that the EC aligns to the parts of the model expected from intuition or process understanding. The alternative situation is where a parameter believed to relate to an EC is perturbed, and potentially adjusted to the outer bounds of its expected value or beyond. The expectation then may be that for that particular ESM, the point moves to the bounds of the EC regression. Such an anomaly confirms that the EC is affected by the part of the climate system the parameter pertains to,
and may indicate that outlier parameter values are not present in the other ESMs.

If spare computational cycles become available, we suggest perturbed physics experiments with ESMs will be beneficial to the development and testing of ECs. 

\section{Emergent constraints found in the Earth system}
\label{sec:ECs}

\newcommand{\rpm}{\raisebox{.2ex}{$\scriptstyle\pm$}}

Numerous ECs have been identified across physical and biogeochemical components of the Earth System (a handful of which have been discussed thus far), with a substantial increase in their number over the past decade. In order to capture the breadth of its application in the geosciences, we discuss previously documented ECs based on their broad geophysical classification here. A list of EC studies is provided in Table \ref{table:ECs}. Note that this list is not exhaustive and may be easily superseded. Moreover, some of the references in Table \ref{table:ECs} offer more than one potential current climate quantity to explain $Y$, in which case we list only one example. For example, \textcite{ref:Sherwood14} provide three metrics that describe various aspects of lower-tropospheric mixing to explain equilibrium climate sensitivity. Rather than list all three metrics, we only discuss the one that appears most robust \cite{ref:Caldwell18,ref:Schlund20}. An important criterion for establishing credibility of an EC is to show robustness to the choice of ensemble. Therefore, Table \ref{table:ECs} also documents the model ensemble(s) for which each EC appears to have value (i.e., a statistically significant emergent relationship is present). In several cases this is only the ensemble from which the EC was derived as many of the more recent examples have not undergone out-of-sample verification at the time of this paper. We expect several upcoming publications to assess the validity of previously published ECs with the newer CMIP6 ensemble (e.g. \textcite{ref:Schlund20,ref:Pendergrass20}).

\begin{figure}[t]
    \centering
    \includegraphics[width=\columnwidth]{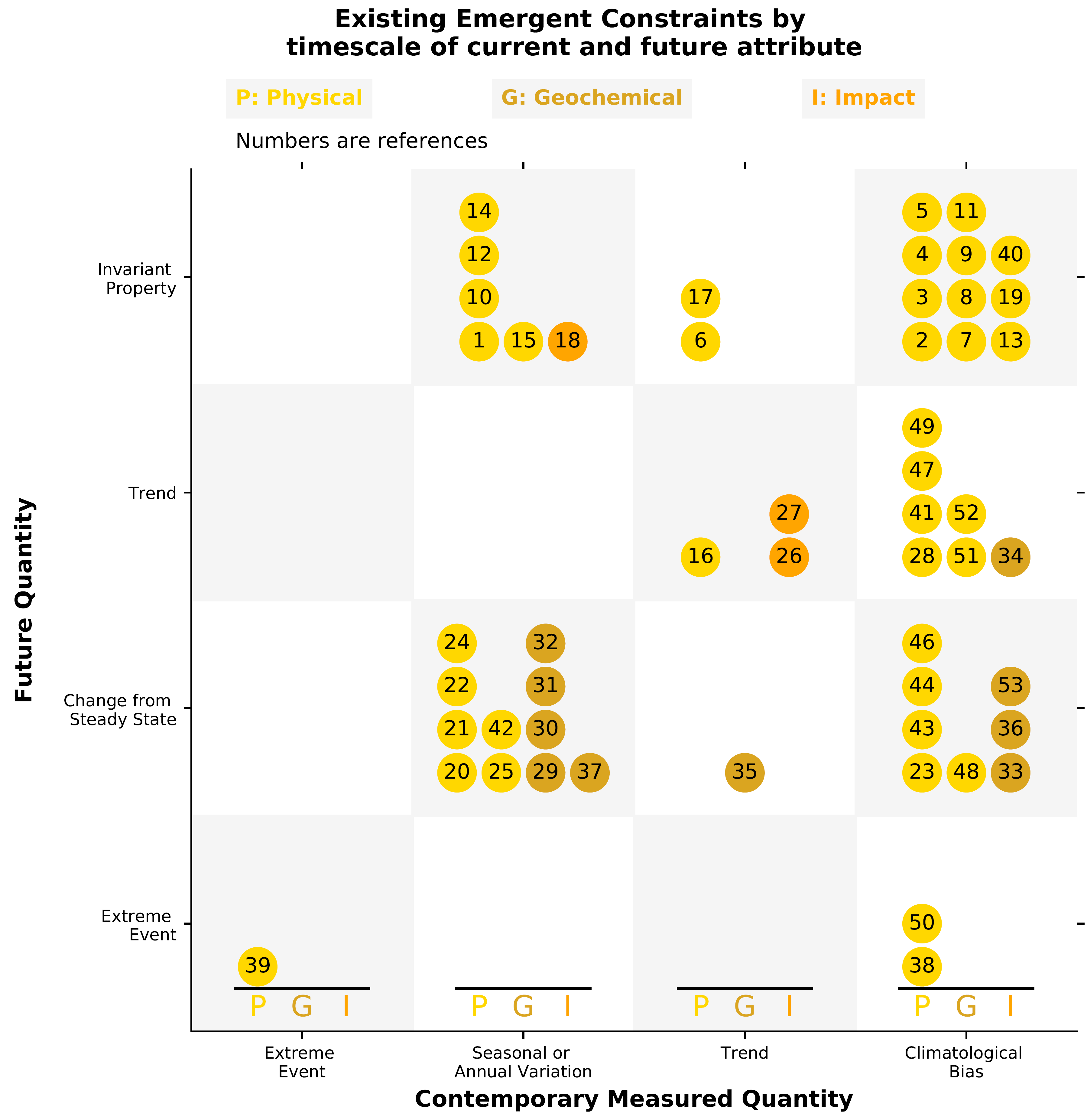}
    \caption{Schematic showing the different types of emergent constraints proposed in the literature to date. Each number corresponds to a study listed in Table 1. ECs are sorted by the timescale of their current climate quantity and the type of future constrained quantity. ECs are also sorted by whether the future quantity relates to an impact or a physical or geochemical quantity.}
    \label{fig:schematic}
\end{figure}

A vast majority of ECs pertain to one of the following general topics: climate sensitivity, cloud feedbacks, cryospheric feedbacks and change, carbon cycle feedbacks, and the hydrologic cycle, but in theory the methodology can be used for any number of applications.  There are also additional examples relating to topics such as radiative forcing \cite{ref:Bowman13}, regional air and sea surface temperature change \cite{ref:Lin17,ref:Selten2020,ref:Sgubin17}, frequency of temperature extremes \cite{ref:Donat18}, and atmospheric circulation changes \cite{ref:Kidston&Gerber10, ref:Simpson&Polvani16}. For the purpose of this review, we will focus primarily on the main applications to date. Along with the differing applications, ECs can also be sorted by the type of constraint (Figure \ref{fig:schematic}). One way to think of this is to group ECs by the time-scale of information that defines their $X$ (e.g., from multi-decadal to extreme events). For example, a rather simple class of ECs relates climatological biases or multi-decadal trends in some quantity ($X$) to the future change in $Y$. Similarly, a collection of ECs use seasonal or interannual variations in $X$ to constrain future changes in $Y$. In many of these instances, variability in some quantity to seasonal or interannual temperature variability is related to sensitivities of that same quantity to future warming. Constraints using climatological biases or short-term sensitivities are also common for attempting to constrain invariant properties such as equilibrium climate sensitivity. Applications to future changes in higher-order climate statistics (e.g., extreme events) are more rare. This highlights a relatively unexplored area of EC research.

\subsection{Equilibrium Climate Sensitivity}

By far the most popular application of the EC approach has been in relation to equilibrium climate sensitivity (ECS) (Table \ref{table:ECs}). ECS is seemingly an ideal candidate for this technique because of its importance to predicting future warming rates and its consistently large spread across various GCMs, which has been a persistent feature of all model generations. (The likely range of ECS according to the IPCC (1.5 to 4.5K) has been largely unchanged for decades). A series of contemporary quantities ($X$) have been identified as being closely linked to ECS, many of which relate to simulated cloud feedbacks or properties. In general, these cloud-based constraints tend to suggest relatively high ECS values (3.5-4.1K) \cite{ref:Knutti2017,ref:Brient2020}. Relevant examples include the height of tropical low clouds \cite{ref:Brient16a}, the sensitivity of the reflection by subtropical low clouds to sea-surface temperature \cite{ref:Brient16}, seasonal sensitivity of low cloud between 20-40°N and 20-40°S to sea-surface temperature \cite{ref:Zhai15}, variability in relative humidity and cloud extent \cite{ref:FasulloTrenberth12}, vertically resolved zonally-average relative humidity and clouds between 45S-40N \cite{ref:Su14}, and the climatological difference between tropical and Southern Hemisphere mid-latitude cloud fraction \cite{ref:Volodin08}.

Another suite of studies relate ECS to intermodel differences in the characteristics of historical radiative fluxes, such as climatological TOA energy balance in the Southern Hemisphere \cite{ref:TrenberthFasullo10}, features of TOA radiation fluxes \citep{ref:Huber11, ref:Tett13, ref:Brown&Caldeira17}, and cloud-sky radiative flux sensitivity to temperature \cite{ref:Lutsko&Takahashi18}. Alternatively, studies have also used temperature characteristics defined over a variety of time-scales to constrain ECS. This approach tends to suggest relatively weaker ECS values than the others \cite{ref:Knutti2017}. Examples of $X$ include the seasonal cycle of temperature \cite{ref:Knutti06}, 20th century warming \cite{ref:Annan&Hargreaves06}, and statistics of interannual variability \cite{ref:Cox18}. Similar to this, there is a collection of studies that have evaluated temperature changes over much longer timescales (e.g., global-mean cooling during the Last Glacial Maximum: 19-23 ka before present) as derived from paleoclimate data \cite{ref:Hargreaves12,ref:Schmidt2014}. These paleoclimate-based constraints suggest ECS of 2.3 and 3.1 K, respectively. However, there is a large amount of uncertainty surrounding the model simulations of the LGM, the proxies used to constrain the intermodel relationship, and the validity of using data from these time periods as proxies for anthropogenic warming \cite{ref:Harrison2015,ref:Brient2020}. A variety of other metrics have also been proposed as being strongly related to ECS with large variability in their constrained predictands. The highest value (4.5K) is derived from a constraint using vertical mixing between the boundary layer and lower troposphere over tropical oceans \cite{ref:Sherwood14}. \textcite{ref:Tian15} use climatological precipitation in the Intertropical Convergence Zone (ITCZ) region to suggest a slightly weaker ECS (4.1K), while the weakest non-paleo estimate uses the climatological latitude of SH Hadley Cell edge in Dec-Feb to suggest an ECS of only 2.5K \cite{ref:Lipat17}.

As discussed in Section \ref{sec:wrong:different_ECs}, having such a large number of ECs for the same quantity can complicate their interpretation when the constraints suggest differing $Y$ values. In this instance, rigorous testing of various proposed ECs is necessary to better understanding their validity. Along these lines, ECs on ECS have been heavily scrutinized over the past few years with assessments tied to their plausibility and robustness across both multimodel ensembles \cite{ref:Grise2015,ref:Caldwell18,ref:Schlund20} and, to a lesser extent, perturbed physics ensembles \cite{ref:Kamae16,ref:Zhao16,ref:Wagman&Jackson18}. \textcite{ref:Caldwell18} provide the most robust assessment to date on the credibility of various constraints on ECS. In doing so, they find that several of the aforementioned constraints lack a physically plausible mechanism connecting the $X$ quantity with ECS \citep{ref:Volodin08,ref:Su14,ref:Tian15,ref:Siler18}. This limits how much faith can be put into the EC itself as the statistical relationship could be fortuitous. Furthermore, many of the constraints are found to be closely related \cite{ref:Caldwell18}. For example, one strongly correlated group relates aspects of present-day Southern Hemisphere cloud cover with ECS, suggesting they may all be capturing a single SH mechanism \citep{ref:Lipat17,ref:TrenberthFasullo10, ref:Volodin08, ref:Siler18}. Additional testing of EC robustness comes in the form of out-of-sample verification (i.e., evaluating the emergent relationship in a different model ensemble to the one from which the EC was originally derived). Although there are some known limitations to this approach related to the existence of common model code across generations (Section  \ref{sec:wrong:common_code}), it is still seen as a valuable exercise for evaluating ECs. For example, \textcite{ref:Grise2015} show that the \textcite{ref:TrenberthFasullo10} EC derived from CMIP3 is not valuable in CMIP5 because the emergent relationship only exists in a subset of models with unrealistic cloud properties in the Southern Hemisphere subtropics. \textcite{ref:Caldwell18} extend this type of analysis to show that \textcite{ref:FasulloTrenberth12} is also non-existent in CMIP5. On the other hand, they find the \textcite{ref:Volodin08} EC to be present in both CMIP3 and CMIP5, despite lacking a physically plausible explanation.  \textcite{ref:Schlund20} recently assessed a number of the constraints on ECS documented here using the CMIP6 ensemble. They find that most of these constraints appear less skilful at predicting ECS in CMIP6, largely tied to differing representation of cloud processes. Three of the ECs discussed here \cite{ref:Brient16,ref:Sherwood14,ref:Volodin08} still exhibit statistically significant skill, while two others no longer have value \cite{ref:Lipat17,ref:Su14}. Lastly, two ECs are somewhat in between, what the authors define as “indeterminate”, including \textcite{ref:Cox18,ref:Tian15} \cite{ref:Schlund20}. Further testing of these ECs has utilised perturbed physics ensembles to better sample the parametric uncertainty space associated with various emergent relationships (section \ref{sec:emerge:common_ensembles}). Of particular interest to these studies is the EC proposed by \textcite{ref:Sherwood14}. Results from three different PPEs show that the physical mechanism proposed by this study may not be valid, while the emergent relationship between lower tropospheric mixing and ECS only exists when certain convection schemes are used \cite{ref:Kamae16,ref:Zhao16,ref:Wagman&Jackson18}. Similar doubts have also been reported for the \textcite{ref:FasulloTrenberth12} EC \cite{ref:Wagman&Jackson18}.

This collection of analyses implies that the most robust constraints on ECS (determined by successful out-of-sample testing and plausible physical mechanisms) are tied to the present-day response of subtropical low clouds to SST variability  \cite{ref:Brient16,ref:Zhai15}. However, one factor that may limit confidence in these constraints is the short observational record of low cloud characteristics \cite{ref:Brient2020}. ECs from \textcite{ref:Sherwood14,ref:Volodin08} are also statistically significant in three ensembles (Table \ref{table:ECs}), but they face credibility issues tied to their physical mechanism (or lack thereof) \cite{ref:Caldwell18,ref:Wagman&Jackson18}. Despite these signs of robustness, the much weaker correlation values associated with these ECs in CMIP6 \cite{ref:Schlund20} supports the belief that it is unlikely that a single predictor will be able to physically explain a large amount of ECS variability \cite{ref:Caldwell18,ref:Hall19}. Rather, several constraints on various feedback components will likely be needed to make a difference for such complex processes. In this fashion, a recent comprehensive assessment of ECS gave little weight to ECs on ECS when attempting to determine its likely value instead favouring observations of actual temperature change and climate feedbacks \cite{ref:Sherwood20}. This is also believed to be true for the global cloud feedback (described in more detail below) \cite{ref:Klein&Hall15} given that the radiative effect of clouds in response to warming is expected to vary by cloud type and regime \cite{ref:Gettelman2016}.

Similar research efforts have recently suggested that historical decadal warming trends are strongly predictive of the transient climate response (TCR; defined as the amount of warming that occurs at the time of atmospheric carbon dioxide doubling, having increased by 1\% each year) \cite{ref:Jimenez-de-la-Cuesta2019,ref:Tokarska2020,ref:Nijsse20}. TCR is of interest because it more closely resembles the way carbon dioxide concentrations have changed in the past. These constraints suggest that the best estimate of TCR is 1.67K in CMIP5 \cite{ref:Jimenez-de-la-Cuesta2019} and between 1.60K and 1.68K in CMIP6 \cite{ref:Tokarska2020,ref:Nijsse20}. The difference between the latter two studies largely stems from the choice of historical time period evaluated \cite{ref:Nijsse20}. All three of these estimates are lower than the raw ensemble statistics (median of 1.95K in CMIP6). Similar constraints on future warming have also been developed for regional temperature changes. Both \textcite{ref:Lin17} and \textcite{ref:Selten2020} find that present-day summer temperatures across central USA and Europe, respectively, are strongly tied to future continental summer warming. The authors hypothesise that the same physical mechanisms operating in the current climate (tied to precipitation biases and soil hydrology) also govern the future climate response. Incorporating observations allows for the future spread in European summer warming to be reduced by nearly 50\% \cite{ref:Selten2020}.

\subsection{Cloud Feedbacks}

One of the primary contributors to intermodel spread in climate sensitivity is the cloud feedback, which encompasses changes in how clouds modulate the radiation budget in response to warming. (For a recent synthesis of cloud feedbacks see Section 3.3 of \textcite{ref:Sherwood20}). The global mean cloud feedback is the most uncertain component of the total climate feedback \cite{ref:Bony2006,ref:Ceppi17,ref:Sherwood20}, varying from -0.36 to 1.17 W m$^{-2}$ K$^{-1}$ in the latest generation of climate models  \cite{ref:Zelinka20}. Strengthening of ensemble mean cloud feedback strength in CMIP6 is also thought to be largely responsible for a recent ECS increase in many models  \cite{ref:Zelinka20}. Thus, many of the aforementioned constraints for ECS include cloud effects either directly or indirectly \cite{ref:Qu2018,ref:Brient16,ref:Brient16a,ref:FasulloTrenberth12,ref:Su14,ref:Volodin08,ref:Zhai15}. This also means that there is great interest in constraining cloud feedbacks as well \cite{ref:Brient&Bony13,ref:Qu14,ref:Gordon14,ref:Zhou15,ref:Siler18}. For example, \textcite{ref:Siler18} propose a constraint on future global mean cloud feedback using the climatological latitudinal gradient in cloud reflectivity (a quantity that can be derived from satellite observations). In general, this EC suggests that ESMs with lower cloud albedo in warm-SST regions and higher cloud albedo in cool-SST regions will exhibit greater cloud feedback (and to a lesser extent ECS). It suggests a global cloud feedback in the upper range of ESM estimates (0.58 \rpm 0.31 W m$^{-2}$ K$^{-1}$). However, this constraint lacks a testable physical explanation \cite{ref:Caldwell18}.  Another study that seeks to constrain the global mean cloud feedback is \textcite{ref:Zhou15}. This paper highlights strong similarities between the global mean cloud feedback and a present-day analogue cloud feedback derived from interannual climate variability \cite{ref:Zhou15}. This follows the assumption of several ECs that processes operating on seasonal to interannual timescales are also likely to be occurring on longer timescales. However, the usefulness of this EC is limited because the observational uncertainty in the present-day quantity is large relative to inter-model spread.

As previously mentioned, it is seen as unlikely that a single current climate quantity can account fully explain variability in the global cloud feedback \cite{ref:Klein&Hall15}. Thus, a number of studies have focused on isolating the individual components of the cloud feedback (high-cloud altitude, tropical marine low-cloud, tropical anvil cloud-area, land cloud amount, mid-latitude marine low cloud amount, and high-latitude low-cloud optical depth). For example, \textcite{ref:Po-Chedley19} narrow in on the changes to upper troposphere clouds and relative humidity across the tropics. They show that future changes in upper troposphere cloud fraction under climate warming are strongly tied to the vertical gradient in climatological mean cloud fraction and relative humidity. Elsewhere, much of this research concerns tropical low clouds as ESMs struggle to represent them, they account for nearly half of the variance in the global mean cloud feedback \citep{ref:Zelinka2016,ref:Klein2017}. Moreover, intermodel spread in the low cloud feedback is closely tied to variability in ECS \cite{ref:Sherwood20}. An early EC along these lines showed that the future low cloud feedback sign ($Y$) is related to current low cloud sensitivity as a result of natural variability ($X$) in CMIP3 \cite{ref:Clement09}. Unlike most other ECs, this example simply suggests the sign of a feedback rather than specific values. Thus, it is not always mentioned in reviews of this nature \cite{ref:Klein&Hall15,ref:Brient2020}. More traditional ECs relating to this topic include the link between low cloud optical depth changes with warming and the optical depth response to temperature anomalies associated with natural variability \cite{ref:Gordon14}. This feedback is of limited importance globally, but makes an important contribution at high latitudes \cite{ref:Klein&Hall15}. The \textcite{ref:Gordon14} EC also has a simple thermodynamic explanation, which makes it a promising example. Low cloud amount changes under future warming play a much larger role in the global cloud feedback so this has naturally been an area of focus \cite{ref:Qu14,ref:Qu15,ref:Brient16}. \textcite{ref:Qu14} identified an EC on this low cloud response through the low cloud response to temperature and stability anomalies derived from interannual variability. Although weaker in CMIP5 than CMIP3 \cite{ref:Klein&Hall15}, this EC has value in both ensembles and features a strong physical basis explaining the relationship \cite{ref:Qu14,ref:Qu15}. This type of process based evidence can then be used in combination to assemble a best estimate of the more complex global mean cloud feedback. Figure 7 of \textcite{ref:Sherwood20} illustrates this concept to get a global cloud feedback of 0.45 \rpm 0.33 W m$^{-2}$ K$^{-1}$. Going forward, it is imperative that we continue to improve our process-based understanding of individual cloud components, where potential physical mechanisms are more likely to be uncovered.

\subsection{Carbon Cycle}

Elsewhere, the EC technique has been used extensively to constrain elements of climate-carbon cycle feedbacks (Table I \#28-36) \citep{ref:Wenzel14, ref:Wenzel16, ref:Cox13, ref:Hoffman14, ref:Mystakidis2017}. Climate-carbon cycle feedbacks are characterised by changes in terrestrial and ocean carbon storage in response to climate change, which modify the atmospheric CO\textsubscript{2} concentration, thus enhancing warming. Intermodel differences in feedback strengths lead to highly variable projections of future atmospheric carbon dioxide (more than 300 ppmv of intermodel spread at the end of the 21st century for a high-emissions scenario \cite{ref:Cox19}). A rather simple relationship was uncovered between the mid-century atmospheric CO\textsubscript{2} concentration in ESMs and the simulated present-day concentration, but it was shown to weaken as the influence of varying climate-carbon feedbacks becomes more apparent later in the 21st century \cite{ref:Hoffman14}. This EC suggests a much narrower likely range for atmospheric CO\textsubscript{2} concentrations at the end of the century (947 \rpm 35 ppm) than the full ensemble (980 \rpm 161 ppm).

Much of the model uncertainty in future atmospheric CO\textsubscript{2} has been shown to stem from differences in the land carbon sink \cite{ref:Cox13, ref:Wenzel14}. Therefore, significant effort has gone into constraining this feature of the carbon cycle. A prominent example uses the historical sensitivity of annual atmospheric carbon dioxide growth-rate to temperature variability (which is strongly impacted by tropical land carbon storage fluctuations) to constrain the future loss of terrestrial tropical carbon \cite{ref:Cox13, ref:Wenzel14}. As with prior examples, the CO\textsubscript{2} growth rate sensitivity is an observable quantity, thus allowing for the EC to suggest future tropical land carbon stability. This EC has been proven robust across both C4MIP and CMIP5 ensembles \cite{ref:Wenzel14}\footnote{It should be noted that \textcite{ref:Wang14} also attempt to evaluate this EC in CMIP5, but their use of RCP8.5 to diagnose the sensitivity of tropical land carbon loss to tropical warming is not comparable to the work of \textcite{ref:Cox13,ref:Wenzel14}. Given that there is land-use change in RCP8.5, the authors conflate changes in tropical land carbon due to climate change, with changes in land carbon due to deforestation (for which there is no physical reason to expect a correlation with X).}, but has yet to undergo testing pertaining to CMIP6. This is an important EC because there was a fourfold simulated spread in tropical land carbon reductions per degree of warming in C\textsuperscript{4}MIP (29-133 GtC K$^{-1}$) \cite{ref:Jones16}, with the higher-end models suggesting potentially catastrophic dieback of the Amazon rainforest \cite{ref:Cox00,ref:Cox04}. However, the \textcite{ref:Cox13} EC suggests tropical land carbon reductions slightly weaker than the ensemble mean (53\rpm 17 GtC K$^{-1}$).

Another carbon cycle EC uses the change in seasonal atmospheric CO\textsubscript{2} amplitude to constrain CO\textsubscript{2} fertilisation of photosynthesis on the extratropics \cite{ref:Wenzel16}. On a more regional basis, \textcite{ref:Winkler2019} highlight a strong relationship between future increases in terrestrial Arctic gross primary productivity and historical increases in leaf area index (greening). These two ECs both suggest that most models are underestimating future changes in gross primary productivity across the high latitudes \cite{ref:Wenzel16, ref:Winkler2019}. While these ECs have yet to undergo out-of-sample verification, \textcite{ref:Winkler2019b} provides detailed evaluation of the factors contributing to uncertainty in the \cite{ref:Winkler2019} constraint. They point to choices relating to the temporal period of the predictor variable, choice of observational dataset, and the rate of CO2 forcing as particularly large sources of uncertainty \cite{ref:Winkler2019b}.  Lastly, additional research identified constraints on terrestrial carbon cycle feedbacks using interannual variability in evapotranspiration, net biome productivity, and gross primary productivity. This EC suggests a 40\% reduction in the climate-carbon feedback and a 30\% reduction in the concentration-carbon feedback \cite{ref:Mystakidis2017}.

When it comes to understanding how the ocean carbon cycle will respond to climate change, one of the main challenges is to reduce uncertainty in estimates of tropical ocean primary productivity. Satellite observations of the interannual variability in ocean productivity resulting from ENSO-driven SST anomalies have been used to constrain highly uncertain projected changes in tropical marine primary productivity with warming \cite{ref:Kwiatkowski17}. There is a strong inverse relationship between NPP and SST anomalies in the observational record, but the number of ENSO events in the satellite record is limited. This EC suggests a substantial reduction in the long-term tropical NPP sensitivity (-3.4\% K$^{-1}$ to -2.4 \% K$^{-1}$) from the unconstrained ensemble average (-4.0 \rpm 2.2 \% K$^{-1}$). Another recently published EC by \textcite{ref:Kessler2016} relates future global ocean carbon uptake to the contemporary carbon uptake by the Southern Ocean. In this example, models with anomalously low uptake in the current climate project low global uptake over the course of the 21st century. Uncertainty in simulated ocean uptake also translates to uncertainty in projections of future ocean acidification, the impacts of which are expected to be greatest in the Arctic ecosystem \cite{ref:Terhaar20}. Recent work uses CMIP5 models to identify an EC on future Arctic ocean acidification using the simulated density of Arctic ocean surface waters in the current climate \cite{ref:Terhaar20}. Observations of sea surface density when coupled with this emergent relationship imply that future Arctic Ocean acidification will be greater than previously expected.

\subsection{High-latitude Processes}

The Earth\textsc{\char13}s high-latitude regions are rapidly warming and there is a great deal of uncertainty in how various components of the cryosphere (Earth\textsc{\char13}s frozen surfaces) will evolve in the future \cite{ref:Mudryk2020,ref:Notz2020,ref:SROCC2019}. Thus, ECs are an intriguing option for reducing intermodel spread here. One of the earliest documented examples of an emergent constraint is for the snow albedo feedback \cite{ref:Hall&Qu06}, which was introduced in Section \ref{sec:intro}. This is a leading example of the EC technique because of the simplicity behind its physical mechanism and its robustness across several generations of models (Figure 2). An interesting point on the SAF EC is that despite substantial time since its initial publication, we have yet to see a meaningful reduction in its current climate quantity ($X$).

Following on from its application to SAF, recent work has identified that the emergent constraint approach can also be applied to a similar process over Arctic sea ice \cite{ref:Thackeray2019}. The sea ice albedo feedback (SIAF) is characterized by the enhancement of future warming through the reduction of surface albedo as a result of decreasing sea ice \cite{ref:Holland03,ref:Hall2004}. An analogue of this process operates in the seasonal cycle during the transition from maximum to minimum ice extent ($X$). These two processes are found to be closely linked across the CMIP5 ensemble, forming the basis of an EC. However, given the projected rapid loss of sea ice in the 21st century, this constraint exhibits a regime-dependence whereby the relationship between seasonal cycle and climate change SIAF begins to break down in the latter half of the 21st century, when most models exhibit an ice-free Arctic \cite{ref:Thackeray2019}. This is similar to the concept of tipping points discussed in Section \ref{sec:wrong:tipping}.

In addition to the aforementioned radiative feedbacks, there is a large amount of uncertainty when it comes to the fate of a number of high-latitude Earth system components, notably sea ice and permafrost \cite{ref:Stroeve2012,ref:Slater2013}. Given this uncertainty and the potential widespread implications associated with these changes, ECs can provide valuable information here. Notably, a number of studies have proposed constraints on projections of future Arctic sea ice \citep{ref:Boe09, ref:Massonnet12, ref:Liu13}. \textcite{ref:Boe09} used the observed historical trend in September Arctic sea ice extent over the satellite era to constrain the time in the 21st century when the Arctic is likely to become ice-free during summer. Similar research by \textcite{ref:Massonnet12} used a series of historical ice characteristics and trends in an attempt to constrain the time period when the Arctic is likely to become seasonally ice-free. Their results suggest ice-free conditions for September to begin somewhere between 2041-2068 under a high-emissions scenario, a significant reduction from the full CMIP5 spread that spans nearly 100 years. The presence of constraints between historical ice properties and the future change in both CMIP3 and CMIP5 is a promising sign, but we are not aware of CMIP6 testing for this EC thus far. Similar historical sea-ice metrics have also been used to inform future changes in high-latitude temperature variability \cite{ref:Borodina2017a}.
Permafrost underlies a significant portion of the Northern Hemisphere high-latitudes and is highly sensitive to warming \cite{ref:Lawrence2012}, with its degradation expected to have detrimental climate and developmental effects \cite{ref:Teufel19}. However, modern ESMs struggle to agree upon future changes in permafrost area \cite{ref:Slater2013}.  \textcite{ref:Chadburn17} use a relationship between mean annual air temperature and permafrost area during the historical period ($X$) to constrain projections of future permafrost thaw ($Y$). This constraint suggests that roughly 20\% more permafrost will be lost per degree of warming than previously expected \cite{ref:Chadburn17}.

\subsection{Hydrologic Cycle}

Future changes to precipitation are of great interest to the climate science community because of wide-ranging impacts on natural and human systems. However, precipitation is highly variable in space and time, thus making it difficult to both observe and predict. These factors contribute to large intermodel differences in future projections of precipitation. Reducing this uncertainty is therefore vital for those making policy and infrastructure decisions based on model projections. ECs have been applied to multiple aspects of the hydrologic cycle across various scales \citep{ref:DeAngelis15,ref:OGorman12,ref:Li17,ref:Rowell2019,ref:Watanabe18}. First, we consider the globally-averaged changes to the hydrologic cycle. In response to rising temperatures and increasing atmospheric water vapour, global-mean precipitation is projected to increase. This is because the atmosphere radiatively cools to space and this radiative cooling must be balanced by latent heat release from precipitation, thus setting the radiative-convective balance \cite{ref:Pendergrass14,ref:DeAngelis15}. Under a high-emission scenario, models project a future increase of between 2 to 10 \% in global-mean precipitation \cite{ref:Kharin2013}. This spread can be partially explained by differing rates of future warming \cite{ref:Flaschner2016}, although a threefold spread still exists when the precipitation change is normalised by warming \cite{ref:Kharin2013}. In an attempt to minimise this spread, \textcite{ref:DeAngelis15} used the sensitivity of clear-sky shortwave radiation absorption to changes in column water vapour to constrain the global mean precipitation increase with warming (termed hydrologic sensitivity). This constraint suggests a 35 \% reduction in the ensemble spread and a 40 \% reduction in the ensemble mean hydrologic sensitivity. This work was followed by \textcite{ref:Watanabe18}, who use the climatological global mean surface longwave cloud radiative effect (CRE) to constrain the future surface longwave CRE feedback. In combination with the most-likely sensitivity of clear-sky shortwave radiative absorption to water vapour \cite{ref:DeAngelis15}, this EC suggests a constrained estimate of 1.8 \% K$^{-1}$ for global hydrologic sensitivity, down from an ensemble mean of 2.6 \% K$^{-1}$ \cite{ref:Watanabe18}. However, a recently published assessment of these constraints in CMIP6 reveals that although the \textcite{ref:DeAngelis15} EC is still present, its strength is weakened \cite{ref:Pendergrass20}. Moreover, the \textcite{ref:Watanabe18} EC on HS is found to be non-existent in the new ensemble. This highlights the importance of applying out-of-sample testing to the remainder of the ECs on hydrologic cycle components discussed below.

Another key component of the global hydrologic cycle is terrestrial evapotranspiration (ET), which is made up of evaporation and biological transpiration (T). The ratio of T to ET is used to estimate land water fluxes, while because similar processes influence T and land-atmosphere carbon exchanges, better estimates of T/ET help to reduce uncertainty in global carbon cycle projections \cite{ref:Lian18}. Climate models exhibit large variability in the global strength of T/ET, spanning a factor of 3. \textcite{ref:Lian18} found a strong relationship ($r^2= 0.93$) between simulated local T/ET (averaged over grid-cells corresponding to in situ measurements) and global mean T/ET. This EC reveals that the constrained global T/ET ratio is significantly higher than the CMIP5 ensemble mean.

Alternatively, several studies have attempted to use this technique to constrain the change in aspects of the hydrologic cycle over specific regions. Regional constraints have the potential for more policy and impacts relevance than globally averaged metrics. As a first attempt, \textcite{ref:OGorman12} proposed an EC on the change in future daily tropical precipitation extremes using the observed sensitivity of tropical precipitation extremes to temperature variability. They found a strong relationship between the sensitivity of very heavy precipitation (99.9th percentile) to both seasonal and future temperature changes. The constrained estimate based on observations (6-14\% K$^{-1}$) was substantially narrower than the intermodel spread across the CMIP3 ensemble (2-23\% K$^{-1}$). Similarly, \textcite{ref:Borodina2017b} found that the scaling of annual maximum one day precipitation (Rx1day) with warming global land temperatures can be used to constrain the future intensification of heavy rainfall across extratropical regions with sufficient data records. This EC suggests that the CMIP5 models are likely underestimating the future change in extreme rainfall where climatological rainfall intensity is high.

Another example links biases in simulated historical climate and future Indian Summer Monsoon (ISM) rainfall \cite{ref:Li17}. In this case, the authors find that models overestimating historical precipitation across the tropical western Pacific also exhibit larger increases in Monsoon rains in the future \cite{ref:Li17}. Using observations of western Pacific rainfall reduces the projected increase in ISM rainfall by nearly 50\%. CMIP5 models also differ greatly in their projections of future total East African Long Rains (March-April-May mean spanning a region from northwest Tanzania to southwest Ethiopia) – from a 20\% decrease to a 120\% increase \cite{ref:Rowell2018}. \textcite{ref:Rowell2019} use an EC between present day interannual SST-low level cloud sensitivity and future SST change over the Indian Ocean to assess the credibility of projections of East African precipitation. They find that one outlier model, which projects a doubling of seasonal precipitation is likely unreliable because of unrealistic SST/low cloud processes. This reduction in intermodel spread is another example of the value that an EC approach can provide.  Along these lines, \textcite{ref:Lehner2019} recently suggested that the sensitivity of historical runoff to temperature and precipitation change across three watersheds in the Western US is closely tied to projections of future runoff across a series of ESMs. It is expected that regional applications of ECs to highly uncertain quantities like precipitation change will be a key area of future research.

\begin{widetext}
\begin{longtable}{p{0.03\textwidth}|p{0.35\textwidth}|p{0.35\textwidth}|p{0.08\textwidth}| p{0.14\textwidth}}
\caption[]{Collection of existing ECs. Note that some of these ECs involve correlations that are lower than those portrayed in Figure 1, with correspondingly less potential for uncertainty reduction. The ensemble(s) for which the EC appears to have value is also listed (note that many have only been tested on the ensemble that they were developed on).}\label{table:ECs}\\
\hline
   & Future constrained quantity (Y)                                                         & Current climate quantity (X)                                                             & Ensemble

   & Reference                                         \\
   \hline \hline
1  & Equilibrium Climate Sensitivity                                                         & Seasonal cycle of temperature                   & CMIP1 PPE                                                       & \textcite{ref:Knutti06,ref:Covey2000}                          \\
\hline
2  & Equilibrium Climate Sensitivity                                                         & Difference in cloud fraction between the tropics and SH midlatitudes                             & CMIP3 CMIP5 CMIP6    & \textcite{ref:Volodin08}                                \\
\hline
3  & Equilibrium Climate Sensitivity                                                         & TOA radiation balance in the SH                                                                 & CMIP3     & \textcite{ref:TrenberthFasullo10}                  \\
\hline
4 & Equilibrium Climate Sensitivity                                                         & Features of TOA radiation fluxes                                                                & CMIP3     & \textcite{ref:Huber11,ref:Tett13}         \\
\hline
5  & Equilibrium Climate Sensitivity                                                         & Variability in climatological May-August relative humidity and cloud extent                                          & CMIP3          & \textcite{ref:FasulloTrenberth12}                  \\
\hline
6 & Equilibrium Climate Sensitivity                                                         & Last Glacial Maximum Cooling                    & CMIP3/
PMIP2 CMIP5/
PMIP3                                                     & \textcite{ref:Hargreaves12, ref:Schmidt2014}  \\
\hline
7 & Equilibrium Climate Sensitivity                                                         & Vertical mixing strength between the boundary layer and lower troposphere over tropical oceans & CMIP3 CMIP5 CMIP6      & \textcite{ref:Sherwood14}                         \\
\hline
8  & Equilibrium Climate Sensitivity                                                         & Vertically resolved relative humidity and clouds between 45\degree S and 40\degree N             & CMIP5                      & \textcite{ref:Su14}                               \\
\hline
9 & Equilibrium Climate Sensitivity                                                         & Precipitation in the double-ITCZ region                                                         & CMIP3 CMIP5     & \textcite{ref:Tian15}                                   \\
\hline
10  & Equilibrium Climate Sensitivity                                                         & Seasonal sensitivity of low cloud to SSTs (20-40\degree latitude)
& CMIP3 CMIP5 CMIP6
& \textcite{ref:Zhai15}                            \\
\hline
11  & Equilibrium Climate Sensitivity                                                         & Height of tropical low clouds                   & CMIP5                                                       & \textcite{ref:Brient16a}                          \\
\hline
12  & Equilibrium Climate Sensitivity                                                         & Sensitivity of subtropical low cloud albedo to SSTs                                            & CMIP5 CMIP6       & \textcite{ref:Brient16}                   \\
\hline
13 & Equilibrium Climate Sensitivity                                                         & Climatological latitude of the SH Hadley Cell edge in winter                                  & CMIP5       & \textcite{ref:Lipat17}                            \\
\hline
14  & Equilibrium Climate Sensitivity                                                         & Statistics of interannual temperature variability                                     & CMIP5                 & \textcite{ref:Cox18}                             \\

\hline
15 & Equilibrium Climate Sensitivity                                                         & Cloudy-sky radiative flux sensitivity to temperature                                     & CMIP5            & Lutsko and Takahashi (2018)                     \\

\hline
16 & Future global warming (mid-century or end of century)                                   & Recent warming trend in CMIP models (1981-2017) & CMIP6                                                      & \textcite{ref:Tokarska2020}                         \\
\hline
17 & Transient climate response  and Equilibrium Climate Sensitivity                                & Recent warming in CMIP models                   & CMIP5 CMIP6                                                      & \textcite{ref:Jimenez-de-la-Cuesta2019,ref:Nijsse20}     \\

\hline
18 & Low cloud feedback sign                                                                 & Low cloud sensitivity to Pacific variability    & CMIP3                                                     & \textcite{ref:Clement09}                         \\
\hline
19 & Global mean cloud feedback and ECS                                                      & Climatological latitudinal gradient in the reflectivity of clouds                          & CMIP5          & \textcite{ref:Siler18}                            \\
\hline
20 & Low cloud optical depth change per degree climate warming                               & Low cloud optical depth response to temperature anomalies                                       & CMIP3 CMIP5     & \textcite{ref:Gordon14}                       \\
\hline
21 & Subtropical low cloud cover change under climate warming                                & Subtropical low cloud cover response to inter-annual temperature and stability anomalies & CMIP3 CMIP5            & \textcite{ref:Qu14,ref:Qu15}                               \\
\hline
22 & Global mean cloud feedback under climate change                                & Global mean cloud feedback derived from interannual variability in current climate              &
CMIP5 &
\textcite{ref:Zhou15}                               \\
\hline

23 & Change in upper troposphere relative humidity and cloud fraction
& Vertical gradient in climatological mean relative humidity and cloud fraction &
CMIP5 &
\textcite{ref:Po-Chedley19}
                \\
\hline
24 & Snow albedo feedback                                                                    & Seasonal snow albedo feedback (spring)          & CMIP3 CMIP5 CMIP6                                                       & Hall and Qu (2006); Qu and Hall (2014)       \\
\hline
25 & Sea ice albedo feedback                                                                 & Seasonal sea ice albedo feedback (summer)       & CMIP5                                                     & \textcite{ref:Thackeray2019}                     \\
\hline
26 & Timing of Arctic ice-free summer                                                        & Historical September sea ice trend              & CMIP3                                                       & \textcite{ref:Boe09}                             \\
\hline
27 & Timing of Arctic ice-free summer                                                        & Historical sea ice characteristics              & CMIP5                                                       & \textcite{ref:Massonnet12}                       \\
\hline
28 & Arctic thermal feedbacks                                                                & Different aspects of Northern Latitude regional temperatures                                    & CMIP3 CMIP5     & \textcite{ref:Bracegirdle13}             \\
\hline
29 & Sensitivity of tropical land carbon storage to warming                                  & Short-term sensitivity of CO2 to interannual temperature variability                         & C4MIP CMIP5        & \textcite{ref:Cox13,ref:Wenzel14}         \\
\hline
30 & CO2 fertilisation of photosynthesis on the extratropics                                 & Seasonal fluctuations in CO2 concentrations     & CMIP5                                                     & \textcite{ref:Wenzel16}                          \\
\hline
31 & Global land carbon feedback strength                                                    & Interannual variations in evapotranspiration, net biome productivity, and gross primary productivity & CMIP5
& \textcite{ref:Mystakidis2017}                       \\
\hline
32 & Change in tropical primary production to temperature anomalies                          & Sensitivity of tropical primary production to interannual SST anomalies                       & CMIP5       & \textcite{ref:Kwiatkowski17}                      \\
\hline
33 & Future permafrost thaw                                                                  & Relationship between mean annual air temperature and permafrost area                             & CMIP5    & \textcite{ref:Chadburn17}                         \\
\hline
34 & Future CO2 concentration                                                                & Simulated CO2 concentration by 2010             & CMIP5                                                     & \textcite{ref:Hoffman14}                          \\
\hline
35 & Change in Arctic Gross Primary Productivity                                             & Sensitivity of annual maximum leaf area index to increasing CO2                                  & CMIP5    & \textcite{ref:Winkler2019}                          \\
\hline
36 & Global Ocean Carbon Uptake                                                              & Present-day Southern Ocean carbon uptake & CMIP5                                                         & \textcite{ref:Kessler2016}                   \\
\hline
37 & Future Arctic Ocean acidification &
Present-day Arctic sea surface density &
CMIP5 &
\textcite{ref:Terhaar20}
\\
\hline

38 & Intensification of heavy rainfall across certain extratropical regions                  & Scaling of annual maximum daily precipitation with global land temperatures                   & CMIP5       & \textcite{ref:Borodina2017b}                         \\
\hline
39 & Change in tropical precipitation extremes under climate warming                         & Sensitivity of tropical precipitation extremes to temperature variability                      & CMIP3 CMIP5      & O{\ttfamily\char'15}Gorman (2012)                               \\
\hline
40 & Indian summer monsoon rainfall increase with warming                                    & Climatological mean precipitation in the Western Tropical Pacific                        & CMIP5            & \textcite{ref:Li17}                              \\
\hline
41 & Future SST change over the Indian Ocean with application for East African precipitation & Present-day interannual SST-low level cloud sensitivity                                     & CMIP5         & \textcite{ref:Rowell2019}                                 \\
\hline
42 & Future runoff projections over the Western US                                           & Historical runoff sensitivity to temperature and precipitation variations
& CMIP5
& \textcite{ref:Lehner2019}                           \\
\hline
43 & Clear-sky shortwave absorption (Global mean precipitation increase)                     & Sensitivity of shortwave radiative absorption to changes in column water vapor
& CMIP5 CMIP6
& \textcite{ref:DeAngelis15}                       \\
\hline
44 & Global hydrologic sensitivity                                                           & Variation between surface longwave cloud radiative effect and its sensitivity
& CMIP5
& \textcite{ref:Watanabe18}                        \\
\hline
45 & Global transpiration to evapotranspiration ratio                                        & Local transpiration to evapotranspiration ratio & CMIP5                                                     & Lian \emph{et al.} (2018)                            \\
\hline
46 & North Atlantic subpolar gyre cooling                                                    & Stratification of the subpolar North Atlantic Ocean                                           & CMIP5       & \textcite{ref:Sgubin17}                           \\
\hline
47 & Summer warming over the Central USA                                                     & Climatological summer temperature over the central USA
& CMIP5
& \textcite{ref:Lin17}                             \\
\hline
48 & Change in high-latitude temperature variability                                         & Historical sea-ice metrics                & CMIP5                                                          & \textcite{ref:Borodina2017a}                        \\
\hline
49 & Future continental warming over Europe                                                  & Present-day climatological summer temperatures over Europe                                     & CMIP5      & \textcite{ref:Selten2020}                           \\
\hline
50 & Frequency of heat extremes                                                              & Seasonal land-atmosphere feedbacks              & CMIP5                                                     & \textcite{ref:Donat18}                            \\
\hline
51 & Poleward jet shifts under warming                                                       & Climatological position of jet stream            & CMIP5                                                    & Simpson and Polvani (2016)                    \\
\hline
52 & Poleward shift of Southern Hemisphere eddy-driven jet stream with climate warming       & Climatological latitudinal position of Southern Hemisphere eddy-driven jet stream       & CMIP3             & Kidston and Gerber (2010)                     \\
\hline
53 & Anthopogenic ozone radiative forcing                                                    & Tropospheric ozone effect on outgoing longwave radiation &
ACCMIP & \textcite{ref:Bowman13}  \\
\hline
\end{longtable}
\end{widetext}



\section{Statistical underpinnings}\label{sec:stats}

Accurately constraining the unknown future value of $Y$ in the real world requires one to include and quantify all the possible sources of uncertainty in each step of the EC procedure. This section will give an overview of four types of uncertainty that have been incorporated into ECs so far: those stemming from uncertainty in the real world observation of $X$, uncertainty in $X$ from internal variability, uncertainty in the functional form of the emergent relationship and uncertainty from ESMs being imperfect replications of the real world. The section concludes with a discussion on how to combine those four types of uncertainty in the resulting in EC. In addition, we discuss how to combine multiple emergent constraints of the same quantity, derived from alternative features of historical climate.


\subsection{Uncertainty in observations}

There are multiple sources of observational error in the real world value of $X$. First, a lack of spatial and/or temporal coverage can be present and this may lead to biases if not taken into account \citep{Cowtan14}. There are two ways to handle missing data: it can be interpolated and extrapolated from existing data, or alternatively, model output can be filtered to reflect only locations and times for which observational data is present (e.g. \citet{ref:AchutaRao2006, ref:Durack2014, ref:Cox18}). Secondly, observational records are of finite length, introducing additional uncertainty from low sample size. Standard errors quantifying finite size effects can be computed, but care should be taken when timeseries are autocorrelated as this increases the standard error by effectively reducing the sample size \citep{ref:Trenberth1984}. For stationary processes, standard equations for autocorrelation errors can be found in \citet{Zhang06}.

Estimates of errors in instrumentation and data gathering are often available from literature e.g.  \citet{ref:ERA5documentation}. If multiple observational data sets are available, these can be used to infer uncertainty \cite{ref:TrenberthFasullo10, ref:Kwiatkowski17}. The errors in these data sets might not be independent; different satellite products might for instance have the same biases. Independent sources of error $\sigma_1$ and $\sigma_2$ can simply be computed by $\sigma_{total}^2 = \sqrt{\sigma_1^2 + \sigma_2^2}$.

If observational uncertainty makes up a large percentage of overall uncertainty, care should be taken to assess whether to use a normal distribution to describe the probability density function (pdf). It may sometimes be possible to estimate a full pdf from measurements. Alternatively, stochastic reduced-form modelling of the system can be used to estimate the shape \citep{ref:Nijsse18, ref:Williamson19}.

Whatever method is used, it is important that this step of capturing observational error is not neglected \citep{ref:Hall19}.

\subsection{Uncertainty from internal variability}

Like the real world, climate models have internal variability. Because of the finite length of the simulation or observed climate record, internal variability can have a significant impact on the estimation of the predictor.  One possibility is to use very long model control simulations to estimate the size of internal variability if it is believed to be independent of forcing \citep{ref:Nijsse19}. Variability may however be dependent on forcing, and consequently, estimating it from a forced initial value ensemble (see section \ref{sec:emerge:common_ensembles}) may be preferable \citep{ref:Tokarska2020}. For instance, global inter-annual variability is expected to decrease in the future \citep{ref:Huntingford2013}. \citet{ref:Jimenez-de-la-Cuesta2019} used the 100-member historical ensemble of MPI-ESM1.1 to quantify the effect of internal variability, whereas \citet{ref:Nijsse20} used all available historical initial value members from each CMIP6 model to estimate the mean model variability. Both used model estimates as a proxy for real internal variability.

\subsection{Uncertainty in the functional form of the relationship}
\begin{figure}[t]
    \centering
    \includegraphics[width=\columnwidth]{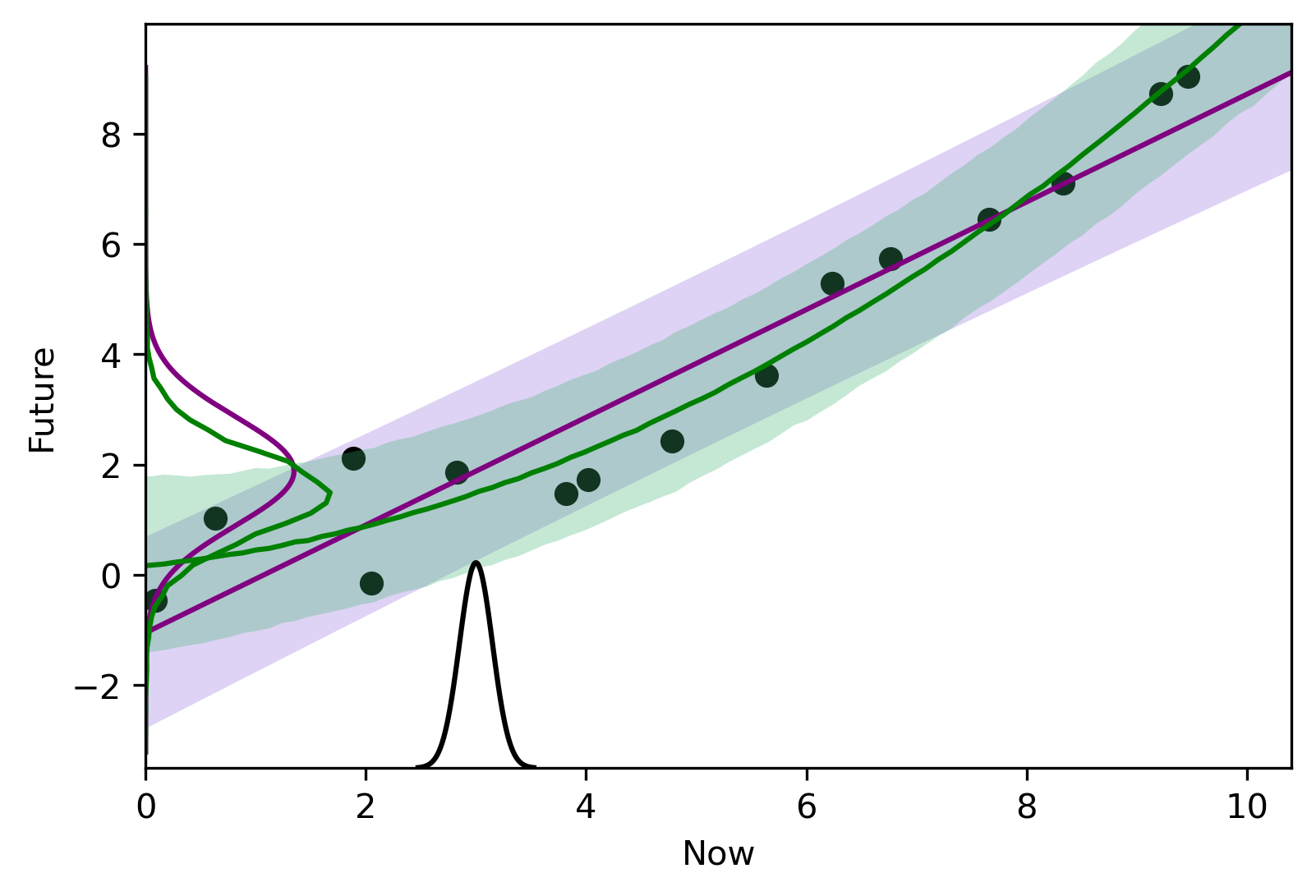}
    \caption{Using a quadratic relationship decreases the uncertainty in $Y$ compared to a linear fit \emph{provided} the observations line up with the shallow section of the quadratic function. Illustration with synthetic data.}
    \label{fig:quadratic_fit}
\end{figure}

Reducing a high-dimensional climate model to a lower dimension brings some uncertainty. Clearly, not all the variance will be explained with only two variables, and performing a regression is a tool to quantify this. While most emergent constraints so far assume linear relationships between $X$ and $Y$ and use linear regression to infer the emergent relationship, the regression does not necessarily have to be linear \citep{ref:Bracegirdle12, ref:Nijsse18}. If a linear relationship is imposed when, in reality, the relationship is nonlinear, additional errors will occur. A nonlinear emergent relationship leads mostly to a non-normal pdf for $Y$ with a standard deviation that is potentially significantly larger or smaller than for a linear fit (see Figure~\ref{fig:quadratic_fit}). Of course, more data or clear prior information on the parameters is needed when fitting additional parameters.

Regression dilution takes place as a consequence of errors in estimating model predictors (e.g. finite simulation length): when there is an error in the modelled explanatory variables, the slope of a linear regression fit will be smaller than without error and the intercept regresses towards the mean \citep{ref:Frost00}. Multiple strategies to reduce this can be employed: taking the mean of a set of initial value simulations \citep{ref:Jimenez-de-la-Cuesta2019}, using orthogonal distance regression, which takes into account both errors in the dependent and independent variable \citep{ref:Jimenez-de-la-Cuesta2019} or using a hierarchical Bayesian method that assume the `true' independent variable is unknown (latent) and using the realizations to infer this true value simultaneously while performing the regression \citep{ref:Nijsse20, ref:Sansom2014}.

Regression confidence intervals also depend on sample size. As climate models often have shared computer code, effective sample size is probably lower than the total number of models \citep{ref:Pennell11, ref:Masson&Knutti12, ref:Herger17}, effectively reducing significance \citep{ref:Knutti13}. This commonality can be partially addressed by selecting only one model per modelling centre \citep{ref:Cox18, Samson19arXiv}, where centres frequently offer multiple versions of alternative spatial resolutions or other small differences in model physics. However, even then effective sample size may be overestimated. As strongly related models in terms of code may have very different values for $X$ and $Y$, it is not always clear how important corrections are \cite{ref:Nijsse20}.

\subsection{Uncertainty from imperfect models}\label{sec:stats:model_UQ}
Almost paradoxically, the technique of emergent constraints \emph{relies} on errors in models. Hence some type of errors are useful, while others simply contribute to widening the confidence interval. Model error comes from parameteric and from structural errors (see also section \ref{sec:emerge:common_ensembles}). Parameter error is often examined by perturbed physics ensembles (PPEs), an ensemble where a single climate model is run repeatedly with systematically varied parameters designed to span a large range of model responses. Care should be taken to only include the physically adequate parameter choices, but not restrict the parameter space too much \citep{ref:Wagman&Jackson18}.

Usually, due to the expense of running climate models, an \emph{ad hoc} ensemble of different models is exploited to establish emergent constraints, the so called `ensemble of opportunity'. In this case, models should be chosen in such a way that they are still comparable: If half the models contain a nitrogen cycle, while the other half does not, they are unlikely to fit on a single regression \citep{ref:Wenzel14}. \textcite{ref:WilliamsonD19} argue that due to structural uncertainty, the derived regression between $X$ and $Y$ should not be regarded as the real relationship, but instead as only informing the real relationship.

\subsection{Combining sources of uncertainty in an EC}

\citet{ref:WilliamsonD19} argue on theoretical grounds that emergent constraints should be performed using a Bayesian framework, instead of the more common frequentists' framework using ordinary least square fitting (e.g. \textcite{ref:Wenzel14}), in which priors are set for the regression parameters but not for the predictand $Y$. \citet{ref:Bowman18} formulated a Bayesian hierarchical statistical framework that combines uncertainty in observations with uncertainty related to the regression itself. An explicit formula for the probability distribution can be approximated incorporating these uncertainties: assuming normality and performing a linear regression with intercept $a$, slope $b$ and confidence around the regression of $\sigma_f$, the probability density function for the predictand $p(y)$ is computed by integrating the conditional probability function $p(y|x)$ with the observational probability density $p(x)$. This formula for $p(y)$, rewritten in the notation of other statistical frameworks and under Gaussian assumptions, turns into

\begin{align}
p(y) &= \int_{-\infty}^{\infty} p(y|x)p(x) d x \nonumber \\
&= \mathcal{N}\left(y\big|a + b X_\text{obs}, \sqrt{\sigma_f^2+b^2\sigma_{O}^2}\right)
\label{eq:normal_regression}
\end{align}
Here $X$ is the predictor, $Y$ the predictand, and $X_\text{obs}$ is the best estimate of the observation. The observational uncertainty is denoted by $\sigma_O$.

To better estimate the regression and incorporate internal climate variability explicitly, a second hierarchical model was developed independently by \citet{ref:Sansom2014} and \citet{ref:Nijsse20}. Internal climate variability is incorporated both as part of the regression, and as an additional term in the uncertainty around observations, by using all initial value simulations $X_{m, j}$ of each model $m$ and each initial value member $j$. Here formulated using Gaussian distributions:

\begin{eqnarray}
  &&\qquad  X_{m,j}\sim \mathcal{N}(X_{m}, \sigma_I)\\
  && \qquad Y_m \sim \mathcal{N}(a + b X_m, \sigma_f)
\end{eqnarray}
The internal variability is denoted by $\sigma_I$ and is assumed to be independent of $Y$ and $X$ (and therefore model independent). The probability density function is sampled from estimates of the observed $X_\text{obs}$ and the parameters of the emergent relationship, giving a similar equation to Eq.~\ref{eq:normal_regression}.
\begin{eqnarray*}
  &&p(y) = \mathcal{N} \left(a+b \mathcal{N}\left(X_\text{obs},\sqrt{\sigma_I^2 + \sigma_{O}^2}\right),  \sigma_f\right)
\end{eqnarray*}

\citet{ref:Hall19} queries the extent to which we should trust in only partially confirmed ECs. This can be addressed formally, by inclusion of a researcher's assessment of how reliable a certain EC is. \citet{ref:WilliamsonD19} describe a method explicitly incorporating the trust that is put in any particular EC, including for instance by how much the EC changes in different ensembles. They introduce an additional parameter $\sigma_R$ that represents the uncertainty even after having taken all model information into account. This parameter is set subjectively, judged on the degree of physical trust the researcher has in the EC and features as $p(y) =  \mathcal{N}(\boldsymbol{\beta}x, \sigma_f^2 + \sigma_R^2)$, where $\boldsymbol{\beta}$ is a covariance matrix of the regression parameters, for which priors are also provided with information about possible biases or errors in the regression parameters.

\subsection{Combining multiple constraints}
A very simple method to combine multiple constraints was used in \citet{ref:Brient2020}. They use a Gaussian kernel density estimation of a histogram of the best values of previous constraints. To account for the variance, the posterior variance of the different ECs was included in a weighted Gaussian kernel density estimation. This method suffers from multiple drawbacks. Most prominently it cannot capture increasing confidence from having multiple independent constraints. Adding a new constraint with the same mean does not automatically lead to a narrower combined constraint, with the choice of kernel bandwidth remaining subjective. It further does not take into account to what extent the different emergent constraints are related to each other.

In \citet{ref:Bretherton2020} multiple emergent constraints were combined using a multivariate Gaussian PDF, which can be viewed as a form of multilinear regression. Their `method C' (for Correlated) includes information about correlations between different emergent relationships. Regularization was applied to deal with the strong collinearity between the emergent relationships. A second `method U' (for Uncorrelated) used a smaller subset of ECs, those regarded as confirmed constraints. As collinearity is less important with fewer ECs this method simplifies C by dropping the covariance ECs. Both variants were extended with a transparent method to account for overconfidence in the EC: they scale the ratio of the explained to unexplained variance with a factor $\alpha^2 \leq 1$, reducing all correlation coefficients.

\begin{figure}[t]
    \centering
    \includegraphics[width=\columnwidth]{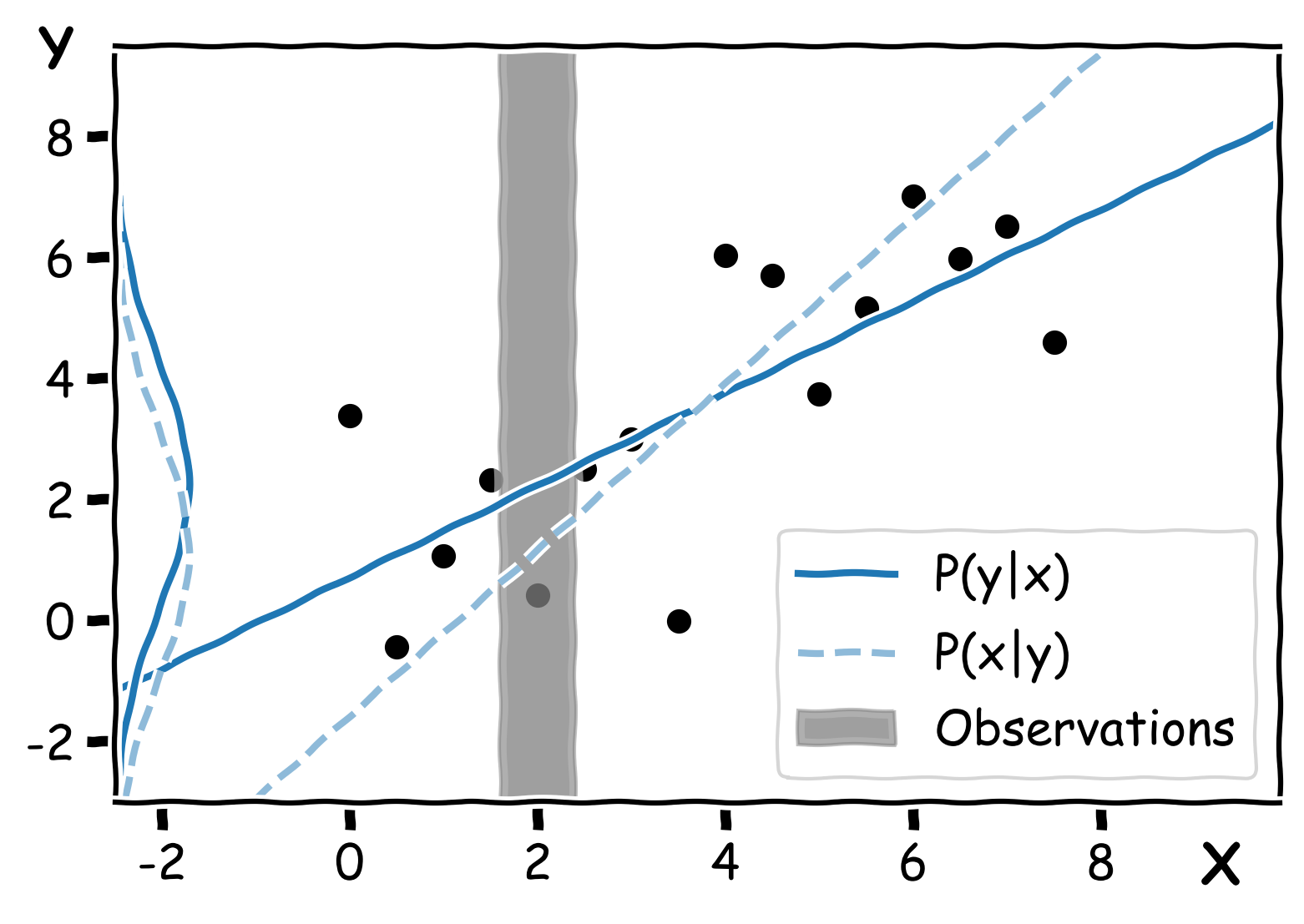}
    \caption{Standard, ordinary least squares linear regression $p(y|x)$ compared with the reverse regression $p(x|y)$. If the latter were to be used as the relationship on which the emergent constraint is founded, the final constraint has a bias.}
    \label{fig:regressions}
\end{figure}

In a discussion paper by \citet{ref:Renoult20}, a simple method was proposed to combine independent ECs to create a tighter estimate for $Y$. Where the regression is normally given as $p(y|x)$, they propose to instead formulate the statistical model as $p(x|y)$, allowing for a prior on $\pi(y)$ to be integrated into the emergent constraint as
\begin{equation}
p(y) = \int_{-\infty}^{\infty} p(x|y)\pi(y)p(x) d x .
\label{eq:combining_ECs}
\end{equation}
$p(y)$ here is the posterior distribution of a previous EC. To make sure the two ECs are indeed independent, the authors state that observations need to be independent, and that insofar possible, the errors in models should also stem from different sources. Their example involved a warm and cold climate state for which temperature change was reconstructed. Temperature change is dominated by different processes in this case, so that model error can be considered independent to first order. This method is not consistent with other methods described above. Linear regression is typically not symmetric; regression where $X$ predicts $Y$, $p(y|x)$, describes a different function than regression where $Y$ predicts $X$, $p(x|y)$ \citep{ref:Smith09}, as illustrated in Fig.~\ref{fig:regressions}.

\section{Outlook}\label{sec:outlook}

In this section we give some directions which we think are promising or exciting for future EC research.

\subsection{Key gaps in ECs to date}
The collection of ECs described in Section \ref{sec:ECs} highlight key areas of focus to date, but this also demonstrates where gaps exist, thus highlighting where the potential for EC growth is greatest going forward. For the most part, ECs have been traditionally focused on globally-aggregated quantities related to the mean state of climate (e.g., climate feedbacks, hydrologic sensitivity). However, recent applications to more regional features of the climate system are particularly encouraging (e.g. informing future monsoonal precipitation change over India; \textcite{ref:Li17}). Seeing as these regional applications have the potential for greater policy impact than most globally-averaged metrics we anticipate much more focus in this area going forwards. Climate extremes are also of great societal importance, but from Figure \ref{fig:schematic} we can see that very few studies have attempted to constrain the uncertainty in their future changes. We anticipate a greater focus on these and other higher order moments of climate statistics going forward. This advance is spurred on by an ever-lengthening satellite record that is now long enough to robustly sample extreme precipitation events. Similarly, improving paleoclimate records offer great potential for constraining emergent relationships pertaining to aspects of the climate system that vary on long timescales. In terms of their Earth system components, proposed constraints are heavily skewed towards atmospheric and biogeochemical fields, while constraints pertaining to oceanography and ice sheets are lacking. The latter is a rather recent addition to ESMs, but as observational records continue to improve there could be great potential for reducing uncertainty in critical metrics like sea level rise. Lastly, most ECs pertain to local relationships, but there are likely many undiscovered remote relationships in the climate system, where the current variability in one region is strongly tied to the future change in another region through atmospheric teleconnections \cite{ref:Rowell2019}. It is also believed that biases in the position of various climate features may be systematically tied to how the features respond to future forcing \cite{ref:Hall19}. Utilising these spatial characteristics of climate for ECs has begun \cite{ref:Kidston&Gerber10,ref:Simpson&Polvani16}, but better detection of these remote constraints likely requires improved analysis techniques e.g. \textcite{ref:Barnes19}.

\subsection{Targeted model development}\label{sec:outlook:spread_reduction}

An appealing yet underused aspect of ECs is targeted model development for bias reduction in a particular aspect of climate change ($Y$), although there are dangers to be aware of too (see section \ref{sec:wrong:spread_reduction}). If an EC has strong physical underpinnings, then we can expect that taking steps to reduce the spread in $X$ will result in a corresponding spread reduction in $Y$. First, we must assess in detail how the structural and parametric makeup of the ESMs influence the spread in $X$. Through this analysis the best parameterizations or parameters for simulating $X$ may be uncovered, thus providing guidelines for modelling centres on how they can efficiently reduce bias in $X$ going forward. This type of analysis is encouraged for all well-established ECs, but has only been completed for ECs on hydrologic cycle intensification \cite{ref:DeAngelis15}, the sensitivity of extratropical cloud reflectivity to temperature \cite{ref:Gordon14}, and snow albedo feedback \cite{ref:Thackeray18}. It is also possible that these targeted development activities will lead to a reduction in spread for related attributes of climate change affected by $Y$.

\subsection{Use of conceptual models as the basis of emergent relationships and understanding of more complex ESMs}

In the earlier years of ECs there may have been a tendency to data mine for variables with high correlations within ESM ensembles, potentially resulting in bad ECs (see section \ref{sec:emerge:data_mine}). However, data mining is not necessarily bad if it guides the search for independently testable mechanisms that may not have been obvious otherwise (section \ref{sec:wrong:data_mine}). 

One way of guarding against fortuitous correlations and hypothesizing new mechanisms is using conceptual, analytically soluble models as the basis of the emergent relationship. Building and solving conceptual models requires one to make assumptions about the real world. These assumptions are often clear, simple to understand and ideally, testable against real world and ESM ensemble data \cite{ref:Williamson18}. Provided the assumptions survive these tests, the solution to the conceptual model can form the basis of the emergent relationship. This testing of simplified theory not only aids understanding and intuition of what might (or might not) be going on in the complex ESMs but also adds a greater degree of rigour to ECs. There are potential dangers one should be aware of however; (i) $p$-hacking might occur (see section \ref{sec:wrong:phack}) and (ii) it might be that a process in the ESM one is trying to constraint is closer to the conceptual model in the spectrum of complexity than it is to reality. The use of conceptual models could therefore be used to justify the relationships arising from slightly less conceptual parameterizations which are implemented in the ESMs.

State-of-the-art ESM spatial resolution has continually increased and more physical and biogeochemical processes have been incorporated making these models ever more comprehensive representations of the Earth as time has gone on. They have become the main theoretical tool in climate science. However, due to their ever increasing complexity, their behaviour is difficult to understand and predict without running the models themselves. They are also frequently treated as one-stop shops and oracles for any question about contemporary and future climate. 

In the past, numerical simulation was more limited by computing power and researchers had to think carefully about which bits of the Earth system were important to include in their models to answer a particular question. This led to wider use of conceptual models specific to that question, analytical techniques and a better intuition due to the reduction of model complexity \cite{ref:Budyko69,ref:Hasselmann76,ref:Wigley&Raper90,ref:SaltzmanBook,ref:Dijkstra13}. 

Today, there is even more of an opportunity for the top down insights of specific conceptual models to meet and complement the comprehensive, bottom up approach of from state-of-the-art ESMs; there are many more high quality observations from satellites \cite{ref:Yang13}, ocean floats \cite{ref:Roemmich09}, and continuous temperature records; the global warming signal also has become clearer through the passage of time; there is also a large archive of past and present ESM simulations. With all the new data, it would be worth revisiting these simple, understandable models. We think a promising way forward to incorporate all the extra information from ESMs, conceptual models and observations is the EC approach.

\subsection{Multidimensional ECs and nonlinear emergent relationships}\label{sec:outlook:multiECs}

Much of the work on ECs has used just two scalar variables, a predictor $X$, and a response or predictand $Y$, related linearly. This could be extended to two or more predictors or predictands \cite{ref:Renoult20}. The predictand space to constrain will become larger however, and this might be difficult with small ESM ensembles. 

Most of the work to date has also assumed a linear emergent relationships, exceptions being the constraints on ECS using global mean temperature change in \textcite{ref:Jimenez-de-la-Cuesta2019,ref:Nijsse20}. This is also not necessary if theoretical reasoning suggests a different relation is more suitable. 

Equivalently to the use of multiple predictors/predictands, one could imagine using vectors for $X$ and $Y$, rather than scalars if theory suggested such a relationship. These vectors could for example encode spatial fields of a climate variable \cite{ref:Brown&Caldeira17}, spatial normal modes of variance, referred to a empirical orthogonal functions (EOFs) in meteorology \cite{ref:vonStorchBook} or the leading linear dynamical modes, also known as principal oscillation patterns (POPs) in meteorology \cite{ref:Hasselmann88,ref:Williamson&Lenton15}.


\subsection{Continued improvement of climate projections and impacts-led requirements}


One prediction we make is that the discovery of new ECs will be led more by those asking questions as to how climate impacts of concern may 
adjust in the future. Many altered meteorological features expected to have the most detrimental effect on societal safety, livelihoods and 
well-being are the magnitude and frequency of extreme weather events. A focus on extremes may result in ECs that are different in composition to 
those discovered to date. Rather than
a fluctuating and observable contemporary quantity ($X$) projecting a `bulk' climate property ($Y$) 
of importance to the future, instead a different format is required. That is, summary climate properties of large-scale features (or trends in them), 
as observable quantity $X$, require investigation to determine if they can estimate the future extreme frequency of quantity $Y$. 
In addition, a search for such ECs will focus attention on three further challenges. 
First, attributes of extremes and their changes might show substantial 
geographical variation. Second, by definition, extremes are rare, and so there are few data points available for investigation in ESM 
diagnostics where meteorological values are above thresholds that may be unsafe. Third, the future connection between observable $X$ 
and extreme feature $Y$ may involve a teleconnection between locations, and possibly with seasonal inertia.


\subsection{Better understanding of the effects of parametric and structural uncertainty on ECs}\label{sec:outlook:PPEs}

Presently, the most commonly used ESM ensembles in climate as well as EC research are multi-model ensembles (MMEs, see section \ref{sec:emerge:common_ensembles}) allowing structural uncertainty to be explored and in more recent MMEs such as the CMIP datasets, initial condition uncertainty. Parametric uncertainty is also captured to some degree although not in a systematic way unlike in PPEs. It is therefore unclear whether all plausible model parameteric configurations are explored. See also sections \ref{sec:wrong:PPEs} and \ref{sec:stats:model_UQ}. 

The strengths of MMEs and PPEs could be combined in a `super-ensemble' of PPEs generated from structurally different models. Apart from better assessment of confidence intervals, questions about the necessity of plausible ESM simulations for good ECs could be answered i.e. is it necessary to filter out unrealistic simulations to strengthen the EC, based on an observation unrelated to the predictor? Statistical emulators \cite{ref:Sacks89} of ESMs may help to give enough effective simulations after filtering unrealistic simulations from a PPE.

\subsection{Machine Learning}

Use of machine learning (ML) techniques in ECs may prove fruitful in future research. Machine Learning (ML) \cite{ref:WebbBook} is a wide definition but encompasses ECs, particularly supervised learning type algorithms. In a supervised ML approach, the algorithm builds a mathematical model from a set of training data that contains both inputs and desired outputs. In the case of ECs, this training data is from an ESM ensemble, the inputs being the observable $X$ and the outputs being the future projection $Y$. The mathematical model is $f$ i.e. $Y=f(X)+\epsilon$. The mathematical model that has been a popular choice in ECs so far has been $f(X)=a X + b$ where $a$ and $b$ are constants  fitted via linear regression. 

Phrasing linear regression as an example of supervised ML could be seen to be an over complicated generalization. However, viewing ECs within this framework potentially allows one to take advantage of the existing tools within ML. For example, the mathematical models fitted by the supervised ML algorithms could be more elaborate with more physical or statistically based parameters that give better explanatory power or more realism. The inputs and outputs equally could be more elaborate having extra spatial dimensions and/or extra variables (section \ref{sec:outlook:multiECs}). This could of course get as complicated as reproducing the original models that the inputs and outputs derive from. However, one ideally would like to walk the line between physical plausibility/understandability of the model and explanatory power to derive a minimal description. 

One also has to be aware of the number of data available to fit to a particular mathematical model. Presently, MMEs consist of around 30 ESMs limiting defensible fits of mathematical models to just a few parameters. However, one may be able to leverage the extra information in the ESMs spatial and temporal degrees of freedom to increase data size and therefore mathematical model complexity.

\subsection{Building connections to other fields}

Though our expertise is in the area of climate and Earth system dynamics, we have found a few examples of EC-like approaches in other fields. 

In the closely related field of weather prediction, an approach known as `Model Output Statistics' is used to enhance the quality and relevance of weather forecasts produced by numerical models (e.g. \textcite{ref:Veenhuis13}). As models are used again and again to make forecasts, an archive of past simulated data accumulates. This archive can be analysed to diagnose systematic statistical relationships between model output and observed quantities of interest. (An example of a quantity of interest might be temperature at a particular weather station.) When the models are then used to make real-time forecasts, these relationships can be applied to the forecast output, generating values for the quantities of interest. This is similar to the EC approach in that the structure of model biases is ascertained and exploited to produce future predictions that are likely more realistic. 

We find another example in the field of cosmology, where models have been created to simulate galaxy development. Like climate models, these models have tunable parameters.  When the model parameters are varied to produce an ensemble of simulations, statistical relationships among galaxy components across the ensemble can be analysed. In the study of \textcite{ref:Terrazas20}, their galaxy model produces strong statistical relationship between the black hole mass and the associated stellar mass (c.f. Figure 7 of that paper). Under certain conditions, observations of black hole and stellar mass mirror the simulated relationship. This suggests that given an observation of black hole mass for a galaxy without a corresponding stellar mass observation, the model could be used to make a meaningful prediction of stellar mass. This example is similar to ECs in the sense that the intrinsic model (and real world) physics connecting key quantities of interest can only be ascertained through analysis of a multi-model ensemble. 

In the field of economics, past forecasts of economic performance using multiple economic models have been analysed and compared to the actual ensuing economic performance. \textcite{ref:Felix18} found that the skewness of the ensemble forecast's distribution is a strong predictor of economic surprises. The implication is that if a current ensemble of economic models produces a skewed forecast distribution, the likelihood of an economic surprise is elevated. Such an outcome might not be detectable from more conventional predictions, such as the ensemble mean forecast. This example has some similarities to the Model Output Statistics example from weather forecasting: Model biases are ascertained from multiple realisations of past forecast performance, and used to improve the current forecast.

This brief survey of a few close cousins to the EC approach reveals that ECs and their variants are most likely to be useful where we are trying to understand complex systems where experimentation with the real system is not possible or is cumbersome. In such systems we often need to simulate dynamical evolution, initialising from a snapshot in time or at best a relatively short period of observations. Multiple models of the system are possible, and which model is the correct one is unknown. Emergent relationships across model variants are a way to reveal the deeper connections between observable elements of simulations and model outcomes of interest. There are likely many more examples of EC-like research being done on complex systems. These disparate communities appear to be unaware of one another, and could benefit from organised efforts to convene in workshops and meetings.

\section{Conclusion}\label{sec:conclusion}








The spread in climate change projections has not reduced substantially despite major advances in model resolution and process understanding and inclusion over recent decades. The lack of progress represents a disappointment for climate science, and hinders society's ability to plan for future impacts. These challenges cannot be overcome by just increasing the resolution of ESMs and including more and more physical and biogeochemical processes. A balanced interplay of observations as well as top down, conceptual and bottom up, comprehensive modelling and theory is required to achieve the required progress and scientific understanding. In this review, we have argued that the EC approach offers a promising way to incorporate all of these elements and ultimately reduce key uncertainties in future climate. 

The development of ECs has arisen from the requirement to reduce often substantial inter-ESM differences in projections of climate. Large model differences makes adaption planning difficult and risks spending funds set aside to help societies cope with climate change. For those tasked with formulating mitigation plans that will reduce greenhouse gas emissions in order to contain global warming, uncertainty prevents the formation of accurate trajectories of `allowable' fossil fuel burning. In this review paper, we have shown existing applications of ECs, discussed theoretical issues surrounding them that likely require further investigation, and characterised a few circumstances where they may start to fail. Nevertheless, although this seemingly points to more research required, ECs are a key methodology to distil the often conflicting information from climate modelling centres across the world. For this reason, we anticipate that the application of ECs will continue to grow, and results from their application will have an increasing role in projecting future change, and with strong representation in future UN reports by the IPCC.

This review has looked in depth at how ECs have become a standard methodology, used by climate researchers to synthesize substantial inter-ESM prediction differences, into projections of change with lower levels of uncertainty. We make a critical assessment of the EC technique, and many of the open questions raised here may lead to exciting avenues of research over the years ahead. It may also be true that other research disciplines will find a role for the EC approach. However, it is always worth recalling that the main scientific issue remains pressing, as without substantial reductions in emissions, the climate is expected to change substantially and to a dangerous state for large fractions of society. For this reason, intense scrutiny of the EC method is highly welcome and appropriate. The relentless need for climate predictions is also likely to trigger the discovery of new ECs, caused instead by impacts led requirement for information taking precedence over more curiosity driven investigation.

\begin{acknowledgments}
This paper benefited from discussions at the Aspen Global Change Institute (AGCI) which hosted a workshop on Earth System Model Evaluation to Improve Process Understanding in August 2017 attended by M.S.W., P.M.C. and A.H. as part of its traditional landmark summer interdisciplinary sessions (\url{http://www.agci.org/event/17s2}). This work was supported by the European Research Council (ERC) ECCLES project, grant agreement number 742472 (M.S.W., P.M.C. and F.J.M.M.N.); the EU Horizon 2020 Research Programme CRESCENDO project, grant agreement number 641816 (M.S.W. and P.M.C.); the National Science Foundation grant (no. 1543268) titled 'Reducing Uncertainty Surrounding Climate Change Using Emergent Constraints' (C.W.T.); the Regional and Global Model Analysis Program for the Office of Science of the U.S. Department of Energy (A.H.); and the NERC CEH National Capability fund (C.H.). We also acknowledge the World Climate Research Programme’s Working Group on Coupled Modelling, which is responsible for CMIP, and we thank the climate modelling groups for producing and making available their model output.
\end{acknowledgments}

\bibliography{masterbib_RMP}

\begin{thebibliography}{205}%
\makeatletter
\providecommand \@ifxundefined [1]{%
 \@ifx{#1\undefined}
}%
\providecommand \@ifnum [1]{%
 \ifnum #1\expandafter \@firstoftwo
 \else \expandafter \@secondoftwo
 \fi
}%
\providecommand \@ifx [1]{%
 \ifx #1\expandafter \@firstoftwo
 \else \expandafter \@secondoftwo
 \fi
}%
\providecommand \natexlab [1]{#1}%
\providecommand \enquote  [1]{``#1''}%
\providecommand \bibnamefont  [1]{#1}%
\providecommand \bibfnamefont [1]{#1}%
\providecommand \citenamefont [1]{#1}%
\providecommand \href@noop [0]{\@secondoftwo}%
\providecommand \href [0]{\begingroup \@sanitize@url \@href}%
\providecommand \@href[1]{\@@startlink{#1}\@@href}%
\providecommand \@@href[1]{\endgroup#1\@@endlink}%
\providecommand \@sanitize@url [0]{\catcode `\\12\catcode `\$12\catcode
  `\&12\catcode `\#12\catcode `\^12\catcode `\_12\catcode `\%12\relax}%
\providecommand \@@startlink[1]{}%
\providecommand \@@endlink[0]{}%
\providecommand \url  [0]{\begingroup\@sanitize@url \@url }%
\providecommand \@url [1]{\endgroup\@href {#1}{\urlprefix }}%
\providecommand \urlprefix  [0]{URL }%
\providecommand \Eprint [0]{\href }%
\providecommand \doibase [0]{http://dx.doi.org/}%
\providecommand \selectlanguage [0]{\@gobble}%
\providecommand \bibinfo  [0]{\@secondoftwo}%
\providecommand \bibfield  [0]{\@secondoftwo}%
\providecommand \translation [1]{[#1]}%
\providecommand \BibitemOpen [0]{}%
\providecommand \bibitemStop [0]{}%
\providecommand \bibitemNoStop [0]{.\EOS\space}%
\providecommand \EOS [0]{\spacefactor3000\relax}%
\providecommand \BibitemShut  [1]{\csname bibitem#1\endcsname}%
\let\auto@bib@innerbib\@empty
\bibitem [{\citenamefont {AchutaRao}\ \emph {et~al.}(2006)\citenamefont
  {AchutaRao}, \citenamefont {Santer}, \citenamefont {Gleckler}, \citenamefont
  {Taylor}, \citenamefont {Pierce}, \citenamefont {Barnett},\ and\
  \citenamefont {Wigley}}]{ref:AchutaRao2006}%
  \BibitemOpen
  \bibfield  {author} {\bibinfo {author} {\bibnamefont {AchutaRao},
  \bibfnamefont {K~M}}, \bibinfo {author} {\bibfnamefont {B.~D.}\ \bibnamefont
  {Santer}}, \bibinfo {author} {\bibfnamefont {P.~J.}\ \bibnamefont
  {Gleckler}}, \bibinfo {author} {\bibfnamefont {K.~E.}\ \bibnamefont
  {Taylor}}, \bibinfo {author} {\bibfnamefont {D.~W.}\ \bibnamefont {Pierce}},
  \bibinfo {author} {\bibfnamefont {T.~P.}\ \bibnamefont {Barnett}}, \ and\
  \bibinfo {author} {\bibfnamefont {T.~M.~L.}\ \bibnamefont {Wigley}}}
  (\bibinfo {year} {2006}),\ \bibfield  {title} {\enquote {\bibinfo {title}
  {Variability of ocean heat uptake: Reconciling observations and models},}\
  }\href {\doibase 10.1029/2005JC003136} {\bibfield  {journal} {\bibinfo
  {journal} {Journal of Geophysical Research: Oceans}\ }\textbf {\bibinfo
  {volume} {111}}~(\bibinfo {number} {C5}),\ 10.1029/2005JC003136},\ \Eprint
  {http://arxiv.org/abs/https://agupubs.onlinelibrary.wiley.com/doi/pdf/10.1029/2005JC003136}
  {https://agupubs.onlinelibrary.wiley.com/doi/pdf/10.1029/2005JC003136}
  \BibitemShut {NoStop}%
\bibitem [{\citenamefont {Annan}\ and\ \citenamefont
  {Hargreaves}(2006)}]{ref:Annan&Hargreaves06}%
  \BibitemOpen
  \bibfield  {author} {\bibinfo {author} {\bibnamefont {Annan}, \bibfnamefont
  {J~D}}, \ and\ \bibinfo {author} {\bibfnamefont {J.~C.}\ \bibnamefont
  {Hargreaves}}} (\bibinfo {year} {2006}),\ \bibfield  {title} {\enquote
  {\bibinfo {title} {Using multiple observationally-based constraints to
  estimate climate sensitivity},}\ }\href {\doibase 10.1029/2005GL025259}
  {\bibfield  {journal} {\bibinfo  {journal} {Geophysical Research Letters}\
  }\textbf {\bibinfo {volume} {33}}~(\bibinfo {number} {6}),\
  10.1029/2005GL025259}\BibitemShut {NoStop}%
\bibitem [{\citenamefont {Barnes}\ \emph {et~al.}(2019)\citenamefont {Barnes},
  \citenamefont {Hurrell},\ and\ \citenamefont {Uphoff}}]{ref:Barnes19}%
  \BibitemOpen
  \bibfield  {author} {\bibinfo {author} {\bibnamefont {Barnes}, \bibfnamefont
  {Elizabeth~A}}, \bibinfo {author} {\bibfnamefont {James~W}\ \bibnamefont
  {Hurrell}}, \ and\ \bibinfo {author} {\bibfnamefont {Imme~Ebert}\
  \bibnamefont {Uphoff}}} (\bibinfo {year} {2019}),\ \bibfield  {title}
  {\enquote {\bibinfo {title} {{Viewing Forced Climate Patterns Through an AI
  Lens Geophysical Research Letters}},}\ }\href {\doibase 10.1029/2019GL084944}
  {\bibfield  {journal} {\bibinfo  {journal} {Geophysical Research Letters}\
  }10.1029/2019GL084944}\BibitemShut {NoStop}%
\bibitem [{\citenamefont {Bell}(1980)}]{ref:Bell80}%
  \BibitemOpen
  \bibfield  {author} {\bibinfo {author} {\bibnamefont {Bell}, \bibfnamefont
  {Thomas~L}}} (\bibinfo {year} {1980}),\ \bibfield  {title} {\enquote
  {\bibinfo {title} {Climate sensitivity from fluctuation dissipation: Some
  simple model tests},}\ }\href {\doibase
  10.1175/1520-0469(1980)037<1700:CSFFDS>2.0.CO;2} {\bibfield  {journal}
  {\bibinfo  {journal} {Journal of the Atmospheric Sciences}\ }\textbf
  {\bibinfo {volume} {37}}~(\bibinfo {number} {8}),\ \bibinfo {pages}
  {1700--1707}}\BibitemShut {NoStop}%
\bibitem [{\citenamefont {Boe}\ \emph {et~al.}(2009)\citenamefont {Boe},
  \citenamefont {Hall},\ and\ \citenamefont {Qu}}]{ref:Boe09}%
  \BibitemOpen
  \bibfield  {author} {\bibinfo {author} {\bibnamefont {Boe}, \bibfnamefont
  {Julien}}, \bibinfo {author} {\bibfnamefont {Alex}\ \bibnamefont {Hall}}, \
  and\ \bibinfo {author} {\bibfnamefont {Xin}\ \bibnamefont {Qu}}} (\bibinfo
  {year} {2009}),\ \bibfield  {title} {\enquote {\bibinfo {title} {September
  sea-ice cover in the arctic ocean projected to vanish by 2100},}\ }\href
  {http://dx.doi.org/10.1038/ngeo467} {\bibfield  {journal} {\bibinfo
  {journal} {Nature Geosci}\ }\textbf {\bibinfo {volume} {2}}~(\bibinfo
  {number} {5}),\ \bibinfo {pages} {341--343}}\BibitemShut {NoStop}%
\bibitem [{\citenamefont {Bond}\ \emph {et~al.}(1992)\citenamefont {Bond},
  \citenamefont {Heinrich}, \citenamefont {Broecker}, \citenamefont {Labeyrie},
  \citenamefont {McManus}, \citenamefont {Andrews}, \citenamefont {Huon},
  \citenamefont {Jantschik}, \citenamefont {Clasen}, \citenamefont {Simet},
  \citenamefont {Tedesco}, \citenamefont {Klas}, \citenamefont {Bonani},\ and\
  \citenamefont {Ivy}}]{ref:Bond92}%
  \BibitemOpen
  \bibfield  {author} {\bibinfo {author} {\bibnamefont {Bond}, \bibfnamefont
  {Gerard}}, \bibinfo {author} {\bibfnamefont {Hartmut}\ \bibnamefont
  {Heinrich}}, \bibinfo {author} {\bibfnamefont {Wallace}\ \bibnamefont
  {Broecker}}, \bibinfo {author} {\bibfnamefont {Laurent}\ \bibnamefont
  {Labeyrie}}, \bibinfo {author} {\bibfnamefont {Jerry}\ \bibnamefont
  {McManus}}, \bibinfo {author} {\bibfnamefont {John}\ \bibnamefont {Andrews}},
  \bibinfo {author} {\bibfnamefont {Sylvain}\ \bibnamefont {Huon}}, \bibinfo
  {author} {\bibfnamefont {Ruediger}\ \bibnamefont {Jantschik}}, \bibinfo
  {author} {\bibfnamefont {Silke}\ \bibnamefont {Clasen}}, \bibinfo {author}
  {\bibfnamefont {Christine}\ \bibnamefont {Simet}}, \bibinfo {author}
  {\bibfnamefont {Kathy}\ \bibnamefont {Tedesco}}, \bibinfo {author}
  {\bibfnamefont {Mieczyslawa}\ \bibnamefont {Klas}}, \bibinfo {author}
  {\bibfnamefont {Georges}\ \bibnamefont {Bonani}}, \ and\ \bibinfo {author}
  {\bibfnamefont {Susan}\ \bibnamefont {Ivy}}} (\bibinfo {year} {1992}),\
  \bibfield  {title} {\enquote {\bibinfo {title} {Evidence for massive
  discharges of icebergs into the north atlantic ocean during the last glacial
  period},}\ }\href {\doibase 10.1038/360245a0} {\bibfield  {journal} {\bibinfo
   {journal} {Nature}\ }\textbf {\bibinfo {volume} {360}}~(\bibinfo {number}
  {6401}),\ \bibinfo {pages} {245--249}}\BibitemShut {NoStop}%
\bibitem [{\citenamefont {Bony}\ \emph {et~al.}(2006)\citenamefont {Bony},
  \citenamefont {Colman}, \citenamefont {Kattsov}, \citenamefont {Allan},
  \citenamefont {Bretherton}, \citenamefont {Dufresne}, \citenamefont {Hall},
  \citenamefont {Hallegatte}, \citenamefont {Holland}, \citenamefont {Ingram},
  \citenamefont {Randall}, \citenamefont {Soden}, \citenamefont {Tselioudis},\
  and\ \citenamefont {Webb}}]{ref:Bony2006}%
  \BibitemOpen
  \bibfield  {author} {\bibinfo {author} {\bibnamefont {Bony}, \bibfnamefont
  {Sandrine}}, \bibinfo {author} {\bibfnamefont {Robert}\ \bibnamefont
  {Colman}}, \bibinfo {author} {\bibfnamefont {Vladimir~M.}\ \bibnamefont
  {Kattsov}}, \bibinfo {author} {\bibfnamefont {Richard~P.}\ \bibnamefont
  {Allan}}, \bibinfo {author} {\bibfnamefont {Christopher~S.}\ \bibnamefont
  {Bretherton}}, \bibinfo {author} {\bibfnamefont {Jean~Louis}\ \bibnamefont
  {Dufresne}}, \bibinfo {author} {\bibfnamefont {Alex}\ \bibnamefont {Hall}},
  \bibinfo {author} {\bibfnamefont {Stephane}\ \bibnamefont {Hallegatte}},
  \bibinfo {author} {\bibfnamefont {Marika~M.}\ \bibnamefont {Holland}},
  \bibinfo {author} {\bibfnamefont {William}\ \bibnamefont {Ingram}}, \bibinfo
  {author} {\bibfnamefont {David~a.}\ \bibnamefont {Randall}}, \bibinfo
  {author} {\bibfnamefont {Brian~J.}\ \bibnamefont {Soden}}, \bibinfo {author}
  {\bibfnamefont {George}\ \bibnamefont {Tselioudis}}, \ and\ \bibinfo {author}
  {\bibfnamefont {Mark~J.}\ \bibnamefont {Webb}}} (\bibinfo {year} {2006}),\
  \bibfield  {title} {\enquote {\bibinfo {title} {{How well do we understand
  and evaluate climate change feedback processes?}}}\ }\href {\doibase
  10.1175/JCLI3819.1} {\bibfield  {journal} {\bibinfo  {journal} {Journal of
  Climate}\ }\textbf {\bibinfo {volume} {19}},\ \bibinfo {pages}
  {3445--3482}}\BibitemShut {NoStop}%
\bibitem [{\citenamefont {Borodina}\ \emph
  {et~al.}(2017{\natexlab{a}})\citenamefont {Borodina}, \citenamefont
  {Fischer},\ and\ \citenamefont {Knutti}}]{ref:Borodina2017a}%
  \BibitemOpen
  \bibfield  {author} {\bibinfo {author} {\bibnamefont {Borodina},
  \bibfnamefont {Aleksandra}}, \bibinfo {author} {\bibfnamefont {Erich~M.}\
  \bibnamefont {Fischer}}, \ and\ \bibinfo {author} {\bibfnamefont {Reto}\
  \bibnamefont {Knutti}}} (\bibinfo {year} {2017}{\natexlab{a}}),\ \bibfield
  {title} {\enquote {\bibinfo {title} {{Emergent constraints in climate
  projections: A case study of changes in high-latitude temperature
  variability}},}\ }\href {\doibase 10.1175/JCLI-D-16-0662.1} {\bibfield
  {journal} {\bibinfo  {journal} {Journal of Climate}\ }\textbf {\bibinfo
  {volume} {30}}~(\bibinfo {number} {10}),\ \bibinfo {pages}
  {3655--3670}}\BibitemShut {NoStop}%
\bibitem [{\citenamefont {Borodina}\ \emph
  {et~al.}(2017{\natexlab{b}})\citenamefont {Borodina}, \citenamefont
  {Fischer},\ and\ \citenamefont {Knutti}}]{ref:Borodina2017b}%
  \BibitemOpen
  \bibfield  {author} {\bibinfo {author} {\bibnamefont {Borodina},
  \bibfnamefont {Aleksandra}}, \bibinfo {author} {\bibfnamefont {Erich~M.}\
  \bibnamefont {Fischer}}, \ and\ \bibinfo {author} {\bibfnamefont {Reto}\
  \bibnamefont {Knutti}}} (\bibinfo {year} {2017}{\natexlab{b}}),\ \bibfield
  {title} {\enquote {\bibinfo {title} {{Models are likely to underestimate
  increase in heavy rainfall in the extratropical regions with high rainfall
  intensity}},}\ }\href {\doibase 10.1002/2017GL074530} {\bibfield  {journal}
  {\bibinfo  {journal} {Geophysical Research Letters}\ }\textbf {\bibinfo
  {volume} {44}}~(\bibinfo {number} {14}),\ \bibinfo {pages}
  {7401--7409}}\BibitemShut {NoStop}%
\bibitem [{\citenamefont {Bowman}\ \emph {et~al.}(2013)\citenamefont {Bowman},
  \citenamefont {Shindell}, \citenamefont {Worden}, \citenamefont {Lamarque},
  \citenamefont {Young}, \citenamefont {Stevenson}, \citenamefont {Qu},
  \citenamefont {de~la Torre}, \citenamefont {Bergmann}, \citenamefont
  {Cameron-Smith}, \citenamefont {Collins}, \citenamefont {Doherty},
  \citenamefont {Dalsoren}, \citenamefont {Faluvegi}, \citenamefont {Folberth},
  \citenamefont {Horowitz}, \citenamefont {Josse}, \citenamefont {Lee},
  \citenamefont {MacKenzie}, \citenamefont {Myhre}, \citenamefont {Nagashima},
  \citenamefont {Naik}, \citenamefont {Plummer}, \citenamefont {Rumbold},
  \citenamefont {Skeie}, \citenamefont {Strode}, \citenamefont {Sudo},
  \citenamefont {Szopa}, \citenamefont {Voulgarakis}, \citenamefont {Zeng},
  \citenamefont {Kulawik}, \citenamefont {Aghedo},\ and\ \citenamefont
  {Worden}}]{ref:Bowman13}%
  \BibitemOpen
  \bibfield  {author} {\bibinfo {author} {\bibnamefont {Bowman}, \bibfnamefont
  {K~W}}, \bibinfo {author} {\bibfnamefont {D.~T.}\ \bibnamefont {Shindell}},
  \bibinfo {author} {\bibfnamefont {H.~M.}\ \bibnamefont {Worden}}, \bibinfo
  {author} {\bibfnamefont {J.~F.}\ \bibnamefont {Lamarque}}, \bibinfo {author}
  {\bibfnamefont {P.~J.}\ \bibnamefont {Young}}, \bibinfo {author}
  {\bibfnamefont {D.~S.}\ \bibnamefont {Stevenson}}, \bibinfo {author}
  {\bibfnamefont {Z.}~\bibnamefont {Qu}}, \bibinfo {author} {\bibfnamefont
  {M.}~\bibnamefont {de~la Torre}}, \bibinfo {author} {\bibfnamefont
  {D.}~\bibnamefont {Bergmann}}, \bibinfo {author} {\bibfnamefont {P.~J.}\
  \bibnamefont {Cameron-Smith}}, \bibinfo {author} {\bibfnamefont {W.~J.}\
  \bibnamefont {Collins}}, \bibinfo {author} {\bibfnamefont {R.}~\bibnamefont
  {Doherty}}, \bibinfo {author} {\bibfnamefont {S.~B.}\ \bibnamefont
  {Dalsoren}}, \bibinfo {author} {\bibfnamefont {G.}~\bibnamefont {Faluvegi}},
  \bibinfo {author} {\bibfnamefont {G.}~\bibnamefont {Folberth}}, \bibinfo
  {author} {\bibfnamefont {L.~W.}\ \bibnamefont {Horowitz}}, \bibinfo {author}
  {\bibfnamefont {B.~M.}\ \bibnamefont {Josse}}, \bibinfo {author}
  {\bibfnamefont {Y.~H.}\ \bibnamefont {Lee}}, \bibinfo {author} {\bibfnamefont
  {I.~A.}\ \bibnamefont {MacKenzie}}, \bibinfo {author} {\bibfnamefont
  {G.}~\bibnamefont {Myhre}}, \bibinfo {author} {\bibfnamefont
  {T.}~\bibnamefont {Nagashima}}, \bibinfo {author} {\bibfnamefont
  {V.}~\bibnamefont {Naik}}, \bibinfo {author} {\bibfnamefont {D.~A.}\
  \bibnamefont {Plummer}}, \bibinfo {author} {\bibfnamefont {S.~T.}\
  \bibnamefont {Rumbold}}, \bibinfo {author} {\bibfnamefont {R.~B.}\
  \bibnamefont {Skeie}}, \bibinfo {author} {\bibfnamefont {S.~A.}\ \bibnamefont
  {Strode}}, \bibinfo {author} {\bibfnamefont {K.}~\bibnamefont {Sudo}},
  \bibinfo {author} {\bibfnamefont {S.}~\bibnamefont {Szopa}}, \bibinfo
  {author} {\bibfnamefont {A.}~\bibnamefont {Voulgarakis}}, \bibinfo {author}
  {\bibfnamefont {G.}~\bibnamefont {Zeng}}, \bibinfo {author} {\bibfnamefont
  {S.~S.}\ \bibnamefont {Kulawik}}, \bibinfo {author} {\bibfnamefont {A.~M.}\
  \bibnamefont {Aghedo}}, \ and\ \bibinfo {author} {\bibfnamefont {J.~R.}\
  \bibnamefont {Worden}}} (\bibinfo {year} {2013}),\ \bibfield  {title}
  {\enquote {\bibinfo {title} {Evaluation of accmip outgoing longwave radiation
  from tropospheric ozone using tes satellite observations},}\ }\href {\doibase
  10.5194/acp-13-4057-2013} {\bibfield  {journal} {\bibinfo  {journal}
  {Atmospheric Chemistry and Physics}\ }\textbf {\bibinfo {volume}
  {13}}~(\bibinfo {number} {8}),\ \bibinfo {pages} {4057--4072}}\BibitemShut
  {NoStop}%
\bibitem [{\citenamefont {Bowman}\ \emph {et~al.}(2018)\citenamefont {Bowman},
  \citenamefont {Cressie}, \citenamefont {Qu},\ and\ \citenamefont
  {Hall}}]{ref:Bowman18}%
  \BibitemOpen
  \bibfield  {author} {\bibinfo {author} {\bibnamefont {Bowman}, \bibfnamefont
  {Kevin~W}}, \bibinfo {author} {\bibfnamefont {Noel}\ \bibnamefont {Cressie}},
  \bibinfo {author} {\bibfnamefont {Xin}\ \bibnamefont {Qu}}, \ and\ \bibinfo
  {author} {\bibfnamefont {Alex}\ \bibnamefont {Hall}}} (\bibinfo {year}
  {2018}),\ \bibfield  {title} {\enquote {\bibinfo {title} {A hierarchical
  statistical framework for emergent constraints: application to snow-albedo
  feedback},}\ }\href {\doibase 10.1029/2018GL080082} {\bibfield  {journal}
  {\bibinfo  {journal} {Geophysical Research Letters}\ }\textbf {\bibinfo
  {volume} {45}}~(\bibinfo {number} {ja}),\ 10.1029/2018GL080082}\BibitemShut
  {NoStop}%
\bibitem [{\citenamefont {Bracegirdle}\ and\ \citenamefont
  {Stephenson}(2012)}]{ref:Bracegirdle12}%
  \BibitemOpen
  \bibfield  {author} {\bibinfo {author} {\bibnamefont {Bracegirdle},
  \bibfnamefont {Thomas~J}}, \ and\ \bibinfo {author} {\bibfnamefont
  {David~B.}\ \bibnamefont {Stephenson}}} (\bibinfo {year} {2012}),\ \bibfield
  {title} {\enquote {\bibinfo {title} {{Higher precision estimates of regional
  polar warming by ensemble regression of climate model projections}},}\ }\href
  {\doibase 10.1007/s00382-012-1330-3} {\bibfield  {journal} {\bibinfo
  {journal} {Climate Dynamics}\ }\textbf {\bibinfo {volume} {39}}~(\bibinfo
  {number} {12}),\ \bibinfo {pages} {2805--2821}}\BibitemShut {NoStop}%
\bibitem [{\citenamefont {Bracegirdle}\ and\ \citenamefont
  {Stephenson}(2013)}]{ref:Bracegirdle13}%
  \BibitemOpen
  \bibfield  {author} {\bibinfo {author} {\bibnamefont {Bracegirdle},
  \bibfnamefont {Thomas~J}}, \ and\ \bibinfo {author} {\bibfnamefont
  {David~B.}\ \bibnamefont {Stephenson}}} (\bibinfo {year} {2013}),\ \bibfield
  {title} {\enquote {\bibinfo {title} {On the robustness of emergent
  constraints used in multimodel climate change projections of arctic
  warming},}\ }\href {\doibase DOI: 10.1175/JCLI-D-12-00537.1} {\bibfield
  {journal} {\bibinfo  {journal} {Journal of Climate}\ }\textbf {\bibinfo
  {volume} {26}},\ \bibinfo {pages} {669 -- 678}}\BibitemShut {NoStop}%
\bibitem [{\citenamefont {Bretherton}\ and\ \citenamefont
  {Caldwell}(2020)}]{ref:Bretherton2020}%
  \BibitemOpen
  \bibfield  {author} {\bibinfo {author} {\bibnamefont {Bretherton},
  \bibfnamefont {Christopher}}, \ and\ \bibinfo {author} {\bibfnamefont
  {Peter}\ \bibnamefont {Caldwell}}} (\bibinfo {year} {2020}),\ \bibfield
  {title} {\enquote {\bibinfo {title} {{Combining Emergent Constraints for
  Climate Sensitivity}},}\ }\href {\doibase 10.1002/ESSOAR.10501381.1}
  {\bibfield  {journal} {\bibinfo  {journal} {Journal of Climate}\ }\textbf
  {\bibinfo {volume} {33}},\ \bibinfo {pages} {7413--7430}}\BibitemShut
  {NoStop}%
\bibitem [{\citenamefont {Brient}\ and\ \citenamefont
  {Schneider}(2016)}]{ref:Brient16}%
  \BibitemOpen
  \bibfield  {author} {\bibinfo {author} {\bibnamefont {Brient}, \bibfnamefont
  {F}}, \ and\ \bibinfo {author} {\bibfnamefont {T.}~\bibnamefont {Schneider}}}
  (\bibinfo {year} {2016}),\ \bibfield  {title} {\enquote {\bibinfo {title}
  {Constraints on climate sensitivity from space-based measurements of
  low-cloud reflection},}\ }\href {\doibase 10.1175/jcli-d-15-0897.1}
  {\bibfield  {journal} {\bibinfo  {journal} {Journal of Climate}\ }\textbf
  {\bibinfo {volume} {29}}~(\bibinfo {number} {16}),\ \bibinfo {pages}
  {5821--5835}}\BibitemShut {NoStop}%
\bibitem [{\citenamefont {Brient}\ \emph {et~al.}(2016)\citenamefont {Brient},
  \citenamefont {Schneider}, \citenamefont {Tan}, \citenamefont {Bony},
  \citenamefont {Qu},\ and\ \citenamefont {Hall}}]{ref:Brient16a}%
  \BibitemOpen
  \bibfield  {author} {\bibinfo {author} {\bibnamefont {Brient}, \bibfnamefont
  {F}}, \bibinfo {author} {\bibfnamefont {T.}~\bibnamefont {Schneider}},
  \bibinfo {author} {\bibfnamefont {Z.~H.}\ \bibnamefont {Tan}}, \bibinfo
  {author} {\bibfnamefont {S.}~\bibnamefont {Bony}}, \bibinfo {author}
  {\bibfnamefont {X.}~\bibnamefont {Qu}}, \ and\ \bibinfo {author}
  {\bibfnamefont {A.}~\bibnamefont {Hall}}} (\bibinfo {year} {2016}),\
  \bibfield  {title} {\enquote {\bibinfo {title} {Shallowness of tropical low
  clouds as a predictor of climate models' response to warming},}\ }\href
  {\doibase 10.1007/s00382-015-2846-0} {\bibfield  {journal} {\bibinfo
  {journal} {Climate Dynamics}\ }\textbf {\bibinfo {volume} {47}}~(\bibinfo
  {number} {1-2}),\ \bibinfo {pages} {433--449}}\BibitemShut {NoStop}%
\bibitem [{\citenamefont {Brient}(2020)}]{ref:Brient2020}%
  \BibitemOpen
  \bibfield  {author} {\bibinfo {author} {\bibnamefont {Brient}, \bibfnamefont
  {Florent}}} (\bibinfo {year} {2020}),\ \bibfield  {title} {\enquote {\bibinfo
  {title} {{Reducing uncertainties in climate projections with emergent
  constraints: Concepts, Examples and Prospects}},}\ }\href@noop {} {\bibfield
  {journal} {\bibinfo  {journal} {Advances in Atmospheric Sciences}\ }\textbf
  {\bibinfo {volume} {37}},\ \bibinfo {pages} {1--15}}\BibitemShut {NoStop}%
\bibitem [{\citenamefont {Brient}\ and\ \citenamefont
  {Bony}(2013)}]{ref:Brient&Bony13}%
  \BibitemOpen
  \bibfield  {author} {\bibinfo {author} {\bibnamefont {Brient}, \bibfnamefont
  {Florent}}, \ and\ \bibinfo {author} {\bibfnamefont {Sandrine}\ \bibnamefont
  {Bony}}} (\bibinfo {year} {2013}),\ \bibfield  {title} {\enquote {\bibinfo
  {title} {Interpretation of the positive low-cloud feedback predicted by a
  climate model under global warming},}\ }\href {\doibase
  10.1007/s00382-011-1279-7} {\bibfield  {journal} {\bibinfo  {journal}
  {Climate Dynamics}\ }\textbf {\bibinfo {volume} {40}}~(\bibinfo {number}
  {9}),\ \bibinfo {pages} {2415--2431}}\BibitemShut {NoStop}%
\bibitem [{\citenamefont {Brown}\ and\ \citenamefont
  {Caldeira}(2017)}]{ref:Brown&Caldeira17}%
  \BibitemOpen
  \bibfield  {author} {\bibinfo {author} {\bibnamefont {Brown}, \bibfnamefont
  {Patrick~T}}, \ and\ \bibinfo {author} {\bibfnamefont {Ken}\ \bibnamefont
  {Caldeira}}} (\bibinfo {year} {2017}),\ \bibfield  {title} {\enquote
  {\bibinfo {title} {Greater future global warming inferred from earth’s
  recent energy budget},}\ }\href {\doibase 10.1038/nature24672
  https://www.nature.com/articles/nature24672#supplementary-information}
  {\bibfield  {journal} {\bibinfo  {journal} {Nature}\ }\textbf {\bibinfo
  {volume} {552}},\ \bibinfo {pages} {45}}\BibitemShut {NoStop}%
\bibitem [{\citenamefont {Brown}\ \emph {et~al.}(2018)\citenamefont {Brown},
  \citenamefont {Stolpe},\ and\ \citenamefont {Caldeira}}]{ref:Brown18}%
  \BibitemOpen
  \bibfield  {author} {\bibinfo {author} {\bibnamefont {Brown}, \bibfnamefont
  {Patrick~T}}, \bibinfo {author} {\bibfnamefont {Martin~B.}\ \bibnamefont
  {Stolpe}}, \ and\ \bibinfo {author} {\bibfnamefont {Ken}\ \bibnamefont
  {Caldeira}}} (\bibinfo {year} {2018}),\ \bibfield  {title} {\enquote
  {\bibinfo {title} {Assumptions for emergent constraints},}\ }\href {\doibase
  10.1038/s41586-018-0638-5} {\bibfield  {journal} {\bibinfo  {journal}
  {Nature}\ }\textbf {\bibinfo {volume} {563}}~(\bibinfo {number} {7729}),\
  \bibinfo {pages} {E1--E3}}\BibitemShut {NoStop}%
\bibitem [{\citenamefont {Budyko}(1969)}]{ref:Budyko69}%
  \BibitemOpen
  \bibfield  {author} {\bibinfo {author} {\bibnamefont {Budyko}, \bibfnamefont
  {M~I}}} (\bibinfo {year} {1969}),\ \bibfield  {title} {\enquote {\bibinfo
  {title} {The effect of solar radiation variations on the climate of the
  {E}arth},}\ }\href@noop {} {\bibfield  {journal} {\bibinfo  {journal}
  {Tellus}\ }\textbf {\bibinfo {volume} {21}},\ \bibinfo {pages}
  {611--619}}\BibitemShut {NoStop}%
\bibitem [{\citenamefont {Caldwell}\ \emph {et~al.}(2018)\citenamefont
  {Caldwell}, \citenamefont {Zelinka},\ and\ \citenamefont
  {Klein}}]{ref:Caldwell18}%
  \BibitemOpen
  \bibfield  {author} {\bibinfo {author} {\bibnamefont {Caldwell},
  \bibfnamefont {P~M}}, \bibinfo {author} {\bibfnamefont {M.~D.}\ \bibnamefont
  {Zelinka}}, \ and\ \bibinfo {author} {\bibfnamefont {S.~A.}\ \bibnamefont
  {Klein}}} (\bibinfo {year} {2018}),\ \bibfield  {title} {\enquote {\bibinfo
  {title} {Evaluating emergent constraints on equilibrium climate
  sensitivity},}\ }\href {\doibase 10.1175/jcli-d-17-0631.1} {\bibfield
  {journal} {\bibinfo  {journal} {Journal of Climate}\ }\textbf {\bibinfo
  {volume} {31}}~(\bibinfo {number} {10}),\ \bibinfo {pages}
  {3921--3942}}\BibitemShut {NoStop}%
\bibitem [{\citenamefont {Caldwell}\ \emph {et~al.}(2014)\citenamefont
  {Caldwell}, \citenamefont {Bretherton}, \citenamefont {Zelinka},
  \citenamefont {Klein}, \citenamefont {Santer},\ and\ \citenamefont
  {Sanderson}}]{ref:Caldwell14}%
  \BibitemOpen
  \bibfield  {author} {\bibinfo {author} {\bibnamefont {Caldwell},
  \bibfnamefont {Peter~M}}, \bibinfo {author} {\bibfnamefont {Christopher~S.}\
  \bibnamefont {Bretherton}}, \bibinfo {author} {\bibfnamefont {Mark~D.}\
  \bibnamefont {Zelinka}}, \bibinfo {author} {\bibfnamefont {Stephen~A.}\
  \bibnamefont {Klein}}, \bibinfo {author} {\bibfnamefont {Benjamin~D.}\
  \bibnamefont {Santer}}, \ and\ \bibinfo {author} {\bibfnamefont
  {Benjamin~M.}\ \bibnamefont {Sanderson}}} (\bibinfo {year} {2014}),\
  \bibfield  {title} {\enquote {\bibinfo {title} {Statistical significance of
  climate sensitivity predictors obtained by data mining},}\ }\href {\doibase
  10.1002/2014GL059205} {\bibfield  {journal} {\bibinfo  {journal} {Geophysical
  Research Letters}\ }\textbf {\bibinfo {volume} {41}}~(\bibinfo {number}
  {5}),\ \bibinfo {pages} {1803--1808}}\BibitemShut {NoStop}%
\bibitem [{\citenamefont {Ceppi}\ \emph {et~al.}(2017)\citenamefont {Ceppi},
  \citenamefont {Brient}, \citenamefont {Zelinka},\ and\ \citenamefont
  {Hartmann}}]{ref:Ceppi17}%
  \BibitemOpen
  \bibfield  {author} {\bibinfo {author} {\bibnamefont {Ceppi}, \bibfnamefont
  {Paulo}}, \bibinfo {author} {\bibfnamefont {Florent}\ \bibnamefont {Brient}},
  \bibinfo {author} {\bibfnamefont {Mark~D.}\ \bibnamefont {Zelinka}}, \ and\
  \bibinfo {author} {\bibfnamefont {Dennis~L.}\ \bibnamefont {Hartmann}}}
  (\bibinfo {year} {2017}),\ \bibfield  {title} {\enquote {\bibinfo {title}
  {{Cloud feedback mechanisms and their representation in global climate
  models}},}\ }\href {\doibase 10.1002/wcc.465} {\bibfield  {journal} {\bibinfo
   {journal} {Wiley Interdisciplinary Reviews: Climate Change}\ }\textbf
  {\bibinfo {volume} {8}}~(\bibinfo {number} {4}),\
  10.1002/wcc.465}\BibitemShut {NoStop}%
\bibitem [{\citenamefont {Chadburn}\ \emph {et~al.}(2017)\citenamefont
  {Chadburn}, \citenamefont {Burke}, \citenamefont {Cox}, \citenamefont
  {Friedlingstein}, \citenamefont {Hugelius},\ and\ \citenamefont
  {Westermann}}]{ref:Chadburn17}%
  \BibitemOpen
  \bibfield  {author} {\bibinfo {author} {\bibnamefont {Chadburn},
  \bibfnamefont {S~E}}, \bibinfo {author} {\bibfnamefont {E.~J.}\ \bibnamefont
  {Burke}}, \bibinfo {author} {\bibfnamefont {P.~M.}\ \bibnamefont {Cox}},
  \bibinfo {author} {\bibfnamefont {P.}~\bibnamefont {Friedlingstein}},
  \bibinfo {author} {\bibfnamefont {G.}~\bibnamefont {Hugelius}}, \ and\
  \bibinfo {author} {\bibfnamefont {S.}~\bibnamefont {Westermann}}} (\bibinfo
  {year} {2017}),\ \bibfield  {title} {\enquote {\bibinfo {title} {An
  observation-based constraint on permafrost loss as a function of global
  warming},}\ }\href {\doibase 10.1038/nclimate3262
  http://www.nature.com/nclimate/journal/v7/n5/abs/nclimate3262.html#supplementary-information}
  {\bibfield  {journal} {\bibinfo  {journal} {Nature Clim. Change}\ }\textbf
  {\bibinfo {volume} {7}}~(\bibinfo {number} {5}),\ \bibinfo {pages}
  {340--344}}\BibitemShut {NoStop}%
\bibitem [{\citenamefont {Clement}\ \emph {et~al.}(2009)\citenamefont
  {Clement}, \citenamefont {Burgman},\ and\ \citenamefont
  {Norris}}]{ref:Clement09}%
  \BibitemOpen
  \bibfield  {author} {\bibinfo {author} {\bibnamefont {Clement}, \bibfnamefont
  {A~C}}, \bibinfo {author} {\bibfnamefont {R.}~\bibnamefont {Burgman}}, \ and\
  \bibinfo {author} {\bibfnamefont {J.~R.}\ \bibnamefont {Norris}}} (\bibinfo
  {year} {2009}),\ \bibfield  {title} {\enquote {\bibinfo {title}
  {Observational and model evidence for positive low-level cloud feedback},}\
  }\href {\doibase 10.1126/science.1171255} {\bibfield  {journal} {\bibinfo
  {journal} {Science}\ }\textbf {\bibinfo {volume} {325}}~(\bibinfo {number}
  {5939}),\ \bibinfo {pages} {460--464}}\BibitemShut {NoStop}%
\bibitem [{\citenamefont {Collins}\ \emph {et~al.}(2013)\citenamefont
  {Collins}, \citenamefont {Knutti}, \citenamefont {Arblaster}, \citenamefont
  {Dufresne}, \citenamefont {Fichefet}, \citenamefont {Friedlingstein},
  \citenamefont {Gao}, \citenamefont {Gutowski}, \citenamefont {Johns},
  \citenamefont {Krinner}, \citenamefont {Shongwe}, \citenamefont {Tebaldi},
  \citenamefont {Weaver},\ and\ \citenamefont
  {Wehner}}]{ref:IPCC_AR5_projections}%
  \BibitemOpen
  \bibfield  {author} {\bibinfo {author} {\bibnamefont {Collins}, \bibfnamefont
  {M}}, \bibinfo {author} {\bibfnamefont {R.}~\bibnamefont {Knutti}}, \bibinfo
  {author} {\bibfnamefont {J.}~\bibnamefont {Arblaster}}, \bibinfo {author}
  {\bibfnamefont {J.~L.}\ \bibnamefont {Dufresne}}, \bibinfo {author}
  {\bibfnamefont {T.}~\bibnamefont {Fichefet}}, \bibinfo {author}
  {\bibfnamefont {P.}~\bibnamefont {Friedlingstein}}, \bibinfo {author}
  {\bibfnamefont {X.}~\bibnamefont {Gao}}, \bibinfo {author} {\bibfnamefont
  {W.~J.}\ \bibnamefont {Gutowski}}, \bibinfo {author} {\bibfnamefont
  {T.}~\bibnamefont {Johns}}, \bibinfo {author} {\bibfnamefont
  {G.}~\bibnamefont {Krinner}}, \bibinfo {author} {\bibfnamefont
  {M.}~\bibnamefont {Shongwe}}, \bibinfo {author} {\bibfnamefont
  {C.}~\bibnamefont {Tebaldi}}, \bibinfo {author} {\bibfnamefont {A.~J.}\
  \bibnamefont {Weaver}}, \ and\ \bibinfo {author} {\bibfnamefont
  {M.}~\bibnamefont {Wehner}}} (\bibinfo {year} {2013}),\ \bibfield  {title}
  {\enquote {\bibinfo {title} {Long-term climate change: projections,
  commitments and irreversibility},}\ }in\ \href@noop {} {\emph {\bibinfo
  {booktitle} {Climate change 2013: the physical science basis. Contribution of
  working group I to the fifth assessment report of the {I}ntergovernmental
  {P}anel on {C}limate {C}hange}}},\ \bibinfo {editor} {edited by\ \bibinfo
  {editor} {\bibfnamefont {T.~F.}\ \bibnamefont {Stocker}}, \bibinfo {editor}
  {\bibfnamefont {D.~Qin}\ \bibnamefont {D}}, \bibinfo {editor} {\bibfnamefont
  {G.~K.}\ \bibnamefont {Plattner}}, \bibinfo {editor} {\bibfnamefont
  {M.}~\bibnamefont {Tignor}}, \bibinfo {editor} {\bibfnamefont {S.~K.}\
  \bibnamefont {Allen}}, \bibinfo {editor} {\bibfnamefont {J.}~\bibnamefont
  {Boschung}}, \bibinfo {editor} {\bibfnamefont {A.}~\bibnamefont {Nauels}},
  \bibinfo {editor} {\bibfnamefont {Y.}~\bibnamefont {Xia}}, \bibinfo {editor}
  {\bibfnamefont {V.~Bex}\ \bibnamefont {V}}, \ and\ \bibinfo {editor}
  {\bibfnamefont {P.~M.}\ \bibnamefont {Midgley}}}\ (\bibinfo  {publisher}
  {Cambridge University Press, Cambridge})\BibitemShut {NoStop}%
\bibitem [{\citenamefont {Collins}\ \emph {et~al.}(2011)\citenamefont
  {Collins}, \citenamefont {Booth}, \citenamefont {Bhaskaran}, \citenamefont
  {Harris}, \citenamefont {Murphy}, \citenamefont {Sexton},\ and\ \citenamefont
  {Webb}}]{ref:Collins11}%
  \BibitemOpen
  \bibfield  {author} {\bibinfo {author} {\bibnamefont {Collins}, \bibfnamefont
  {Matthew}}, \bibinfo {author} {\bibfnamefont {Ben B.~B.}\ \bibnamefont
  {Booth}}, \bibinfo {author} {\bibfnamefont {B.}~\bibnamefont {Bhaskaran}},
  \bibinfo {author} {\bibfnamefont {Glen~R.}\ \bibnamefont {Harris}}, \bibinfo
  {author} {\bibfnamefont {James~M.}\ \bibnamefont {Murphy}}, \bibinfo {author}
  {\bibfnamefont {David M.~H.}\ \bibnamefont {Sexton}}, \ and\ \bibinfo
  {author} {\bibfnamefont {Mark~J.}\ \bibnamefont {Webb}}} (\bibinfo {year}
  {2011}),\ \bibfield  {title} {\enquote {\bibinfo {title} {Climate model
  errors, feedbacks and forcings: a comparison of perturbed physics and
  multi-model ensembles},}\ }\href {\doibase 10.1007/s00382-010-0808-0}
  {\bibfield  {journal} {\bibinfo  {journal} {Climate Dynamics}\ }\textbf
  {\bibinfo {volume} {36}}~(\bibinfo {number} {9}),\ \bibinfo {pages}
  {1737--1766}}\BibitemShut {NoStop}%
\bibitem [{\citenamefont {Covey}\ \emph {et~al.}(2000)\citenamefont {Covey},
  \citenamefont {Guilyardi}, \citenamefont {Jiang}, \citenamefont {Johns},
  \citenamefont {Treut}, \citenamefont {Madec}, \citenamefont {Meehl},
  \citenamefont {Miller}, \citenamefont {Power}, \citenamefont {Roeckner},\
  and\ \citenamefont {Russell}}]{ref:Covey2000}%
  \BibitemOpen
  \bibfield  {author} {\bibinfo {author} {\bibnamefont {Covey}, \bibfnamefont
  {C}}, \bibinfo {author} {\bibfnamefont {E}~\bibnamefont {Guilyardi}},
  \bibinfo {author} {\bibfnamefont {X}~\bibnamefont {Jiang}}, \bibinfo {author}
  {\bibfnamefont {T~C}\ \bibnamefont {Johns}}, \bibinfo {author} {\bibfnamefont
  {H~Le}\ \bibnamefont {Treut}}, \bibinfo {author} {\bibfnamefont
  {G}~\bibnamefont {Madec}}, \bibinfo {author} {\bibfnamefont {G~a}\
  \bibnamefont {Meehl}}, \bibinfo {author} {\bibfnamefont {R}~\bibnamefont
  {Miller}}, \bibinfo {author} {\bibfnamefont {S~B}\ \bibnamefont {Power}},
  \bibinfo {author} {\bibfnamefont {E}~\bibnamefont {Roeckner}}, \ and\
  \bibinfo {author} {\bibfnamefont {G}~\bibnamefont {Russell}}} (\bibinfo
  {year} {2000}),\ \bibfield  {title} {\enquote {\bibinfo {title} {{The
  seasonal cycle in coupled ocean-atmosphere general circulation models}},}\
  }\href@noop {} {\bibfield  {journal} {\bibinfo  {journal} {Climate Dynamics}\
  }\textbf {\bibinfo {volume} {16}},\ \bibinfo {pages} {775--787}}\BibitemShut
  {NoStop}%
\bibitem [{\citenamefont {Cowtan}\ and\ \citenamefont {Way}(2014)}]{Cowtan14}%
  \BibitemOpen
  \bibfield  {author} {\bibinfo {author} {\bibnamefont {Cowtan}, \bibfnamefont
  {Kevin}}, \ and\ \bibinfo {author} {\bibfnamefont {Robert~G.}\ \bibnamefont
  {Way}}} (\bibinfo {year} {2014}),\ \bibfield  {title} {\enquote {\bibinfo
  {title} {Coverage bias in the hadcrut4 temperature series and its impact on
  recent temperature trends},}\ }\href@noop {} {\bibfield  {journal} {\bibinfo
  {journal} {Quarterly Journal of the Royal Meteorological Society}\ }\textbf
  {\bibinfo {volume} {140}}~(\bibinfo {number} {683}),\ \bibinfo {pages}
  {1935--1944}}\BibitemShut {NoStop}%
\bibitem [{\citenamefont {Cox}\ \emph {et~al.}(2004)\citenamefont {Cox},
  \citenamefont {Betts}, \citenamefont {Collins}, \citenamefont {Harris},
  \citenamefont {Huntingford},\ and\ \citenamefont {Jones}}]{ref:Cox04}%
  \BibitemOpen
  \bibfield  {author} {\bibinfo {author} {\bibnamefont {Cox}, \bibfnamefont
  {P~M}}, \bibinfo {author} {\bibfnamefont {R.~A.}\ \bibnamefont {Betts}},
  \bibinfo {author} {\bibfnamefont {M.}~\bibnamefont {Collins}}, \bibinfo
  {author} {\bibfnamefont {P.~P.}\ \bibnamefont {Harris}}, \bibinfo {author}
  {\bibfnamefont {C.}~\bibnamefont {Huntingford}}, \ and\ \bibinfo {author}
  {\bibfnamefont {C.~D.}\ \bibnamefont {Jones}}} (\bibinfo {year} {2004}),\
  \bibfield  {title} {\enquote {\bibinfo {title} {Amazonian forest dieback
  under climate-carbon cycle projections for the 21st century},}\ }\href
  {\doibase 10.1007/s00704-004-0049-4} {\bibfield  {journal} {\bibinfo
  {journal} {Theoretical and Applied Climatology}\ }\textbf {\bibinfo {volume}
  {78}}~(\bibinfo {number} {1-3}),\ \bibinfo {pages} {137--156}}\BibitemShut
  {NoStop}%
\bibitem [{\citenamefont {Cox}\ \emph {et~al.}(2000)\citenamefont {Cox},
  \citenamefont {Betts}, \citenamefont {Jones}, \citenamefont {Spall},\ and\
  \citenamefont {Totterdell}}]{ref:Cox00}%
  \BibitemOpen
  \bibfield  {author} {\bibinfo {author} {\bibnamefont {Cox}, \bibfnamefont
  {P~M}}, \bibinfo {author} {\bibfnamefont {R.~A.}\ \bibnamefont {Betts}},
  \bibinfo {author} {\bibfnamefont {C.~D.}\ \bibnamefont {Jones}}, \bibinfo
  {author} {\bibfnamefont {S.~A.}\ \bibnamefont {Spall}}, \ and\ \bibinfo
  {author} {\bibfnamefont {I.~J.}\ \bibnamefont {Totterdell}}} (\bibinfo {year}
  {2000}),\ \bibfield  {title} {\enquote {\bibinfo {title} {Acceleration of
  global warming due to carbon-cycle feedbacks in a coupled climate model (vol
  408, pg 184, 2000)},}\ }\href {\doibase 10.1038/35047138} {\bibfield
  {journal} {\bibinfo  {journal} {Nature}\ }\textbf {\bibinfo {volume}
  {408}}~(\bibinfo {number} {6813}),\ \bibinfo {pages} {750--750}}\BibitemShut
  {NoStop}%
\bibitem [{\citenamefont {Cox}(2019)}]{ref:Cox19}%
  \BibitemOpen
  \bibfield  {author} {\bibinfo {author} {\bibnamefont {Cox}, \bibfnamefont
  {Peter~M}}} (\bibinfo {year} {2019}),\ \bibfield  {title} {\enquote {\bibinfo
  {title} {Emergent constraints on climate-carbon cycle feedbacks},}\ }\href
  {\doibase 10.1007/s40641-019-00141-y} {\bibfield  {journal} {\bibinfo
  {journal} {Current Climate Change Reports}\ }\textbf {\bibinfo {volume}
  {5}}~(\bibinfo {number} {4}),\ \bibinfo {pages} {275--281}}\BibitemShut
  {NoStop}%
\bibitem [{\citenamefont {Cox}\ \emph {et~al.}(2018{\natexlab{a}})\citenamefont
  {Cox}, \citenamefont {Huntingford},\ and\ \citenamefont
  {Williamson}}]{ref:Cox18}%
  \BibitemOpen
  \bibfield  {author} {\bibinfo {author} {\bibnamefont {Cox}, \bibfnamefont
  {Peter~M}}, \bibinfo {author} {\bibfnamefont {Chris}\ \bibnamefont
  {Huntingford}}, \ and\ \bibinfo {author} {\bibfnamefont {Mark~S.}\
  \bibnamefont {Williamson}}} (\bibinfo {year} {2018}{\natexlab{a}}),\
  \bibfield  {title} {\enquote {\bibinfo {title} {Emergent constraint on
  equilibrium climate sensitivity from global temperature variability},}\
  }\href {\doibase 10.1038/nature25450} {\bibfield  {journal} {\bibinfo
  {journal} {Nature}\ }\textbf {\bibinfo {volume} {553}},\ \bibinfo {pages}
  {319}}\BibitemShut {NoStop}%
\bibitem [{\citenamefont {Cox}\ \emph {et~al.}(2013)\citenamefont {Cox},
  \citenamefont {Pearson}, \citenamefont {Booth}, \citenamefont
  {Friedlingstein}, \citenamefont {Huntingford}, \citenamefont {Jones},\ and\
  \citenamefont {Luke}}]{ref:Cox13}%
  \BibitemOpen
  \bibfield  {author} {\bibinfo {author} {\bibnamefont {Cox}, \bibfnamefont
  {Peter~M}}, \bibinfo {author} {\bibfnamefont {David}\ \bibnamefont
  {Pearson}}, \bibinfo {author} {\bibfnamefont {Ben~B.}\ \bibnamefont {Booth}},
  \bibinfo {author} {\bibfnamefont {Pierre}\ \bibnamefont {Friedlingstein}},
  \bibinfo {author} {\bibfnamefont {Chris}\ \bibnamefont {Huntingford}},
  \bibinfo {author} {\bibfnamefont {Chris~D.}\ \bibnamefont {Jones}}, \ and\
  \bibinfo {author} {\bibfnamefont {Catherine~M.}\ \bibnamefont {Luke}}}
  (\bibinfo {year} {2013}),\ \bibfield  {title} {\enquote {\bibinfo {title}
  {Sensitivity of tropical carbon to climate change constrained by carbon
  dioxide variability},}\ }\href {\doibase 10.1038/nature11882
  https://www.nature.com/articles/nature11882#supplementary-information}
  {\bibfield  {journal} {\bibinfo  {journal} {Nature}\ }\textbf {\bibinfo
  {volume} {494}},\ \bibinfo {pages} {341}}\BibitemShut {NoStop}%
\bibitem [{\citenamefont {Cox}\ \emph {et~al.}(2018{\natexlab{b}})\citenamefont
  {Cox}, \citenamefont {Williamson}, \citenamefont {Nijsse},\ and\
  \citenamefont {Huntingford}}]{ref:Cox18a}%
  \BibitemOpen
  \bibfield  {author} {\bibinfo {author} {\bibnamefont {Cox}, \bibfnamefont
  {Peter~M}}, \bibinfo {author} {\bibfnamefont {Mark~S.}\ \bibnamefont
  {Williamson}}, \bibinfo {author} {\bibfnamefont {Femke J. M.~M.}\
  \bibnamefont {Nijsse}}, \ and\ \bibinfo {author} {\bibfnamefont {Chris}\
  \bibnamefont {Huntingford}}} (\bibinfo {year} {2018}{\natexlab{b}}),\
  \bibfield  {title} {\enquote {\bibinfo {title} {Cox et al. reply},}\ }\href
  {\doibase 10.1038/s41586-018-0641-x} {\bibfield  {journal} {\bibinfo
  {journal} {Nature}\ }\textbf {\bibinfo {volume} {563}}~(\bibinfo {number}
  {7729}),\ \bibinfo {pages} {E10--E15}}\BibitemShut {NoStop}%
\bibitem [{\citenamefont {Cubasch}\ \emph {et~al.}(2001)\citenamefont
  {Cubasch}, \citenamefont {Meehl}, \citenamefont {Boer}, \citenamefont
  {Stouffer}, \citenamefont {Dix}, \citenamefont {Noda}, \citenamefont
  {Senior}, \citenamefont {Raper},\ and\ \citenamefont
  {Yap}}]{ref:IPCC_AR3_projections}%
  \BibitemOpen
  \bibfield  {author} {\bibinfo {author} {\bibnamefont {Cubasch}, \bibfnamefont
  {U}}, \bibinfo {author} {\bibfnamefont {GA}~\bibnamefont {Meehl}}, \bibinfo
  {author} {\bibfnamefont {GJ}~\bibnamefont {Boer}}, \bibinfo {author}
  {\bibfnamefont {RJ}~\bibnamefont {Stouffer}}, \bibinfo {author}
  {\bibfnamefont {M}~\bibnamefont {Dix}}, \bibinfo {author} {\bibfnamefont
  {A}~\bibnamefont {Noda}}, \bibinfo {author} {\bibfnamefont {CA}~\bibnamefont
  {Senior}}, \bibinfo {author} {\bibfnamefont {S}~\bibnamefont {Raper}}, \ and\
  \bibinfo {author} {\bibfnamefont {KS}~\bibnamefont {Yap}}} (\bibinfo {year}
  {2001}),\ \enquote {\bibinfo {title} {Projections of future climate
  change},}\ in\ \href@noop {} {\emph {\bibinfo {booktitle} {Climate Change
  2001: The scientific basis. Contribution of WG1 to the Third Assessment
  Report of the IPCC (TAR)}}}\ (\bibinfo  {publisher} {Cambridge University
  Press})\ pp.\ \bibinfo {pages} {525--582}\BibitemShut {NoStop}%
\bibitem [{\citenamefont {Jim{\'{e}}nez-de-la Cuesta}\ and\ \citenamefont
  {Mauritsen}(2019)}]{ref:Jimenez-de-la-Cuesta2019}%
  \BibitemOpen
  \bibfield  {author} {\bibinfo {author} {\bibnamefont {Jim{\'{e}}nez-de-la
  Cuesta}, \bibfnamefont {Diego}}, \ and\ \bibinfo {author} {\bibfnamefont
  {Thorsten}\ \bibnamefont {Mauritsen}}} (\bibinfo {year} {2019}),\ \bibfield
  {title} {\enquote {\bibinfo {title} {{Emergent constraints on Earth's
  transient and equilibrium response to doubled CO2 from post-1970s global
  warming}},}\ }\href {\doibase 10.1038/s41561-019-0463-y} {\bibfield
  {journal} {\bibinfo  {journal} {Nature Geoscience}\ }\textbf {\bibinfo
  {volume} {12}}~(\bibinfo {number} {11}),\ \bibinfo {pages}
  {902--905}}\BibitemShut {NoStop}%
\bibitem [{\citenamefont {DeAngelis}\ \emph {et~al.}(2015)\citenamefont
  {DeAngelis}, \citenamefont {Qu}, \citenamefont {Zelinka},\ and\ \citenamefont
  {Hall}}]{ref:DeAngelis15}%
  \BibitemOpen
  \bibfield  {author} {\bibinfo {author} {\bibnamefont {DeAngelis},
  \bibfnamefont {Anthony~M}}, \bibinfo {author} {\bibfnamefont {Xin}\
  \bibnamefont {Qu}}, \bibinfo {author} {\bibfnamefont {Mark~D.}\ \bibnamefont
  {Zelinka}}, \ and\ \bibinfo {author} {\bibfnamefont {Alex}\ \bibnamefont
  {Hall}}} (\bibinfo {year} {2015}),\ \bibfield  {title} {\enquote {\bibinfo
  {title} {An observational radiative constraint on hydrologic cycle
  intensification},}\ }\href {\doibase 10.1038/nature15770} {\bibfield
  {journal} {\bibinfo  {journal} {Nature}\ }\textbf {\bibinfo {volume}
  {528}}~(\bibinfo {number} {7581}),\ \bibinfo {pages} {249--253}}\BibitemShut
  {NoStop}%
\bibitem [{\citenamefont {Dijkstra}(2013)}]{ref:Dijkstra13}%
  \BibitemOpen
  \bibfield  {author} {\bibinfo {author} {\bibnamefont {Dijkstra},
  \bibfnamefont {Henk~A}}} (\bibinfo {year} {2013}),\ \href {\doibase
  10.1017/CBO9781139034135} {\emph {\bibinfo {title} {Nonlinear Climate
  Dynamics}}}\ (\bibinfo  {publisher} {Cambridge University Press})\BibitemShut
  {NoStop}%
\bibitem [{\citenamefont {Donat}\ \emph {et~al.}(2018)\citenamefont {Donat},
  \citenamefont {Pitman},\ and\ \citenamefont {Ang{\'{e}}lil}}]{ref:Donat18}%
  \BibitemOpen
  \bibfield  {author} {\bibinfo {author} {\bibnamefont {Donat}, \bibfnamefont
  {Markus~G}}, \bibinfo {author} {\bibfnamefont {Andrew~J}\ \bibnamefont
  {Pitman}}, \ and\ \bibinfo {author} {\bibfnamefont {Oliver}\ \bibnamefont
  {Ang{\'{e}}lil}}} (\bibinfo {year} {2018}),\ \bibfield  {title} {\enquote
  {\bibinfo {title} {{Understanding and Reducing Future Uncertainty in
  Midlatitude Daily Heat Extremes Via Land Surface Feedback Constraints}},}\
  }\href {\doibase 10.1029/2018GL079128} {\bibfield  {journal} {\bibinfo
  {journal} {Geophysical Research Letters}\ }\textbf {\bibinfo {volume}
  {45}}~(\bibinfo {number} {19}),\ \bibinfo {pages}
  {10,610--627,636}}\BibitemShut {NoStop}%
\bibitem [{\citenamefont {Donner}\ \emph {et~al.}(2011)\citenamefont {Donner},
  \citenamefont {Wyman}, \citenamefont {Hemler}, \citenamefont {Horowitz},
  \citenamefont {Ming}, \citenamefont {Zhao}, \citenamefont {Golaz},
  \citenamefont {Ginoux}, \citenamefont {Lin}, \citenamefont {Schwarzkopf},
  \citenamefont {Austin}, \citenamefont {Alaka}, \citenamefont {Cooke},
  \citenamefont {Delworth}, \citenamefont {Freidenreich}, \citenamefont
  {Gordon}, \citenamefont {Griffies}, \citenamefont {Held}, \citenamefont
  {Hurlin}, \citenamefont {Klein}, \citenamefont {Knutson}, \citenamefont
  {Langenhorst}, \citenamefont {Lee}, \citenamefont {Lin}, \citenamefont
  {Magi}, \citenamefont {Malyshev}, \citenamefont {Milly}, \citenamefont
  {Naik}, \citenamefont {Nath}, \citenamefont {Pincus}, \citenamefont
  {Ploshay}, \citenamefont {Ramaswamy}, \citenamefont {Seman}, \citenamefont
  {Shevliakova}, \citenamefont {Sirutis}, \citenamefont {Stern}, \citenamefont
  {Stouffer}, \citenamefont {Wilson}, \citenamefont {Winton}, \citenamefont
  {Wittenberg},\ and\ \citenamefont {Zeng}}]{ref:Donner11}%
  \BibitemOpen
  \bibfield  {author} {\bibinfo {author} {\bibnamefont {Donner}, \bibfnamefont
  {Leo~J}}, \bibinfo {author} {\bibfnamefont {Bruce~L.}\ \bibnamefont {Wyman}},
  \bibinfo {author} {\bibfnamefont {Richard~S.}\ \bibnamefont {Hemler}},
  \bibinfo {author} {\bibfnamefont {Larry~W.}\ \bibnamefont {Horowitz}},
  \bibinfo {author} {\bibfnamefont {Yi}~\bibnamefont {Ming}}, \bibinfo {author}
  {\bibfnamefont {Ming}\ \bibnamefont {Zhao}}, \bibinfo {author} {\bibfnamefont
  {Jean-Christophe}\ \bibnamefont {Golaz}}, \bibinfo {author} {\bibfnamefont
  {Paul}\ \bibnamefont {Ginoux}}, \bibinfo {author} {\bibfnamefont {S.~J.}\
  \bibnamefont {Lin}}, \bibinfo {author} {\bibfnamefont {M.~Daniel}\
  \bibnamefont {Schwarzkopf}}, \bibinfo {author} {\bibfnamefont {John}\
  \bibnamefont {Austin}}, \bibinfo {author} {\bibfnamefont {Ghassan}\
  \bibnamefont {Alaka}}, \bibinfo {author} {\bibfnamefont {William~F.}\
  \bibnamefont {Cooke}}, \bibinfo {author} {\bibfnamefont {Thomas~L.}\
  \bibnamefont {Delworth}}, \bibinfo {author} {\bibfnamefont {Stuart~M.}\
  \bibnamefont {Freidenreich}}, \bibinfo {author} {\bibfnamefont {C.~T.}\
  \bibnamefont {Gordon}}, \bibinfo {author} {\bibfnamefont {Stephen~M.}\
  \bibnamefont {Griffies}}, \bibinfo {author} {\bibfnamefont {Isaac~M.}\
  \bibnamefont {Held}}, \bibinfo {author} {\bibfnamefont {William~J.}\
  \bibnamefont {Hurlin}}, \bibinfo {author} {\bibfnamefont {Stephen~A.}\
  \bibnamefont {Klein}}, \bibinfo {author} {\bibfnamefont {Thomas~R.}\
  \bibnamefont {Knutson}}, \bibinfo {author} {\bibfnamefont {Amy~R.}\
  \bibnamefont {Langenhorst}}, \bibinfo {author} {\bibfnamefont {Hyun-Chul}\
  \bibnamefont {Lee}}, \bibinfo {author} {\bibfnamefont {Yanluan}\ \bibnamefont
  {Lin}}, \bibinfo {author} {\bibfnamefont {Brian~I.}\ \bibnamefont {Magi}},
  \bibinfo {author} {\bibfnamefont {Sergey~L.}\ \bibnamefont {Malyshev}},
  \bibinfo {author} {\bibfnamefont {P.~C.~D.}\ \bibnamefont {Milly}}, \bibinfo
  {author} {\bibfnamefont {Vaishali}\ \bibnamefont {Naik}}, \bibinfo {author}
  {\bibfnamefont {Mary~J.}\ \bibnamefont {Nath}}, \bibinfo {author}
  {\bibfnamefont {Robert}\ \bibnamefont {Pincus}}, \bibinfo {author}
  {\bibfnamefont {Jeffrey~J.}\ \bibnamefont {Ploshay}}, \bibinfo {author}
  {\bibfnamefont {V.}~\bibnamefont {Ramaswamy}}, \bibinfo {author}
  {\bibfnamefont {Charles~J.}\ \bibnamefont {Seman}}, \bibinfo {author}
  {\bibfnamefont {Elena}\ \bibnamefont {Shevliakova}}, \bibinfo {author}
  {\bibfnamefont {Joseph~J.}\ \bibnamefont {Sirutis}}, \bibinfo {author}
  {\bibfnamefont {William~F.}\ \bibnamefont {Stern}}, \bibinfo {author}
  {\bibfnamefont {Ronald~J.}\ \bibnamefont {Stouffer}}, \bibinfo {author}
  {\bibfnamefont {R.~John}\ \bibnamefont {Wilson}}, \bibinfo {author}
  {\bibfnamefont {Michael}\ \bibnamefont {Winton}}, \bibinfo {author}
  {\bibfnamefont {Andrew~T.}\ \bibnamefont {Wittenberg}}, \ and\ \bibinfo
  {author} {\bibfnamefont {Fanrong}\ \bibnamefont {Zeng}}} (\bibinfo {year}
  {2011}),\ \bibfield  {title} {\enquote {\bibinfo {title} {The dynamical core,
  physical parameterizations, and basic simulation characteristics of the
  atmospheric component am3 of the gfdl global coupled model cm3},}\ }\href
  {\doibase 10.1175/2011JCLI3955.1} {\bibfield  {journal} {\bibinfo  {journal}
  {Journal of Climate}\ }\textbf {\bibinfo {volume} {24}}~(\bibinfo {number}
  {13}),\ \bibinfo {pages} {3484--3519}}\BibitemShut {NoStop}%
\bibitem [{\citenamefont {Drijfhout}\ \emph {et~al.}(2015)\citenamefont
  {Drijfhout}, \citenamefont {Bathiany}, \citenamefont {Beaulieu},
  \citenamefont {Brovkin}, \citenamefont {Claussen}, \citenamefont
  {Huntingford}, \citenamefont {Scheffer}, \citenamefont {Sgubin},\ and\
  \citenamefont {Swingedouw}}]{ref:Drijfhout15a}%
  \BibitemOpen
  \bibfield  {author} {\bibinfo {author} {\bibnamefont {Drijfhout},
  \bibfnamefont {Sybren}}, \bibinfo {author} {\bibfnamefont {Sebastian}\
  \bibnamefont {Bathiany}}, \bibinfo {author} {\bibfnamefont {Claudie}\
  \bibnamefont {Beaulieu}}, \bibinfo {author} {\bibfnamefont {Victor}\
  \bibnamefont {Brovkin}}, \bibinfo {author} {\bibfnamefont {Martin}\
  \bibnamefont {Claussen}}, \bibinfo {author} {\bibfnamefont {Chris}\
  \bibnamefont {Huntingford}}, \bibinfo {author} {\bibfnamefont {Marten}\
  \bibnamefont {Scheffer}}, \bibinfo {author} {\bibfnamefont {Giovanni}\
  \bibnamefont {Sgubin}}, \ and\ \bibinfo {author} {\bibfnamefont {Didier}\
  \bibnamefont {Swingedouw}}} (\bibinfo {year} {2015}),\ \bibfield  {title}
  {\enquote {\bibinfo {title} {Catalogue of abrupt shifts in intergovernmental
  panel on climate change climate models},}\ }\href {\doibase
  10.1073/pnas.1511451112} {\bibfield  {journal} {\bibinfo  {journal}
  {Proceedings of the National Academy of Sciences}\ }\textbf {\bibinfo
  {volume} {112}}~(\bibinfo {number} {43}),\ \bibinfo {pages} {E5777--E5786}},\
  \Eprint
  {http://arxiv.org/abs/https://www.pnas.org/content/112/43/E5777.full.pdf}
  {https://www.pnas.org/content/112/43/E5777.full.pdf} \BibitemShut {NoStop}%
\bibitem [{\citenamefont {Durack}\ \emph {et~al.}(2014)\citenamefont {Durack},
  \citenamefont {Gleckler}, \citenamefont {Landerer},\ and\ \citenamefont
  {Taylor}}]{ref:Durack2014}%
  \BibitemOpen
  \bibfield  {author} {\bibinfo {author} {\bibnamefont {Durack}, \bibfnamefont
  {Paul~J}}, \bibinfo {author} {\bibfnamefont {Peter~J.}\ \bibnamefont
  {Gleckler}}, \bibinfo {author} {\bibfnamefont {Felix~W.}\ \bibnamefont
  {Landerer}}, \ and\ \bibinfo {author} {\bibfnamefont {Karl~E.}\ \bibnamefont
  {Taylor}}} (\bibinfo {year} {2014}),\ \bibfield  {title} {\enquote {\bibinfo
  {title} {Quantifying underestimates of long-term upper-ocean warming},}\
  }\href {\doibase 10.1038/nclimate2389} {\bibfield  {journal} {\bibinfo
  {journal} {Nature Climate Change}\ }\textbf {\bibinfo {volume} {4}}~(\bibinfo
  {number} {11}),\ \bibinfo {pages} {999--1005}}\BibitemShut {NoStop}%
\bibitem [{\citenamefont {Einstein}(1905)}]{ref:Einstein05}%
  \BibitemOpen
  \bibfield  {author} {\bibinfo {author} {\bibnamefont {Einstein},
  \bibfnamefont {A}}} (\bibinfo {year} {1905}),\ \bibfield  {title} {\enquote
  {\bibinfo {title} {Über die von der molekularkinetischen theorie der wärme
  geforderte bewegung von in ruhenden flüssigkeiten suspendierten teilchen},}\
  }\href {\doibase 10.1002/andp.19053220806} {\bibfield  {journal} {\bibinfo
  {journal} {Annalen der Physik}\ }\textbf {\bibinfo {volume} {322}}~(\bibinfo
  {number} {8}),\ \bibinfo {pages} {549--560}}\BibitemShut {NoStop}%
\bibitem [{\citenamefont {Epstein}(1969)}]{ref:Epstein69}%
  \BibitemOpen
  \bibfield  {author} {\bibinfo {author} {\bibnamefont {Epstein}, \bibfnamefont
  {Edward~S}}} (\bibinfo {year} {1969}),\ \bibfield  {title} {\enquote
  {\bibinfo {title} {Stochastic dynamic prediction},}\ }\href {\doibase
  10.1111/j.2153-3490.1969.tb00483.x} {\bibfield  {journal} {\bibinfo
  {journal} {Tellus}\ }\textbf {\bibinfo {volume} {21}}~(\bibinfo {number}
  {6}),\ \bibinfo {pages} {739--759}}\BibitemShut {NoStop}%
\bibitem [{\citenamefont {Eyring}\ \emph {et~al.}(2016)\citenamefont {Eyring},
  \citenamefont {Bony}, \citenamefont {Meehl}, \citenamefont {Senior},
  \citenamefont {Stevens}, \citenamefont {Stouffer},\ and\ \citenamefont
  {Taylor}}]{ref:CMIP6}%
  \BibitemOpen
  \bibfield  {author} {\bibinfo {author} {\bibnamefont {Eyring}, \bibfnamefont
  {V}}, \bibinfo {author} {\bibfnamefont {S.}~\bibnamefont {Bony}}, \bibinfo
  {author} {\bibfnamefont {G.~A.}\ \bibnamefont {Meehl}}, \bibinfo {author}
  {\bibfnamefont {C.~A.}\ \bibnamefont {Senior}}, \bibinfo {author}
  {\bibfnamefont {B.}~\bibnamefont {Stevens}}, \bibinfo {author} {\bibfnamefont
  {R.~J.}\ \bibnamefont {Stouffer}}, \ and\ \bibinfo {author} {\bibfnamefont
  {K.~E.}\ \bibnamefont {Taylor}}} (\bibinfo {year} {2016}),\ \bibfield
  {title} {\enquote {\bibinfo {title} {Overview of the coupled model
  intercomparison project phase 6 (cmip6) experimental design and
  organization},}\ }\href {\doibase 10.5194/gmd-9-1937-2016} {\bibfield
  {journal} {\bibinfo  {journal} {Geoscientific Model Development}\ }\textbf
  {\bibinfo {volume} {9}}~(\bibinfo {number} {5}),\ \bibinfo {pages}
  {1937--1958}}\BibitemShut {NoStop}%
\bibitem [{\citenamefont {Eyring}\ \emph {et~al.}(2019)\citenamefont {Eyring},
  \citenamefont {Cox}, \citenamefont {Flato}, \citenamefont {Gleckler},
  \citenamefont {Abramowitz}, \citenamefont {Caldwell}, \citenamefont
  {Collins}, \citenamefont {Gier}, \citenamefont {Hall}, \citenamefont
  {Hoffman}, \citenamefont {Hurtt}, \citenamefont {Jahn}, \citenamefont
  {Jones}, \citenamefont {Klein}, \citenamefont {Krasting}, \citenamefont
  {Kwiatkowski}, \citenamefont {Lorenz}, \citenamefont {Maloney}, \citenamefont
  {Meehl}, \citenamefont {Pendergrass}, \citenamefont {Pincus}, \citenamefont
  {Ruane}, \citenamefont {Russell}, \citenamefont {Sanderson}, \citenamefont
  {Santer}, \citenamefont {Sherwood}, \citenamefont {Simpson}, \citenamefont
  {Stouffer},\ and\ \citenamefont {Williamson}}]{ref:Eyring19}%
  \BibitemOpen
  \bibfield  {author} {\bibinfo {author} {\bibnamefont {Eyring}, \bibfnamefont
  {Veronika}}, \bibinfo {author} {\bibfnamefont {Peter~M.}\ \bibnamefont
  {Cox}}, \bibinfo {author} {\bibfnamefont {Gregory~M.}\ \bibnamefont {Flato}},
  \bibinfo {author} {\bibfnamefont {Peter~J.}\ \bibnamefont {Gleckler}},
  \bibinfo {author} {\bibfnamefont {Gab}\ \bibnamefont {Abramowitz}}, \bibinfo
  {author} {\bibfnamefont {Peter}\ \bibnamefont {Caldwell}}, \bibinfo {author}
  {\bibfnamefont {William~D.}\ \bibnamefont {Collins}}, \bibinfo {author}
  {\bibfnamefont {Bettina~K.}\ \bibnamefont {Gier}}, \bibinfo {author}
  {\bibfnamefont {Alex~D.}\ \bibnamefont {Hall}}, \bibinfo {author}
  {\bibfnamefont {Forrest~M.}\ \bibnamefont {Hoffman}}, \bibinfo {author}
  {\bibfnamefont {George~C.}\ \bibnamefont {Hurtt}}, \bibinfo {author}
  {\bibfnamefont {Alexandra}\ \bibnamefont {Jahn}}, \bibinfo {author}
  {\bibfnamefont {Chris~D.}\ \bibnamefont {Jones}}, \bibinfo {author}
  {\bibfnamefont {Stephen~A.}\ \bibnamefont {Klein}}, \bibinfo {author}
  {\bibfnamefont {John~P.}\ \bibnamefont {Krasting}}, \bibinfo {author}
  {\bibfnamefont {Lester}\ \bibnamefont {Kwiatkowski}}, \bibinfo {author}
  {\bibfnamefont {Ruth}\ \bibnamefont {Lorenz}}, \bibinfo {author}
  {\bibfnamefont {Eric}\ \bibnamefont {Maloney}}, \bibinfo {author}
  {\bibfnamefont {Gerald~A.}\ \bibnamefont {Meehl}}, \bibinfo {author}
  {\bibfnamefont {Angeline~G.}\ \bibnamefont {Pendergrass}}, \bibinfo {author}
  {\bibfnamefont {Robert}\ \bibnamefont {Pincus}}, \bibinfo {author}
  {\bibfnamefont {Alex~C.}\ \bibnamefont {Ruane}}, \bibinfo {author}
  {\bibfnamefont {Joellen~L.}\ \bibnamefont {Russell}}, \bibinfo {author}
  {\bibfnamefont {Benjamin~M.}\ \bibnamefont {Sanderson}}, \bibinfo {author}
  {\bibfnamefont {Benjamin~D.}\ \bibnamefont {Santer}}, \bibinfo {author}
  {\bibfnamefont {Steven~C.}\ \bibnamefont {Sherwood}}, \bibinfo {author}
  {\bibfnamefont {Isla~R.}\ \bibnamefont {Simpson}}, \bibinfo {author}
  {\bibfnamefont {Ronald~J.}\ \bibnamefont {Stouffer}}, \ and\ \bibinfo
  {author} {\bibfnamefont {Mark~S.}\ \bibnamefont {Williamson}}} (\bibinfo
  {year} {2019}),\ \bibfield  {title} {\enquote {\bibinfo {title} {Taking
  climate model evaluation to the next level},}\ }\href {\doibase
  10.1038/s41558-018-0355-y} {\bibfield  {journal} {\bibinfo  {journal} {Nature
  Climate Change}\ }\textbf {\bibinfo {volume} {9}}~(\bibinfo {number} {2}),\
  \bibinfo {pages} {102--110}}\BibitemShut {NoStop}%
\bibitem [{\citenamefont {Fasullo}\ and\ \citenamefont
  {Trenberth}(2012)}]{ref:FasulloTrenberth12}%
  \BibitemOpen
  \bibfield  {author} {\bibinfo {author} {\bibnamefont {Fasullo}, \bibfnamefont
  {J~T}}, \ and\ \bibinfo {author} {\bibfnamefont {K.~E.}\ \bibnamefont
  {Trenberth}}} (\bibinfo {year} {2012}),\ \bibfield  {title} {\enquote
  {\bibinfo {title} {A less cloudy future: The role of subtropical subsidence
  in climate sensitivity},}\ }\href {\doibase 10.1126/science.1227465}
  {\bibfield  {journal} {\bibinfo  {journal} {Science}\ }\textbf {\bibinfo
  {volume} {338}}~(\bibinfo {number} {6108}),\ \bibinfo {pages}
  {792--794}}\BibitemShut {NoStop}%
\bibitem [{\citenamefont {Feldstein}(2000)}]{ref:Feldstein00}%
  \BibitemOpen
  \bibfield  {author} {\bibinfo {author} {\bibnamefont {Feldstein},
  \bibfnamefont {Steven~B}}} (\bibinfo {year} {2000}),\ \bibfield  {title}
  {\enquote {\bibinfo {title} {The timescale, power spectra, and climate noise
  properties of teleconnection patterns},}\ }\href {\doibase
  10.1175/1520-0442(2000)013<4430:TTPSAC>2.0.CO;2} {\bibfield  {journal}
  {\bibinfo  {journal} {Journal of Climate}\ }\textbf {\bibinfo {volume}
  {13}}~(\bibinfo {number} {24}),\ \bibinfo {pages} {4430--4440}}\BibitemShut
  {NoStop}%
\bibitem [{\citenamefont {Felix}\ \emph {et~al.}(2018)\citenamefont {Felix},
  \citenamefont {Kraeussl},\ and\ \citenamefont {Stork}}]{ref:Felix18}%
  \BibitemOpen
  \bibfield  {author} {\bibinfo {author} {\bibnamefont {Felix}, \bibfnamefont
  {Luiz F~F}}, \bibinfo {author} {\bibfnamefont {Roman}\ \bibnamefont
  {Kraeussl}}, \ and\ \bibinfo {author} {\bibfnamefont {Philip~A.}\
  \bibnamefont {Stork}}} (\bibinfo {year} {2018}),\ \bibfield  {title}
  {\enquote {\bibinfo {title} {Predictable biases in macroeconomic forecasts
  and their impact across asset classes (november 20, 2018).}}\ }\href
  {\doibase http://dx.doi.org/10.2139/ssrn.3008976} {\bibfield  {journal}
  {\bibinfo  {journal} {SSRN}\
  }http://dx.doi.org/10.2139/ssrn.3008976}\BibitemShut {NoStop}%
\bibitem [{\citenamefont {Fl{\"{a}}schner}\ \emph {et~al.}(2016)\citenamefont
  {Fl{\"{a}}schner}, \citenamefont {Mauritsen},\ and\ \citenamefont
  {Stevens}}]{ref:Flaschner2016}%
  \BibitemOpen
  \bibfield  {author} {\bibinfo {author} {\bibnamefont {Fl{\"{a}}schner},
  \bibfnamefont {Dagmar}}, \bibinfo {author} {\bibfnamefont {Thorsten}\
  \bibnamefont {Mauritsen}}, \ and\ \bibinfo {author} {\bibfnamefont {Bjorn}\
  \bibnamefont {Stevens}}} (\bibinfo {year} {2016}),\ \bibfield  {title}
  {\enquote {\bibinfo {title} {{Understanding the Intermodel Spread in
  Global-Mean Hydrological Sensitivity*}},}\ }\href {\doibase
  10.1175/JCLI-D-15-0351.1} {\bibfield  {journal} {\bibinfo  {journal} {Journal
  of Climate}\ }\textbf {\bibinfo {volume} {29}},\ \bibinfo {pages}
  {801--817}}\BibitemShut {NoStop}%
\bibitem [{\citenamefont {Forster}\ \emph {et~al.}(2019)\citenamefont
  {Forster}, \citenamefont {Maycock}, \citenamefont {McKenna},\ and\
  \citenamefont {Smith}}]{ref:Forster19}%
  \BibitemOpen
  \bibfield  {author} {\bibinfo {author} {\bibnamefont {Forster}, \bibfnamefont
  {Piers~M}}, \bibinfo {author} {\bibfnamefont {Amanda~C.}\ \bibnamefont
  {Maycock}}, \bibinfo {author} {\bibfnamefont {Christine~M.}\ \bibnamefont
  {McKenna}}, \ and\ \bibinfo {author} {\bibfnamefont {Christopher~J.}\
  \bibnamefont {Smith}}} (\bibinfo {year} {2019}),\ \bibfield  {title}
  {\enquote {\bibinfo {title} {{Latest climate models confirm need for urgent
  mitigation}},}\ }\href {\doibase 10.1038/s41558-019-0660-0} {\bibinfo
  {journal} {Nature Climate Change}\ ,\ \bibinfo {pages} {1--4}}\BibitemShut
  {NoStop}%
\bibitem [{\citenamefont {Frost}\ and\ \citenamefont
  {Thompson}(2000)}]{ref:Frost00}%
  \BibitemOpen
\bibfield  {journal} {  }\bibfield  {author} {\bibinfo {author} {\bibnamefont
  {Frost}, \bibfnamefont {Chris}}, \ and\ \bibinfo {author} {\bibfnamefont
  {Simon~G.}\ \bibnamefont {Thompson}}} (\bibinfo {year} {2000}),\ \bibfield
  {title} {\enquote {\bibinfo {title} {Correcting for regression dilution bias:
  Comparison of methods for a single predictor variable},}\ }\href
  {http://www.jstor.org/stable/2680496} {\bibfield  {journal} {\bibinfo
  {journal} {Journal of the Royal Statistical Society. Series A (Statistics in
  Society)}\ }\textbf {\bibinfo {volume} {163}}~(\bibinfo {number} {2}),\
  \bibinfo {pages} {173--189}}\BibitemShut {NoStop}%
\bibitem [{\citenamefont {Gettelman}\ and\ \citenamefont
  {Sherwood}(2016)}]{ref:Gettelman2016}%
  \BibitemOpen
  \bibfield  {author} {\bibinfo {author} {\bibnamefont {Gettelman},
  \bibfnamefont {A}}, \ and\ \bibinfo {author} {\bibfnamefont {S.~C.}\
  \bibnamefont {Sherwood}}} (\bibinfo {year} {2016}),\ \bibfield  {title}
  {\enquote {\bibinfo {title} {{Processes Responsible for Cloud Feedback}},}\
  }\href {\doibase 10.1007/s40641-016-0052-8} {\bibfield  {journal} {\bibinfo
  {journal} {Current Climate Change Reports}\ }\textbf {\bibinfo {volume}
  {2}}~(\bibinfo {number} {4}),\ \bibinfo {pages} {179--189}}\BibitemShut
  {NoStop}%
\bibitem [{\citenamefont {Ghil}\ and\ \citenamefont
  {Lucarini}(2020)}]{ref:Ghil&Lucarini20}%
  \BibitemOpen
  \bibfield  {author} {\bibinfo {author} {\bibnamefont {Ghil}, \bibfnamefont
  {Michael}}, \ and\ \bibinfo {author} {\bibfnamefont {Valerio}\ \bibnamefont
  {Lucarini}}} (\bibinfo {year} {2020}),\ \bibfield  {title} {\enquote
  {\bibinfo {title} {The physics of climate variability and climate change},}\
  }\href {\doibase 10.1103/RevModPhys.92.035002} {\bibfield  {journal}
  {\bibinfo  {journal} {Reviews of Modern Physics}\ }\textbf {\bibinfo {volume}
  {92}}~(\bibinfo {number} {3}),\ \bibinfo {pages} {035002}}\BibitemShut
  {NoStop}%
\bibitem [{\citenamefont {Gordon}\ and\ \citenamefont
  {Klein}(2014)}]{ref:Gordon14}%
  \BibitemOpen
  \bibfield  {author} {\bibinfo {author} {\bibnamefont {Gordon}, \bibfnamefont
  {N~D}}, \ and\ \bibinfo {author} {\bibfnamefont {S.~A.}\ \bibnamefont
  {Klein}}} (\bibinfo {year} {2014}),\ \bibfield  {title} {\enquote {\bibinfo
  {title} {Low-cloud optical depth feedback in climate models},}\ }\href
  {\doibase 10.1002/2013jd021052} {\bibfield  {journal} {\bibinfo  {journal}
  {Journal of Geophysical Research-Atmospheres}\ }\textbf {\bibinfo {volume}
  {119}}~(\bibinfo {number} {10}),\ \bibinfo {pages} {6052--6065}}\BibitemShut
  {NoStop}%
\bibitem [{\citenamefont {Gregory}(2000)}]{ref:Gregory00}%
  \BibitemOpen
  \bibfield  {author} {\bibinfo {author} {\bibnamefont {Gregory}, \bibfnamefont
  {J~M}}} (\bibinfo {year} {2000}),\ \bibfield  {title} {\enquote {\bibinfo
  {title} {Vertical heat transports in the ocean and their effect on
  time-dependent climate change},}\ }\href {\doibase 10.1007/s003820000059}
  {\bibfield  {journal} {\bibinfo  {journal} {Climate Dynamics}\ }\textbf
  {\bibinfo {volume} {16}}~(\bibinfo {number} {7}),\ \bibinfo {pages}
  {501--515}}\BibitemShut {NoStop}%
\bibitem [{\citenamefont {Grise}\ \emph {et~al.}(2015)\citenamefont {Grise},
  \citenamefont {Polvani},\ and\ \citenamefont {Fasullo}}]{ref:Grise2015}%
  \BibitemOpen
  \bibfield  {author} {\bibinfo {author} {\bibnamefont {Grise}, \bibfnamefont
  {Kevin~M}}, \bibinfo {author} {\bibfnamefont {Lorenzo~M.}\ \bibnamefont
  {Polvani}}, \ and\ \bibinfo {author} {\bibfnamefont {John~T.}\ \bibnamefont
  {Fasullo}}} (\bibinfo {year} {2015}),\ \bibfield  {title} {\enquote {\bibinfo
  {title} {{Reexamining the relationship between climate sensitivity and the
  Southern Hemisphere radiation budget in CMIP models}},}\ }\href {\doibase
  10.1175/JCLI-D-15-0031.1} {\bibfield  {journal} {\bibinfo  {journal} {Journal
  of Climate}\ }\textbf {\bibinfo {volume} {28}}~(\bibinfo {number} {23}),\
  \bibinfo {pages} {9298--9312}}\BibitemShut {NoStop}%
\bibitem [{\citenamefont {Hall}\ and\ \citenamefont
  {Qu}(2006)}]{ref:Hall&Qu06}%
  \BibitemOpen
  \bibfield  {author} {\bibinfo {author} {\bibnamefont {Hall}, \bibfnamefont
  {A}}, \ and\ \bibinfo {author} {\bibfnamefont {X.}~\bibnamefont {Qu}}}
  (\bibinfo {year} {2006}),\ \bibfield  {title} {\enquote {\bibinfo {title}
  {Using the current seasonal cycle to constrain snow albedo feedback in future
  climate change},}\ }\href {\doibase 10.1029/2005gl025127} {\bibfield
  {journal} {\bibinfo  {journal} {Geophysical Research Letters}\ }\textbf
  {\bibinfo {volume} {33}}~(\bibinfo {number} {3}),\
  10.1029/2005gl025127}\BibitemShut {NoStop}%
\bibitem [{\citenamefont {Hall}(2004)}]{ref:Hall2004}%
  \BibitemOpen
  \bibfield  {author} {\bibinfo {author} {\bibnamefont {Hall}, \bibfnamefont
  {Alex}}} (\bibinfo {year} {2004}),\ \bibfield  {title} {\enquote {\bibinfo
  {title} {{The Role of Surface Albedo Feedback in Climate}},}\ }\href
  {\doibase 10.1175/1520-0442(2004)017 <1550:TROSAF>2.0.CO;2} {\bibfield
  {journal} {\bibinfo  {journal} {Journal of Climate}\ }\textbf {\bibinfo
  {volume} {17}},\ \bibinfo {pages} {1550--1568}}\BibitemShut {NoStop}%
\bibitem [{\citenamefont {Hall}\ \emph {et~al.}(2019)\citenamefont {Hall},
  \citenamefont {Cox}, \citenamefont {Huntingford},\ and\ \citenamefont
  {Klein}}]{ref:Hall19}%
  \BibitemOpen
  \bibfield  {author} {\bibinfo {author} {\bibnamefont {Hall}, \bibfnamefont
  {Alex}}, \bibinfo {author} {\bibfnamefont {Peter}\ \bibnamefont {Cox}},
  \bibinfo {author} {\bibfnamefont {Chris}\ \bibnamefont {Huntingford}}, \ and\
  \bibinfo {author} {\bibfnamefont {Stephen}\ \bibnamefont {Klein}}} (\bibinfo
  {year} {2019}),\ \bibfield  {title} {\enquote {\bibinfo {title} {{Progressing
  emergent constraints on future climate change}},}\ }\href {\doibase
  10.1038/s41558-019-0436-6} {\bibfield  {journal} {\bibinfo  {journal} {Nature
  Climate Change}\ }\textbf {\bibinfo {volume} {9}}~(\bibinfo {number} {4}),\
  \bibinfo {pages} {269--278}}\BibitemShut {NoStop}%
\bibitem [{\citenamefont {Hargreaves}\ \emph {et~al.}(2012)\citenamefont
  {Hargreaves}, \citenamefont {Annan}, \citenamefont {Yoshimori},\ and\
  \citenamefont {Abe-Ouchi}}]{ref:Hargreaves12}%
  \BibitemOpen
  \bibfield  {author} {\bibinfo {author} {\bibnamefont {Hargreaves},
  \bibfnamefont {J~C}}, \bibinfo {author} {\bibfnamefont {J.~D.}\ \bibnamefont
  {Annan}}, \bibinfo {author} {\bibfnamefont {M.}~\bibnamefont {Yoshimori}}, \
  and\ \bibinfo {author} {\bibfnamefont {A.}~\bibnamefont {Abe-Ouchi}}}
  (\bibinfo {year} {2012}),\ \bibfield  {title} {\enquote {\bibinfo {title}
  {{Can the Last Glacial Maximum constrain climate sensitivity?}}}\ }\href
  {\doibase 10.1029/2012GL053872} {\bibfield  {journal} {\bibinfo  {journal}
  {Geophysical Research Letters}\ }\textbf {\bibinfo {volume} {39}}~(\bibinfo
  {number} {24}),\ \bibinfo {pages} {1--5}}\BibitemShut {NoStop}%
\bibitem [{\citenamefont {Harrison}\ \emph {et~al.}(2015)\citenamefont
  {Harrison}, \citenamefont {Bartlein}, \citenamefont {Izumi}, \citenamefont
  {Li}, \citenamefont {Annan}, \citenamefont {Hargreaves}, \citenamefont
  {Braconnot},\ and\ \citenamefont {Kageyama}}]{ref:Harrison2015}%
  \BibitemOpen
  \bibfield  {author} {\bibinfo {author} {\bibnamefont {Harrison},
  \bibfnamefont {S~P}}, \bibinfo {author} {\bibfnamefont {P.~J.}\ \bibnamefont
  {Bartlein}}, \bibinfo {author} {\bibfnamefont {K.}~\bibnamefont {Izumi}},
  \bibinfo {author} {\bibfnamefont {G.}~\bibnamefont {Li}}, \bibinfo {author}
  {\bibfnamefont {J.}~\bibnamefont {Annan}}, \bibinfo {author} {\bibfnamefont
  {J.}~\bibnamefont {Hargreaves}}, \bibinfo {author} {\bibfnamefont
  {P.}~\bibnamefont {Braconnot}}, \ and\ \bibinfo {author} {\bibfnamefont
  {M.}~\bibnamefont {Kageyama}}} (\bibinfo {year} {2015}),\ \bibfield  {title}
  {\enquote {\bibinfo {title} {{Evaluation of CMIP5 palaeo-simulations to
  improve climate projections}},}\ }\href {\doibase 10.1038/nclimate2649}
  {\bibfield  {journal} {\bibinfo  {journal} {Nature Climate Change}\ }\textbf
  {\bibinfo {volume} {5}}~(\bibinfo {number} {8}),\ \bibinfo {pages}
  {735--743}}\BibitemShut {NoStop}%
\bibitem [{\citenamefont {Hasselmann}(1976)}]{ref:Hasselmann76}%
  \BibitemOpen
  \bibfield  {author} {\bibinfo {author} {\bibnamefont {Hasselmann},
  \bibfnamefont {K}}} (\bibinfo {year} {1976}),\ \bibfield  {title} {\enquote
  {\bibinfo {title} {Stochastic climate models. part {I}. theory},}\
  }\href@noop {} {\bibfield  {journal} {\bibinfo  {journal} {Tellus}\ }\textbf
  {\bibinfo {volume} {28}},\ \bibinfo {pages} {473--484}}\BibitemShut {NoStop}%
\bibitem [{\citenamefont {Hasselmann}(1988)}]{ref:Hasselmann88}%
  \BibitemOpen
  \bibfield  {author} {\bibinfo {author} {\bibnamefont {Hasselmann},
  \bibfnamefont {K}}} (\bibinfo {year} {1988}),\ \bibfield  {title} {\enquote
  {\bibinfo {title} {Pips and pops: The reduction of complex dynamical systems
  using principal interaction and oscillation patterns},}\ }\href {\doibase
  10.1029/JD093iD09p11015} {\bibfield  {journal} {\bibinfo  {journal} {Journal
  of Geophysical Research: Atmospheres}\ }\textbf {\bibinfo {volume}
  {93}}~(\bibinfo {number} {D9}),\ \bibinfo {pages} {11015--11021}}\BibitemShut
  {NoStop}%
\bibitem [{\citenamefont {Hawkins}\ and\ \citenamefont
  {Sutton}(2009)}]{ref:Hawkins&Sutton09}%
  \BibitemOpen
  \bibfield  {author} {\bibinfo {author} {\bibnamefont {Hawkins}, \bibfnamefont
  {Ed}}, \ and\ \bibinfo {author} {\bibfnamefont {Rowan}\ \bibnamefont
  {Sutton}}} (\bibinfo {year} {2009}),\ \bibfield  {title} {\enquote {\bibinfo
  {title} {The potential to narrow uncertainty in regional climate
  predictions},}\ }\href {\doibase 10.1175/2009BAMS2607.1} {\bibfield
  {journal} {\bibinfo  {journal} {Bulletin of the American Meteorological
  Society}\ }\textbf {\bibinfo {volume} {90}}~(\bibinfo {number} {8}),\
  \bibinfo {pages} {1095--1108}}\BibitemShut {NoStop}%
\bibitem [{\citenamefont {Hawkins}\ and\ \citenamefont
  {Sutton}(2011)}]{ref:Hawkins&Sutton11}%
  \BibitemOpen
  \bibfield  {author} {\bibinfo {author} {\bibnamefont {Hawkins}, \bibfnamefont
  {Ed}}, \ and\ \bibinfo {author} {\bibfnamefont {Rowan}\ \bibnamefont
  {Sutton}}} (\bibinfo {year} {2011}),\ \bibfield  {title} {\enquote {\bibinfo
  {title} {The potential to narrow uncertainty in projections of regional
  precipitation change},}\ }\href {\doibase 10.1007/s00382-010-0810-6}
  {\bibfield  {journal} {\bibinfo  {journal} {Climate Dynamics}\ }\textbf
  {\bibinfo {volume} {37}}~(\bibinfo {number} {1}),\ \bibinfo {pages}
  {407--418}}\BibitemShut {NoStop}%
\bibitem [{\citenamefont {Hennermann}(last edited July
  2018)}]{ref:ERA5documentation}%
  \BibitemOpen
  \bibfield  {author} {\bibinfo {author} {\bibnamefont {Hennermann},
  \bibfnamefont {Karl}}} (\bibinfo {year} {last edited July 2018}),\ \href
  {https://confluence.ecmwf.int/display/CKB/ERA5%3A+uncertainty+estimation}
  {\enquote {\bibinfo {title} {Era5 data documentation, era5: uncertainty
  estimation v14},}\ }\BibitemShut {NoStop}%
\bibitem [{\citenamefont {Herger}\ \emph {et~al.}(2018)\citenamefont {Herger},
  \citenamefont {Abramowitz}, \citenamefont {Knutti}, \citenamefont
  {Ang\'elil}, \citenamefont {Lehmann},\ and\ \citenamefont
  {Sanderson}}]{ref:Herger17}%
  \BibitemOpen
  \bibfield  {author} {\bibinfo {author} {\bibnamefont {Herger}, \bibfnamefont
  {N}}, \bibinfo {author} {\bibfnamefont {G.}~\bibnamefont {Abramowitz}},
  \bibinfo {author} {\bibfnamefont {R.}~\bibnamefont {Knutti}}, \bibinfo
  {author} {\bibfnamefont {O.}~\bibnamefont {Ang\'elil}}, \bibinfo {author}
  {\bibfnamefont {K.}~\bibnamefont {Lehmann}}, \ and\ \bibinfo {author}
  {\bibfnamefont {B.~M.}\ \bibnamefont {Sanderson}}} (\bibinfo {year} {2018}),\
  \bibfield  {title} {\enquote {\bibinfo {title} {Selecting a climate model
  subset to optimise key ensemble properties},}\ }\href {\doibase
  10.5194/esd-9-135-2018} {\bibfield  {journal} {\bibinfo  {journal} {Earth
  System Dynamics}\ }\textbf {\bibinfo {volume} {9}}~(\bibinfo {number} {1}),\
  \bibinfo {pages} {135--151}}\BibitemShut {NoStop}%
\bibitem [{\citenamefont {Hoffman}\ \emph {et~al.}(2014)\citenamefont
  {Hoffman}, \citenamefont {Randerson}, \citenamefont {Arora}, \citenamefont
  {Bao}, \citenamefont {Cadule}, \citenamefont {Ji}, \citenamefont {Jones},
  \citenamefont {Kawamiya}, \citenamefont {Khatiwala}, \citenamefont {Lindsay},
  \citenamefont {Obata}, \citenamefont {Shevliakova}, \citenamefont {Six},
  \citenamefont {Tjiputra}, \citenamefont {Volodin},\ and\ \citenamefont
  {Wu}}]{ref:Hoffman14}%
  \BibitemOpen
  \bibfield  {author} {\bibinfo {author} {\bibnamefont {Hoffman}, \bibfnamefont
  {F~M}}, \bibinfo {author} {\bibfnamefont {J.~T.}\ \bibnamefont {Randerson}},
  \bibinfo {author} {\bibfnamefont {V.~K.}\ \bibnamefont {Arora}}, \bibinfo
  {author} {\bibfnamefont {Q.}~\bibnamefont {Bao}}, \bibinfo {author}
  {\bibfnamefont {P.}~\bibnamefont {Cadule}}, \bibinfo {author} {\bibfnamefont
  {D.}~\bibnamefont {Ji}}, \bibinfo {author} {\bibfnamefont {C.~D.}\
  \bibnamefont {Jones}}, \bibinfo {author} {\bibfnamefont {M.}~\bibnamefont
  {Kawamiya}}, \bibinfo {author} {\bibfnamefont {S.}~\bibnamefont {Khatiwala}},
  \bibinfo {author} {\bibfnamefont {K.}~\bibnamefont {Lindsay}}, \bibinfo
  {author} {\bibfnamefont {A.}~\bibnamefont {Obata}}, \bibinfo {author}
  {\bibfnamefont {E.}~\bibnamefont {Shevliakova}}, \bibinfo {author}
  {\bibfnamefont {K.~D.}\ \bibnamefont {Six}}, \bibinfo {author} {\bibfnamefont
  {J.~F.}\ \bibnamefont {Tjiputra}}, \bibinfo {author} {\bibfnamefont {E.~M.}\
  \bibnamefont {Volodin}}, \ and\ \bibinfo {author} {\bibfnamefont
  {T.}~\bibnamefont {Wu}}} (\bibinfo {year} {2014}),\ \bibfield  {title}
  {\enquote {\bibinfo {title} {Causes and implications of persistent
  atmospheric carbon dioxide biases in {Earth} system models},}\ }\href
  {\doibase 10.1002/2013JG002381} {\bibfield  {journal} {\bibinfo  {journal}
  {Journal of Geophysical Research: Biogeosciences}\ }\textbf {\bibinfo
  {volume} {119}}~(\bibinfo {number} {2}),\ \bibinfo {pages}
  {141--162}}\BibitemShut {NoStop}%
\bibitem [{\citenamefont {Holland}\ and\ \citenamefont
  {Bitz}(2003)}]{ref:Holland03}%
  \BibitemOpen
  \bibfield  {author} {\bibinfo {author} {\bibnamefont {Holland}, \bibfnamefont
  {M~M}}, \ and\ \bibinfo {author} {\bibfnamefont {C.~M.}\ \bibnamefont
  {Bitz}}} (\bibinfo {year} {2003}),\ \bibfield  {title} {\enquote {\bibinfo
  {title} {{Polar amplification of climate change in coupled models}},}\ }\href
  {\doibase 10.1007/s00382-003-0332-6} {\bibfield  {journal} {\bibinfo
  {journal} {Climate Dynamics}\ }\textbf {\bibinfo {volume} {21}},\ \bibinfo
  {pages} {221--232}}\BibitemShut {NoStop}%
\bibitem [{\citenamefont {Huber}\ \emph {et~al.}(2011)\citenamefont {Huber},
  \citenamefont {Mahlstein}, \citenamefont {Wild}, \citenamefont {Fasullo},\
  and\ \citenamefont {Knutti}}]{ref:Huber11}%
  \BibitemOpen
  \bibfield  {author} {\bibinfo {author} {\bibnamefont {Huber}, \bibfnamefont
  {Markus}}, \bibinfo {author} {\bibfnamefont {Irina}\ \bibnamefont
  {Mahlstein}}, \bibinfo {author} {\bibfnamefont {Martin}\ \bibnamefont
  {Wild}}, \bibinfo {author} {\bibfnamefont {John}\ \bibnamefont {Fasullo}}, \
  and\ \bibinfo {author} {\bibfnamefont {Reto}\ \bibnamefont {Knutti}}}
  (\bibinfo {year} {2011}),\ \bibfield  {title} {\enquote {\bibinfo {title}
  {Constraints on climate sensitivity from radiation patterns in climate
  models},}\ }\href {\doibase 10.1175/2010JCLI3403.1} {\bibfield  {journal}
  {\bibinfo  {journal} {Journal of Climate}\ }\textbf {\bibinfo {volume}
  {24}}~(\bibinfo {number} {4}),\ \bibinfo {pages} {1034--1052}}\BibitemShut
  {NoStop}%
\bibitem [{\citenamefont {Huntingford}\ \emph {et~al.}(2013)\citenamefont
  {Huntingford}, \citenamefont {Jones}, \citenamefont {Livina}, \citenamefont
  {Lenton},\ and\ \citenamefont {Cox}}]{ref:Huntingford2013}%
  \BibitemOpen
  \bibfield  {author} {\bibinfo {author} {\bibnamefont {Huntingford},
  \bibfnamefont {Chris}}, \bibinfo {author} {\bibfnamefont {Philip~D.}\
  \bibnamefont {Jones}}, \bibinfo {author} {\bibfnamefont {Valerie~N.}\
  \bibnamefont {Livina}}, \bibinfo {author} {\bibfnamefont {Timothy~M.}\
  \bibnamefont {Lenton}}, \ and\ \bibinfo {author} {\bibfnamefont {Peter~M.}\
  \bibnamefont {Cox}}} (\bibinfo {year} {2013}),\ \bibfield  {title} {\enquote
  {\bibinfo {title} {{No increase in global temperature variability despite
  changing regional patterns}},}\ }\href {\doibase 10.1038/nature12310}
  {\bibfield  {journal} {\bibinfo  {journal} {Nature}\ }\textbf {\bibinfo
  {volume} {500}}~(\bibinfo {number} {7462}),\ \bibinfo {pages}
  {327--330}}\BibitemShut {NoStop}%
\bibitem [{\citenamefont {IPCC}(2013)}]{ref:IPCC_AR5}%
  \BibitemOpen
  \bibfield  {author} {\bibinfo {author} {\bibnamefont {IPCC},}} (\bibinfo
  {year} {2013}),\ \href {\doibase 10.1017/CBO9781107415324} {\emph {\bibinfo
  {title} {Climate Change 2013: The Physical Science Basis. Contribution of
  Working Group I to the Fifth Assessment Report of the Intergovernmental Panel
  on Climate Change}}}\ (\bibinfo  {publisher} {Cambridge University Press},\
  \bibinfo {address} {Cambridge, United Kingdom and New York, NY,
  USA})\BibitemShut {NoStop}%
\bibitem [{\citenamefont {Johnson}(1928)}]{ref:Johnson28}%
  \BibitemOpen
  \bibfield  {author} {\bibinfo {author} {\bibnamefont {Johnson}, \bibfnamefont
  {J~B}}} (\bibinfo {year} {1928}),\ \bibfield  {title} {\enquote {\bibinfo
  {title} {Thermal agitation of electricity in conductors},}\ }\href {\doibase
  10.1103/PhysRev.32.97} {\bibfield  {journal} {\bibinfo  {journal} {Phys.
  Rev.}\ }\textbf {\bibinfo {volume} {32}},\ \bibinfo {pages}
  {97--109}}\BibitemShut {NoStop}%
\bibitem [{\citenamefont {Jones}\ \emph {et~al.}(2016)\citenamefont {Jones},
  \citenamefont {Arora}, \citenamefont {Friedlingstein}, \citenamefont {Bopp},
  \citenamefont {Brovkin}, \citenamefont {Dunne}, \citenamefont {Graven},
  \citenamefont {Hoffman}, \citenamefont {Ilyina}, \citenamefont {John},
  \citenamefont {Jung}, \citenamefont {Kawamiya}, \citenamefont {Koven},
  \citenamefont {Pongratz}, \citenamefont {Raddatz}, \citenamefont
  {Randerson},\ and\ \citenamefont {Zaehle}}]{ref:Jones16}%
  \BibitemOpen
  \bibfield  {author} {\bibinfo {author} {\bibnamefont {Jones}, \bibfnamefont
  {C~D}}, \bibinfo {author} {\bibfnamefont {V.}~\bibnamefont {Arora}}, \bibinfo
  {author} {\bibfnamefont {P.}~\bibnamefont {Friedlingstein}}, \bibinfo
  {author} {\bibfnamefont {L.}~\bibnamefont {Bopp}}, \bibinfo {author}
  {\bibfnamefont {V.}~\bibnamefont {Brovkin}}, \bibinfo {author} {\bibfnamefont
  {J.}~\bibnamefont {Dunne}}, \bibinfo {author} {\bibfnamefont
  {H.}~\bibnamefont {Graven}}, \bibinfo {author} {\bibfnamefont
  {F.}~\bibnamefont {Hoffman}}, \bibinfo {author} {\bibfnamefont
  {T.}~\bibnamefont {Ilyina}}, \bibinfo {author} {\bibfnamefont {J.~G.}\
  \bibnamefont {John}}, \bibinfo {author} {\bibfnamefont {M.}~\bibnamefont
  {Jung}}, \bibinfo {author} {\bibfnamefont {M.}~\bibnamefont {Kawamiya}},
  \bibinfo {author} {\bibfnamefont {C.}~\bibnamefont {Koven}}, \bibinfo
  {author} {\bibfnamefont {J.}~\bibnamefont {Pongratz}}, \bibinfo {author}
  {\bibfnamefont {T.}~\bibnamefont {Raddatz}}, \bibinfo {author} {\bibfnamefont
  {J.~T.}\ \bibnamefont {Randerson}}, \ and\ \bibinfo {author} {\bibfnamefont
  {S.}~\bibnamefont {Zaehle}}} (\bibinfo {year} {2016}),\ \bibfield  {title}
  {\enquote {\bibinfo {title} {C4mip -- the coupled climate--carbon cycle model
  intercomparison project: experimental protocol for cmip6},}\ }\href {\doibase
  10.5194/gmd-9-2853-2016} {\bibfield  {journal} {\bibinfo  {journal}
  {Geoscientific Model Development}\ }\textbf {\bibinfo {volume} {9}}~(\bibinfo
  {number} {8}),\ \bibinfo {pages} {2853--2880}}\BibitemShut {NoStop}%
\bibitem [{\citenamefont {Kamae}\ \emph {et~al.}(2016)\citenamefont {Kamae},
  \citenamefont {Shiogama}, \citenamefont {Watanabe}, \citenamefont {Ogura},
  \citenamefont {Yokohata},\ and\ \citenamefont {Kimoto}}]{ref:Kamae16}%
  \BibitemOpen
  \bibfield  {author} {\bibinfo {author} {\bibnamefont {Kamae}, \bibfnamefont
  {Y}}, \bibinfo {author} {\bibfnamefont {H.}~\bibnamefont {Shiogama}},
  \bibinfo {author} {\bibfnamefont {M.}~\bibnamefont {Watanabe}}, \bibinfo
  {author} {\bibfnamefont {T.}~\bibnamefont {Ogura}}, \bibinfo {author}
  {\bibfnamefont {T.}~\bibnamefont {Yokohata}}, \ and\ \bibinfo {author}
  {\bibfnamefont {M.}~\bibnamefont {Kimoto}}} (\bibinfo {year} {2016}),\
  \bibfield  {title} {\enquote {\bibinfo {title} {Lower-tropospheric mixing as
  a constraint on cloud feedback in a multiparameter multiphysics ensemble},}\
  }\href {\doibase 10.1175/jcli-d-16-0042.1} {\bibfield  {journal} {\bibinfo
  {journal} {Journal of Climate}\ }\textbf {\bibinfo {volume} {29}}~(\bibinfo
  {number} {17}),\ \bibinfo {pages} {6259--6275}}\BibitemShut {NoStop}%
\bibitem [{\citenamefont {Kay}\ \emph {et~al.}(2015)\citenamefont {Kay},
  \citenamefont {Deser}, \citenamefont {Phillips}, \citenamefont {Mai},
  \citenamefont {Hannay}, \citenamefont {Strand}, \citenamefont {Arblaster},
  \citenamefont {Bates}, \citenamefont {Danabasoglu}, \citenamefont {Edwards},
  \citenamefont {Holland}, \citenamefont {Kushner}, \citenamefont {Lamarque},
  \citenamefont {Lawrence}, \citenamefont {Lindsay}, \citenamefont {Middleton},
  \citenamefont {Munoz}, \citenamefont {Neale}, \citenamefont {Oleson},
  \citenamefont {Polvani},\ and\ \citenamefont {Vertenstein}}]{ref:Kay15}%
  \BibitemOpen
  \bibfield  {author} {\bibinfo {author} {\bibnamefont {Kay}, \bibfnamefont
  {J~E}}, \bibinfo {author} {\bibfnamefont {C.}~\bibnamefont {Deser}}, \bibinfo
  {author} {\bibfnamefont {A.}~\bibnamefont {Phillips}}, \bibinfo {author}
  {\bibfnamefont {A.}~\bibnamefont {Mai}}, \bibinfo {author} {\bibfnamefont
  {C.}~\bibnamefont {Hannay}}, \bibinfo {author} {\bibfnamefont
  {G.}~\bibnamefont {Strand}}, \bibinfo {author} {\bibfnamefont {J.~M.}\
  \bibnamefont {Arblaster}}, \bibinfo {author} {\bibfnamefont {S.~C.}\
  \bibnamefont {Bates}}, \bibinfo {author} {\bibfnamefont {G.}~\bibnamefont
  {Danabasoglu}}, \bibinfo {author} {\bibfnamefont {J.}~\bibnamefont
  {Edwards}}, \bibinfo {author} {\bibfnamefont {M.}~\bibnamefont {Holland}},
  \bibinfo {author} {\bibfnamefont {P.}~\bibnamefont {Kushner}}, \bibinfo
  {author} {\bibfnamefont {J.~F.}\ \bibnamefont {Lamarque}}, \bibinfo {author}
  {\bibfnamefont {D.}~\bibnamefont {Lawrence}}, \bibinfo {author}
  {\bibfnamefont {K.}~\bibnamefont {Lindsay}}, \bibinfo {author} {\bibfnamefont
  {A.}~\bibnamefont {Middleton}}, \bibinfo {author} {\bibfnamefont
  {E.}~\bibnamefont {Munoz}}, \bibinfo {author} {\bibfnamefont
  {R.}~\bibnamefont {Neale}}, \bibinfo {author} {\bibfnamefont
  {K.}~\bibnamefont {Oleson}}, \bibinfo {author} {\bibfnamefont
  {L.}~\bibnamefont {Polvani}}, \ and\ \bibinfo {author} {\bibfnamefont
  {M.}~\bibnamefont {Vertenstein}}} (\bibinfo {year} {2015}),\ \bibfield
  {title} {\enquote {\bibinfo {title} {The community earth system model (cesm)
  large ensemble project: A community resource for studying climate change in
  the presence of internal climate variability},}\ }\href {\doibase
  10.1175/BAMS-D-13-00255.1} {\bibfield  {journal} {\bibinfo  {journal}
  {Bulletin of the American Meteorological Society}\ }\textbf {\bibinfo
  {volume} {96}}~(\bibinfo {number} {8}),\ \bibinfo {pages}
  {1333--1349}}\BibitemShut {NoStop}%
\bibitem [{\citenamefont {Kessler}\ and\ \citenamefont
  {Tjiputra}(2016)}]{ref:Kessler2016}%
  \BibitemOpen
  \bibfield  {author} {\bibinfo {author} {\bibnamefont {Kessler}, \bibfnamefont
  {A}}, \ and\ \bibinfo {author} {\bibfnamefont {J.}~\bibnamefont {Tjiputra}}}
  (\bibinfo {year} {2016}),\ \bibfield  {title} {\enquote {\bibinfo {title}
  {{The Southern Ocean as a constraint to reduce uncertainty in future ocean
  carbon sinks}},}\ }\href {\doibase 10.5194/esd-7-295-2016} {\bibfield
  {journal} {\bibinfo  {journal} {Earth System Dynamics}\ }\textbf {\bibinfo
  {volume} {7}}~(\bibinfo {number} {2}),\ \bibinfo {pages}
  {295--312}}\BibitemShut {NoStop}%
\bibitem [{\citenamefont {Kharin}\ \emph {et~al.}(2013)\citenamefont {Kharin},
  \citenamefont {Zwiers}, \citenamefont {Zhang},\ and\ \citenamefont
  {Wehner}}]{ref:Kharin2013}%
  \BibitemOpen
  \bibfield  {author} {\bibinfo {author} {\bibnamefont {Kharin}, \bibfnamefont
  {V~V}}, \bibinfo {author} {\bibfnamefont {F~W}\ \bibnamefont {Zwiers}},
  \bibinfo {author} {\bibfnamefont {X}~\bibnamefont {Zhang}}, \ and\ \bibinfo
  {author} {\bibfnamefont {M}~\bibnamefont {Wehner}}} (\bibinfo {year}
  {2013}),\ \bibfield  {title} {\enquote {\bibinfo {title} {{Changes in
  temperature and precipitation extremes in the CMIP5 ensemble}},}\ }\href
  {\doibase 10.1007/s10584-013-0705-8} {\bibfield  {journal} {\bibinfo
  {journal} {Climatic Change}\ }\textbf {\bibinfo {volume} {119}},\ \bibinfo
  {pages} {345--357}}\BibitemShut {NoStop}%
\bibitem [{\citenamefont {Kidston}\ and\ \citenamefont
  {Gerber}(2010)}]{ref:Kidston&Gerber10}%
  \BibitemOpen
  \bibfield  {author} {\bibinfo {author} {\bibnamefont {Kidston}, \bibfnamefont
  {J}}, \ and\ \bibinfo {author} {\bibfnamefont {E.~P.}\ \bibnamefont
  {Gerber}}} (\bibinfo {year} {2010}),\ \bibfield  {title} {\enquote {\bibinfo
  {title} {Intermodel variability of the poleward shift of the austral jet
  stream in the {CMIP3} integrations linked to biases in 20th century
  climatology},}\ }\href {\doibase 10.1029/2010GL042873} {\bibfield  {journal}
  {\bibinfo  {journal} {Geophysical Research Letters}\ }\textbf {\bibinfo
  {volume} {37}}~(\bibinfo {number} {9}),\ 10.1029/2010GL042873}\BibitemShut
  {NoStop}%
\bibitem [{\citenamefont {Kirk-Davidoff}(2009)}]{ref:Kirk-Davidoff09}%
  \BibitemOpen
  \bibfield  {author} {\bibinfo {author} {\bibnamefont {Kirk-Davidoff},
  \bibfnamefont {D~B}}} (\bibinfo {year} {2009}),\ \bibfield  {title} {\enquote
  {\bibinfo {title} {On the diagnosis of climate sensitivity using observations
  of fluctuations},}\ }\href {\doibase 10.5194/acp-9-813-2009} {\bibfield
  {journal} {\bibinfo  {journal} {Atmospheric Chemistry and Physics}\ }\textbf
  {\bibinfo {volume} {9}}~(\bibinfo {number} {3}),\ \bibinfo {pages}
  {813--822}}\BibitemShut {NoStop}%
\bibitem [{\citenamefont {Klein}\ and\ \citenamefont
  {Hall}(2015)}]{ref:Klein&Hall15}%
  \BibitemOpen
  \bibfield  {author} {\bibinfo {author} {\bibnamefont {Klein}, \bibfnamefont
  {Stephen~A}}, \ and\ \bibinfo {author} {\bibfnamefont {Alex}\ \bibnamefont
  {Hall}}} (\bibinfo {year} {2015}),\ \bibfield  {title} {\enquote {\bibinfo
  {title} {Emergent constraints for cloud feedbacks},}\ }\href {\doibase
  10.1007/s40641-015-0027-1} {\bibfield  {journal} {\bibinfo  {journal}
  {Current Climate Change Reports}\ }\textbf {\bibinfo {volume} {1}}~(\bibinfo
  {number} {4}),\ \bibinfo {pages} {276--287}}\BibitemShut {NoStop}%
\bibitem [{\citenamefont {Klein}\ \emph {et~al.}(2017)\citenamefont {Klein},
  \citenamefont {Hall}, \citenamefont {Norris},\ and\ \citenamefont
  {Pincus}}]{ref:Klein2017}%
  \BibitemOpen
  \bibfield  {author} {\bibinfo {author} {\bibnamefont {Klein}, \bibfnamefont
  {Stephen~A}}, \bibinfo {author} {\bibfnamefont {Alex}\ \bibnamefont {Hall}},
  \bibinfo {author} {\bibfnamefont {Joel~R.}\ \bibnamefont {Norris}}, \ and\
  \bibinfo {author} {\bibfnamefont {Robert}\ \bibnamefont {Pincus}}} (\bibinfo
  {year} {2017}),\ \bibfield  {title} {\enquote {\bibinfo {title} {{Low-Cloud
  Feedbacks from Cloud-Controlling Factors: A Review}},}\ }\href {\doibase
  10.1007/s10712-017-9433-3} {\bibfield  {journal} {\bibinfo  {journal}
  {Surveys in Geophysics}\ }\textbf {\bibinfo {volume} {38}}~(\bibinfo {number}
  {6}),\ \bibinfo {pages} {1307--1329}}\BibitemShut {NoStop}%
\bibitem [{\citenamefont {Klein}\ \emph {et~al.}(2013)\citenamefont {Klein},
  \citenamefont {Zhang}, \citenamefont {Zelinka}, \citenamefont {Pincus},
  \citenamefont {Boyle},\ and\ \citenamefont {Gleckler}}]{ref:Klein13}%
  \BibitemOpen
  \bibfield  {author} {\bibinfo {author} {\bibnamefont {Klein}, \bibfnamefont
  {Stephen~A}}, \bibinfo {author} {\bibfnamefont {Yuying}\ \bibnamefont
  {Zhang}}, \bibinfo {author} {\bibfnamefont {Mark~D.}\ \bibnamefont
  {Zelinka}}, \bibinfo {author} {\bibfnamefont {Robert}\ \bibnamefont
  {Pincus}}, \bibinfo {author} {\bibfnamefont {James}\ \bibnamefont {Boyle}}, \
  and\ \bibinfo {author} {\bibfnamefont {Peter~J.}\ \bibnamefont {Gleckler}}}
  (\bibinfo {year} {2013}),\ \bibfield  {title} {\enquote {\bibinfo {title}
  {Are climate model simulations of clouds improving? an evaluation using the
  isccp simulator},}\ }\href {\doibase 10.1002/jgrd.50141} {\bibfield
  {journal} {\bibinfo  {journal} {Journal of Geophysical Research:
  Atmospheres}\ }\textbf {\bibinfo {volume} {118}}~(\bibinfo {number} {3}),\
  \bibinfo {pages} {1329--1342}}\BibitemShut {NoStop}%
\bibitem [{\citenamefont {Knutti}\ \emph {et~al.}(2013)\citenamefont {Knutti},
  \citenamefont {Masson},\ and\ \citenamefont {Gettelman}}]{ref:Knutti13}%
  \BibitemOpen
  \bibfield  {author} {\bibinfo {author} {\bibnamefont {Knutti}, \bibfnamefont
  {R}}, \bibinfo {author} {\bibfnamefont {D.}~\bibnamefont {Masson}}, \ and\
  \bibinfo {author} {\bibfnamefont {A.}~\bibnamefont {Gettelman}}} (\bibinfo
  {year} {2013}),\ \bibfield  {title} {\enquote {\bibinfo {title} {Climate
  model genealogy: Generation cmip5 and how we got there},}\ }\href {\doibase
  10.1002/grl.50256} {\bibfield  {journal} {\bibinfo  {journal} {Geophysical
  Research Letters}\ }\textbf {\bibinfo {volume} {40}}~(\bibinfo {number}
  {6}),\ \bibinfo {pages} {1194--1199}}\BibitemShut {NoStop}%
\bibitem [{\citenamefont {Knutti}\ \emph {et~al.}(2006)\citenamefont {Knutti},
  \citenamefont {Meehl}, \citenamefont {Allen},\ and\ \citenamefont
  {Stainforth}}]{ref:Knutti06}%
  \BibitemOpen
  \bibfield  {author} {\bibinfo {author} {\bibnamefont {Knutti}, \bibfnamefont
  {R}}, \bibinfo {author} {\bibfnamefont {G.~A.}\ \bibnamefont {Meehl}},
  \bibinfo {author} {\bibfnamefont {M.~R.}\ \bibnamefont {Allen}}, \ and\
  \bibinfo {author} {\bibfnamefont {D.~A.}\ \bibnamefont {Stainforth}}}
  (\bibinfo {year} {2006}),\ \bibfield  {title} {\enquote {\bibinfo {title}
  {Constraining climate sensitivity from the seasonal cycle in surface
  temperature},}\ }\href {\doibase 10.1175/jcli3865.1} {\bibfield  {journal}
  {\bibinfo  {journal} {Journal of Climate}\ }\textbf {\bibinfo {volume}
  {19}}~(\bibinfo {number} {17}),\ \bibinfo {pages} {4224--4233}}\BibitemShut
  {NoStop}%
\bibitem [{\citenamefont {Knutti}\ \emph {et~al.}(2017)\citenamefont {Knutti},
  \citenamefont {Rugenstein},\ and\ \citenamefont {Hegerl}}]{ref:Knutti2017}%
  \BibitemOpen
  \bibfield  {author} {\bibinfo {author} {\bibnamefont {Knutti}, \bibfnamefont
  {Reto}}, \bibinfo {author} {\bibfnamefont {Maria A.~A.}\ \bibnamefont
  {Rugenstein}}, \ and\ \bibinfo {author} {\bibfnamefont {Gabriele~C.}\
  \bibnamefont {Hegerl}}} (\bibinfo {year} {2017}),\ \bibfield  {title}
  {\enquote {\bibinfo {title} {{Beyond equilibrium climate sensitivity}},}\
  }\href {\doibase 10.1038/ngeo3017} {\bibfield  {journal} {\bibinfo  {journal}
  {Nature Geoscience}\ }\textbf {\bibinfo {volume} {10}}~(\bibinfo {number}
  {10}),\ \bibinfo {pages} {727--736}}\BibitemShut {NoStop}%
\bibitem [{\citenamefont {Kubo}(1966)}]{ref:Kubo66}%
  \BibitemOpen
  \bibfield  {author} {\bibinfo {author} {\bibnamefont {Kubo}, \bibfnamefont
  {R}}} (\bibinfo {year} {1966}),\ \bibfield  {title} {\enquote {\bibinfo
  {title} {The fluctuation-dissipation theorem},}\ }\href
  {http://stacks.iop.org/0034-4885/29/i=1/a=306} {\bibfield  {journal}
  {\bibinfo  {journal} {Reports on Progress in Physics}\ }\textbf {\bibinfo
  {volume} {29}}~(\bibinfo {number} {1}),\ \bibinfo {pages} {255}}\BibitemShut
  {NoStop}%
\bibitem [{\citenamefont {Kwiatkowski}\ \emph {et~al.}(2017)\citenamefont
  {Kwiatkowski}, \citenamefont {Bopp}, \citenamefont {Aumont}, \citenamefont
  {Ciais}, \citenamefont {Cox}, \citenamefont {Laufkötter}, \citenamefont
  {Li},\ and\ \citenamefont {Séférian}}]{ref:Kwiatkowski17}%
  \BibitemOpen
  \bibfield  {author} {\bibinfo {author} {\bibnamefont {Kwiatkowski},
  \bibfnamefont {Lester}}, \bibinfo {author} {\bibfnamefont {Laurent}\
  \bibnamefont {Bopp}}, \bibinfo {author} {\bibfnamefont {Olivier}\
  \bibnamefont {Aumont}}, \bibinfo {author} {\bibfnamefont {Philippe}\
  \bibnamefont {Ciais}}, \bibinfo {author} {\bibfnamefont {Peter~M.}\
  \bibnamefont {Cox}}, \bibinfo {author} {\bibfnamefont {Charlotte}\
  \bibnamefont {Laufkötter}}, \bibinfo {author} {\bibfnamefont {Yue}\
  \bibnamefont {Li}}, \ and\ \bibinfo {author} {\bibfnamefont {Roland}\
  \bibnamefont {Séférian}}} (\bibinfo {year} {2017}),\ \bibfield  {title}
  {\enquote {\bibinfo {title} {Emergent constraints on projections of declining
  primary production in the tropical oceans},}\ }\href {\doibase
  10.1038/nclimate3265
  https://www.nature.com/articles/nclimate3265#supplementary-information}
  {\bibfield  {journal} {\bibinfo  {journal} {Nature Climate Change}\ }\textbf
  {\bibinfo {volume} {7}},\ \bibinfo {pages} {355}}\BibitemShut {NoStop}%
\bibitem [{\citenamefont {Lambert}\ \emph {et~al.}(2013)\citenamefont
  {Lambert}, \citenamefont {Harris}, \citenamefont {Collins}, \citenamefont
  {Murphy}, \citenamefont {Sexton},\ and\ \citenamefont
  {Booth}}]{ref:Lambert13}%
  \BibitemOpen
  \bibfield  {author} {\bibinfo {author} {\bibnamefont {Lambert}, \bibfnamefont
  {F~Hugo}}, \bibinfo {author} {\bibfnamefont {Glen~R.}\ \bibnamefont
  {Harris}}, \bibinfo {author} {\bibfnamefont {Matthew}\ \bibnamefont
  {Collins}}, \bibinfo {author} {\bibfnamefont {James~M.}\ \bibnamefont
  {Murphy}}, \bibinfo {author} {\bibfnamefont {David M.~H.}\ \bibnamefont
  {Sexton}}, \ and\ \bibinfo {author} {\bibfnamefont {Ben B.~B.}\ \bibnamefont
  {Booth}}} (\bibinfo {year} {2013}),\ \bibfield  {title} {\enquote {\bibinfo
  {title} {Interactions between perturbations to different earth system
  components simulated by a fully-coupled climate model},}\ }\href {\doibase
  10.1007/s00382-012-1618-3} {\bibfield  {journal} {\bibinfo  {journal}
  {Climate Dynamics}\ }\textbf {\bibinfo {volume} {41}}~(\bibinfo {number}
  {11}),\ \bibinfo {pages} {3055--3072}}\BibitemShut {NoStop}%
\bibitem [{\citenamefont {Lawrence}\ \emph {et~al.}(2012)\citenamefont
  {Lawrence}, \citenamefont {Slater},\ and\ \citenamefont
  {Swenson}}]{ref:Lawrence2012}%
  \BibitemOpen
  \bibfield  {author} {\bibinfo {author} {\bibnamefont {Lawrence},
  \bibfnamefont {David~M}}, \bibinfo {author} {\bibfnamefont {Andrew~G}\
  \bibnamefont {Slater}}, \ and\ \bibinfo {author} {\bibfnamefont {Sean~C.}\
  \bibnamefont {Swenson}}} (\bibinfo {year} {2012}),\ \bibfield  {title}
  {\enquote {\bibinfo {title} {{Simulation of Present-Day and Future Permafrost
  and Seasonally Frozen Ground Conditions in CCSM4}},}\ }\href {\doibase
  10.1175/JCLI-D-11-00334.1} {\bibfield  {journal} {\bibinfo  {journal}
  {Journal of Climate}\ }\textbf {\bibinfo {volume} {25}},\ \bibinfo {pages}
  {2207--2225}}\BibitemShut {NoStop}%
\bibitem [{\citenamefont {Lehner}\ \emph {et~al.}(2019)\citenamefont {Lehner},
  \citenamefont {Wood}, \citenamefont {Vano}, \citenamefont {Lawrence},
  \citenamefont {Clark},\ and\ \citenamefont {Mankin}}]{ref:Lehner2019}%
  \BibitemOpen
  \bibfield  {author} {\bibinfo {author} {\bibnamefont {Lehner}, \bibfnamefont
  {Flavio}}, \bibinfo {author} {\bibfnamefont {Andrew~W.}\ \bibnamefont
  {Wood}}, \bibinfo {author} {\bibfnamefont {Julie~A.}\ \bibnamefont {Vano}},
  \bibinfo {author} {\bibfnamefont {David~M.}\ \bibnamefont {Lawrence}},
  \bibinfo {author} {\bibfnamefont {Martyn~P.}\ \bibnamefont {Clark}}, \ and\
  \bibinfo {author} {\bibfnamefont {Justin~S.}\ \bibnamefont {Mankin}}}
  (\bibinfo {year} {2019}),\ \bibfield  {title} {\enquote {\bibinfo {title}
  {{The potential to reduce uncertainty in regional runoff projections from
  climate models}},}\ }\href {\doibase 10.1038/s41558-019-0639-x} {\bibfield
  {journal} {\bibinfo  {journal} {Nature Climate Change}\ }\textbf {\bibinfo
  {volume} {9}}~(\bibinfo {number} {12}),\ \bibinfo {pages}
  {926--933}}\BibitemShut {NoStop}%
\bibitem [{\citenamefont {Leith}(1974)}]{ref:Leith74}%
  \BibitemOpen
  \bibfield  {author} {\bibinfo {author} {\bibnamefont {Leith}, \bibfnamefont
  {C~E}}} (\bibinfo {year} {1974}),\ \bibfield  {title} {\enquote {\bibinfo
  {title} {Theoretical skill of monte carlo forecasts},}\ }\href {\doibase
  10.1175/1520-0493(1974)102<0409:TSOMCF>2.0.CO;2} {\bibfield  {journal}
  {\bibinfo  {journal} {Monthly Weather Review}\ }\textbf {\bibinfo {volume}
  {102}}~(\bibinfo {number} {6}),\ \bibinfo {pages} {409--418}}\BibitemShut
  {NoStop}%
\bibitem [{\citenamefont {Leith}(1975)}]{ref:Leith75}%
  \BibitemOpen
  \bibfield  {author} {\bibinfo {author} {\bibnamefont {Leith}, \bibfnamefont
  {C~E}}} (\bibinfo {year} {1975}),\ \bibfield  {title} {\enquote {\bibinfo
  {title} {Climate response and fluctuation dissipation},}\ }\href {\doibase
  10.1175/1520-0469(1975)032<2022:CRAFD>2.0.CO;2} {\bibfield  {journal}
  {\bibinfo  {journal} {Journal of the Atmospheric Sciences}\ }\textbf
  {\bibinfo {volume} {32}}~(\bibinfo {number} {10}),\ \bibinfo {pages}
  {2022--2026}}\BibitemShut {NoStop}%
\bibitem [{\citenamefont {Lenton}\ \emph {et~al.}(2008)\citenamefont {Lenton},
  \citenamefont {Held}, \citenamefont {Kriegler}, \citenamefont {Hall},
  \citenamefont {Lucht}, \citenamefont {Rahmstorf},\ and\ \citenamefont
  {Schellnhuber}}]{ref:Lenton08}%
  \BibitemOpen
  \bibfield  {author} {\bibinfo {author} {\bibnamefont {Lenton}, \bibfnamefont
  {T~M}}, \bibinfo {author} {\bibfnamefont {H.}~\bibnamefont {Held}}, \bibinfo
  {author} {\bibfnamefont {E.}~\bibnamefont {Kriegler}}, \bibinfo {author}
  {\bibfnamefont {J.~W.}\ \bibnamefont {Hall}}, \bibinfo {author}
  {\bibfnamefont {W.}~\bibnamefont {Lucht}}, \bibinfo {author} {\bibfnamefont
  {S.}~\bibnamefont {Rahmstorf}}, \ and\ \bibinfo {author} {\bibfnamefont
  {H.~J.}\ \bibnamefont {Schellnhuber}}} (\bibinfo {year} {2008}),\ \bibfield
  {title} {\enquote {\bibinfo {title} {Tipping elements in the {E}arth's
  climate system},}\ }\href@noop {} {\bibfield  {journal} {\bibinfo  {journal}
  {Proc. Natl. Acad. Sci. USA}\ }\textbf {\bibinfo {volume} {105}},\ \bibinfo
  {pages} {1786--93}}\BibitemShut {NoStop}%
\bibitem [{\citenamefont {Li}\ \emph {et~al.}(2017)\citenamefont {Li},
  \citenamefont {Xie}, \citenamefont {He},\ and\ \citenamefont
  {Chen}}]{ref:Li17}%
  \BibitemOpen
  \bibfield  {author} {\bibinfo {author} {\bibnamefont {Li}, \bibfnamefont
  {G}}, \bibinfo {author} {\bibfnamefont {S.~P.}\ \bibnamefont {Xie}}, \bibinfo
  {author} {\bibfnamefont {C.}~\bibnamefont {He}}, \ and\ \bibinfo {author}
  {\bibfnamefont {Z.~S.}\ \bibnamefont {Chen}}} (\bibinfo {year} {2017}),\
  \bibfield  {title} {\enquote {\bibinfo {title} {Western pacific emergent
  constraint lowers projected increase in indian summer monsoon rainfall},}\
  }\href {\doibase 10.1038/nclimate3387} {\bibfield  {journal} {\bibinfo
  {journal} {Nature Climate Change}\ }\textbf {\bibinfo {volume} {7}}~(\bibinfo
  {number} {10}),\ \bibinfo {pages} {708}}\BibitemShut {NoStop}%
\bibitem [{\citenamefont {Lian}\ \emph {et~al.}(2018)\citenamefont {Lian},
  \citenamefont {Piao}, \citenamefont {Huntingford}, \citenamefont {Li},
  \citenamefont {Zeng}, \citenamefont {Wang}, \citenamefont {Ciais},
  \citenamefont {McVicar}, \citenamefont {Peng}, \citenamefont {Ottlé},
  \citenamefont {Yang}, \citenamefont {Yang}, \citenamefont {Zhang},\ and\
  \citenamefont {Wang}}]{ref:Lian18}%
  \BibitemOpen
  \bibfield  {author} {\bibinfo {author} {\bibnamefont {Lian}, \bibfnamefont
  {Xu}}, \bibinfo {author} {\bibfnamefont {Shilong}\ \bibnamefont {Piao}},
  \bibinfo {author} {\bibfnamefont {Chris}\ \bibnamefont {Huntingford}},
  \bibinfo {author} {\bibfnamefont {Yue}\ \bibnamefont {Li}}, \bibinfo {author}
  {\bibfnamefont {Zhenzhong}\ \bibnamefont {Zeng}}, \bibinfo {author}
  {\bibfnamefont {Xuhui}\ \bibnamefont {Wang}}, \bibinfo {author}
  {\bibfnamefont {Philippe}\ \bibnamefont {Ciais}}, \bibinfo {author}
  {\bibfnamefont {Tim~R.}\ \bibnamefont {McVicar}}, \bibinfo {author}
  {\bibfnamefont {Shushi}\ \bibnamefont {Peng}}, \bibinfo {author}
  {\bibfnamefont {Catherine}\ \bibnamefont {Ottlé}}, \bibinfo {author}
  {\bibfnamefont {Hui}\ \bibnamefont {Yang}}, \bibinfo {author} {\bibfnamefont
  {Yuting}\ \bibnamefont {Yang}}, \bibinfo {author} {\bibfnamefont {Yongqiang}\
  \bibnamefont {Zhang}}, \ and\ \bibinfo {author} {\bibfnamefont {Tao}\
  \bibnamefont {Wang}}} (\bibinfo {year} {2018}),\ \bibfield  {title} {\enquote
  {\bibinfo {title} {Partitioning global land evapotranspiration using cmip5
  models constrained by observations},}\ }\href {\doibase
  10.1038/s41558-018-0207-9} {\bibfield  {journal} {\bibinfo  {journal} {Nature
  Climate Change}\ }\textbf {\bibinfo {volume} {8}}~(\bibinfo {number} {7}),\
  \bibinfo {pages} {640--646}}\BibitemShut {NoStop}%
\bibitem [{\citenamefont {Lin}\ \emph {et~al.}(2017)\citenamefont {Lin},
  \citenamefont {Dong}, \citenamefont {Zhang}, \citenamefont {Xie},
  \citenamefont {Xue}, \citenamefont {Huang},\ and\ \citenamefont
  {Luo}}]{ref:Lin17}%
  \BibitemOpen
  \bibfield  {author} {\bibinfo {author} {\bibnamefont {Lin}, \bibfnamefont
  {Yanluan}}, \bibinfo {author} {\bibfnamefont {Wenhao}\ \bibnamefont {Dong}},
  \bibinfo {author} {\bibfnamefont {Minghua}\ \bibnamefont {Zhang}}, \bibinfo
  {author} {\bibfnamefont {Yuanyu}\ \bibnamefont {Xie}}, \bibinfo {author}
  {\bibfnamefont {Wei}\ \bibnamefont {Xue}}, \bibinfo {author} {\bibfnamefont
  {Jianbin}\ \bibnamefont {Huang}}, \ and\ \bibinfo {author} {\bibfnamefont
  {Yong}\ \bibnamefont {Luo}}} (\bibinfo {year} {2017}),\ \bibfield  {title}
  {\enquote {\bibinfo {title} {Causes of model dry and warm bias over central
  u.s. and impact on climate projections},}\ }\href {\doibase
  10.1038/s41467-017-01040-2} {\bibfield  {journal} {\bibinfo  {journal}
  {Nature Communications}\ }\textbf {\bibinfo {volume} {8}}~(\bibinfo {number}
  {1}),\ \bibinfo {pages} {881}}\BibitemShut {NoStop}%
\bibitem [{\citenamefont {Lipat}\ \emph {et~al.}(2017)\citenamefont {Lipat},
  \citenamefont {Tselioudis}, \citenamefont {Grise},\ and\ \citenamefont
  {Polvani}}]{ref:Lipat17}%
  \BibitemOpen
  \bibfield  {author} {\bibinfo {author} {\bibnamefont {Lipat}, \bibfnamefont
  {Bernard~R}}, \bibinfo {author} {\bibfnamefont {George}\ \bibnamefont
  {Tselioudis}}, \bibinfo {author} {\bibfnamefont {Kevin~M.}\ \bibnamefont
  {Grise}}, \ and\ \bibinfo {author} {\bibfnamefont {Lorenzo~M.}\ \bibnamefont
  {Polvani}}} (\bibinfo {year} {2017}),\ \bibfield  {title} {\enquote {\bibinfo
  {title} {{CMIP}5 models' shortwave cloud radiative response and climate
  sensitivity linked to the climatological {Hadley} cell extent},}\ }\href
  {\doibase 10.1002/2017GL073151} {\bibfield  {journal} {\bibinfo  {journal}
  {Geophysical Research Letters}\ }\textbf {\bibinfo {volume} {44}}~(\bibinfo
  {number} {11}),\ \bibinfo {pages} {5739--5748}}\BibitemShut {NoStop}%
\bibitem [{\citenamefont {Liu}\ \emph {et~al.}(2013)\citenamefont {Liu},
  \citenamefont {Song}, \citenamefont {Horton},\ and\ \citenamefont
  {Hu}}]{ref:Liu13}%
  \BibitemOpen
  \bibfield  {author} {\bibinfo {author} {\bibnamefont {Liu}, \bibfnamefont
  {Jiping}}, \bibinfo {author} {\bibfnamefont {Mirong}\ \bibnamefont {Song}},
  \bibinfo {author} {\bibfnamefont {Radley~M.}\ \bibnamefont {Horton}}, \ and\
  \bibinfo {author} {\bibfnamefont {Yongyun}\ \bibnamefont {Hu}}} (\bibinfo
  {year} {2013}),\ \bibfield  {title} {\enquote {\bibinfo {title} {Reducing
  spread in climate model projections of a september ice-free arctic},}\ }\href
  {\doibase 10.1073/pnas.1219716110} {\bibfield  {journal} {\bibinfo  {journal}
  {Proceedings of the National Academy of Sciences}\ }\textbf {\bibinfo
  {volume} {110}}~(\bibinfo {number} {31}),\ \bibinfo {pages}
  {12571}}\BibitemShut {NoStop}%
\bibitem [{\citenamefont {Lutsko}\ and\ \citenamefont
  {Takahashi}(2018)}]{ref:Lutsko&Takahashi18}%
  \BibitemOpen
  \bibfield  {author} {\bibinfo {author} {\bibnamefont {Lutsko}, \bibfnamefont
  {N~J}}, \ and\ \bibinfo {author} {\bibfnamefont {K.}~\bibnamefont
  {Takahashi}}} (\bibinfo {year} {2018}),\ \bibfield  {title} {\enquote
  {\bibinfo {title} {What can the internal variability of cmip5 models tell us
  about their climate sensitivity?}}\ }\href {\doibase
  10.1175/jcli-d-17-0736.1} {\bibfield  {journal} {\bibinfo  {journal} {Journal
  of Climate}\ }\textbf {\bibinfo {volume} {31}}~(\bibinfo {number} {13}),\
  \bibinfo {pages} {5051--5069}}\BibitemShut {NoStop}%
\bibitem [{\citenamefont {Maher}\ \emph {et~al.}(2019)\citenamefont {Maher},
  \citenamefont {Milinski}, \citenamefont {Suarez-Gutierrez}, \citenamefont
  {Botzet}, \citenamefont {Dobrynin}, \citenamefont {Kornblueh}, \citenamefont
  {Kröger}, \citenamefont {Takano}, \citenamefont {Ghosh}, \citenamefont
  {Hedemann}, \citenamefont {Li}, \citenamefont {Li}, \citenamefont {Manzini},
  \citenamefont {Notz}, \citenamefont {Putrasahan}, \citenamefont {Boysen},
  \citenamefont {Claussen}, \citenamefont {Ilyina}, \citenamefont {Olonscheck},
  \citenamefont {Raddatz}, \citenamefont {Stevens},\ and\ \citenamefont
  {Marotzke}}]{ref:Maher19}%
  \BibitemOpen
  \bibfield  {author} {\bibinfo {author} {\bibnamefont {Maher}, \bibfnamefont
  {Nicola}}, \bibinfo {author} {\bibfnamefont {Sebastian}\ \bibnamefont
  {Milinski}}, \bibinfo {author} {\bibfnamefont {Laura}\ \bibnamefont
  {Suarez-Gutierrez}}, \bibinfo {author} {\bibfnamefont {Michael}\ \bibnamefont
  {Botzet}}, \bibinfo {author} {\bibfnamefont {Mikhail}\ \bibnamefont
  {Dobrynin}}, \bibinfo {author} {\bibfnamefont {Luis}\ \bibnamefont
  {Kornblueh}}, \bibinfo {author} {\bibfnamefont {Jürgen}\ \bibnamefont
  {Kröger}}, \bibinfo {author} {\bibfnamefont {Yohei}\ \bibnamefont {Takano}},
  \bibinfo {author} {\bibfnamefont {Rohit}\ \bibnamefont {Ghosh}}, \bibinfo
  {author} {\bibfnamefont {Christopher}\ \bibnamefont {Hedemann}}, \bibinfo
  {author} {\bibfnamefont {Chao}\ \bibnamefont {Li}}, \bibinfo {author}
  {\bibfnamefont {Hongmei}\ \bibnamefont {Li}}, \bibinfo {author}
  {\bibfnamefont {Elisa}\ \bibnamefont {Manzini}}, \bibinfo {author}
  {\bibfnamefont {Dirk}\ \bibnamefont {Notz}}, \bibinfo {author} {\bibfnamefont
  {Dian}\ \bibnamefont {Putrasahan}}, \bibinfo {author} {\bibfnamefont {Lena}\
  \bibnamefont {Boysen}}, \bibinfo {author} {\bibfnamefont {Martin}\
  \bibnamefont {Claussen}}, \bibinfo {author} {\bibfnamefont {Tatiana}\
  \bibnamefont {Ilyina}}, \bibinfo {author} {\bibfnamefont {Dirk}\ \bibnamefont
  {Olonscheck}}, \bibinfo {author} {\bibfnamefont {Thomas}\ \bibnamefont
  {Raddatz}}, \bibinfo {author} {\bibfnamefont {Bjorn}\ \bibnamefont
  {Stevens}}, \ and\ \bibinfo {author} {\bibfnamefont {Jochem}\ \bibnamefont
  {Marotzke}}} (\bibinfo {year} {2019}),\ \bibfield  {title} {\enquote
  {\bibinfo {title} {The max planck institute grand ensemble: Enabling the
  exploration of climate system variability},}\ }\href {\doibase
  10.1029/2019MS001639} {\bibfield  {journal} {\bibinfo  {journal} {Journal of
  Advances in Modeling Earth Systems}\ }\textbf {\bibinfo {volume}
  {11}}~(\bibinfo {number} {7}),\ \bibinfo {pages} {2050--2069}}\BibitemShut
  {NoStop}%
\bibitem [{\citenamefont {Majda}\ \emph {et~al.}(2010)\citenamefont {Majda},
  \citenamefont {Abramov},\ and\ \citenamefont {Gershgorin}}]{ref:Majda10}%
  \BibitemOpen
  \bibfield  {author} {\bibinfo {author} {\bibnamefont {Majda}, \bibfnamefont
  {A~J}}, \bibinfo {author} {\bibfnamefont {R.}~\bibnamefont {Abramov}}, \ and\
  \bibinfo {author} {\bibfnamefont {B.}~\bibnamefont {Gershgorin}}} (\bibinfo
  {year} {2010}),\ \bibfield  {title} {\enquote {\bibinfo {title} {High skill
  in low-frequency climate response through fluctuation dissipation theorems
  despite structural instability},}\ }\href {\doibase 10.1073/pnas.0912997107}
  {\bibfield  {journal} {\bibinfo  {journal} {Proceedings of the National
  Academy of Sciences of the United States of America}\ }\textbf {\bibinfo
  {volume} {107}}~(\bibinfo {number} {2}),\ \bibinfo {pages}
  {581--586}}\BibitemShut {NoStop}%
\bibitem [{\citenamefont {Manabe}\ and\ \citenamefont
  {Wetherald}(1975)}]{ref:Manabe&Wetherald75}%
  \BibitemOpen
  \bibfield  {author} {\bibinfo {author} {\bibnamefont {Manabe}, \bibfnamefont
  {Syukuro}}, \ and\ \bibinfo {author} {\bibfnamefont {Richard~T.}\
  \bibnamefont {Wetherald}}} (\bibinfo {year} {1975}),\ \bibfield  {title}
  {\enquote {\bibinfo {title} {The effects of doubling the co2 concentration on
  the climate of a general circulation model},}\ }\href {\doibase
  10.1175/1520-0469(1975)032<0003:TEODTC>2.0.CO;2} {\bibfield  {journal}
  {\bibinfo  {journal} {Journal of the Atmospheric Sciences}\ }\textbf
  {\bibinfo {volume} {32}}~(\bibinfo {number} {1}),\ \bibinfo {pages}
  {3--15}}\BibitemShut {NoStop}%
\bibitem [{\citenamefont {Masson}\ and\ \citenamefont
  {Knutti}(2012)}]{ref:Masson&Knutti12}%
  \BibitemOpen
  \bibfield  {author} {\bibinfo {author} {\bibnamefont {Masson}, \bibfnamefont
  {David}}, \ and\ \bibinfo {author} {\bibfnamefont {Reto}\ \bibnamefont
  {Knutti}}} (\bibinfo {year} {2012}),\ \bibfield  {title} {\enquote {\bibinfo
  {title} {Predictor screening, calibration, and observational constraints in
  climate model ensembles: An illustration using climate sensitivity},}\ }\href
  {\doibase 10.1175/JCLI-D-11-00540.1} {\bibfield  {journal} {\bibinfo
  {journal} {Journal of Climate}\ }\textbf {\bibinfo {volume} {26}}~(\bibinfo
  {number} {3}),\ \bibinfo {pages} {887--898}}\BibitemShut {NoStop}%
\bibitem [{\citenamefont {Massonnet}\ \emph {et~al.}(2012)\citenamefont
  {Massonnet}, \citenamefont {Fichefet}, \citenamefont {Goosse}, \citenamefont
  {Bitz}, \citenamefont {Philippon-Berthier}, \citenamefont {Holland},\ and\
  \citenamefont {Barriat}}]{ref:Massonnet12}%
  \BibitemOpen
  \bibfield  {author} {\bibinfo {author} {\bibnamefont {Massonnet},
  \bibfnamefont {F}}, \bibinfo {author} {\bibfnamefont {T.}~\bibnamefont
  {Fichefet}}, \bibinfo {author} {\bibfnamefont {H.}~\bibnamefont {Goosse}},
  \bibinfo {author} {\bibfnamefont {C.~M.}\ \bibnamefont {Bitz}}, \bibinfo
  {author} {\bibfnamefont {G.}~\bibnamefont {Philippon-Berthier}}, \bibinfo
  {author} {\bibfnamefont {M.~M.}\ \bibnamefont {Holland}}, \ and\ \bibinfo
  {author} {\bibfnamefont {P.~Y.}\ \bibnamefont {Barriat}}} (\bibinfo {year}
  {2012}),\ \bibfield  {title} {\enquote {\bibinfo {title} {Constraining
  projections of summer arctic sea ice},}\ }\href {\doibase
  10.5194/tc-6-1383-2012} {\bibfield  {journal} {\bibinfo  {journal}
  {Cryosphere}\ }\textbf {\bibinfo {volume} {6}}~(\bibinfo {number} {6}),\
  \bibinfo {pages} {1383--1394}}\BibitemShut {NoStop}%
\bibitem [{\citenamefont {Meehl}\ \emph
  {et~al.}(2007{\natexlab{a}})\citenamefont {Meehl}, \citenamefont {Covey},
  \citenamefont {Delworth}, \citenamefont {Latif}, \citenamefont {Mc{A}vaney},
  \citenamefont {Mitchell}, \citenamefont {Stouffer},\ and\ \citenamefont
  {Taylor}}]{ref:CMIP3}%
  \BibitemOpen
  \bibfield  {author} {\bibinfo {author} {\bibnamefont {Meehl}, \bibfnamefont
  {G~A}}, \bibinfo {author} {\bibfnamefont {C.}~\bibnamefont {Covey}}, \bibinfo
  {author} {\bibfnamefont {T.}~\bibnamefont {Delworth}}, \bibinfo {author}
  {\bibfnamefont {M.}~\bibnamefont {Latif}}, \bibinfo {author} {\bibfnamefont
  {B.}~\bibnamefont {Mc{A}vaney}}, \bibinfo {author} {\bibfnamefont {J.~F.~B.}\
  \bibnamefont {Mitchell}}, \bibinfo {author} {\bibfnamefont {R.~J.}\
  \bibnamefont {Stouffer}}, \ and\ \bibinfo {author} {\bibfnamefont {K.~E.}\
  \bibnamefont {Taylor}}} (\bibinfo {year} {2007}{\natexlab{a}}),\ \bibfield
  {title} {\enquote {\bibinfo {title} {The {WCRP} {CMIP}3 multimodel
  dataset},}\ }\href@noop {} {\bibfield  {journal} {\bibinfo  {journal} {Bull.
  Am. Meteor. Soc.}\ }\textbf {\bibinfo {volume} {88}},\ \bibinfo {pages}
  {1383--1394}}\BibitemShut {NoStop}%
\bibitem [{\citenamefont {Meehl}\ \emph
  {et~al.}(2007{\natexlab{b}})\citenamefont {Meehl}, \citenamefont {Stocker},
  \citenamefont {Collins}, \citenamefont {Friedlingstein}, \citenamefont
  {Gaye}, \citenamefont {Gregory}, \citenamefont {Kitoh}, \citenamefont
  {Knutti}, \citenamefont {Murphy}, \citenamefont {Noda}, \citenamefont
  {Raper}, \citenamefont {Watterson}, \citenamefont {Weaver},\ and\
  \citenamefont {Zhao}}]{ref:IPCC_AR4_projections}%
  \BibitemOpen
  \bibfield  {author} {\bibinfo {author} {\bibnamefont {Meehl}, \bibfnamefont
  {G~A}}, \bibinfo {author} {\bibfnamefont {T.F.}\ \bibnamefont {Stocker}},
  \bibinfo {author} {\bibfnamefont {W.D.}\ \bibnamefont {Collins}}, \bibinfo
  {author} {\bibfnamefont {P.}~\bibnamefont {Friedlingstein}}, \bibinfo
  {author} {\bibfnamefont {A.T.}\ \bibnamefont {Gaye}}, \bibinfo {author}
  {\bibfnamefont {J.M.}\ \bibnamefont {Gregory}}, \bibinfo {author}
  {\bibfnamefont {A.}~\bibnamefont {Kitoh}}, \bibinfo {author} {\bibfnamefont
  {R.}~\bibnamefont {Knutti}}, \bibinfo {author} {\bibfnamefont {J.M.}\
  \bibnamefont {Murphy}}, \bibinfo {author} {\bibfnamefont {A.}~\bibnamefont
  {Noda}}, \bibinfo {author} {\bibfnamefont {S.C.B.}\ \bibnamefont {Raper}},
  \bibinfo {author} {\bibfnamefont {I.G.}\ \bibnamefont {Watterson}}, \bibinfo
  {author} {\bibfnamefont {A.J.}\ \bibnamefont {Weaver}}, \ and\ \bibinfo
  {author} {\bibfnamefont {Z.-C.}\ \bibnamefont {Zhao}}} (\bibinfo {year}
  {2007}{\natexlab{b}}),\ \bibfield  {title} {\enquote {\bibinfo {title}
  {Global climate projections.}}\ }in\ \href@noop {} {\emph {\bibinfo
  {booktitle} {Climate Change 2007: The Physical Science Basis. Contribution of
  Working Group I to the Fourth Assessment Report of the Intergovernmental
  Panel on Climate Change}}},\ \bibinfo {editor} {edited by\ \bibinfo {editor}
  {\bibfnamefont {S.}~\bibnamefont {Solomon}}, \bibinfo {editor} {\bibfnamefont
  {D.}~\bibnamefont {Qin}}, \bibinfo {editor} {\bibfnamefont {M.}~\bibnamefont
  {Manning}}, \bibinfo {editor} {\bibfnamefont {Z.}~\bibnamefont {Chen}},
  \bibinfo {editor} {\bibfnamefont {M.}~\bibnamefont {Marquis}}, \bibinfo
  {editor} {\bibfnamefont {K.B.}\ \bibnamefont {Averyt}}, \bibinfo {editor}
  {\bibfnamefont {M.}~\bibnamefont {Tignor}}, \ and\ \bibinfo {editor}
  {\bibfnamefont {H.L.}\ \bibnamefont {Miller}}}\ (\bibinfo  {publisher}
  {Cambridge University Press, Cambridge, United Kingdom and New York, NY,
  USA})\BibitemShut {NoStop}%
\bibitem [{\citenamefont {Meinshausen}\ \emph {et~al.}(2011)\citenamefont
  {Meinshausen}, \citenamefont {Smith}, \citenamefont {Calvin}, \citenamefont
  {Daniel}, \citenamefont {Kainuma}, \citenamefont {Lamarque}, \citenamefont
  {Matsumoto}, \citenamefont {Montzka}, \citenamefont {Raper}, \citenamefont
  {Riahi}, \citenamefont {Thomson}, \citenamefont {Velders},\ and\
  \citenamefont {van Vuuren}}]{ref:Meinshausen11}%
  \BibitemOpen
  \bibfield  {author} {\bibinfo {author} {\bibnamefont {Meinshausen},
  \bibfnamefont {Malte}}, \bibinfo {author} {\bibfnamefont {S.~J.}\
  \bibnamefont {Smith}}, \bibinfo {author} {\bibfnamefont {K.}~\bibnamefont
  {Calvin}}, \bibinfo {author} {\bibfnamefont {J.~S.}\ \bibnamefont {Daniel}},
  \bibinfo {author} {\bibfnamefont {M.~L.~T.}\ \bibnamefont {Kainuma}},
  \bibinfo {author} {\bibfnamefont {J.~F.}\ \bibnamefont {Lamarque}}, \bibinfo
  {author} {\bibfnamefont {K.}~\bibnamefont {Matsumoto}}, \bibinfo {author}
  {\bibfnamefont {S.~A.}\ \bibnamefont {Montzka}}, \bibinfo {author}
  {\bibfnamefont {S.~C.~B.}\ \bibnamefont {Raper}}, \bibinfo {author}
  {\bibfnamefont {K.}~\bibnamefont {Riahi}}, \bibinfo {author} {\bibfnamefont
  {A.}~\bibnamefont {Thomson}}, \bibinfo {author} {\bibfnamefont {G.~J.~M.}\
  \bibnamefont {Velders}}, \ and\ \bibinfo {author} {\bibfnamefont {D.~P.~P.}\
  \bibnamefont {van Vuuren}}} (\bibinfo {year} {2011}),\ \bibfield  {title}
  {\enquote {\bibinfo {title} {The rcp greenhouse gas concentrations and their
  extensions from 1765 to 2300},}\ }\href {\doibase 10.1007/s10584-011-0156-z}
  {\bibfield  {journal} {\bibinfo  {journal} {Climatic Change}\ }\textbf
  {\bibinfo {volume} {109}}~(\bibinfo {number} {1}),\ \bibinfo {pages}
  {213}}\BibitemShut {NoStop}%
\bibitem [{\citenamefont {Meredith}\ \emph {et~al.}(2019)\citenamefont
  {Meredith}, \citenamefont {Sommerkorn}, \citenamefont {Cassotta},
  \citenamefont {Derksen}, \citenamefont {Ekaykin}, \citenamefont {Hollowed},
  \citenamefont {Kofinas}, \citenamefont {Mackintosh}, \citenamefont
  {Melbourne-Thomas}, \citenamefont {Muelbert}, \citenamefont {Ottersen},
  \citenamefont {Pritchard},\ and\ \citenamefont {Schuur}}]{ref:SROCC2019}%
  \BibitemOpen
  \bibfield  {author} {\bibinfo {author} {\bibnamefont {Meredith},
  \bibfnamefont {M}}, \bibinfo {author} {\bibfnamefont {M}~\bibnamefont
  {Sommerkorn}}, \bibinfo {author} {\bibfnamefont {S}~\bibnamefont {Cassotta}},
  \bibinfo {author} {\bibfnamefont {Chris}\ \bibnamefont {Derksen}}, \bibinfo
  {author} {\bibfnamefont {A}~\bibnamefont {Ekaykin}}, \bibinfo {author}
  {\bibfnamefont {A}~\bibnamefont {Hollowed}}, \bibinfo {author} {\bibfnamefont
  {G}~\bibnamefont {Kofinas}}, \bibinfo {author} {\bibfnamefont
  {A}~\bibnamefont {Mackintosh}}, \bibinfo {author} {\bibfnamefont
  {J}~\bibnamefont {Melbourne-Thomas}}, \bibinfo {author} {\bibfnamefont
  {M~M~C}\ \bibnamefont {Muelbert}}, \bibinfo {author} {\bibfnamefont
  {G}~\bibnamefont {Ottersen}}, \bibinfo {author} {\bibfnamefont
  {H}~\bibnamefont {Pritchard}}, \ and\ \bibinfo {author} {\bibfnamefont
  {E~A~G}\ \bibnamefont {Schuur}}} (\bibinfo {year} {2019}),\ \bibfield
  {title} {\enquote {\bibinfo {title} {{Polar regions}},}\ }in\ \href {\doibase
  10.1016/S1366-7017(01)00066-6} {\emph {\bibinfo {booktitle} {IPCC Special
  Report on the Ocean and Cryosphere in a Changing Climate}}},\ Chap.~\bibinfo
  {chapter} {3},\ pp.\ \bibinfo {pages} {203--320}\BibitemShut {NoStop}%
\bibitem [{\citenamefont {Molteni}\ \emph {et~al.}(1996)\citenamefont
  {Molteni}, \citenamefont {Buizza}, \citenamefont {Palmer},\ and\
  \citenamefont {Petroliagis}}]{ref:Molteni96}%
  \BibitemOpen
  \bibfield  {author} {\bibinfo {author} {\bibnamefont {Molteni}, \bibfnamefont
  {F}}, \bibinfo {author} {\bibfnamefont {R.}~\bibnamefont {Buizza}}, \bibinfo
  {author} {\bibfnamefont {T.~N.}\ \bibnamefont {Palmer}}, \ and\ \bibinfo
  {author} {\bibfnamefont {T.}~\bibnamefont {Petroliagis}}} (\bibinfo {year}
  {1996}),\ \bibfield  {title} {\enquote {\bibinfo {title} {The ecmwf ensemble
  prediction system: Methodology and validation},}\ }\href {\doibase
  10.1002/qj.49712252905} {\bibfield  {journal} {\bibinfo  {journal} {Quarterly
  Journal of the Royal Meteorological Society}\ }\textbf {\bibinfo {volume}
  {122}}~(\bibinfo {number} {529}),\ \bibinfo {pages} {73--119}}\BibitemShut
  {NoStop}%
\bibitem [{\citenamefont {Moss}\ \emph {et~al.}(2010)\citenamefont {Moss},
  \citenamefont {Edmonds}, \citenamefont {Hibbard}, \citenamefont {Manning},
  \citenamefont {Rose}, \citenamefont {{Van Vuuren}}, \citenamefont {Carter},
  \citenamefont {Emori}, \citenamefont {Kainuma}, \citenamefont {Kram},
  \citenamefont {Meehl}, \citenamefont {Mitchell}, \citenamefont {Nakicenovic},
  \citenamefont {Riahi}, \citenamefont {Smith}, \citenamefont {Stouffer},
  \citenamefont {Thomson}, \citenamefont {Weyant},\ and\ \citenamefont
  {Wilbanks}}]{ref:Moss2010}%
  \BibitemOpen
  \bibfield  {author} {\bibinfo {author} {\bibnamefont {Moss}, \bibfnamefont
  {Richard~H}}, \bibinfo {author} {\bibfnamefont {Jae~A.}\ \bibnamefont
  {Edmonds}}, \bibinfo {author} {\bibfnamefont {Kathy~A.}\ \bibnamefont
  {Hibbard}}, \bibinfo {author} {\bibfnamefont {Martin~R.}\ \bibnamefont
  {Manning}}, \bibinfo {author} {\bibfnamefont {Steven~K.}\ \bibnamefont
  {Rose}}, \bibinfo {author} {\bibfnamefont {Detlef~P.}\ \bibnamefont {{Van
  Vuuren}}}, \bibinfo {author} {\bibfnamefont {Timothy~R.}\ \bibnamefont
  {Carter}}, \bibinfo {author} {\bibfnamefont {Seita}\ \bibnamefont {Emori}},
  \bibinfo {author} {\bibfnamefont {Mikiko}\ \bibnamefont {Kainuma}}, \bibinfo
  {author} {\bibfnamefont {Tom}\ \bibnamefont {Kram}}, \bibinfo {author}
  {\bibfnamefont {Gerald~A.}\ \bibnamefont {Meehl}}, \bibinfo {author}
  {\bibfnamefont {John~F.B.}\ \bibnamefont {Mitchell}}, \bibinfo {author}
  {\bibfnamefont {Nebojsa}\ \bibnamefont {Nakicenovic}}, \bibinfo {author}
  {\bibfnamefont {Keywan}\ \bibnamefont {Riahi}}, \bibinfo {author}
  {\bibfnamefont {Steven~J.}\ \bibnamefont {Smith}}, \bibinfo {author}
  {\bibfnamefont {Ronald~J.}\ \bibnamefont {Stouffer}}, \bibinfo {author}
  {\bibfnamefont {Allison~M.}\ \bibnamefont {Thomson}}, \bibinfo {author}
  {\bibfnamefont {John~P.}\ \bibnamefont {Weyant}}, \ and\ \bibinfo {author}
  {\bibfnamefont {Thomas~J.}\ \bibnamefont {Wilbanks}}} (\bibinfo {year}
  {2010}),\ \bibfield  {title} {\enquote {\bibinfo {title} {{The next
  generation of scenarios for climate change research and assessment}},}\
  }\href {\doibase 10.1038/nature08823} {\bibfield  {journal} {\bibinfo
  {journal} {Nature}\ }\textbf {\bibinfo {volume} {463}}~(\bibinfo {number}
  {7282}),\ \bibinfo {pages} {747--756}}\BibitemShut {NoStop}%
\bibitem [{\citenamefont {Mudryk}\ \emph {et~al.}(2020)\citenamefont {Mudryk},
  \citenamefont {Santolaria-Ot{\'{i}}n}, \citenamefont {Krinner}, \citenamefont
  {M{\'{e}}n{\'{e}}goz}, \citenamefont {Derksen}, \citenamefont
  {Brutel-Vuilmet}, \citenamefont {Brady},\ and\ \citenamefont
  {Essery}}]{ref:Mudryk2020}%
  \BibitemOpen
  \bibfield  {author} {\bibinfo {author} {\bibnamefont {Mudryk}, \bibfnamefont
  {Lawrence}}, \bibinfo {author} {\bibfnamefont {Mar{\'{i}}a}\ \bibnamefont
  {Santolaria-Ot{\'{i}}n}}, \bibinfo {author} {\bibfnamefont {Gerhard}\
  \bibnamefont {Krinner}}, \bibinfo {author} {\bibfnamefont {Martin}\
  \bibnamefont {M{\'{e}}n{\'{e}}goz}}, \bibinfo {author} {\bibfnamefont
  {Chris}\ \bibnamefont {Derksen}}, \bibinfo {author} {\bibfnamefont {Claire}\
  \bibnamefont {Brutel-Vuilmet}}, \bibinfo {author} {\bibfnamefont {Mike}\
  \bibnamefont {Brady}}, \ and\ \bibinfo {author} {\bibfnamefont {Richard}\
  \bibnamefont {Essery}}} (\bibinfo {year} {2020}),\ \bibfield  {title}
  {\enquote {\bibinfo {title} {{Historical Northern Hemisphere snow cover
  trends and projected changes in the CMIP6 multi-model ensemble}},}\ }\href
  {\doibase 10.5194/tc-14-2495-2020} {\bibfield  {journal} {\bibinfo  {journal}
  {The Cryosphere}\ }\textbf {\bibinfo {volume} {14}}~(\bibinfo {number} {7}),\
  \bibinfo {pages} {2495--2514}}\BibitemShut {NoStop}%
\bibitem [{\citenamefont {Murphy}\ \emph {et~al.}(2004)\citenamefont {Murphy},
  \citenamefont {Sexton}, \citenamefont {Barnett}, \citenamefont {Jones},
  \citenamefont {Webb}, \citenamefont {Collins},\ and\ \citenamefont
  {Stainforth}}]{ref:Murphy04}%
  \BibitemOpen
  \bibfield  {author} {\bibinfo {author} {\bibnamefont {Murphy}, \bibfnamefont
  {James~M}}, \bibinfo {author} {\bibfnamefont {David M.~H.}\ \bibnamefont
  {Sexton}}, \bibinfo {author} {\bibfnamefont {David~N.}\ \bibnamefont
  {Barnett}}, \bibinfo {author} {\bibfnamefont {Gareth~S.}\ \bibnamefont
  {Jones}}, \bibinfo {author} {\bibfnamefont {Mark~J.}\ \bibnamefont {Webb}},
  \bibinfo {author} {\bibfnamefont {Matthew}\ \bibnamefont {Collins}}, \ and\
  \bibinfo {author} {\bibfnamefont {David~A.}\ \bibnamefont {Stainforth}}}
  (\bibinfo {year} {2004}),\ \bibfield  {title} {\enquote {\bibinfo {title}
  {Quantification of modelling uncertainties in a large ensemble of climate
  change simulations},}\ }\href {\doibase 10.1038/nature02771} {\bibfield
  {journal} {\bibinfo  {journal} {Nature}\ }\textbf {\bibinfo {volume}
  {430}}~(\bibinfo {number} {7001}),\ \bibinfo {pages} {768--772}}\BibitemShut
  {NoStop}%
\bibitem [{\citenamefont {Mystakidis}\ \emph {et~al.}(2017)\citenamefont
  {Mystakidis}, \citenamefont {Seneviratne}, \citenamefont {Gruber},\ and\
  \citenamefont {Davin}}]{ref:Mystakidis2017}%
  \BibitemOpen
  \bibfield  {author} {\bibinfo {author} {\bibnamefont {Mystakidis},
  \bibfnamefont {Stefanos}}, \bibinfo {author} {\bibfnamefont {Sonia~I.}\
  \bibnamefont {Seneviratne}}, \bibinfo {author} {\bibfnamefont {Nicolas}\
  \bibnamefont {Gruber}}, \ and\ \bibinfo {author} {\bibfnamefont {Edouard~L.}\
  \bibnamefont {Davin}}} (\bibinfo {year} {2017}),\ \bibfield  {title}
  {\enquote {\bibinfo {title} {{Hydrological and biogeochemical constraints on
  terrestrial carbon cycle feedbacks}},}\ }\href {\doibase
  10.1088/1748-9326/12/1/014009} {\bibfield  {journal} {\bibinfo  {journal}
  {Environmental Research Letters}\ }\textbf {\bibinfo {volume} {12}},\
  \bibinfo {pages} {014009}}\BibitemShut {NoStop}%
\bibitem [{\citenamefont {Nigam}\ and\ \citenamefont
  {Baxter}(2015)}]{ref:Nigam&Baxter15}%
  \BibitemOpen
  \bibfield  {author} {\bibinfo {author} {\bibnamefont {Nigam}, \bibfnamefont
  {S}}, \ and\ \bibinfo {author} {\bibfnamefont {S.}~\bibnamefont {Baxter}}}
  (\bibinfo {year} {2015}),\ \enquote {\bibinfo {title} {General circulation of
  the atmosphere | teleconnections},}\ in\ \href {\doibase
  https://doi.org/10.1016/B978-0-12-382225-3.00400-X} {\emph {\bibinfo
  {booktitle} {Encyclopedia of Atmospheric Sciences (Second Edition)}}},\
  \bibinfo {editor} {edited by\ \bibinfo {editor} {\bibfnamefont {Gerald~R.}\
  \bibnamefont {North}}, \bibinfo {editor} {\bibfnamefont {John}\ \bibnamefont
  {Pyle}}, \ and\ \bibinfo {editor} {\bibfnamefont {Fuqing}\ \bibnamefont
  {Zhang}}}\ (\bibinfo  {publisher} {Academic Press},\ \bibinfo {address}
  {Oxford})\ pp.\ \bibinfo {pages} {90--109}\BibitemShut {NoStop}%
\bibitem [{\citenamefont {Nijsse}\ \emph {et~al.}(2020)\citenamefont {Nijsse},
  \citenamefont {Cox},\ and\ \citenamefont {Williamson}}]{ref:Nijsse20}%
  \BibitemOpen
  \bibfield  {author} {\bibinfo {author} {\bibnamefont {Nijsse}, \bibfnamefont
  {F~J M~M}}, \bibinfo {author} {\bibfnamefont {P.~M.}\ \bibnamefont {Cox}}, \
  and\ \bibinfo {author} {\bibfnamefont {M.~S.}\ \bibnamefont {Williamson}}}
  (\bibinfo {year} {2020}),\ \bibfield  {title} {\enquote {\bibinfo {title}
  {Emergent constraints on transient climate response (tcr) and equilibrium
  climate sensitivity (ecs) from historical warming in cmip5 and cmip6
  models},}\ }\href {\doibase 10.5194/esd-11-737-2020} {\bibfield  {journal}
  {\bibinfo  {journal} {Earth System Dynamics}\ }\textbf {\bibinfo {volume}
  {11}}~(\bibinfo {number} {3}),\ \bibinfo {pages} {737--750}}\BibitemShut
  {NoStop}%
\bibitem [{\citenamefont {Nijsse}\ \emph {et~al.}(2019)\citenamefont {Nijsse},
  \citenamefont {Cox}, \citenamefont {Huntingford},\ and\ \citenamefont
  {Williamson}}]{ref:Nijsse19}%
  \BibitemOpen
  \bibfield  {author} {\bibinfo {author} {\bibnamefont {Nijsse}, \bibfnamefont
  {Femke~JMM}}, \bibinfo {author} {\bibfnamefont {Peter~M.}\ \bibnamefont
  {Cox}}, \bibinfo {author} {\bibfnamefont {Chris}\ \bibnamefont
  {Huntingford}}, \ and\ \bibinfo {author} {\bibfnamefont {Mark~S.}\
  \bibnamefont {Williamson}}} (\bibinfo {year} {2019}),\ \bibfield  {title}
  {\enquote {\bibinfo {title} {{Decadal global temperature variability
  increases strongly with climate sensitivity}},}\ }\href@noop {} {\bibfield
  {journal} {\bibinfo  {journal} {Nature Climate Change}\ }\textbf {\bibinfo
  {volume} {9}}~(\bibinfo {number} {8}),\ \bibinfo {pages}
  {598--601}}\BibitemShut {NoStop}%
\bibitem [{\citenamefont {Nijsse}\ and\ \citenamefont
  {Dijkstra}(2018)}]{ref:Nijsse18}%
  \BibitemOpen
  \bibfield  {author} {\bibinfo {author} {\bibnamefont {Nijsse}, \bibfnamefont
  {Femke~JMM}}, \ and\ \bibinfo {author} {\bibfnamefont {Henk~A.}\ \bibnamefont
  {Dijkstra}}} (\bibinfo {year} {2018}),\ \bibfield  {title} {\enquote
  {\bibinfo {title} {{A mathematical approach to understanding emergent
  constraints}},}\ }\href@noop {} {\bibfield  {journal} {\bibinfo  {journal}
  {Earth System Dynamics}\ }\textbf {\bibinfo {volume} {9}}~(\bibinfo {number}
  {3}),\ \bibinfo {pages} {999--1012}}\BibitemShut {NoStop}%
\bibitem [{\citenamefont {Notz}\ \emph {et~al.}(2020)\citenamefont {Notz},
  \citenamefont {Dorr}, \citenamefont {Bailey}, \citenamefont {Blockley},
  \citenamefont {Bushuk}, \citenamefont {Debernard}, \citenamefont
  {Derepentigny}, \citenamefont {Docquier}, \citenamefont {Fuckar},
  \citenamefont {Fyfe}, \citenamefont {Jahn}, \citenamefont {Holland},
  \citenamefont {Hunke}, \citenamefont {Iovino}, \citenamefont {Khosravi},
  \citenamefont {Madec}, \citenamefont {Farrell}, \citenamefont {Petty},
  \citenamefont {Rana}, \citenamefont {Roach}, \citenamefont {Rosenblum},
  \citenamefont {Rousset}, \citenamefont {Semmler}, \citenamefont {Stroeve},
  \citenamefont {Tremblay}, \citenamefont {Toyoda}, \citenamefont {Tsujino},\
  and\ \citenamefont {Vancoppenolle}}]{ref:Notz2020}%
  \BibitemOpen
  \bibfield  {author} {\bibinfo {author} {\bibnamefont {Notz}, \bibfnamefont
  {Dirk}}, \bibinfo {author} {\bibfnamefont {Jakob}\ \bibnamefont {Dorr}},
  \bibinfo {author} {\bibfnamefont {David~A}\ \bibnamefont {Bailey}}, \bibinfo
  {author} {\bibfnamefont {Ed}~\bibnamefont {Blockley}}, \bibinfo {author}
  {\bibfnamefont {Mitchell}\ \bibnamefont {Bushuk}}, \bibinfo {author}
  {\bibfnamefont {Jens~Boldingh}\ \bibnamefont {Debernard}}, \bibinfo {author}
  {\bibfnamefont {Patricia}\ \bibnamefont {Derepentigny}}, \bibinfo {author}
  {\bibfnamefont {David}\ \bibnamefont {Docquier}}, \bibinfo {author}
  {\bibfnamefont {Neven~S}\ \bibnamefont {Fuckar}}, \bibinfo {author}
  {\bibfnamefont {John~C}\ \bibnamefont {Fyfe}}, \bibinfo {author}
  {\bibfnamefont {Alexandra}\ \bibnamefont {Jahn}}, \bibinfo {author}
  {\bibfnamefont {Marika}\ \bibnamefont {Holland}}, \bibinfo {author}
  {\bibfnamefont {Elizabeth}\ \bibnamefont {Hunke}}, \bibinfo {author}
  {\bibfnamefont {Doroteaciro}\ \bibnamefont {Iovino}}, \bibinfo {author}
  {\bibfnamefont {Narges}\ \bibnamefont {Khosravi}}, \bibinfo {author}
  {\bibfnamefont {Gurvan}\ \bibnamefont {Madec}}, \bibinfo {author}
  {\bibfnamefont {Siobhan~O}\ \bibnamefont {Farrell}}, \bibinfo {author}
  {\bibfnamefont {Alek}\ \bibnamefont {Petty}}, \bibinfo {author}
  {\bibfnamefont {Arun}\ \bibnamefont {Rana}}, \bibinfo {author} {\bibfnamefont
  {Lettie}\ \bibnamefont {Roach}}, \bibinfo {author} {\bibfnamefont {Erica}\
  \bibnamefont {Rosenblum}}, \bibinfo {author} {\bibfnamefont {Clement}\
  \bibnamefont {Rousset}}, \bibinfo {author} {\bibfnamefont {Tido}\
  \bibnamefont {Semmler}}, \bibinfo {author} {\bibfnamefont {Julienne}\
  \bibnamefont {Stroeve}}, \bibinfo {author} {\bibfnamefont {Bruno}\
  \bibnamefont {Tremblay}}, \bibinfo {author} {\bibfnamefont {Takahiro}\
  \bibnamefont {Toyoda}}, \bibinfo {author} {\bibfnamefont {Hiroyuki}\
  \bibnamefont {Tsujino}}, \ and\ \bibinfo {author} {\bibfnamefont {Martin}\
  \bibnamefont {Vancoppenolle}}} (\bibinfo {year} {2020}),\ \bibfield  {title}
  {\enquote {\bibinfo {title} {{Arctic Sea Ice in CMIP6}},}\ }\href {\doibase
  10.1029/2019GL086749} {\bibfield  {journal} {\bibinfo  {journal} {Geohpysical
  Research Letters}\ }\textbf {\bibinfo {volume} {47}},\ \bibinfo {pages}
  {e2019GL086749}}\BibitemShut {NoStop}%
\bibitem [{\citenamefont {Nuzzo}(2014)}]{ref:Nuzzo14}%
  \BibitemOpen
  \bibfield  {author} {\bibinfo {author} {\bibnamefont {Nuzzo}, \bibfnamefont
  {Regina}}} (\bibinfo {year} {2014}),\ \bibfield  {title} {\enquote {\bibinfo
  {title} {Statistical errors: p values, the 'gold standard' of statistical
  validity, are not as reliable as many scientists assume},}\ }\href@noop {}
  {\bibfield  {journal} {\bibinfo  {journal} {Nature}\ }\textbf {\bibinfo
  {volume} {130}}}\BibitemShut {NoStop}%
\bibitem [{\citenamefont {Nyquist}(1928)}]{ref:Nyquist28}%
  \BibitemOpen
  \bibfield  {author} {\bibinfo {author} {\bibnamefont {Nyquist}, \bibfnamefont
  {H}}} (\bibinfo {year} {1928}),\ \bibfield  {title} {\enquote {\bibinfo
  {title} {Thermal agitation of electric charge in conductors},}\ }\href
  {\doibase 10.1103/PhysRev.32.110} {\bibfield  {journal} {\bibinfo  {journal}
  {Phys. Rev.}\ }\textbf {\bibinfo {volume} {32}},\ \bibinfo {pages}
  {110--113}}\BibitemShut {NoStop}%
\bibitem [{\citenamefont {O{'}Gorman}(2012)}]{ref:OGorman12}%
  \BibitemOpen
  \bibfield  {author} {\bibinfo {author} {\bibnamefont {O{'}Gorman},
  \bibfnamefont {Paul~A}}} (\bibinfo {year} {2012}),\ \bibfield  {title}
  {\enquote {\bibinfo {title} {Sensitivity of tropical precipitation extremes
  to climate change},}\ }\href {\doibase
  http://www.nature.com/ngeo/journal/v5/n10/abs/ngeo1568.html#supplementary-information}
  {\bibfield  {journal} {\bibinfo  {journal} {Nature Geosci}\ }\textbf
  {\bibinfo {volume} {5}}~(\bibinfo {number} {10}),\ \bibinfo {pages}
  {697--700}}\BibitemShut {NoStop}%
\bibitem [{\citenamefont {Palmer}(2019)}]{ref:Palmer19}%
  \BibitemOpen
  \bibfield  {author} {\bibinfo {author} {\bibnamefont {Palmer}, \bibfnamefont
  {Tim}}} (\bibinfo {year} {2019}),\ \bibfield  {title} {\enquote {\bibinfo
  {title} {The ecmwf ensemble prediction system: Looking back (more than) 25
  years and projecting forward 25 years},}\ }\href {\doibase 10.1002/qj.3383}
  {\bibfield  {journal} {\bibinfo  {journal} {Quarterly Journal of the Royal
  Meteorological Society}\ }\textbf {\bibinfo {volume} {145}}~(\bibinfo
  {number} {S1}),\ \bibinfo {pages} {12--24}}\BibitemShut {NoStop}%
\bibitem [{\citenamefont {Pendergrass}(2020)}]{ref:Pendergrass20}%
  \BibitemOpen
  \bibfield  {author} {\bibinfo {author} {\bibnamefont {Pendergrass},
  \bibfnamefont {A~G}}} (\bibinfo {year} {2020}),\ \bibfield  {title} {\enquote
  {\bibinfo {title} {{The global-mean precipitation response to CO2-induced
  warming in CMIP6 models}},}\ }\href {\doibase 10.1029/2020GL089964}
  {\bibfield  {journal} {\bibinfo  {journal} {Geophysical Research Letters}\
  }\textbf {\bibinfo {volume} {n/a}}~(\bibinfo {number} {n/a}),\ \bibinfo
  {pages} {e2020GL089964}}\BibitemShut {NoStop}%
\bibitem [{\citenamefont {Pendergrass}\ and\ \citenamefont
  {Hartmann}(2014)}]{ref:Pendergrass14}%
  \BibitemOpen
  \bibfield  {author} {\bibinfo {author} {\bibnamefont {Pendergrass},
  \bibfnamefont {Angeline~G}}, \ and\ \bibinfo {author} {\bibfnamefont
  {Dennis~L.}\ \bibnamefont {Hartmann}}} (\bibinfo {year} {2014}),\ \bibfield
  {title} {\enquote {\bibinfo {title} {{The atmospheric energy constraint on
  global-mean precipitation change}},}\ }\href {\doibase
  10.1175/JCLI-D-13-00163.1} {\bibfield  {journal} {\bibinfo  {journal}
  {Journal of Climate}\ }\textbf {\bibinfo {volume} {27}}~(\bibinfo {number}
  {2}),\ \bibinfo {pages} {757--768}}\BibitemShut {NoStop}%
\bibitem [{\citenamefont {Pennell}\ and\ \citenamefont
  {Reichler}(2010)}]{ref:Pennell11}%
  \BibitemOpen
  \bibfield  {author} {\bibinfo {author} {\bibnamefont {Pennell}, \bibfnamefont
  {C}}, \ and\ \bibinfo {author} {\bibfnamefont {T.}~\bibnamefont {Reichler}}}
  (\bibinfo {year} {2010}),\ \bibfield  {title} {\enquote {\bibinfo {title} {On
  the effective number of climate models},}\ }\href {\doibase
  10.1175/2010JCLI3814.1} {\bibfield  {journal} {\bibinfo  {journal} {Journal
  of Climate}\ }\textbf {\bibinfo {volume} {24}}~(\bibinfo {number} {9}),\
  \bibinfo {pages} {2358--2367}}\BibitemShut {NoStop}%
\bibitem [{\citenamefont {Po-Chedley}\ \emph {et~al.}(2018)\citenamefont
  {Po-Chedley}, \citenamefont {Proistosescu}, \citenamefont {Armour},\ and\
  \citenamefont {Santer}}]{ref:Po-Chedley18}%
  \BibitemOpen
  \bibfield  {author} {\bibinfo {author} {\bibnamefont {Po-Chedley},
  \bibfnamefont {Stephen}}, \bibinfo {author} {\bibfnamefont {Cristian}\
  \bibnamefont {Proistosescu}}, \bibinfo {author} {\bibfnamefont {Kyle~C.}\
  \bibnamefont {Armour}}, \ and\ \bibinfo {author} {\bibfnamefont
  {Benjamin~D.}\ \bibnamefont {Santer}}} (\bibinfo {year} {2018}),\ \bibfield
  {title} {\enquote {\bibinfo {title} {Climate constraint reflects forced
  signal},}\ }\href {\doibase 10.1038/s41586-018-0640-y} {\bibfield  {journal}
  {\bibinfo  {journal} {Nature}\ }\textbf {\bibinfo {volume} {563}}~(\bibinfo
  {number} {7729}),\ \bibinfo {pages} {E6--E9}}\BibitemShut {NoStop}%
\bibitem [{\citenamefont {Po-Chedley}\ \emph {et~al.}(2019)\citenamefont
  {Po-Chedley}, \citenamefont {Zelinka}, \citenamefont {Jeevanjee},
  \citenamefont {Thorsen},\ and\ \citenamefont {Santer}}]{ref:Po-Chedley19}%
  \BibitemOpen
  \bibfield  {author} {\bibinfo {author} {\bibnamefont {Po-Chedley},
  \bibfnamefont {Stephen}}, \bibinfo {author} {\bibfnamefont {Mark~D.}\
  \bibnamefont {Zelinka}}, \bibinfo {author} {\bibfnamefont {Nadir}\
  \bibnamefont {Jeevanjee}}, \bibinfo {author} {\bibfnamefont {Tyler~J.}\
  \bibnamefont {Thorsen}}, \ and\ \bibinfo {author} {\bibfnamefont
  {Benjamin~D.}\ \bibnamefont {Santer}}} (\bibinfo {year} {2019}),\ \bibfield
  {title} {\enquote {\bibinfo {title} {Climatology explains intermodel spread
  in tropical upper tropospheric cloud and relative humidity response to
  greenhouse warming},}\ }\href {\doibase https://doi.org/10.1029/2019GL084786}
  {\bibfield  {journal} {\bibinfo  {journal} {Geophysical Research Letters}\
  }\textbf {\bibinfo {volume} {46}}~(\bibinfo {number} {22}),\ \bibinfo {pages}
  {13399--13409}}\BibitemShut {NoStop}%
\bibitem [{\citenamefont {Qu}\ and\ \citenamefont
  {Hall}(2014)}]{ref:Qu&Hall14}%
  \BibitemOpen
  \bibfield  {author} {\bibinfo {author} {\bibnamefont {Qu}, \bibfnamefont
  {X}}, \ and\ \bibinfo {author} {\bibfnamefont {A.}~\bibnamefont {Hall}}}
  (\bibinfo {year} {2014}),\ \bibfield  {title} {\enquote {\bibinfo {title} {On
  the persistent spread in snow-albedo feedback},}\ }\href {\doibase
  10.1007/s00382-013-1774-0} {\bibfield  {journal} {\bibinfo  {journal}
  {Climate Dynamics}\ }\textbf {\bibinfo {volume} {42}}~(\bibinfo {number}
  {1-2}),\ \bibinfo {pages} {69--81}}\BibitemShut {NoStop}%
\bibitem [{\citenamefont {Qu}\ \emph {et~al.}(2015)\citenamefont {Qu},
  \citenamefont {Hall}, \citenamefont {Klein},\ and\ \citenamefont
  {DeAngelis}}]{ref:Qu15}%
  \BibitemOpen
  \bibfield  {author} {\bibinfo {author} {\bibnamefont {Qu}, \bibfnamefont
  {X}}, \bibinfo {author} {\bibfnamefont {A.}~\bibnamefont {Hall}}, \bibinfo
  {author} {\bibfnamefont {S.~A.}\ \bibnamefont {Klein}}, \ and\ \bibinfo
  {author} {\bibfnamefont {A.~M.}\ \bibnamefont {DeAngelis}}} (\bibinfo {year}
  {2015}),\ \bibfield  {title} {\enquote {\bibinfo {title} {Positive tropical
  marine low-cloud cover feedback inferred from cloud-controlling factors},}\
  }\href {\doibase 10.1002/2015gl065627} {\bibfield  {journal} {\bibinfo
  {journal} {Geophysical Research Letters}\ }\textbf {\bibinfo {volume}
  {42}}~(\bibinfo {number} {18}),\ \bibinfo {pages} {7767--7775}}\BibitemShut
  {NoStop}%
\bibitem [{\citenamefont {Qu}\ and\ \citenamefont
  {Hall}(2007)}]{ref:Qu&Hall07}%
  \BibitemOpen
  \bibfield  {author} {\bibinfo {author} {\bibnamefont {Qu}, \bibfnamefont
  {Xin}}, \ and\ \bibinfo {author} {\bibfnamefont {Alex}\ \bibnamefont {Hall}}}
  (\bibinfo {year} {2007}),\ \bibfield  {title} {\enquote {\bibinfo {title}
  {What controls the strength of snow-albedo feedback?}}\ }\href {\doibase
  10.1175/JCLI4186.1} {\bibfield  {journal} {\bibinfo  {journal} {Journal of
  Climate}\ }\textbf {\bibinfo {volume} {20}}~(\bibinfo {number} {15}),\
  \bibinfo {pages} {3971--3981}},\ \Eprint
  {http://arxiv.org/abs/https://doi.org/10.1175/JCLI4186.1}
  {https://doi.org/10.1175/JCLI4186.1} \BibitemShut {NoStop}%
\bibitem [{\citenamefont {Qu}\ \emph {et~al.}(2018)\citenamefont {Qu},
  \citenamefont {Hall}, \citenamefont {DeAngelis}, \citenamefont {Zelinka},
  \citenamefont {Klein}, \citenamefont {Su}, \citenamefont {Tian},\ and\
  \citenamefont {Zhai}}]{ref:Qu2018}%
  \BibitemOpen
  \bibfield  {author} {\bibinfo {author} {\bibnamefont {Qu}, \bibfnamefont
  {Xin}}, \bibinfo {author} {\bibfnamefont {Alex}\ \bibnamefont {Hall}},
  \bibinfo {author} {\bibfnamefont {Anthony~M.}\ \bibnamefont {DeAngelis}},
  \bibinfo {author} {\bibfnamefont {Mark~D.}\ \bibnamefont {Zelinka}}, \bibinfo
  {author} {\bibfnamefont {Stephen~A.}\ \bibnamefont {Klein}}, \bibinfo
  {author} {\bibfnamefont {Hui}\ \bibnamefont {Su}}, \bibinfo {author}
  {\bibfnamefont {Baijun}\ \bibnamefont {Tian}}, \ and\ \bibinfo {author}
  {\bibfnamefont {Chengxing}\ \bibnamefont {Zhai}}} (\bibinfo {year} {2018}),\
  \bibfield  {title} {\enquote {\bibinfo {title} {{On the emergent constraints
  of climate sensitivity}},}\ }\href {\doibase 10.1175/JCLI-D-17-0482.1}
  {\bibfield  {journal} {\bibinfo  {journal} {Journal of Climate}\ }\textbf
  {\bibinfo {volume} {31}}~(\bibinfo {number} {2}),\ \bibinfo {pages}
  {863--875}}\BibitemShut {NoStop}%
\bibitem [{\citenamefont {Qu}\ \emph {et~al.}(2014)\citenamefont {Qu},
  \citenamefont {Hall}, \citenamefont {Klein},\ and\ \citenamefont
  {Caldwell}}]{ref:Qu14}%
  \BibitemOpen
  \bibfield  {author} {\bibinfo {author} {\bibnamefont {Qu}, \bibfnamefont
  {Xin}}, \bibinfo {author} {\bibfnamefont {Alex}\ \bibnamefont {Hall}},
  \bibinfo {author} {\bibfnamefont {Stephen~A.}\ \bibnamefont {Klein}}, \ and\
  \bibinfo {author} {\bibfnamefont {Peter~M.}\ \bibnamefont {Caldwell}}}
  (\bibinfo {year} {2014}),\ \bibfield  {title} {\enquote {\bibinfo {title} {On
  the spread of changes in marine low cloud cover in climate model simulations
  of the 21st century},}\ }\href {\doibase 10.1007/s00382-013-1945-z}
  {\bibfield  {journal} {\bibinfo  {journal} {Climate Dynamics}\ }\textbf
  {\bibinfo {volume} {42}}~(\bibinfo {number} {9}),\ \bibinfo {pages}
  {2603--2626}}\BibitemShut {NoStop}%
\bibitem [{\citenamefont {Renoult}\ \emph {et~al.}(2020)\citenamefont
  {Renoult}, \citenamefont {Annan}, \citenamefont {Hargreaves}, \citenamefont
  {Sagoo}, \citenamefont {Flynn}, \citenamefont {Kapsch}, \citenamefont
  {Mikolajewicz}, \citenamefont {Ohgaito},\ and\ \citenamefont
  {Mauritsen}}]{ref:Renoult20}%
  \BibitemOpen
  \bibfield  {author} {\bibinfo {author} {\bibnamefont {Renoult}, \bibfnamefont
  {Martin}}, \bibinfo {author} {\bibfnamefont {James~Douglas}\ \bibnamefont
  {Annan}}, \bibinfo {author} {\bibfnamefont {Julia~Catherine}\ \bibnamefont
  {Hargreaves}}, \bibinfo {author} {\bibfnamefont {Navjit}\ \bibnamefont
  {Sagoo}}, \bibinfo {author} {\bibfnamefont {Clare}\ \bibnamefont {Flynn}},
  \bibinfo {author} {\bibfnamefont {Marie-Luise}\ \bibnamefont {Kapsch}},
  \bibinfo {author} {\bibfnamefont {Uwe}\ \bibnamefont {Mikolajewicz}},
  \bibinfo {author} {\bibfnamefont {Rumi}\ \bibnamefont {Ohgaito}}, \ and\
  \bibinfo {author} {\bibfnamefont {Thorsten}\ \bibnamefont {Mauritsen}}}
  (\bibinfo {year} {2020}),\ \bibfield  {title} {\enquote {\bibinfo {title} {{A
  Bayesian framework for emergent constraints: case studies of climate
  sensitivity with PMIP}},}\ }\href {\doibase 10.5194/cp-2019-162} {\bibinfo
  {journal} {Climate of the Past Discussions}\ ,\ \bibinfo {pages}
  {1--29}}\BibitemShut {NoStop}%
\bibitem [{\citenamefont {Riahi}\ \emph {et~al.}(2017)\citenamefont {Riahi},
  \citenamefont {van Vuuren}, \citenamefont {Kriegler}, \citenamefont
  {Edmonds}, \citenamefont {O'Neill}, \citenamefont {Fujimori}, \citenamefont
  {Bauer}, \citenamefont {Calvin}, \citenamefont {Dellink}, \citenamefont
  {Fricko}, \citenamefont {Lutz}, \citenamefont {Popp}, \citenamefont
  {Cuaresma}, \citenamefont {KC}, \citenamefont {Leimbach}, \citenamefont
  {Jiang}, \citenamefont {Kram}, \citenamefont {Rao}, \citenamefont
  {Emmerling}, \citenamefont {Ebi}, \citenamefont {Hasegawa}, \citenamefont
  {Havlik}, \citenamefont {Humpen{\"{o}}der}, \citenamefont {{Da Silva}},
  \citenamefont {Smith}, \citenamefont {Stehfest}, \citenamefont {Bosetti},
  \citenamefont {Eom}, \citenamefont {Gernaat}, \citenamefont {Masui},
  \citenamefont {Rogelj}, \citenamefont {Strefler}, \citenamefont {Drouet},
  \citenamefont {Krey}, \citenamefont {Luderer}, \citenamefont {Harmsen},
  \citenamefont {Takahashi}, \citenamefont {Baumstark}, \citenamefont
  {Doelman}, \citenamefont {Kainuma}, \citenamefont {Klimont}, \citenamefont
  {Marangoni}, \citenamefont {Lotze-Campen}, \citenamefont {Obersteiner},
  \citenamefont {Tabeau},\ and\ \citenamefont {Tavoni}}]{ref:Riahi17}%
  \BibitemOpen
\bibfield  {journal} {  }\bibfield  {author} {\bibinfo {author} {\bibnamefont
  {Riahi}, \bibfnamefont {Keywan}}, \bibinfo {author} {\bibfnamefont
  {Detlef~P.}\ \bibnamefont {van Vuuren}}, \bibinfo {author} {\bibfnamefont
  {Elmar}\ \bibnamefont {Kriegler}}, \bibinfo {author} {\bibfnamefont {Jae}\
  \bibnamefont {Edmonds}}, \bibinfo {author} {\bibfnamefont {Brian~C.}\
  \bibnamefont {O'Neill}}, \bibinfo {author} {\bibfnamefont {Shinichiro}\
  \bibnamefont {Fujimori}}, \bibinfo {author} {\bibfnamefont {Nico}\
  \bibnamefont {Bauer}}, \bibinfo {author} {\bibfnamefont {Katherine}\
  \bibnamefont {Calvin}}, \bibinfo {author} {\bibfnamefont {Rob}\ \bibnamefont
  {Dellink}}, \bibinfo {author} {\bibfnamefont {Oliver}\ \bibnamefont
  {Fricko}}, \bibinfo {author} {\bibfnamefont {Wolfgang}\ \bibnamefont {Lutz}},
  \bibinfo {author} {\bibfnamefont {Alexander}\ \bibnamefont {Popp}}, \bibinfo
  {author} {\bibfnamefont {Jesus~Crespo}\ \bibnamefont {Cuaresma}}, \bibinfo
  {author} {\bibfnamefont {Samir}\ \bibnamefont {KC}}, \bibinfo {author}
  {\bibfnamefont {Marian}\ \bibnamefont {Leimbach}}, \bibinfo {author}
  {\bibfnamefont {Leiwen}\ \bibnamefont {Jiang}}, \bibinfo {author}
  {\bibfnamefont {Tom}\ \bibnamefont {Kram}}, \bibinfo {author} {\bibfnamefont
  {Shilpa}\ \bibnamefont {Rao}}, \bibinfo {author} {\bibfnamefont {Johannes}\
  \bibnamefont {Emmerling}}, \bibinfo {author} {\bibfnamefont {Kristie}\
  \bibnamefont {Ebi}}, \bibinfo {author} {\bibfnamefont {Tomoko}\ \bibnamefont
  {Hasegawa}}, \bibinfo {author} {\bibfnamefont {Petr}\ \bibnamefont {Havlik}},
  \bibinfo {author} {\bibfnamefont {Florian}\ \bibnamefont {Humpen{\"{o}}der}},
  \bibinfo {author} {\bibfnamefont {Lara~Aleluia}\ \bibnamefont {{Da Silva}}},
  \bibinfo {author} {\bibfnamefont {Steve}\ \bibnamefont {Smith}}, \bibinfo
  {author} {\bibfnamefont {Elke}\ \bibnamefont {Stehfest}}, \bibinfo {author}
  {\bibfnamefont {Valentina}\ \bibnamefont {Bosetti}}, \bibinfo {author}
  {\bibfnamefont {Jiyong}\ \bibnamefont {Eom}}, \bibinfo {author}
  {\bibfnamefont {David}\ \bibnamefont {Gernaat}}, \bibinfo {author}
  {\bibfnamefont {Toshihiko}\ \bibnamefont {Masui}}, \bibinfo {author}
  {\bibfnamefont {Joeri}\ \bibnamefont {Rogelj}}, \bibinfo {author}
  {\bibfnamefont {Jessica}\ \bibnamefont {Strefler}}, \bibinfo {author}
  {\bibfnamefont {Laurent}\ \bibnamefont {Drouet}}, \bibinfo {author}
  {\bibfnamefont {Volker}\ \bibnamefont {Krey}}, \bibinfo {author}
  {\bibfnamefont {Gunnar}\ \bibnamefont {Luderer}}, \bibinfo {author}
  {\bibfnamefont {Mathijs}\ \bibnamefont {Harmsen}}, \bibinfo {author}
  {\bibfnamefont {Kiyoshi}\ \bibnamefont {Takahashi}}, \bibinfo {author}
  {\bibfnamefont {Lavinia}\ \bibnamefont {Baumstark}}, \bibinfo {author}
  {\bibfnamefont {Jonathan~C.}\ \bibnamefont {Doelman}}, \bibinfo {author}
  {\bibfnamefont {Mikiko}\ \bibnamefont {Kainuma}}, \bibinfo {author}
  {\bibfnamefont {Zbigniew}\ \bibnamefont {Klimont}}, \bibinfo {author}
  {\bibfnamefont {Giacomo}\ \bibnamefont {Marangoni}}, \bibinfo {author}
  {\bibfnamefont {Hermann}\ \bibnamefont {Lotze-Campen}}, \bibinfo {author}
  {\bibfnamefont {Michael}\ \bibnamefont {Obersteiner}}, \bibinfo {author}
  {\bibfnamefont {Andrzej}\ \bibnamefont {Tabeau}}, \ and\ \bibinfo {author}
  {\bibfnamefont {Massimo}\ \bibnamefont {Tavoni}}} (\bibinfo {year} {2017}),\
  \bibfield  {title} {\enquote {\bibinfo {title} {{The Shared Socioeconomic
  Pathways and their energy, land use, and greenhouse gas emissions
  implications: An overview}},}\ }\href {\doibase
  10.1016/j.gloenvcha.2016.05.009} {\bibfield  {journal} {\bibinfo  {journal}
  {Global Environmental Change}\ }\textbf {\bibinfo {volume} {42}},\ \bibinfo
  {pages} {153--168}}\BibitemShut {NoStop}%
\bibitem [{\citenamefont {Roemmich}\ \emph {et~al.}(2019)\citenamefont
  {Roemmich}, \citenamefont {Alford}, \citenamefont {Claustre}, \citenamefont
  {Johnson}, \citenamefont {King}, \citenamefont {Moum}, \citenamefont {Oke},
  \citenamefont {Owens}, \citenamefont {Pouliquen}, \citenamefont {Purkey},
  \citenamefont {Scanderbeg}, \citenamefont {Suga}, \citenamefont {Wijffels},
  \citenamefont {Zilberman}, \citenamefont {Bakker}, \citenamefont {Baringer},
  \citenamefont {Belbeoch}, \citenamefont {Bittig}, \citenamefont {Boss},
  \citenamefont {Calil}, \citenamefont {Carse}, \citenamefont {Carval},
  \citenamefont {Chai}, \citenamefont {Conchubhair}, \citenamefont
  {d’Ortenzio}, \citenamefont {Dall’Olmo}, \citenamefont {Desbruyeres},
  \citenamefont {Fennel}, \citenamefont {Fer}, \citenamefont {Ferrari},
  \citenamefont {Forget}, \citenamefont {Freeland}, \citenamefont {Fujiki},
  \citenamefont {Gehlen}, \citenamefont {Greenan}, \citenamefont {Hallberg},
  \citenamefont {Hibiya}, \citenamefont {Hosoda}, \citenamefont {Jayne},
  \citenamefont {Jochum}, \citenamefont {Johnson}, \citenamefont {Kang},
  \citenamefont {Kolodziejczyk}, \citenamefont {Körtzinger}, \citenamefont
  {Traon}, \citenamefont {Lenn}, \citenamefont {Maze}, \citenamefont {Mork},
  \citenamefont {Morris}, \citenamefont {Nagai}, \citenamefont {Nash},
  \citenamefont {Garabato}, \citenamefont {Olsen}, \citenamefont {Pattabhi},
  \citenamefont {Prakash}, \citenamefont {Riser}, \citenamefont {Schmechtig},
  \citenamefont {Schmid}, \citenamefont {Shroyer}, \citenamefont {Sterl},
  \citenamefont {Sutton}, \citenamefont {Talley}, \citenamefont {Tanhua},
  \citenamefont {Thierry}, \citenamefont {Thomalla}, \citenamefont {Toole},
  \citenamefont {Troisi}, \citenamefont {Trull}, \citenamefont {Turton},
  \citenamefont {Velez-Belchi}, \citenamefont {Walczowski}, \citenamefont
  {Wang}, \citenamefont {Wanninkhof}, \citenamefont {Waterhouse}, \citenamefont
  {Waterman}, \citenamefont {Watson}, \citenamefont {Wilson}, \citenamefont
  {Wong}, \citenamefont {Xu},\ and\ \citenamefont {Yasuda}}]{ref:Roemmich09}%
  \BibitemOpen
  \bibfield  {author} {\bibinfo {author} {\bibnamefont {Roemmich},
  \bibfnamefont {Dean}}, \bibinfo {author} {\bibfnamefont {Matthew~H.}\
  \bibnamefont {Alford}}, \bibinfo {author} {\bibfnamefont {Hervé}\
  \bibnamefont {Claustre}}, \bibinfo {author} {\bibfnamefont {Kenneth}\
  \bibnamefont {Johnson}}, \bibinfo {author} {\bibfnamefont {Brian}\
  \bibnamefont {King}}, \bibinfo {author} {\bibfnamefont {James}\ \bibnamefont
  {Moum}}, \bibinfo {author} {\bibfnamefont {Peter}\ \bibnamefont {Oke}},
  \bibinfo {author} {\bibfnamefont {W.~Brechner}\ \bibnamefont {Owens}},
  \bibinfo {author} {\bibfnamefont {Sylvie}\ \bibnamefont {Pouliquen}},
  \bibinfo {author} {\bibfnamefont {Sarah}\ \bibnamefont {Purkey}}, \bibinfo
  {author} {\bibfnamefont {Megan}\ \bibnamefont {Scanderbeg}}, \bibinfo
  {author} {\bibfnamefont {Toshio}\ \bibnamefont {Suga}}, \bibinfo {author}
  {\bibfnamefont {Susan}\ \bibnamefont {Wijffels}}, \bibinfo {author}
  {\bibfnamefont {Nathalie}\ \bibnamefont {Zilberman}}, \bibinfo {author}
  {\bibfnamefont {Dorothee}\ \bibnamefont {Bakker}}, \bibinfo {author}
  {\bibfnamefont {Molly}\ \bibnamefont {Baringer}}, \bibinfo {author}
  {\bibfnamefont {Mathieu}\ \bibnamefont {Belbeoch}}, \bibinfo {author}
  {\bibfnamefont {Henry~C.}\ \bibnamefont {Bittig}}, \bibinfo {author}
  {\bibfnamefont {Emmanuel}\ \bibnamefont {Boss}}, \bibinfo {author}
  {\bibfnamefont {Paulo}\ \bibnamefont {Calil}}, \bibinfo {author}
  {\bibfnamefont {Fiona}\ \bibnamefont {Carse}}, \bibinfo {author}
  {\bibfnamefont {Thierry}\ \bibnamefont {Carval}}, \bibinfo {author}
  {\bibfnamefont {Fei}\ \bibnamefont {Chai}}, \bibinfo {author} {\bibfnamefont
  {Diarmuid~Ó.}\ \bibnamefont {Conchubhair}}, \bibinfo {author} {\bibfnamefont
  {Fabrizio}\ \bibnamefont {d’Ortenzio}}, \bibinfo {author} {\bibfnamefont
  {Giorgio}\ \bibnamefont {Dall’Olmo}}, \bibinfo {author} {\bibfnamefont
  {Damien}\ \bibnamefont {Desbruyeres}}, \bibinfo {author} {\bibfnamefont
  {Katja}\ \bibnamefont {Fennel}}, \bibinfo {author} {\bibfnamefont {Ilker}\
  \bibnamefont {Fer}}, \bibinfo {author} {\bibfnamefont {Raffaele}\
  \bibnamefont {Ferrari}}, \bibinfo {author} {\bibfnamefont {Gael}\
  \bibnamefont {Forget}}, \bibinfo {author} {\bibfnamefont {Howard}\
  \bibnamefont {Freeland}}, \bibinfo {author} {\bibfnamefont {Tetsuichi}\
  \bibnamefont {Fujiki}}, \bibinfo {author} {\bibfnamefont {Marion}\
  \bibnamefont {Gehlen}}, \bibinfo {author} {\bibfnamefont {Blair}\
  \bibnamefont {Greenan}}, \bibinfo {author} {\bibfnamefont {Robert}\
  \bibnamefont {Hallberg}}, \bibinfo {author} {\bibfnamefont {Toshiyuki}\
  \bibnamefont {Hibiya}}, \bibinfo {author} {\bibfnamefont {Shigeki}\
  \bibnamefont {Hosoda}}, \bibinfo {author} {\bibfnamefont {Steven}\
  \bibnamefont {Jayne}}, \bibinfo {author} {\bibfnamefont {Markus}\
  \bibnamefont {Jochum}}, \bibinfo {author} {\bibfnamefont {Gregory~C.}\
  \bibnamefont {Johnson}}, \bibinfo {author} {\bibfnamefont {KiRyong}\
  \bibnamefont {Kang}}, \bibinfo {author} {\bibfnamefont {Nicolas}\
  \bibnamefont {Kolodziejczyk}}, \bibinfo {author} {\bibfnamefont {Arne}\
  \bibnamefont {Körtzinger}}, \bibinfo {author} {\bibfnamefont
  {Pierre-Yves~Le}\ \bibnamefont {Traon}}, \bibinfo {author} {\bibfnamefont
  {Yueng-Djern}\ \bibnamefont {Lenn}}, \bibinfo {author} {\bibfnamefont
  {Guillaume}\ \bibnamefont {Maze}}, \bibinfo {author} {\bibfnamefont
  {Kjell~Arne}\ \bibnamefont {Mork}}, \bibinfo {author} {\bibfnamefont
  {Tamaryn}\ \bibnamefont {Morris}}, \bibinfo {author} {\bibfnamefont
  {Takeyoshi}\ \bibnamefont {Nagai}}, \bibinfo {author} {\bibfnamefont
  {Jonathan}\ \bibnamefont {Nash}}, \bibinfo {author} {\bibfnamefont
  {Alberto~Naveira}\ \bibnamefont {Garabato}}, \bibinfo {author} {\bibfnamefont
  {Are}\ \bibnamefont {Olsen}}, \bibinfo {author} {\bibfnamefont {Rama~Rao}\
  \bibnamefont {Pattabhi}}, \bibinfo {author} {\bibfnamefont {Satya}\
  \bibnamefont {Prakash}}, \bibinfo {author} {\bibfnamefont {Stephen}\
  \bibnamefont {Riser}}, \bibinfo {author} {\bibfnamefont {Catherine}\
  \bibnamefont {Schmechtig}}, \bibinfo {author} {\bibfnamefont {Claudia}\
  \bibnamefont {Schmid}}, \bibinfo {author} {\bibfnamefont {Emily}\
  \bibnamefont {Shroyer}}, \bibinfo {author} {\bibfnamefont {Andreas}\
  \bibnamefont {Sterl}}, \bibinfo {author} {\bibfnamefont {Philip}\
  \bibnamefont {Sutton}}, \bibinfo {author} {\bibfnamefont {Lynne}\
  \bibnamefont {Talley}}, \bibinfo {author} {\bibfnamefont {Toste}\
  \bibnamefont {Tanhua}}, \bibinfo {author} {\bibfnamefont {Virginie}\
  \bibnamefont {Thierry}}, \bibinfo {author} {\bibfnamefont {Sandy}\
  \bibnamefont {Thomalla}}, \bibinfo {author} {\bibfnamefont {John}\
  \bibnamefont {Toole}}, \bibinfo {author} {\bibfnamefont {Ariel}\ \bibnamefont
  {Troisi}}, \bibinfo {author} {\bibfnamefont {Thomas~W.}\ \bibnamefont
  {Trull}}, \bibinfo {author} {\bibfnamefont {Jon}\ \bibnamefont {Turton}},
  \bibinfo {author} {\bibfnamefont {Pedro~Joaquin}\ \bibnamefont
  {Velez-Belchi}}, \bibinfo {author} {\bibfnamefont {Waldemar}\ \bibnamefont
  {Walczowski}}, \bibinfo {author} {\bibfnamefont {Haili}\ \bibnamefont
  {Wang}}, \bibinfo {author} {\bibfnamefont {Rik}\ \bibnamefont {Wanninkhof}},
  \bibinfo {author} {\bibfnamefont {Amy~F.}\ \bibnamefont {Waterhouse}},
  \bibinfo {author} {\bibfnamefont {Stephanie}\ \bibnamefont {Waterman}},
  \bibinfo {author} {\bibfnamefont {Andrew}\ \bibnamefont {Watson}}, \bibinfo
  {author} {\bibfnamefont {Cara}\ \bibnamefont {Wilson}}, \bibinfo {author}
  {\bibfnamefont {Annie P.~S.}\ \bibnamefont {Wong}}, \bibinfo {author}
  {\bibfnamefont {Jianping}\ \bibnamefont {Xu}}, \ and\ \bibinfo {author}
  {\bibfnamefont {Ichiro}\ \bibnamefont {Yasuda}}} (\bibinfo {year} {2019}),\
  \bibfield  {title} {\enquote {\bibinfo {title} {On the future of argo: A
  global, full-depth, multi-disciplinary array},}\ }\href {\doibase
  10.3389/fmars.2019.00439} {\bibfield  {journal} {\bibinfo  {journal}
  {Frontiers in Marine Science}\ }\textbf {\bibinfo {volume} {6}},\ \bibinfo
  {pages} {439}}\BibitemShut {NoStop}%
\bibitem [{\citenamefont {Rowell}(2019)}]{ref:Rowell2019}%
  \BibitemOpen
  \bibfield  {author} {\bibinfo {author} {\bibnamefont {Rowell}, \bibfnamefont
  {David~P}}} (\bibinfo {year} {2019}),\ \bibfield  {title} {\enquote {\bibinfo
  {title} {{An Observational Constraint on CMIP5 Projections of the East
  African Long Rains and Southern Indian Ocean Warming}},}\ }\href {\doibase
  10.1029/2019GL082847} {\bibfield  {journal} {\bibinfo  {journal} {Geophysical
  Research Letters}\ }\textbf {\bibinfo {volume} {46}},\ \bibinfo {pages}
  {6050--6058}}\BibitemShut {NoStop}%
\bibitem [{\citenamefont {Rowell}\ and\ \citenamefont
  {Chadwick}(2018)}]{ref:Rowell2018}%
  \BibitemOpen
  \bibfield  {author} {\bibinfo {author} {\bibnamefont {Rowell}, \bibfnamefont
  {David~P}}, \ and\ \bibinfo {author} {\bibfnamefont {Robin}\ \bibnamefont
  {Chadwick}}} (\bibinfo {year} {2018}),\ \bibfield  {title} {\enquote
  {\bibinfo {title} {{Causes of the uncertainty in projections of tropical
  terrestrial rainfall change: East Africa}},}\ }\href {\doibase
  10.1175/JCLI-D-17-0830.1} {\bibfield  {journal} {\bibinfo  {journal} {Journal
  of Climate}\ }\textbf {\bibinfo {volume} {31}}~(\bibinfo {number} {15}),\
  \bibinfo {pages} {5977--5995}}\BibitemShut {NoStop}%
\bibitem [{\citenamefont {Rugenstein}\ \emph {et~al.}(2020)\citenamefont
  {Rugenstein}, \citenamefont {Bloch-Johnson}, \citenamefont {Gregory},
  \citenamefont {Andrews}, \citenamefont {Mauritsen}, \citenamefont {Li},
  \citenamefont {Frölicher}, \citenamefont {Paynter}, \citenamefont
  {Danabasoglu}, \citenamefont {Yang}, \citenamefont {Dufresne}, \citenamefont
  {Cao}, \citenamefont {Schmidt}, \citenamefont {Abe-Ouchi}, \citenamefont
  {Geoffroy},\ and\ \citenamefont {Knutti}}]{ref:Rugenstein20}%
  \BibitemOpen
  \bibfield  {author} {\bibinfo {author} {\bibnamefont {Rugenstein},
  \bibfnamefont {Maria}}, \bibinfo {author} {\bibfnamefont {Jonah}\
  \bibnamefont {Bloch-Johnson}}, \bibinfo {author} {\bibfnamefont {Jonathan}\
  \bibnamefont {Gregory}}, \bibinfo {author} {\bibfnamefont {Timothy}\
  \bibnamefont {Andrews}}, \bibinfo {author} {\bibfnamefont {Thorsten}\
  \bibnamefont {Mauritsen}}, \bibinfo {author} {\bibfnamefont {Chao}\
  \bibnamefont {Li}}, \bibinfo {author} {\bibfnamefont {Thomas~L.}\
  \bibnamefont {Frölicher}}, \bibinfo {author} {\bibfnamefont {David}\
  \bibnamefont {Paynter}}, \bibinfo {author} {\bibfnamefont {Gokhan}\
  \bibnamefont {Danabasoglu}}, \bibinfo {author} {\bibfnamefont {Shuting}\
  \bibnamefont {Yang}}, \bibinfo {author} {\bibfnamefont {Jean-Louis}\
  \bibnamefont {Dufresne}}, \bibinfo {author} {\bibfnamefont {Long}\
  \bibnamefont {Cao}}, \bibinfo {author} {\bibfnamefont {Gavin~A.}\
  \bibnamefont {Schmidt}}, \bibinfo {author} {\bibfnamefont {Ayako}\
  \bibnamefont {Abe-Ouchi}}, \bibinfo {author} {\bibfnamefont {Olivier}\
  \bibnamefont {Geoffroy}}, \ and\ \bibinfo {author} {\bibfnamefont {Reto}\
  \bibnamefont {Knutti}}} (\bibinfo {year} {2020}),\ \bibfield  {title}
  {\enquote {\bibinfo {title} {Equilibrium climate sensitivity estimated by
  equilibrating climate models},}\ }\href {\doibase 10.1029/2019GL083898}
  {\bibfield  {journal} {\bibinfo  {journal} {Geophysical Research Letters}\
  }\textbf {\bibinfo {volume} {47}}~(\bibinfo {number} {4}),\ \bibinfo {pages}
  {e2019GL083898}}\BibitemShut {NoStop}%
\bibitem [{\citenamefont {Rypdal}\ \emph {et~al.}(2018)\citenamefont {Rypdal},
  \citenamefont {Fredriksen}, \citenamefont {Rypdal},\ and\ \citenamefont
  {Steene}}]{ref:Rypdal18}%
  \BibitemOpen
  \bibfield  {author} {\bibinfo {author} {\bibnamefont {Rypdal}, \bibfnamefont
  {Martin}}, \bibinfo {author} {\bibfnamefont {Hege-Beate}\ \bibnamefont
  {Fredriksen}}, \bibinfo {author} {\bibfnamefont {Kristoffer}\ \bibnamefont
  {Rypdal}}, \ and\ \bibinfo {author} {\bibfnamefont {Rebekka~J.}\ \bibnamefont
  {Steene}}} (\bibinfo {year} {2018}),\ \bibfield  {title} {\enquote {\bibinfo
  {title} {Emergent constraints on climate sensitivity},}\ }\href {\doibase
  10.1038/s41586-018-0639-4} {\bibfield  {journal} {\bibinfo  {journal}
  {Nature}\ }\textbf {\bibinfo {volume} {563}}~(\bibinfo {number} {7729}),\
  \bibinfo {pages} {E4--E5}}\BibitemShut {NoStop}%
\bibitem [{\citenamefont {Sacks}\ \emph {et~al.}(1989)\citenamefont {Sacks},
  \citenamefont {Welch}, \citenamefont {Mitchell},\ and\ \citenamefont
  {Wynn}}]{ref:Sacks89}%
  \BibitemOpen
  \bibfield  {author} {\bibinfo {author} {\bibnamefont {Sacks}, \bibfnamefont
  {Jerome}}, \bibinfo {author} {\bibfnamefont {William~J.}\ \bibnamefont
  {Welch}}, \bibinfo {author} {\bibfnamefont {Toby~J.}\ \bibnamefont
  {Mitchell}}, \ and\ \bibinfo {author} {\bibfnamefont {Henry~P.}\ \bibnamefont
  {Wynn}}} (\bibinfo {year} {1989}),\ \bibfield  {title} {\enquote {\bibinfo
  {title} {Design and analysis of computer experiments},}\ }\href {\doibase
  10.1214/ss/1177012413} {\bibfield  {journal} {\bibinfo  {journal} {Statist.
  Sci.}\ }\textbf {\bibinfo {volume} {4}}~(\bibinfo {number} {4}),\ \bibinfo
  {pages} {409--423}}\BibitemShut {NoStop}%
\bibitem [{\citenamefont {Saltzman}(2002)}]{ref:SaltzmanBook}%
  \BibitemOpen
  \bibfield  {author} {\bibinfo {author} {\bibnamefont {Saltzman},
  \bibfnamefont {B}}} (\bibinfo {year} {2002}),\ \href@noop {} {\emph {\bibinfo
  {title} {Dynamical Paleoclimatology: Generalized theory of global climate
  change}}}\ (\bibinfo  {publisher} {Academic Press})\BibitemShut {NoStop}%
\bibitem [{\citenamefont {{Sansom}}\ \emph {et~al.}(2019)\citenamefont
  {{Sansom}}, \citenamefont {{Stephenson}},\ and\ \citenamefont
  {{Bracegirdle}}}]{Samson19arXiv}%
  \BibitemOpen
  \bibfield  {author} {\bibinfo {author} {\bibnamefont {{Sansom}},
  \bibfnamefont {Philip~G}}, \bibinfo {author} {\bibfnamefont {David~B.}\
  \bibnamefont {{Stephenson}}}, \ and\ \bibinfo {author} {\bibfnamefont
  {Thomas~J.}\ \bibnamefont {{Bracegirdle}}}} (\bibinfo {year} {2019}),\
  \bibfield  {title} {\enquote {\bibinfo {title} {{On constraining projections
  of future climate using observations and simulations from multiple climate
  models}},}\ }\href@noop {} {\bibfield  {journal} {\bibinfo  {journal} {arXiv
  e-prints}\ ,\ \bibinfo {eid} {arXiv:1711.04139v3}}}\Eprint
  {http://arxiv.org/abs/1711.04139v3} {arXiv:1711.04139v3 [stat.AP]}
  \BibitemShut {NoStop}%
\bibitem [{\citenamefont {Sansom}(2014)}]{ref:Sansom2014}%
  \BibitemOpen
  \bibfield  {author} {\bibinfo {author} {\bibnamefont {Sansom}, \bibfnamefont
  {Philip~George}}} (\bibinfo {year} {2014}),\ \emph {\bibinfo {title}
  {{Statistical methods for quantifying uncertainty in climate projections from
  ensembles of climate models}}},\ \href
  {https://ore.exeter.ac.uk/repository/handle/10871/15292} {Ph.D. thesis}\
  (\bibinfo  {school} {University of Exeter})\BibitemShut {NoStop}%
\bibitem [{\citenamefont {Schekochihin}(2015)}]{ref:SchekochihinNotes}%
  \BibitemOpen
  \bibfield  {author} {\bibinfo {author} {\bibnamefont {Schekochihin},
  \bibfnamefont {A~A}}} (\bibinfo {year} {2015}),\ \href@noop {} {\enquote
  {\bibinfo {title} {University of oxford 2nd year undergraduate physics:
  Lectures on kinetic theory of gases and statistical physics},}\ }\bibinfo
  {howpublished}
  {\url{http://www-thphys.physics.ox.ac.uk/people/AlexanderSchekochihin/A1/2014/A1LectureNotes.pdf}}\BibitemShut
  {NoStop}%
\bibitem [{\citenamefont {Schlund}\ \emph {et~al.}(2020)\citenamefont
  {Schlund}, \citenamefont {Lauer}, \citenamefont {Gentine}, \citenamefont
  {Sherwood},\ and\ \citenamefont {Eyring}}]{ref:Schlund20}%
  \BibitemOpen
  \bibfield  {author} {\bibinfo {author} {\bibnamefont {Schlund}, \bibfnamefont
  {M}}, \bibinfo {author} {\bibfnamefont {A}~\bibnamefont {Lauer}}, \bibinfo
  {author} {\bibfnamefont {P}~\bibnamefont {Gentine}}, \bibinfo {author}
  {\bibfnamefont {S~C}\ \bibnamefont {Sherwood}}, \ and\ \bibinfo {author}
  {\bibfnamefont {V}~\bibnamefont {Eyring}}} (\bibinfo {year} {2020}),\
  \bibfield  {title} {\enquote {\bibinfo {title} {{Emergent constraints on
  Equilibrium Climate Sensitivity in CMIP5: do they hold for CMIP6?}}}\ }\href
  {\doibase 10.5194/esd-2020-49} {\bibinfo  {journal} {Earth System Dynamics
  Discussions}\ ,\ \bibinfo {pages} {1--40}}\BibitemShut {NoStop}%
\bibitem [{\citenamefont {Schmidt}\ \emph {et~al.}(2014)\citenamefont
  {Schmidt}, \citenamefont {Annan}, \citenamefont {Bartlein}, \citenamefont
  {Cook}, \citenamefont {Guilyardi}, \citenamefont {Hargreaves}, \citenamefont
  {Harrison}, \citenamefont {Kageyama}, \citenamefont {Legrande}, \citenamefont
  {Konecky}, \citenamefont {Lovejoy}, \citenamefont {Mann}, \citenamefont
  {Masson-Delmotte}, \citenamefont {Risi}, \citenamefont {Thompson},
  \citenamefont {Timmermann},\ and\ \citenamefont {Yiou}}]{ref:Schmidt2014}%
  \BibitemOpen
\bibfield  {journal} {  }\bibfield  {author} {\bibinfo {author} {\bibnamefont
  {Schmidt}, \bibfnamefont {G~A}}, \bibinfo {author} {\bibfnamefont {J.~D.}\
  \bibnamefont {Annan}}, \bibinfo {author} {\bibfnamefont {P.~J.}\ \bibnamefont
  {Bartlein}}, \bibinfo {author} {\bibfnamefont {B.~I.}\ \bibnamefont {Cook}},
  \bibinfo {author} {\bibfnamefont {E.}~\bibnamefont {Guilyardi}}, \bibinfo
  {author} {\bibfnamefont {J.~C.}\ \bibnamefont {Hargreaves}}, \bibinfo
  {author} {\bibfnamefont {S.~P.}\ \bibnamefont {Harrison}}, \bibinfo {author}
  {\bibfnamefont {M.}~\bibnamefont {Kageyama}}, \bibinfo {author}
  {\bibfnamefont {A.~N.}\ \bibnamefont {Legrande}}, \bibinfo {author}
  {\bibfnamefont {B.}~\bibnamefont {Konecky}}, \bibinfo {author} {\bibfnamefont
  {S.}~\bibnamefont {Lovejoy}}, \bibinfo {author} {\bibfnamefont {M.~E.}\
  \bibnamefont {Mann}}, \bibinfo {author} {\bibfnamefont {V.}~\bibnamefont
  {Masson-Delmotte}}, \bibinfo {author} {\bibfnamefont {C.}~\bibnamefont
  {Risi}}, \bibinfo {author} {\bibfnamefont {D.}~\bibnamefont {Thompson}},
  \bibinfo {author} {\bibfnamefont {A.}~\bibnamefont {Timmermann}}, \ and\
  \bibinfo {author} {\bibfnamefont {P.}~\bibnamefont {Yiou}}} (\bibinfo {year}
  {2014}),\ \bibfield  {title} {\enquote {\bibinfo {title} {{Using
  palaeo-climate comparisons to constrain future projections in CMIP5}},}\
  }\href {\doibase 10.5194/cp-10-221-2014} {\bibfield  {journal} {\bibinfo
  {journal} {Climate of the Past}\ }\textbf {\bibinfo {volume} {10}}~(\bibinfo
  {number} {1}),\ \bibinfo {pages} {221--250}}\BibitemShut {NoStop}%
\bibitem [{\citenamefont {Schwartz}(2007)}]{ref:Schwartz07}%
  \BibitemOpen
  \bibfield  {author} {\bibinfo {author} {\bibnamefont {Schwartz},
  \bibfnamefont {Stephen~E}}} (\bibinfo {year} {2007}),\ \bibfield  {title}
  {\enquote {\bibinfo {title} {Heat capacity, time constant, and sensitivity of
  earth's climate system},}\ }\href {\doibase 10.1029/2007JD008746} {\bibfield
  {journal} {\bibinfo  {journal} {Journal of Geophysical Research:
  Atmospheres}\ }\textbf {\bibinfo {volume} {112}}~(\bibinfo {number} {D24}),\
  \bibinfo {pages} {n/a--n/a}}\BibitemShut {NoStop}%
\bibitem [{\citenamefont {Selten}\ \emph {et~al.}(2020)\citenamefont {Selten},
  \citenamefont {Bintanja}, \citenamefont {Vautard},\ and\ \citenamefont
  {van~den Hurk}}]{ref:Selten2020}%
  \BibitemOpen
  \bibfield  {author} {\bibinfo {author} {\bibnamefont {Selten}, \bibfnamefont
  {F~M}}, \bibinfo {author} {\bibfnamefont {R.}~\bibnamefont {Bintanja}},
  \bibinfo {author} {\bibfnamefont {R.}~\bibnamefont {Vautard}}, \ and\
  \bibinfo {author} {\bibfnamefont {B.~J.J.M.}\ \bibnamefont {van~den Hurk}}}
  (\bibinfo {year} {2020}),\ \bibfield  {title} {\enquote {\bibinfo {title}
  {{Future continental summer warming constrained by the present-day seasonal
  cycle of surface hydrology}},}\ }\href {\doibase 10.1038/s41598-020-61721-9}
  {\bibfield  {journal} {\bibinfo  {journal} {Scientific Reports}\ }\textbf
  {\bibinfo {volume} {10}}~(\bibinfo {number} {1}),\ \bibinfo {pages}
  {1--7}}\BibitemShut {NoStop}%
\bibitem [{\citenamefont {Sgubin}\ \emph {et~al.}(2017)\citenamefont {Sgubin},
  \citenamefont {Swingedouw}, \citenamefont {Drijfhout}, \citenamefont {Mary},\
  and\ \citenamefont {Bennabi}}]{ref:Sgubin17}%
  \BibitemOpen
  \bibfield  {author} {\bibinfo {author} {\bibnamefont {Sgubin}, \bibfnamefont
  {Giovanni}}, \bibinfo {author} {\bibfnamefont {Didier}\ \bibnamefont
  {Swingedouw}}, \bibinfo {author} {\bibfnamefont {Sybren}\ \bibnamefont
  {Drijfhout}}, \bibinfo {author} {\bibfnamefont {Yannick}\ \bibnamefont
  {Mary}}, \ and\ \bibinfo {author} {\bibfnamefont {Amine}\ \bibnamefont
  {Bennabi}}} (\bibinfo {year} {2017}),\ \bibfield  {title} {\enquote {\bibinfo
  {title} {{Abrupt cooling over the North Atlantic in modern climate
  models}},}\ }\href {\doibase 10.1038/ncomms14375} {\bibfield  {journal}
  {\bibinfo  {journal} {Nature Communications}\ }\textbf {\bibinfo {volume}
  {8}},\ \bibinfo {pages} {1--12}}\BibitemShut {NoStop}%
\bibitem [{\citenamefont {Sherwood}\ \emph {et~al.}(2020)\citenamefont
  {Sherwood}, \citenamefont {Webb}, \citenamefont {Annan}, \citenamefont
  {Armour}, \citenamefont {Forster}, \citenamefont {Hargreaves}, \citenamefont
  {Hegerl}, \citenamefont {Klein}, \citenamefont {Marvel}, \citenamefont
  {Rohling}, \citenamefont {Watanabe}, \citenamefont {Andrews}, \citenamefont
  {Braconnot}, \citenamefont {Bretherton}, \citenamefont {Foster},
  \citenamefont {Hausfather}, \citenamefont {von~der Heydt}, \citenamefont
  {Knutti}, \citenamefont {Mauritsen}, \citenamefont {Norris}, \citenamefont
  {Proistosescu}, \citenamefont {Rugenstein}, \citenamefont {Schmidt},
  \citenamefont {Tokarska},\ and\ \citenamefont {Zelinka}}]{ref:Sherwood20}%
  \BibitemOpen
  \bibfield  {author} {\bibinfo {author} {\bibnamefont {Sherwood},
  \bibfnamefont {S}}, \bibinfo {author} {\bibfnamefont {M~J}\ \bibnamefont
  {Webb}}, \bibinfo {author} {\bibfnamefont {J~D}\ \bibnamefont {Annan}},
  \bibinfo {author} {\bibfnamefont {K~C}\ \bibnamefont {Armour}}, \bibinfo
  {author} {\bibfnamefont {P~M}\ \bibnamefont {Forster}}, \bibinfo {author}
  {\bibfnamefont {J~C}\ \bibnamefont {Hargreaves}}, \bibinfo {author}
  {\bibfnamefont {G}~\bibnamefont {Hegerl}}, \bibinfo {author} {\bibfnamefont
  {S~A}\ \bibnamefont {Klein}}, \bibinfo {author} {\bibfnamefont {K~D}\
  \bibnamefont {Marvel}}, \bibinfo {author} {\bibfnamefont {E~J}\ \bibnamefont
  {Rohling}}, \bibinfo {author} {\bibfnamefont {M}~\bibnamefont {Watanabe}},
  \bibinfo {author} {\bibfnamefont {T}~\bibnamefont {Andrews}}, \bibinfo
  {author} {\bibfnamefont {P}~\bibnamefont {Braconnot}}, \bibinfo {author}
  {\bibfnamefont {C~S}\ \bibnamefont {Bretherton}}, \bibinfo {author}
  {\bibfnamefont {G~L}\ \bibnamefont {Foster}}, \bibinfo {author}
  {\bibfnamefont {Z}~\bibnamefont {Hausfather}}, \bibinfo {author}
  {\bibfnamefont {A~S}\ \bibnamefont {von~der Heydt}}, \bibinfo {author}
  {\bibfnamefont {R}~\bibnamefont {Knutti}}, \bibinfo {author} {\bibfnamefont
  {T}~\bibnamefont {Mauritsen}}, \bibinfo {author} {\bibfnamefont {J~R}\
  \bibnamefont {Norris}}, \bibinfo {author} {\bibfnamefont {C}~\bibnamefont
  {Proistosescu}}, \bibinfo {author} {\bibfnamefont {M}~\bibnamefont
  {Rugenstein}}, \bibinfo {author} {\bibfnamefont {G~A}\ \bibnamefont
  {Schmidt}}, \bibinfo {author} {\bibfnamefont {K~B}\ \bibnamefont {Tokarska}},
  \ and\ \bibinfo {author} {\bibfnamefont {M~D}\ \bibnamefont {Zelinka}}}
  (\bibinfo {year} {2020}),\ \bibfield  {title} {\enquote {\bibinfo {title}
  {{An assessment of Earth's climate sensitivity using multiple lines of
  evidence}},}\ }\href {\doibase 10.1029/2019RG000678} {\bibfield  {journal}
  {\bibinfo  {journal} {Reviews of Geophysics}\ }\textbf {\bibinfo {volume}
  {n/a}}~(\bibinfo {number} {n/a}),\ \bibinfo {pages}
  {e2019RG000678}}\BibitemShut {NoStop}%
\bibitem [{\citenamefont {Sherwood}\ \emph {et~al.}(2014)\citenamefont
  {Sherwood}, \citenamefont {Bony},\ and\ \citenamefont
  {Dufresne}}]{ref:Sherwood14}%
  \BibitemOpen
  \bibfield  {author} {\bibinfo {author} {\bibnamefont {Sherwood},
  \bibfnamefont {S~C}}, \bibinfo {author} {\bibfnamefont {S.}~\bibnamefont
  {Bony}}, \ and\ \bibinfo {author} {\bibfnamefont {J.~L.}\ \bibnamefont
  {Dufresne}}} (\bibinfo {year} {2014}),\ \bibfield  {title} {\enquote
  {\bibinfo {title} {Spread in model climate sensitivity traced to atmospheric
  convective mixing},}\ }\href {\doibase 10.1038/nature12829} {\bibfield
  {journal} {\bibinfo  {journal} {Nature}\ }\textbf {\bibinfo {volume}
  {505}}~(\bibinfo {number} {7481}),\ \bibinfo {pages} {37--42}}\BibitemShut
  {NoStop}%
\bibitem [{\citenamefont {Siler}\ \emph {et~al.}(2018)\citenamefont {Siler},
  \citenamefont {Po-Chedley},\ and\ \citenamefont {Bretherton}}]{ref:Siler18}%
  \BibitemOpen
  \bibfield  {author} {\bibinfo {author} {\bibnamefont {Siler}, \bibfnamefont
  {N}}, \bibinfo {author} {\bibfnamefont {S.}~\bibnamefont {Po-Chedley}}, \
  and\ \bibinfo {author} {\bibfnamefont {C.~S.}\ \bibnamefont {Bretherton}}}
  (\bibinfo {year} {2018}),\ \bibfield  {title} {\enquote {\bibinfo {title}
  {Variability in modeled cloud feedback tied to differences in the
  climatological spatial pattern of clouds},}\ }\href {\doibase
  10.1007/s00382-017-3673-2} {\bibfield  {journal} {\bibinfo  {journal}
  {Climate Dynamics}\ }\textbf {\bibinfo {volume} {50}}~(\bibinfo {number}
  {3-4}),\ \bibinfo {pages} {1209--1220}}\BibitemShut {NoStop}%
\bibitem [{\citenamefont {Simpson}\ and\ \citenamefont
  {Polvani}(2016)}]{ref:Simpson&Polvani16}%
  \BibitemOpen
  \bibfield  {author} {\bibinfo {author} {\bibnamefont {Simpson}, \bibfnamefont
  {I~R}}, \ and\ \bibinfo {author} {\bibfnamefont {L.~M.}\ \bibnamefont
  {Polvani}}} (\bibinfo {year} {2016}),\ \bibfield  {title} {\enquote {\bibinfo
  {title} {Revisiting the relationship between jet position, forced response,
  and annular mode variability in the southern midlatitudes},}\ }\href
  {\doibase 10.1002/2016gl067989} {\bibfield  {journal} {\bibinfo  {journal}
  {Geophysical Research Letters}\ }\textbf {\bibinfo {volume} {43}}~(\bibinfo
  {number} {6}),\ \bibinfo {pages} {2896--2903}}\BibitemShut {NoStop}%
\bibitem [{\citenamefont {Slater}\ and\ \citenamefont
  {Lawrence}(2013)}]{ref:Slater2013}%
  \BibitemOpen
  \bibfield  {author} {\bibinfo {author} {\bibnamefont {Slater}, \bibfnamefont
  {Andrew~G}}, \ and\ \bibinfo {author} {\bibfnamefont {David~M.}\ \bibnamefont
  {Lawrence}}} (\bibinfo {year} {2013}),\ \bibfield  {title} {\enquote
  {\bibinfo {title} {{Diagnosing present and future permafrost from climate
  models}},}\ }\href {\doibase 10.1175/JCLI-D-12-00341.1} {\bibfield  {journal}
  {\bibinfo  {journal} {Journal of Climate}\ }\textbf {\bibinfo {volume}
  {26}}~(\bibinfo {number} {15}),\ \bibinfo {pages} {5608--5623}}\BibitemShut
  {NoStop}%
\bibitem [{\citenamefont {Smith}(2009)}]{ref:Smith09}%
  \BibitemOpen
  \bibfield  {author} {\bibinfo {author} {\bibnamefont {Smith}, \bibfnamefont
  {Richard~J}}} (\bibinfo {year} {2009}),\ \bibfield  {title} {\enquote
  {\bibinfo {title} {Use and misuse of the reduced major axis for
  line-fitting},}\ }\href {\doibase 10.1002/ajpa.21090} {\bibfield  {journal}
  {\bibinfo  {journal} {American Journal of Physical Anthropology}\ }\textbf
  {\bibinfo {volume} {140}}~(\bibinfo {number} {3}),\ \bibinfo {pages}
  {476--486}}\BibitemShut {NoStop}%
\bibitem [{\citenamefont {Stainforth}\ \emph {et~al.}(2005)\citenamefont
  {Stainforth}, \citenamefont {Aina}, \citenamefont {Christensen},
  \citenamefont {Collins}, \citenamefont {Faull}, \citenamefont {Frame},
  \citenamefont {Kettleborough}, \citenamefont {Knight}, \citenamefont
  {Martin}, \citenamefont {Murphy}, \citenamefont {Piani}, \citenamefont
  {Sexton}, \citenamefont {Smith}, \citenamefont {Spicer}, \citenamefont
  {Thorpe},\ and\ \citenamefont {Allen}}]{ref:Stainforth05}%
  \BibitemOpen
  \bibfield  {author} {\bibinfo {author} {\bibnamefont {Stainforth},
  \bibfnamefont {D~A}}, \bibinfo {author} {\bibfnamefont {T.}~\bibnamefont
  {Aina}}, \bibinfo {author} {\bibfnamefont {C.}~\bibnamefont {Christensen}},
  \bibinfo {author} {\bibfnamefont {M.}~\bibnamefont {Collins}}, \bibinfo
  {author} {\bibfnamefont {N.}~\bibnamefont {Faull}}, \bibinfo {author}
  {\bibfnamefont {D.~J.}\ \bibnamefont {Frame}}, \bibinfo {author}
  {\bibfnamefont {J.~A.}\ \bibnamefont {Kettleborough}}, \bibinfo {author}
  {\bibfnamefont {S.}~\bibnamefont {Knight}}, \bibinfo {author} {\bibfnamefont
  {A.}~\bibnamefont {Martin}}, \bibinfo {author} {\bibfnamefont {J.~M.}\
  \bibnamefont {Murphy}}, \bibinfo {author} {\bibfnamefont {C.}~\bibnamefont
  {Piani}}, \bibinfo {author} {\bibfnamefont {D.}~\bibnamefont {Sexton}},
  \bibinfo {author} {\bibfnamefont {L.~A.}\ \bibnamefont {Smith}}, \bibinfo
  {author} {\bibfnamefont {R.~A.}\ \bibnamefont {Spicer}}, \bibinfo {author}
  {\bibfnamefont {A.~J.}\ \bibnamefont {Thorpe}}, \ and\ \bibinfo {author}
  {\bibfnamefont {M.~R.}\ \bibnamefont {Allen}}} (\bibinfo {year} {2005}),\
  \bibfield  {title} {\enquote {\bibinfo {title} {Uncertainty in predictions of
  the climate response to rising levels of greenhouse gases},}\ }\href
  {\doibase 10.1038/nature03301} {\bibfield  {journal} {\bibinfo  {journal}
  {Nature}\ }\textbf {\bibinfo {volume} {433}}~(\bibinfo {number} {7024}),\
  \bibinfo {pages} {403--406}}\BibitemShut {NoStop}%
\bibitem [{\citenamefont {Stocker}\ and\ \citenamefont
  {Wright}(1991)}]{ref:Stocker&Wright91}%
  \BibitemOpen
  \bibfield  {author} {\bibinfo {author} {\bibnamefont {Stocker}, \bibfnamefont
  {Thomas~F}}, \ and\ \bibinfo {author} {\bibfnamefont {Daniel~G.}\
  \bibnamefont {Wright}}} (\bibinfo {year} {1991}),\ \bibfield  {title}
  {\enquote {\bibinfo {title} {Rapid transitions of the ocean's deep
  circulation induced by changes in surface water fluxes},}\ }\href {\doibase
  10.1038/351729a0} {\bibfield  {journal} {\bibinfo  {journal} {Nature}\
  }\textbf {\bibinfo {volume} {351}}~(\bibinfo {number} {6329}),\ \bibinfo
  {pages} {729--732}}\BibitemShut {NoStop}%
\bibitem [{\citenamefont {von Storch}\ and\ \citenamefont
  {Zwiers}(1999)}]{ref:vonStorchBook}%
  \BibitemOpen
  \bibfield  {author} {\bibinfo {author} {\bibnamefont {von Storch},
  \bibfnamefont {H}}, \ and\ \bibinfo {author} {\bibfnamefont {F.~W.}\
  \bibnamefont {Zwiers}}} (\bibinfo {year} {1999}),\ \href@noop {} {\emph
  {\bibinfo {title} {Statistical Analysis in Climate Research}}}\ (\bibinfo
  {publisher} {Cambridge University Press})\BibitemShut {NoStop}%
\bibitem [{\citenamefont {Stroeve}\ \emph {et~al.}(2012)\citenamefont
  {Stroeve}, \citenamefont {Kattsov}, \citenamefont {Barrett}, \citenamefont
  {Serreze}, \citenamefont {Pavlova}, \citenamefont {Holland},\ and\
  \citenamefont {Meier}}]{ref:Stroeve2012}%
  \BibitemOpen
  \bibfield  {author} {\bibinfo {author} {\bibnamefont {Stroeve}, \bibfnamefont
  {Julienne~C}}, \bibinfo {author} {\bibfnamefont {Vladimir}\ \bibnamefont
  {Kattsov}}, \bibinfo {author} {\bibfnamefont {Andrew}\ \bibnamefont
  {Barrett}}, \bibinfo {author} {\bibfnamefont {Mark}\ \bibnamefont {Serreze}},
  \bibinfo {author} {\bibfnamefont {Tatiana}\ \bibnamefont {Pavlova}}, \bibinfo
  {author} {\bibfnamefont {Marika}\ \bibnamefont {Holland}}, \ and\ \bibinfo
  {author} {\bibfnamefont {Walter~N.}\ \bibnamefont {Meier}}} (\bibinfo {year}
  {2012}),\ \bibfield  {title} {\enquote {\bibinfo {title} {{Trends in Arctic
  sea ice extent from CMIP5, CMIP3 and observations}},}\ }\href {\doibase
  10.1029/2012GL052676} {\bibfield  {journal} {\bibinfo  {journal} {Geophysical
  Research Letters}\ }\textbf {\bibinfo {volume} {39}},\ \bibinfo {pages}
  {1--7}}\BibitemShut {NoStop}%
\bibitem [{\citenamefont {Su}\ \emph {et~al.}(2014)\citenamefont {Su},
  \citenamefont {Jiang}, \citenamefont {Zhai}, \citenamefont {Shen},
  \citenamefont {Neelin}, \citenamefont {Stephens},\ and\ \citenamefont
  {Yung}}]{ref:Su14}%
  \BibitemOpen
  \bibfield  {author} {\bibinfo {author} {\bibnamefont {Su}, \bibfnamefont
  {H}}, \bibinfo {author} {\bibfnamefont {J.~H.}\ \bibnamefont {Jiang}},
  \bibinfo {author} {\bibfnamefont {C.~X.}\ \bibnamefont {Zhai}}, \bibinfo
  {author} {\bibfnamefont {T.~J.}\ \bibnamefont {Shen}}, \bibinfo {author}
  {\bibfnamefont {J.~D.}\ \bibnamefont {Neelin}}, \bibinfo {author}
  {\bibfnamefont {G.~L.}\ \bibnamefont {Stephens}}, \ and\ \bibinfo {author}
  {\bibfnamefont {Y.~L.}\ \bibnamefont {Yung}}} (\bibinfo {year} {2014}),\
  \bibfield  {title} {\enquote {\bibinfo {title} {Weakening and strengthening
  structures in the hadley circulation change under global warming and
  implications for cloud response and climate sensitivity},}\ }\href {\doibase
  10.1002/2014jd021642} {\bibfield  {journal} {\bibinfo  {journal} {Journal of
  Geophysical Research-Atmospheres}\ }\textbf {\bibinfo {volume}
  {119}}~(\bibinfo {number} {10}),\ \bibinfo {pages} {5787--5805}}\BibitemShut
  {NoStop}%
\bibitem [{\citenamefont {Taylor}\ \emph {et~al.}(2011)\citenamefont {Taylor},
  \citenamefont {Stouffer},\ and\ \citenamefont {Meehl}}]{ref:CMIP5}%
  \BibitemOpen
  \bibfield  {author} {\bibinfo {author} {\bibnamefont {Taylor}, \bibfnamefont
  {Karl~E}}, \bibinfo {author} {\bibfnamefont {Ronald~J.}\ \bibnamefont
  {Stouffer}}, \ and\ \bibinfo {author} {\bibfnamefont {Gerald~A.}\
  \bibnamefont {Meehl}}} (\bibinfo {year} {2011}),\ \bibfield  {title}
  {\enquote {\bibinfo {title} {An overview of {CMIP5} and the experiment
  design},}\ }\href {\doibase 10.1175/BAMS-D-11-00094.1} {\bibfield  {journal}
  {\bibinfo  {journal} {Bulletin of the American Meteorological Society}\
  }\textbf {\bibinfo {volume} {93}}~(\bibinfo {number} {4}),\ \bibinfo {pages}
  {485--498}}\BibitemShut {NoStop}%
\bibitem [{\citenamefont {Terhaar}\ \emph {et~al.}(2020)\citenamefont
  {Terhaar}, \citenamefont {Kwiatkowski},\ and\ \citenamefont
  {Bopp}}]{ref:Terhaar20}%
  \BibitemOpen
  \bibfield  {author} {\bibinfo {author} {\bibnamefont {Terhaar}, \bibfnamefont
  {Jens}}, \bibinfo {author} {\bibfnamefont {Lester}\ \bibnamefont
  {Kwiatkowski}}, \ and\ \bibinfo {author} {\bibfnamefont {Laurent}\
  \bibnamefont {Bopp}}} (\bibinfo {year} {2020}),\ \bibfield  {title} {\enquote
  {\bibinfo {title} {{Emergent constraint on Arctic Ocean acidification in the
  twenty-first century}},}\ }\href {\doibase 10.1038/s41586-020-2360-3}
  {\bibfield  {journal} {\bibinfo  {journal} {Nature}\ }\textbf {\bibinfo
  {volume} {582}}~(\bibinfo {number} {7812}),\ \bibinfo {pages}
  {379--383}}\BibitemShut {NoStop}%
\bibitem [{\citenamefont {Terrazas}\ \emph {et~al.}(2020)\citenamefont
  {Terrazas}, \citenamefont {Bell}, \citenamefont {Pillepich}, \citenamefont
  {Nelson}, \citenamefont {Somerville}, \citenamefont {Genel}, \citenamefont
  {Weinberger}, \citenamefont {Habouzit}, \citenamefont {Li}, \citenamefont
  {Hernquist},\ and\ \citenamefont {Vogelsberger}}]{ref:Terrazas20}%
  \BibitemOpen
  \bibfield  {author} {\bibinfo {author} {\bibnamefont {Terrazas},
  \bibfnamefont {Bryan~A}}, \bibinfo {author} {\bibfnamefont {Eric~F}\
  \bibnamefont {Bell}}, \bibinfo {author} {\bibfnamefont {Annalisa}\
  \bibnamefont {Pillepich}}, \bibinfo {author} {\bibfnamefont {Dylan}\
  \bibnamefont {Nelson}}, \bibinfo {author} {\bibfnamefont {Rachel~S}\
  \bibnamefont {Somerville}}, \bibinfo {author} {\bibfnamefont {Shy}\
  \bibnamefont {Genel}}, \bibinfo {author} {\bibfnamefont {Rainer}\
  \bibnamefont {Weinberger}}, \bibinfo {author} {\bibfnamefont {MÃ©lanie}\
  \bibnamefont {Habouzit}}, \bibinfo {author} {\bibfnamefont {Yuan}\
  \bibnamefont {Li}}, \bibinfo {author} {\bibfnamefont {Lars}\ \bibnamefont
  {Hernquist}}, \ and\ \bibinfo {author} {\bibfnamefont {Mark}\ \bibnamefont
  {Vogelsberger}}} (\bibinfo {year} {2020}),\ \bibfield  {title} {\enquote
  {\bibinfo {title} {{The relationship between black hole mass and galaxy
  properties: examining the black hole feedback model in IllustrisTNG}},}\
  }\href {\doibase 10.1093/mnras/staa374} {\bibfield  {journal} {\bibinfo
  {journal} {Monthly Notices of the Royal Astronomical Society}\ }\textbf
  {\bibinfo {volume} {493}}~(\bibinfo {number} {2}),\ \bibinfo {pages}
  {1888--1906}}\BibitemShut {NoStop}%
\bibitem [{\citenamefont {Tett}\ \emph {et~al.}(2013)\citenamefont {Tett},
  \citenamefont {Rowlands}, \citenamefont {Mineter},\ and\ \citenamefont
  {Cartis}}]{ref:Tett13}%
  \BibitemOpen
  \bibfield  {author} {\bibinfo {author} {\bibnamefont {Tett}, \bibfnamefont
  {S~F~B}}, \bibinfo {author} {\bibfnamefont {D.~J.}\ \bibnamefont {Rowlands}},
  \bibinfo {author} {\bibfnamefont {M.~J.}\ \bibnamefont {Mineter}}, \ and\
  \bibinfo {author} {\bibfnamefont {C.}~\bibnamefont {Cartis}}} (\bibinfo
  {year} {2013}),\ \bibfield  {title} {\enquote {\bibinfo {title} {Can
  top-of-atmosphere radiation measurements constrain climate predictions? part
  ii: Climate sensitivity},}\ }\href {\doibase 10.1175/jcli-d-12-00596.1}
  {\bibfield  {journal} {\bibinfo  {journal} {Journal of Climate}\ }\textbf
  {\bibinfo {volume} {26}}~(\bibinfo {number} {23}),\ \bibinfo {pages}
  {9367--9383}}\BibitemShut {NoStop}%
\bibitem [{\citenamefont {Teufel}\ and\ \citenamefont
  {Sushama}(2019)}]{ref:Teufel19}%
  \BibitemOpen
  \bibfield  {author} {\bibinfo {author} {\bibnamefont {Teufel}, \bibfnamefont
  {B}}, \ and\ \bibinfo {author} {\bibfnamefont {L}~\bibnamefont {Sushama}}}
  (\bibinfo {year} {2019}),\ \bibfield  {title} {\enquote {\bibinfo {title}
  {{Abrupt changes across the Arctic permafrost region endanger northern
  development}},}\ }\href {\doibase 10.1038/s41558-019-0614-6} {\bibfield
  {journal} {\bibinfo  {journal} {Nature Climate Change}\ }\textbf {\bibinfo
  {volume} {9}}~(\bibinfo {number} {November}),\ \bibinfo {pages}
  {858--862}}\BibitemShut {NoStop}%
\bibitem [{\citenamefont {Thackeray}\ \emph {et~al.}(2018)\citenamefont
  {Thackeray}, \citenamefont {Qu},\ and\ \citenamefont
  {Hall}}]{ref:Thackeray18}%
  \BibitemOpen
  \bibfield  {author} {\bibinfo {author} {\bibnamefont {Thackeray},
  \bibfnamefont {C~W}}, \bibinfo {author} {\bibfnamefont {X.}~\bibnamefont
  {Qu}}, \ and\ \bibinfo {author} {\bibfnamefont {A.}~\bibnamefont {Hall}}}
  (\bibinfo {year} {2018}),\ \bibfield  {title} {\enquote {\bibinfo {title}
  {Why do models produce spread in snow albedo feedback?}}\ }\href {\doibase
  10.1029/2018gl078493} {\bibfield  {journal} {\bibinfo  {journal} {Geophysical
  Research Letters}\ }\textbf {\bibinfo {volume} {45}}~(\bibinfo {number}
  {12}),\ \bibinfo {pages} {6223--6231}}\BibitemShut {NoStop}%
\bibitem [{\citenamefont {Thackeray}\ and\ \citenamefont
  {Fletcher}(2016)}]{ref:Thackeray16}%
  \BibitemOpen
  \bibfield  {author} {\bibinfo {author} {\bibnamefont {Thackeray},
  \bibfnamefont {Chad~W}}, \ and\ \bibinfo {author} {\bibfnamefont
  {Christopher~G.}\ \bibnamefont {Fletcher}}} (\bibinfo {year} {2016}),\
  \bibfield  {title} {\enquote {\bibinfo {title} {{Snow albedo feedback:
  Current knowledge, importance, outstanding issues and future directions}},}\
  }\href {\doibase 10.1177/0309133315620999} {\bibfield  {journal} {\bibinfo
  {journal} {Progress in Physical Geography}\ }\textbf {\bibinfo {volume}
  {40}}~(\bibinfo {number} {3}),\ \bibinfo {pages} {392--408}}\BibitemShut
  {NoStop}%
\bibitem [{\citenamefont {Thackeray}\ and\ \citenamefont
  {Hall}(2019)}]{ref:Thackeray2019}%
  \BibitemOpen
  \bibfield  {author} {\bibinfo {author} {\bibnamefont {Thackeray},
  \bibfnamefont {Chad~W}}, \ and\ \bibinfo {author} {\bibfnamefont {Alex}\
  \bibnamefont {Hall}}} (\bibinfo {year} {2019}),\ \bibfield  {title} {\enquote
  {\bibinfo {title} {{An emergent constraint on future Arctic sea-ice albedo
  feedback}},}\ }\href {\doibase 10.1038/s41558-019-0619-1} {\bibfield
  {journal} {\bibinfo  {journal} {Nature Climate Change}\ }\textbf {\bibinfo
  {volume} {9}},\ \bibinfo {pages} {972--978}}\BibitemShut {NoStop}%
\bibitem [{\citenamefont {Thackeray}\ \emph {et~al.}(submitted)\citenamefont
  {Thackeray}, \citenamefont {Hall}, \citenamefont {Zelinka},\ and\
  \citenamefont {Fletcher}}]{ref:Thackeray2020}%
  \BibitemOpen
  \bibfield  {author} {\bibinfo {author} {\bibnamefont {Thackeray},
  \bibfnamefont {Chad~W}}, \bibinfo {author} {\bibfnamefont {Alex}\
  \bibnamefont {Hall}}, \bibinfo {author} {\bibfnamefont {Mark}\ \bibnamefont
  {Zelinka}}, \ and\ \bibinfo {author} {\bibfnamefont {Christopher}\
  \bibnamefont {Fletcher}}} (\bibinfo {year} {submitted}),\ \bibfield  {title}
  {\enquote {\bibinfo {title} {{Assessing prior emergent constraints on surface
  albedo feedback in CMIP6}},}\ }\href@noop {} {\bibinfo  {journal} {Journal of
  Climate}\ }\BibitemShut {NoStop}%
\bibitem [{\citenamefont {Thomas}\ \emph {et~al.}(2015)\citenamefont {Thomas},
  \citenamefont {Brookshire},\ and\ \citenamefont {Gerber}}]{ref:Quinn15}%
  \BibitemOpen
\bibfield  {journal} {  }\bibfield  {author} {\bibinfo {author} {\bibnamefont
  {Thomas}, \bibfnamefont {R~Quinn}}, \bibinfo {author} {\bibfnamefont
  {E.~N.~Jack}\ \bibnamefont {Brookshire}}, \ and\ \bibinfo {author}
  {\bibfnamefont {Stefan}\ \bibnamefont {Gerber}}} (\bibinfo {year} {2015}),\
  \bibfield  {title} {\enquote {\bibinfo {title} {Nitrogen limitation on land:
  how can it occur in earth system models?}}\ }\href {\doibase
  10.1111/gcb.12813} {\bibfield  {journal} {\bibinfo  {journal} {Global Change
  Biology}\ }\textbf {\bibinfo {volume} {21}}~(\bibinfo {number} {5}),\
  \bibinfo {pages} {1777--1793}}\BibitemShut {NoStop}%
\bibitem [{\citenamefont {Tian}(2015)}]{ref:Tian15}%
  \BibitemOpen
  \bibfield  {author} {\bibinfo {author} {\bibnamefont {Tian}, \bibfnamefont
  {Baijun}}} (\bibinfo {year} {2015}),\ \bibfield  {title} {\enquote {\bibinfo
  {title} {Spread of model climate sensitivity linked to double-intertropical
  convergence zone bias},}\ }\href {\doibase 10.1002/2015GL064119} {\bibfield
  {journal} {\bibinfo  {journal} {Geophysical Research Letters}\ }\textbf
  {\bibinfo {volume} {42}}~(\bibinfo {number} {10}),\ \bibinfo {pages}
  {4133--4141}}\BibitemShut {NoStop}%
\bibitem [{\citenamefont {Tokarska}\ \emph {et~al.}(2020)\citenamefont
  {Tokarska}, \citenamefont {Stolpe}, \citenamefont {Sippel}, \citenamefont
  {Fischer}, \citenamefont {Smith}, \citenamefont {Lehner},\ and\ \citenamefont
  {Knutti}}]{ref:Tokarska2020}%
  \BibitemOpen
  \bibfield  {author} {\bibinfo {author} {\bibnamefont {Tokarska},
  \bibfnamefont {Katarzyna~B}}, \bibinfo {author} {\bibfnamefont {Martin~B.}\
  \bibnamefont {Stolpe}}, \bibinfo {author} {\bibfnamefont {Sebastian}\
  \bibnamefont {Sippel}}, \bibinfo {author} {\bibfnamefont {Erich~M.}\
  \bibnamefont {Fischer}}, \bibinfo {author} {\bibfnamefont {Christopher~J.}\
  \bibnamefont {Smith}}, \bibinfo {author} {\bibfnamefont {Flavio}\
  \bibnamefont {Lehner}}, \ and\ \bibinfo {author} {\bibfnamefont {Reto}\
  \bibnamefont {Knutti}}} (\bibinfo {year} {2020}),\ \bibfield  {title}
  {\enquote {\bibinfo {title} {{Past warming trend constrains future warming in
  CMIP6 models}},}\ }\href {\doibase 10.1126/sciadv.aaz9549} {\bibfield
  {journal} {\bibinfo  {journal} {Science Advances}\ }\textbf {\bibinfo
  {volume} {6}}~(\bibinfo {number} {12}),\ \bibinfo {pages}
  {eaaz9549}}\BibitemShut {NoStop}%
\bibitem [{\citenamefont {Toniazzo}\ \emph {et~al.}(2004)\citenamefont
  {Toniazzo}, \citenamefont {Gregory},\ and\ \citenamefont
  {Huybrechts}}]{ref:Toniazzo04}%
  \BibitemOpen
  \bibfield  {author} {\bibinfo {author} {\bibnamefont {Toniazzo},
  \bibfnamefont {T}}, \bibinfo {author} {\bibfnamefont {J.~M.}\ \bibnamefont
  {Gregory}}, \ and\ \bibinfo {author} {\bibfnamefont {P.}~\bibnamefont
  {Huybrechts}}} (\bibinfo {year} {2004}),\ \bibfield  {title} {\enquote
  {\bibinfo {title} {Climatic impact of a greenland deglaciation and its
  possible irreversibility},}\ }\href {\doibase
  10.1175/1520-0442(2004)017<0021:CIOAGD>2.0.CO;2} {\bibfield  {journal}
  {\bibinfo  {journal} {Journal of Climate}\ }\textbf {\bibinfo {volume}
  {17}}~(\bibinfo {number} {1}),\ \bibinfo {pages} {21--33}}\BibitemShut
  {NoStop}%
\bibitem [{\citenamefont {Toth}\ and\ \citenamefont
  {Kalnay}(1993)}]{ref:Toth&Kalnay93}%
  \BibitemOpen
  \bibfield  {author} {\bibinfo {author} {\bibnamefont {Toth}, \bibfnamefont
  {Zoltan}}, \ and\ \bibinfo {author} {\bibfnamefont {Eugenia}\ \bibnamefont
  {Kalnay}}} (\bibinfo {year} {1993}),\ \bibfield  {title} {\enquote {\bibinfo
  {title} {Ensemble forecasting at nmc: The generation of perturbations},}\
  }\href {\doibase 10.1175/1520-0477(1993)074<2317:EFANTG>2.0.CO;2} {\bibfield
  {journal} {\bibinfo  {journal} {Bulletin of the American Meteorological
  Society}\ }\textbf {\bibinfo {volume} {74}}~(\bibinfo {number} {12}),\
  \bibinfo {pages} {2317--2330}}\BibitemShut {NoStop}%
\bibitem [{\citenamefont {Trenberth}\ and\ \citenamefont
  {Fasullo}(2010)}]{ref:TrenberthFasullo10}%
  \BibitemOpen
  \bibfield  {author} {\bibinfo {author} {\bibnamefont {Trenberth},
  \bibfnamefont {K~E}}, \ and\ \bibinfo {author} {\bibfnamefont {J.~T.}\
  \bibnamefont {Fasullo}}} (\bibinfo {year} {2010}),\ \bibfield  {title}
  {\enquote {\bibinfo {title} {Simulation of present-day and
  twenty-first-century energy budgets of the southern oceans},}\ }\href
  {\doibase 10.1175/2009jcli3152.1} {\bibfield  {journal} {\bibinfo  {journal}
  {Journal of Climate}\ }\textbf {\bibinfo {volume} {23}}~(\bibinfo {number}
  {2}),\ \bibinfo {pages} {440--454}}\BibitemShut {NoStop}%
\bibitem [{\citenamefont {Trenberth}(1984)}]{ref:Trenberth1984}%
  \BibitemOpen
  \bibfield  {author} {\bibinfo {author} {\bibnamefont {Trenberth},
  \bibfnamefont {Kevin~E}}} (\bibinfo {year} {1984}),\ \bibfield  {title}
  {\enquote {\bibinfo {title} {{Some Effects of Finite Sample Size and
  Persistence on Meteorological Statistics. Part I: Autocorrelations}},}\
  }\href {\doibase 10.1175/1520-0493(1984)112<2359:SEOFSS>2.0.CO;2} {\bibfield
  {journal} {\bibinfo  {journal} {Monthly Weather Review}\ }\textbf {\bibinfo
  {volume} {112}}~(\bibinfo {number} {12}),\ \bibinfo {pages}
  {2359--2368}}\BibitemShut {NoStop}%
\bibitem [{\citenamefont {Uhlenbeck}\ and\ \citenamefont
  {Ornstein}(1930)}]{ref:Uhlenbeck&Ornstein30}%
  \BibitemOpen
  \bibfield  {author} {\bibinfo {author} {\bibnamefont {Uhlenbeck},
  \bibfnamefont {G~E}}, \ and\ \bibinfo {author} {\bibfnamefont {L.~S.}\
  \bibnamefont {Ornstein}}} (\bibinfo {year} {1930}),\ \bibfield  {title}
  {\enquote {\bibinfo {title} {On the theory of the brownian motion},}\ }\href
  {\doibase 10.1103/PhysRev.36.823} {\bibfield  {journal} {\bibinfo  {journal}
  {Phys. Rev.}\ }\textbf {\bibinfo {volume} {36}},\ \bibinfo {pages}
  {823--841}}\BibitemShut {NoStop}%
\bibitem [{\citenamefont {Veenhuis}(2013)}]{ref:Veenhuis13}%
  \BibitemOpen
  \bibfield  {author} {\bibinfo {author} {\bibnamefont {Veenhuis},
  \bibfnamefont {Bruce~A}}} (\bibinfo {year} {2013}),\ \bibfield  {title}
  {\enquote {\bibinfo {title} {Spread calibration of ensemble mos forecasts},}\
  }\href {\doibase 10.1175/MWR-D-12-00191.1} {\bibfield  {journal} {\bibinfo
  {journal} {Monthly Weather Review}\ }\textbf {\bibinfo {volume}
  {141}}~(\bibinfo {number} {7}),\ \bibinfo {pages} {2467--2482}},\ \Eprint
  {http://arxiv.org/abs/https://doi.org/10.1175/MWR-D-12-00191.1}
  {https://doi.org/10.1175/MWR-D-12-00191.1} \BibitemShut {NoStop}%
\bibitem [{\citenamefont {Volodin}(2008)}]{ref:Volodin08}%
  \BibitemOpen
  \bibfield  {author} {\bibinfo {author} {\bibnamefont {Volodin}, \bibfnamefont
  {E~M}}} (\bibinfo {year} {2008}),\ \bibfield  {title} {\enquote {\bibinfo
  {title} {Methane cycle in the inm ras climate model},}\ }\href {\doibase
  10.1134/s0001433808020023} {\bibfield  {journal} {\bibinfo  {journal}
  {Izvestiya Atmospheric and Oceanic Physics}\ }\textbf {\bibinfo {volume}
  {44}}~(\bibinfo {number} {2}),\ \bibinfo {pages} {153--159}}\BibitemShut
  {NoStop}%
\bibitem [{\citenamefont {Wagman}\ and\ \citenamefont
  {Jackson}(2018)}]{ref:Wagman&Jackson18}%
  \BibitemOpen
  \bibfield  {author} {\bibinfo {author} {\bibnamefont {Wagman}, \bibfnamefont
  {B~M}}, \ and\ \bibinfo {author} {\bibfnamefont {C.~S.}\ \bibnamefont
  {Jackson}}} (\bibinfo {year} {2018}),\ \bibfield  {title} {\enquote {\bibinfo
  {title} {A test of emergent constraints on cloud feedback and climate
  sensitivity using a calibrated single-model ensemble},}\ }\href {\doibase
  10.1175/jcli-d-17-0682.1} {\bibfield  {journal} {\bibinfo  {journal} {Journal
  of Climate}\ }\textbf {\bibinfo {volume} {31}}~(\bibinfo {number} {18}),\
  \bibinfo {pages} {7515--7532}}\BibitemShut {NoStop}%
\bibitem [{\citenamefont {Wang}\ \emph {et~al.}(2014)\citenamefont {Wang},
  \citenamefont {Zeng}, \citenamefont {Liu},\ and\ \citenamefont
  {Bao}}]{ref:Wang14}%
  \BibitemOpen
  \bibfield  {author} {\bibinfo {author} {\bibnamefont {Wang}, \bibfnamefont
  {Jun}}, \bibinfo {author} {\bibfnamefont {Ning}\ \bibnamefont {Zeng}},
  \bibinfo {author} {\bibfnamefont {Yimin}\ \bibnamefont {Liu}}, \ and\
  \bibinfo {author} {\bibfnamefont {Qing}\ \bibnamefont {Bao}}} (\bibinfo
  {year} {2014}),\ \bibfield  {title} {\enquote {\bibinfo {title} {{To what
  extent can interannual CO2 variability constrain carbon cycle sensitivity to
  climate change in CMIP5 Earth System Models?}}}\ }\href {\doibase
  10.1002/2014GL061184} {\bibfield  {journal} {\bibinfo  {journal} {Geophysical
  Research Letters}\ }\textbf {\bibinfo {volume} {41}},\ \bibinfo {pages}
  {3535--3544}}\BibitemShut {NoStop}%
\bibitem [{\citenamefont {Watanabe}\ \emph {et~al.}(2018)\citenamefont
  {Watanabe}, \citenamefont {Kamae}, \citenamefont {Shiogama}, \citenamefont
  {DeAngelis},\ and\ \citenamefont {Suzuki}}]{ref:Watanabe18}%
  \BibitemOpen
  \bibfield  {author} {\bibinfo {author} {\bibnamefont {Watanabe},
  \bibfnamefont {Masahiro}}, \bibinfo {author} {\bibfnamefont {Youichi}\
  \bibnamefont {Kamae}}, \bibinfo {author} {\bibfnamefont {Hideo}\ \bibnamefont
  {Shiogama}}, \bibinfo {author} {\bibfnamefont {Anthony~M.}\ \bibnamefont
  {DeAngelis}}, \ and\ \bibinfo {author} {\bibfnamefont {Kentaroh}\
  \bibnamefont {Suzuki}}} (\bibinfo {year} {2018}),\ \bibfield  {title}
  {\enquote {\bibinfo {title} {Low clouds link equilibrium climate sensitivity
  to hydrological sensitivity},}\ }\href {\doibase 10.1038/s41558-018-0272-0}
  {\bibfield  {journal} {\bibinfo  {journal} {Nature Climate Change}\ }\textbf
  {\bibinfo {volume} {8}}~(\bibinfo {number} {10}),\ \bibinfo {pages}
  {901--906}}\BibitemShut {NoStop}%
\bibitem [{\citenamefont {Webb}\ and\ \citenamefont
  {Copsey}(2011)}]{ref:WebbBook}%
  \BibitemOpen
  \bibfield  {author} {\bibinfo {author} {\bibnamefont {Webb}, \bibfnamefont
  {AR}}, \ and\ \bibinfo {author} {\bibfnamefont {K.D.}\ \bibnamefont
  {Copsey}}} (\bibinfo {year} {2011}),\ \href@noop {} {\emph {\bibinfo {title}
  {Statistical Pattern Recognition}}},\ \bibinfo {edition} {3rd}\ ed.\
  (\bibinfo  {publisher} {Wiley})\BibitemShut {NoStop}%
\bibitem [{\citenamefont {Wenzel}\ \emph {et~al.}(2014)\citenamefont {Wenzel},
  \citenamefont {Cox}, \citenamefont {Eyring},\ and\ \citenamefont
  {Friedlingstein}}]{ref:Wenzel14}%
  \BibitemOpen
  \bibfield  {author} {\bibinfo {author} {\bibnamefont {Wenzel}, \bibfnamefont
  {Sabrina}}, \bibinfo {author} {\bibfnamefont {Peter~M.}\ \bibnamefont {Cox}},
  \bibinfo {author} {\bibfnamefont {Veronika}\ \bibnamefont {Eyring}}, \ and\
  \bibinfo {author} {\bibfnamefont {Pierre}\ \bibnamefont {Friedlingstein}}}
  (\bibinfo {year} {2014}),\ \bibfield  {title} {\enquote {\bibinfo {title}
  {Emergent constraints on climate-carbon cycle feedbacks in the {CMIP}5
  {E}arth system models},}\ }\href {\doibase 10.1002/2013JG002591} {\bibfield
  {journal} {\bibinfo  {journal} {Journal of Geophysical Research:
  Biogeosciences}\ }\textbf {\bibinfo {volume} {119}}~(\bibinfo {number} {5}),\
  \bibinfo {pages} {2013JG002591}}\BibitemShut {NoStop}%
\bibitem [{\citenamefont {Wenzel}\ \emph {et~al.}(2016)\citenamefont {Wenzel},
  \citenamefont {Cox}, \citenamefont {Eyring},\ and\ \citenamefont
  {Friedlingstein}}]{ref:Wenzel16}%
  \BibitemOpen
  \bibfield  {author} {\bibinfo {author} {\bibnamefont {Wenzel}, \bibfnamefont
  {Sabrina}}, \bibinfo {author} {\bibfnamefont {Peter~M.}\ \bibnamefont {Cox}},
  \bibinfo {author} {\bibfnamefont {Veronika}\ \bibnamefont {Eyring}}, \ and\
  \bibinfo {author} {\bibfnamefont {Pierre}\ \bibnamefont {Friedlingstein}}}
  (\bibinfo {year} {2016}),\ \bibfield  {title} {\enquote {\bibinfo {title}
  {Projected land photosynthesis constrained by changes in the seasonal cycle
  of atmospheric {CO2}},}\ }\href {\doibase 10.1038/nature19772} {\bibfield
  {journal} {\bibinfo  {journal} {Nature}\ }\textbf {\bibinfo {volume}
  {538}}~(\bibinfo {number} {7626}),\ \bibinfo {pages} {499--501}}\BibitemShut
  {NoStop}%
\bibitem [{\citenamefont {Wigley}\ and\ \citenamefont
  {Raper}(1990)}]{ref:Wigley&Raper90}%
  \BibitemOpen
  \bibfield  {author} {\bibinfo {author} {\bibnamefont {Wigley}, \bibfnamefont
  {T~M~L}}, \ and\ \bibinfo {author} {\bibfnamefont {S.~C.~B.}\ \bibnamefont
  {Raper}}} (\bibinfo {year} {1990}),\ \bibfield  {title} {\enquote {\bibinfo
  {title} {Natural variability of the climate system and detection of the
  greenhouse effect},}\ }\href {\doibase 10.1038/344324a0} {\bibfield
  {journal} {\bibinfo  {journal} {Nature}\ }\textbf {\bibinfo {volume}
  {344}}~(\bibinfo {number} {6264}),\ \bibinfo {pages} {324--327}}\BibitemShut
  {NoStop}%
\bibitem [{\citenamefont {Williamson}\ and\ \citenamefont
  {Sansom}(2019)}]{ref:WilliamsonD19}%
  \BibitemOpen
  \bibfield  {author} {\bibinfo {author} {\bibnamefont {Williamson},
  \bibfnamefont {Daniel~B}}, \ and\ \bibinfo {author} {\bibfnamefont
  {Philip~G.}\ \bibnamefont {Sansom}}} (\bibinfo {year} {2019}),\ \bibfield
  {title} {\enquote {\bibinfo {title} {How are emergent constraints quantifying
  uncertainty and what do they leave behind?}}\ }\href {\doibase
  10.1175/BAMS-D-19-0131.1} {\bibfield  {journal} {\bibinfo  {journal}
  {Bulletin of the American Meteorological Society}\ }\textbf {\bibinfo
  {volume} {100}}~(\bibinfo {number} {12}),\ \bibinfo {pages}
  {2571--2588}}\BibitemShut {NoStop}%
\bibitem [{\citenamefont {Williamson}\ \emph {et~al.}(2018)\citenamefont
  {Williamson}, \citenamefont {Collins}, \citenamefont {Drijfhout},
  \citenamefont {Kahana}, \citenamefont {Mecking},\ and\ \citenamefont
  {Lenton}}]{ref:Williamson18}%
  \BibitemOpen
  \bibfield  {author} {\bibinfo {author} {\bibnamefont {Williamson},
  \bibfnamefont {M~S}}, \bibinfo {author} {\bibfnamefont {M.}~\bibnamefont
  {Collins}}, \bibinfo {author} {\bibfnamefont {S.}~\bibnamefont {Drijfhout}},
  \bibinfo {author} {\bibfnamefont {R.}~\bibnamefont {Kahana}}, \bibinfo
  {author} {\bibfnamefont {J.~V.}\ \bibnamefont {Mecking}}, \ and\ \bibinfo
  {author} {\bibfnamefont {T.~M.}\ \bibnamefont {Lenton}}} (\bibinfo {year}
  {2018}),\ \bibfield  {title} {\enquote {\bibinfo {title} {Effect of {AMOC}
  collapse on {ENSO} in a high resolution general circulation model},}\
  }\href@noop {} {\bibfield  {journal} {\bibinfo  {journal} {Climate Dynamics}\
  }\textbf {\bibinfo {volume} {50}}~(\bibinfo {number} {7-8}),\ \bibinfo
  {pages} {2537--2552}}\BibitemShut {NoStop}%
\bibitem [{\citenamefont {Williamson}\ and\ \citenamefont
  {Lenton}(2015)}]{ref:Williamson&Lenton15}%
  \BibitemOpen
  \bibfield  {author} {\bibinfo {author} {\bibnamefont {Williamson},
  \bibfnamefont {M~S}}, \ and\ \bibinfo {author} {\bibfnamefont {T.~M.}\
  \bibnamefont {Lenton}}} (\bibinfo {year} {2015}),\ \bibfield  {title}
  {\enquote {\bibinfo {title} {Detection of bifurcations in noisy coupled
  systems from multiple time series},}\ }\href@noop {} {\bibfield  {journal}
  {\bibinfo  {journal} {Chaos}\ }\textbf {\bibinfo {volume} {25}},\ \bibinfo
  {pages} {036407}}\BibitemShut {NoStop}%
\bibitem [{\citenamefont {Williamson}\ \emph {et~al.}(2019)\citenamefont
  {Williamson}, \citenamefont {Cox},\ and\ \citenamefont
  {Nijsse}}]{ref:Williamson19}%
  \BibitemOpen
  \bibfield  {author} {\bibinfo {author} {\bibnamefont {Williamson},
  \bibfnamefont {Mark~S}}, \bibinfo {author} {\bibfnamefont {Peter~M.}\
  \bibnamefont {Cox}}, \ and\ \bibinfo {author} {\bibfnamefont {Femke J.
  M.~M.}\ \bibnamefont {Nijsse}}} (\bibinfo {year} {2019}),\ \bibfield  {title}
  {\enquote {\bibinfo {title} {Theoretical foundations of emergent constraints:
  relationships between climate sensitivity and global temperature variability
  in conceptual models},}\ }\href {\doibase 10.1093/climsys/dzy006} {\bibfield
  {journal} {\bibinfo  {journal} {Dynamics and Statistics of the Climate
  System}\ }\textbf {\bibinfo {volume} {3}}~(\bibinfo {number} {1}),\
  10.1093/climsys/dzy006}\BibitemShut {NoStop}%
\bibitem [{\citenamefont {Winkler}\ \emph
  {et~al.}(2019{\natexlab{a}})\citenamefont {Winkler}, \citenamefont {Myneni},
  \citenamefont {Alexandrov},\ and\ \citenamefont {Brovkin}}]{ref:Winkler2019}%
  \BibitemOpen
  \bibfield  {author} {\bibinfo {author} {\bibnamefont {Winkler}, \bibfnamefont
  {Alexander~J}}, \bibinfo {author} {\bibfnamefont {Ranga~B.}\ \bibnamefont
  {Myneni}}, \bibinfo {author} {\bibfnamefont {Georgii~A.}\ \bibnamefont
  {Alexandrov}}, \ and\ \bibinfo {author} {\bibfnamefont {Victor}\ \bibnamefont
  {Brovkin}}} (\bibinfo {year} {2019}{\natexlab{a}}),\ \bibfield  {title}
  {\enquote {\bibinfo {title} {{Earth system models underestimate carbon
  fixation by plants in the high latitudes}},}\ }\href {\doibase
  10.1038/s41467-019-08633-z} {\bibfield  {journal} {\bibinfo  {journal}
  {Nature Communications}\ }\textbf {\bibinfo {volume} {10}}~(\bibinfo {number}
  {1}),\ 10.1038/s41467-019-08633-z}\BibitemShut {NoStop}%
\bibitem [{\citenamefont {Winkler}\ \emph
  {et~al.}(2019{\natexlab{b}})\citenamefont {Winkler}, \citenamefont {Myneni},\
  and\ \citenamefont {Brovkin}}]{ref:Winkler2019b}%
  \BibitemOpen
  \bibfield  {author} {\bibinfo {author} {\bibnamefont {Winkler}, \bibfnamefont
  {Alexander~J}}, \bibinfo {author} {\bibfnamefont {Ranga~B.}\ \bibnamefont
  {Myneni}}, \ and\ \bibinfo {author} {\bibfnamefont {Victor}\ \bibnamefont
  {Brovkin}}} (\bibinfo {year} {2019}{\natexlab{b}}),\ \bibfield  {title}
  {\enquote {\bibinfo {title} {{Investigating the applicability of emergent
  constraints}},}\ }\href {\doibase 10.5194/esd-10-501-2019} {\bibfield
  {journal} {\bibinfo  {journal} {Earth System Dynamics}\ }\textbf {\bibinfo
  {volume} {10}}~(\bibinfo {number} {3}),\ \bibinfo {pages}
  {501--523}}\BibitemShut {NoStop}%
\bibitem [{\citenamefont {Yang}\ \emph {et~al.}(2018)\citenamefont {Yang},
  \citenamefont {Piao}, \citenamefont {Huntingford}, \citenamefont {Ciais},
  \citenamefont {Li}, \citenamefont {Wang}, \citenamefont {Peng}, \citenamefont
  {Yang}, \citenamefont {Yang},\ and\ \citenamefont {Chang}}]{ref:Yang18}%
  \BibitemOpen
  \bibfield  {author} {\bibinfo {author} {\bibnamefont {Yang}, \bibfnamefont
  {Hui}}, \bibinfo {author} {\bibfnamefont {Shilong}\ \bibnamefont {Piao}},
  \bibinfo {author} {\bibfnamefont {Chris}\ \bibnamefont {Huntingford}},
  \bibinfo {author} {\bibfnamefont {Philippe}\ \bibnamefont {Ciais}}, \bibinfo
  {author} {\bibfnamefont {Yue}\ \bibnamefont {Li}}, \bibinfo {author}
  {\bibfnamefont {Tao}\ \bibnamefont {Wang}}, \bibinfo {author} {\bibfnamefont
  {Shushi}\ \bibnamefont {Peng}}, \bibinfo {author} {\bibfnamefont {Yuting}\
  \bibnamefont {Yang}}, \bibinfo {author} {\bibfnamefont {Dawen}\ \bibnamefont
  {Yang}}, \ and\ \bibinfo {author} {\bibfnamefont {Jinfeng}\ \bibnamefont
  {Chang}}} (\bibinfo {year} {2018}),\ \bibfield  {title} {\enquote {\bibinfo
  {title} {Changing the retention properties of catchments and their influence
  on runoff under climate change},}\ }\href {\doibase 10.1088/1748-9326/aadd32}
  {\bibfield  {journal} {\bibinfo  {journal} {Environmental Research Letters}\
  }\textbf {\bibinfo {volume} {13}}~(\bibinfo {number} {9}),\ \bibinfo {pages}
  {094019}}\BibitemShut {NoStop}%
\bibitem [{\citenamefont {Yang}\ \emph {et~al.}(2013)\citenamefont {Yang},
  \citenamefont {Gong}, \citenamefont {Fu}, \citenamefont {Zhang},
  \citenamefont {Chen}, \citenamefont {Liang}, \citenamefont {Xu},
  \citenamefont {Shi},\ and\ \citenamefont {Dickinson}}]{ref:Yang13}%
  \BibitemOpen
  \bibfield  {author} {\bibinfo {author} {\bibnamefont {Yang}, \bibfnamefont
  {Jun}}, \bibinfo {author} {\bibfnamefont {Peng}\ \bibnamefont {Gong}},
  \bibinfo {author} {\bibfnamefont {Rong}\ \bibnamefont {Fu}}, \bibinfo
  {author} {\bibfnamefont {Minghua}\ \bibnamefont {Zhang}}, \bibinfo {author}
  {\bibfnamefont {Jingming}\ \bibnamefont {Chen}}, \bibinfo {author}
  {\bibfnamefont {Shunlin}\ \bibnamefont {Liang}}, \bibinfo {author}
  {\bibfnamefont {Bing}\ \bibnamefont {Xu}}, \bibinfo {author} {\bibfnamefont
  {Jiancheng}\ \bibnamefont {Shi}}, \ and\ \bibinfo {author} {\bibfnamefont
  {Robert}\ \bibnamefont {Dickinson}}} (\bibinfo {year} {2013}),\ \bibfield
  {title} {\enquote {\bibinfo {title} {The role of satellite remote sensing in
  climate change studies},}\ }\href {\doibase 10.1038/nclimate1908} {\bibfield
  {journal} {\bibinfo  {journal} {Nature Climate Change}\ }\textbf {\bibinfo
  {volume} {3}}~(\bibinfo {number} {10}),\ \bibinfo {pages}
  {875--883}}\BibitemShut {NoStop}%
\bibitem [{\citenamefont {Yokohata}\ \emph {et~al.}(2010)\citenamefont
  {Yokohata}, \citenamefont {Webb}, \citenamefont {Collins}, \citenamefont
  {Williams}, \citenamefont {Yoshimori}, \citenamefont {Hargreaves},\ and\
  \citenamefont {Annan}}]{ref:Yokohata10}%
  \BibitemOpen
  \bibfield  {author} {\bibinfo {author} {\bibnamefont {Yokohata},
  \bibfnamefont {Tokuta}}, \bibinfo {author} {\bibfnamefont {Mark~J.}\
  \bibnamefont {Webb}}, \bibinfo {author} {\bibfnamefont {Matthew}\
  \bibnamefont {Collins}}, \bibinfo {author} {\bibfnamefont {Keith~D.}\
  \bibnamefont {Williams}}, \bibinfo {author} {\bibfnamefont {Masakazu}\
  \bibnamefont {Yoshimori}}, \bibinfo {author} {\bibfnamefont {Julia~C.}\
  \bibnamefont {Hargreaves}}, \ and\ \bibinfo {author} {\bibfnamefont
  {James~D.}\ \bibnamefont {Annan}}} (\bibinfo {year} {2010}),\ \bibfield
  {title} {\enquote {\bibinfo {title} {Structural similarities and differences
  in climate responses to co2 increase between two perturbed physics
  ensembles},}\ }\href {\doibase 10.1175/2009JCLI2917.1} {\bibfield  {journal}
  {\bibinfo  {journal} {Journal of Climate}\ }\textbf {\bibinfo {volume}
  {23}}~(\bibinfo {number} {6}),\ \bibinfo {pages} {1392--1410}}\BibitemShut
  {NoStop}%
\bibitem [{\citenamefont {Zelinka}\ \emph {et~al.}(2020)\citenamefont
  {Zelinka}, \citenamefont {Myers}, \citenamefont {McCoy}, \citenamefont
  {Po-Chedley}, \citenamefont {Caldwell}, \citenamefont {Ceppi}, \citenamefont
  {Klein},\ and\ \citenamefont {Taylor}}]{ref:Zelinka20}%
  \BibitemOpen
  \bibfield  {author} {\bibinfo {author} {\bibnamefont {Zelinka}, \bibfnamefont
  {Mark~D}}, \bibinfo {author} {\bibfnamefont {Timothy~A}\ \bibnamefont
  {Myers}}, \bibinfo {author} {\bibfnamefont {Daniel~T}\ \bibnamefont {McCoy}},
  \bibinfo {author} {\bibfnamefont {Stephen}\ \bibnamefont {Po-Chedley}},
  \bibinfo {author} {\bibfnamefont {Peter~M}\ \bibnamefont {Caldwell}},
  \bibinfo {author} {\bibfnamefont {Paulo}\ \bibnamefont {Ceppi}}, \bibinfo
  {author} {\bibfnamefont {Stephen~A}\ \bibnamefont {Klein}}, \ and\ \bibinfo
  {author} {\bibfnamefont {Karl~E}\ \bibnamefont {Taylor}}} (\bibinfo {year}
  {2020}),\ \bibfield  {title} {\enquote {\bibinfo {title} {{Causes of Higher
  Climate Sensitivity in CMIP6 Models}},}\ }\href {\doibase
  10.1029/2019GL085782} {\bibfield  {journal} {\bibinfo  {journal} {Geophysical
  Research Letters}\ }\textbf {\bibinfo {volume} {47}}~(\bibinfo {number}
  {1}),\ \bibinfo {pages} {e2019GL085782}}\BibitemShut {NoStop}%
\bibitem [{\citenamefont {Zelinka}\ \emph {et~al.}(2016)\citenamefont
  {Zelinka}, \citenamefont {Zhou},\ and\ \citenamefont
  {Klein}}]{ref:Zelinka2016}%
  \BibitemOpen
  \bibfield  {author} {\bibinfo {author} {\bibnamefont {Zelinka}, \bibfnamefont
  {Mark~D}}, \bibinfo {author} {\bibfnamefont {Chen}\ \bibnamefont {Zhou}}, \
  and\ \bibinfo {author} {\bibfnamefont {Stephen~A.}\ \bibnamefont {Klein}}}
  (\bibinfo {year} {2016}),\ \bibfield  {title} {\enquote {\bibinfo {title}
  {{Insights from a refined decomposition of cloud feedbacks}},}\ }\href
  {\doibase 10.1002/2016GL069917} {\bibfield  {journal} {\bibinfo  {journal}
  {Geophysical Research Letters}\ }\textbf {\bibinfo {volume} {43}}~(\bibinfo
  {number} {17}),\ \bibinfo {pages} {9259--9269}}\BibitemShut {NoStop}%
\bibitem [{\citenamefont {Zhai}\ \emph {et~al.}(2015)\citenamefont {Zhai},
  \citenamefont {Jiang},\ and\ \citenamefont {Su}}]{ref:Zhai15}%
  \BibitemOpen
  \bibfield  {author} {\bibinfo {author} {\bibnamefont {Zhai}, \bibfnamefont
  {C~X}}, \bibinfo {author} {\bibfnamefont {J.~H.}\ \bibnamefont {Jiang}}, \
  and\ \bibinfo {author} {\bibfnamefont {H.}~\bibnamefont {Su}}} (\bibinfo
  {year} {2015}),\ \bibfield  {title} {\enquote {\bibinfo {title} {Long-term
  cloud change imprinted in seasonal cloud variation: More evidence of high
  climate sensitivity},}\ }\href {\doibase 10.1002/2015gl065911} {\bibfield
  {journal} {\bibinfo  {journal} {Geophysical Research Letters}\ }\textbf
  {\bibinfo {volume} {42}}~(\bibinfo {number} {20}),\ \bibinfo {pages}
  {8729--8737}}\BibitemShut {NoStop}%
\bibitem [{\citenamefont {Zhang}(2006)}]{Zhang06}%
  \BibitemOpen
  \bibfield  {author} {\bibinfo {author} {\bibnamefont {Zhang}, \bibfnamefont
  {Nien~Fan}}} (\bibinfo {year} {2006}),\ \bibfield  {title} {\enquote
  {\bibinfo {title} {{Calculation of the uncertainty of the mean of
  autocorrelated measurements}},}\ }\href {\doibase 10.1088/0026-1394/43/4/S15}
  {\bibfield  {journal} {\bibinfo  {journal} {Metrologia}\ }\textbf {\bibinfo
  {volume} {43}}~(\bibinfo {number} {4}),\
  10.1088/0026-1394/43/4/S15}\BibitemShut {NoStop}%
\bibitem [{\citenamefont {Zhao}\ \emph {et~al.}(2016)\citenamefont {Zhao},
  \citenamefont {Golaz}, \citenamefont {Held}, \citenamefont {Ramaswamy},
  \citenamefont {Lin}, \citenamefont {Ming}, \citenamefont {Ginoux},
  \citenamefont {Wyman}, \citenamefont {Donner}, \citenamefont {Paynter},\ and\
  \citenamefont {Guo}}]{ref:Zhao16}%
  \BibitemOpen
  \bibfield  {author} {\bibinfo {author} {\bibnamefont {Zhao}, \bibfnamefont
  {Ming}}, \bibinfo {author} {\bibfnamefont {J.~C.}\ \bibnamefont {Golaz}},
  \bibinfo {author} {\bibfnamefont {I.~M.}\ \bibnamefont {Held}}, \bibinfo
  {author} {\bibfnamefont {V.}~\bibnamefont {Ramaswamy}}, \bibinfo {author}
  {\bibfnamefont {S.~J.}\ \bibnamefont {Lin}}, \bibinfo {author} {\bibfnamefont
  {Y.}~\bibnamefont {Ming}}, \bibinfo {author} {\bibfnamefont {P.}~\bibnamefont
  {Ginoux}}, \bibinfo {author} {\bibfnamefont {B.}~\bibnamefont {Wyman}},
  \bibinfo {author} {\bibfnamefont {L.~J.}\ \bibnamefont {Donner}}, \bibinfo
  {author} {\bibfnamefont {D.}~\bibnamefont {Paynter}}, \ and\ \bibinfo
  {author} {\bibfnamefont {H.}~\bibnamefont {Guo}}} (\bibinfo {year} {2016}),\
  \bibfield  {title} {\enquote {\bibinfo {title} {{Uncertainty in model climate
  sensitivity traced to representations of cumulus precipitation
  microphysics}},}\ }\href {\doibase 10.1175/JCLI-D-15-0191.1} {\bibfield
  {journal} {\bibinfo  {journal} {Journal of Climate}\ }\textbf {\bibinfo
  {volume} {29}}~(\bibinfo {number} {2}),\ \bibinfo {pages}
  {543--560}}\BibitemShut {NoStop}%
\bibitem [{\citenamefont {Zhou}\ \emph {et~al.}(2015)\citenamefont {Zhou},
  \citenamefont {Zelinka}, \citenamefont {Dessler},\ and\ \citenamefont
  {Klein}}]{ref:Zhou15}%
  \BibitemOpen
  \bibfield  {author} {\bibinfo {author} {\bibnamefont {Zhou}, \bibfnamefont
  {Chen}}, \bibinfo {author} {\bibfnamefont {Mark~D.}\ \bibnamefont {Zelinka}},
  \bibinfo {author} {\bibfnamefont {Andrew~E.}\ \bibnamefont {Dessler}}, \ and\
  \bibinfo {author} {\bibfnamefont {Stephen~A.}\ \bibnamefont {Klein}}}
  (\bibinfo {year} {2015}),\ \bibfield  {title} {\enquote {\bibinfo {title}
  {{The relationship between interannual and long-term cloud feedbacks}},}\
  }\href {\doibase 10.1002/2015GL066698} {\bibfield  {journal} {\bibinfo
  {journal} {Geophysical Research Letters}\ }\textbf {\bibinfo {volume}
  {42}}~(\bibinfo {number} {23}),\ \bibinfo {pages} {10463--10469}}\BibitemShut
  {NoStop}%
\end{thebibliography}%

\end{document}